\documentclass[12pt]{article}

\usepackage{amstext}
\usepackage{latexsym}
\usepackage{amssymb}
\usepackage{amsmath}
\usepackage{tensor}
\usepackage{bm}
\usepackage{xcolor}
\usepackage{import}
\usepackage{comment}
\usepackage{enumitem}

\def\R{\mathbb R}

\newcommand{\calT}{\mathcal{T}}

\newcommand{\eps}{\varepsilon}
\newcommand{\bbR}{\mathbb{R}}

\newcommand\blfootnote[1]{%
  \begingroup
  \renewcommand\thefootnote{}\footnote{#1}%
  \addtocounter{footnote}{-1}%
  \endgroup
}

\allowdisplaybreaks

\usepackage[many]{tcolorbox}
\tcbuselibrary{theorems}
\tcbuselibrary{skins}

\definecolor{li_red}{rgb}{0.9,0,0}
\definecolor{dk_red}{rgb}{0.4,0,0}

\newtheorem{theorem}{Theorem}[section]

\newtheorem{lemma}[theorem]{Lemma}

\newtheorem{remark}[theorem]{Remark}

\newtheorem{cor}[theorem]{Corollary}

\newcommand{\rem}[1]{({\rm rem}_{#1})}

\begin{document}


\title{Proof of the Einstein quadrupole formula for solutions 
of the Einstein-Vlasov system close to Minkowski spacetime}

\author{\'{E}rik Amorim\\
        Mathematisches Institut\\
        Universit\"at K\"oln\\
        Weyertal 86-90\\
        D-50931 K\"oln, Germany\\
        email: eamorim@uni-koeln.de\\
        \ \\
        H{\aa}kan Andr\'{e}asson\\
        Mathematical Sciences\\
        Chalmers University of Technology\\
        G\"{o}teborg University\\
        S-41296 G\"oteborg, Sweden\\
        email: hand@chalmers.se\\
        \ \\
        Markus Kunze\\
        Mathematisches Institut\\
        Universit\"at K\"oln\\
        Weyertal 86-90\\
        D-50931 K\"oln, Germany\\
        email: mkunze1@uni-koeln.de}

\maketitle



\begin{abstract}
We rigorously derive the quadrupole formula within the context of the Einstein-Vlasov system. 
The main contribution of this work is an estimate of the remainder terms, derived from well-defined assumptions, 
with explicitly stated error terms that depend on the solution's boundedness and decay properties, 
and the distance to the source. The assumptions are linked to established properties of global solutions 
of the Einstein-Vlasov system as in \cite{LT}. Prior derivations of the quadrupole formula have relied 
on post-Newtonian analysis and lacked comparisons with global solution properties. The importance 
of the no-incoming-radiation condition is emphasized underscoring the need for solutions 
satisfying this condition. This work thus addresses the limitations of existing results 
and provides motivation for further research on global solution properties of the Einstein-Vlasov system. 
\end{abstract}


\tableofcontents

\newpage


\section{Introduction and main result}
\setcounter{equation}{0}
Einstein's theory of general relativity, formulated more than a century ago, revolutionized 
our understanding of gravity and predicted the existence of 
gravitational waves-a groundbreaking achievement that remained unconfirmed until 
the historic detection by LIGO in 2015. Among the various aspects of general relativity, 
the quadrupole formula stands as a fundamental cornerstone, providing a theoretical framework 
to understand the emission of gravitational waves from massive, accelerating objects. 

First introduced by Einstein in 1918 as a prediction of his general theory of relativity, 
the quadrupole formula provides a crucial link between the dynamics of massive, 
accelerating objects and the emission of gravitational radiation. This formula gained 
renewed attention with the discovery of the Hulse-Taylor binary pulsar system, 
leading to the 1993 Nobel Prize in physics awarded to Russell Hulse and Joseph Taylor 
for their seminal confirmation of Einstein's theory through the observation 
of the system's orbital decay due to gravitational wave emission.

However, despite its fundamental significance and successful application to explain 
the Hulse-Taylor binary pulsar, the quadrupole formula has been accompanied 
by a lingering controversy surrounding its derivation. The theoretical foundation 
of the formula was initially met with skepticism, as some researchers questioned the rigor 
of its derivation and the validity of the underlying assumptions, cf.~\cite{PoWi,ERGH}. 
This controversy led to extensive debates within the gravitational physics community, 
prompting experts to seek a more comprehensive and rigorous derivation that 
would not only resolve the existing concerns but also reaffirm the formula's relevance 
in the era of high precision gravitational wave astronomy.

Our goal in the present work is to rigorously derive the quadrupole formula in the context 
of the Einstein-Vlasov system for a compactly supported distribution of collisionless matter. 
We start from a set of physically and mathematically motivated assumptions, 
and we prove the validity of the approximation by providing an estimate 
for its corresponding error term. An important aspect about the assumptions we make 
is how these are related to established properties of global solutions of the Einstein-Vlasov system. 
More precisely, global existence of solutions for initial data close to Minkowski space 
has recently been shown in two independent works \cite{LT,FJS}. The set-up in \cite{LT}, 
which relies on harmonic coordinates, is the framework that we are going to use in this paper. 
Hence, it is natural to ask if the result in \cite{LT} gives sufficient information 
on the solutions to validate the assumptions that we make. 

Previous results on the rigorous derivation of the quadrupole formula exist in the literature, 
cf.~\cite{B_LRR,PoWi}. These works are based on a post-Newtonian analysis. 
The assumptions on the solutions in these results cannot be compared to mathematically established 
global properties of solutions of any Einstein-matter system, e.g.~to decay properties 
specified in well-defined norms. In fact, no specific matter model is considered there. 
A standard application of the quadrupole formula is to binary systems such as the Hulse-Taylor 
binary pulsar. Such systems are naturally described by a phenomenological matter model, 
in which case global existence results are rare. Vlasov matter constitutes a unique 
case of a phenomenological matter model for which global existence has been shown 
for small initial data in the case of asymptotically flat spacetimes \cite{LT,FJS}. 
(Clearly, for a binary system the assumption about small initial data is not justified 
from a physics point of view, but since we aim for rigorous mathematical results 
we have to accept this limitation at this point.) We thus find it motivated 
to ask what are the mathematical assumptions required on the solutions 
to the Einstein-Vlasov system to rigorously derive the quadrupole formula. 
Although the result by Lindblad and Taylor \cite{LT} provides a motivation for our study, 
it is not sufficient for our goal. One reason for this is related to the well-known 
\textit{no-incoming-radiation condition}. The purpose of this condition is to make sure 
that the matter source is isolated and does not receive any radiation from infinity. 
Without this condition there are terms in the expansion used to prove the quadrupole formula 
that are infinite. This implies, strictly speaking, that solutions of the Einstein-Vlasov system 
have to be established that satisfy the no-incoming-radiation condition; 
this is a highly non-trivial problem. Hence, the conditions we impose motivate 
further studies on the global properties of solutions to the Einstein-Vlasov system 
that go beyond the results obtained in \cite{LT,FJS}. The no-incoming-radiation condition 
will be used in a form suggested by Blanchet \cite[equ.~(19)]{B_LRR}, and it is 
discussed in more detail in Section \ref{swnir}. 

The main result in this work is an estimate for the quadrupole formula derived 
from mathematically well-defined assumptions. The error terms in the estimate 
are explicitly given and depend on the boundedness and decay assumptions 
made on the solution and the distance to the source. Hence, it is possible 
to read off the contribution of each error term and to understand its origin. 

The paper is organized as follows. In the next two subsections 
the Einstein-Vlasov system and the quadrupole formula are introduced. 
In Section \ref{subsec_apriori} the assumptions on the solutions that we make are specified; 
the absence of incoming radiation is discussed in Section \ref{swnir} 
and the decay and boundedness conditions are stated in Section \ref{LT_intro}. 
The relation between the assumptions and the global properties of the solutions 
that are obtained in \cite{LT} are also discussed. In Section \ref{sub_main} 
we formulate our main result and discuss its meaning. In Section \ref{sec_expansion}, 
and its subsections, we carry out the expansion of the quantities that are needed 
to derive our main result. The estimates of the terms in the expansion are carried 
out in Section \ref{sec_estimating}. Some technicalities are brought up in the appendix, 
in particular the no-incoming-radiation condition is discussed in Appendix \ref{Blanch_discuss}. 

\subsection{The Einstein-Vlasov system}
\label{EV_system}

The Einstein-Vlasov system (with factors of the speed of light $c$ and of Newton's constant 
$G$ included where they belong) is given by Einstein's equation,
\begin{equation}\label{einst} 
   G_{\alpha\beta}=R_{\alpha\beta}-\frac{1}{2}\,R\,g_{\alpha\beta}
   =\frac{8\pi G}{c^4}\,T_{\alpha\beta},
\end{equation} 
coupled to the Vlasov equation for a non-negative particle-density function $f$,
\begin{equation}\label{vlas} 
   p^\alpha\partial_{x^\alpha}f-\Gamma_{\beta\gamma}^\alpha\,p^\beta p^\gamma\,\partial_{p^\alpha} f=0
\end{equation} 
and to the equation defining the energy-momentum tensor $T_{\alpha\beta}$,
\begin{equation}\label{tabori} 
   T_{\alpha\beta}=c^{-1}\,|g|^{1/2}\int_{\R^4} p_\alpha p_\beta\,f\,
   \frac{dp^0\,dp^1\,dp^2\,dp^3}{m}.
\end{equation}    
Here $(x^\alpha, p^\beta)$ are coordinates on the tangent bundle $T{\cal M}$ 
of a four-dimensional spacetime manifold $({\cal M}, (g_{\alpha\beta}))$, where $(g_{\alpha\beta})$ 
is a Lorentzian metric of signature $(-, +, +, +)$ expressed as 
\[ -c^2\,d\tau^2=ds^2=g_{\alpha\beta}\,dx^\alpha\,dx^\beta, \] 
with $\tau$ denoting proper time and $t = x^0/c$ coordinate time. The scalar $g<0$ 
in~\eqref{tabori} denotes the metric determinant, and indices are raised 
and lowered using the metric; thus for instance $p_\alpha=g_{\alpha\beta} p^\beta$. 
The scalar $f$ in (\ref{vlas}) depends on $(x^\alpha, p^\beta)$, i.e., $f=f(x^\alpha, p^\beta)$, 
and the Christoffel symbols are given by
\[ \Gamma^{\alpha}_{\beta\gamma}=\frac{1}{2}\,g^{\alpha\delta}(\partial_{x^\beta} g_{\gamma\delta}
   +\partial_{x^\gamma} g_{\beta\delta}-\partial_{x^\delta} g_{\beta\gamma}). \] 
They are used not only in (\ref{vlas}), but also in the definition 
\[ \tensor{R}{_\alpha_\beta_\gamma^\delta}
   =\partial_{x^\beta}\Gamma^{\delta}_{\alpha\gamma}
   -\partial_{x^\alpha}\Gamma^{\delta}_{\beta\gamma}
   +\Gamma^{\delta}_{\beta\eps}\,\Gamma^{\eps}_{\alpha\gamma}
   -\Gamma^{\delta}_{\alpha\eps}\,\Gamma^{\eps}_{\beta\gamma} \] 
of the Riemann curvature tensor. Then 
\[ R_{\alpha\beta}=\tensor{R}{_\alpha_\gamma_\beta^\gamma} \] 
is the Ricci tensor, whereas $R=\tensor{R}{_\alpha^\alpha}$ 
denotes the scalar curvature (i.e., the trace) of the metric. The Vlasov equation (\ref{vlas}) 
expresses the fact that $f$ is constant along solutions of the geodesic equations 
\begin{equation}\label{geod} 
   \frac{dx^\alpha}{d\tau}=p^\alpha,\quad\frac{dp^\alpha}{d\tau}
   =-\Gamma_{\beta\gamma}^\alpha\,p^\beta p^\gamma. 
\end{equation}
The density function $f$ is restricted to the mass shell
\begin{equation} \label{aa_mass_shell}
    P{\cal M}=\{(x^\mu, p^\nu)\in T{\cal M}: p^0>0, \ g_{\alpha\beta} p^\alpha p^\beta=-m^2c^2\}
\end{equation}
that is a seven-dimensional geodesically invariant submanifold 
of the tangent bundle. We assume that all particles have the same rest mass, i.e. $m=1$. The expression (\ref{tabori}) can then be simplified. Indeed, let us change variables from $(p^0,p^1,p^2,p^3)$ to $(m,\tilde{p}^1,\tilde{p}^2,\tilde{p}^3)$ by
\begin{eqnarray*}
   & &p^0=p^0(m,\tilde{p}^1,\tilde{p}^2,\tilde{p}^3),
   \\ & &p^a=\tilde{p}^a,\; a=1,2,3. 
\end{eqnarray*}
Here $p^0$ is given implicitly by the mass shell relation
\[ g_{\alpha\beta}p^{\alpha}p^{\beta}=-m^2c^2, \]
and we remark that this is always possible in our situation since the solution 
we consider is close to Minkowski space where $(p^0)^2=m^2c^2+(p^1)^2+(p^2)^2+(p^3)^2$. 
The Jacobian of the map above is simply $|\partial p^0/\partial m|$, 
and we get by differentiating the mass shell relation with respect to $m$: 
\[ -2mc^2=2g_{\alpha 0}p^{\alpha}\frac{\partial p^0}{\partial m}
   =2p_0\frac{\partial p_0}{\partial m}, \]
so that
\[ \Big|\frac{\partial p_0}{\partial m}\Big|=\frac{mc^2}{-p_0}, \]
where we notice that $-p_0>0$. Hence, we find that (\ref{tabori}) takes the form
\begin{equation}\label{Tfull} 
   T_{\alpha\beta}=c\,|g|^{1/2}\int_{\R^4} p_\alpha p_\beta\,f\,
   \frac{dm\,dp^1\,dp^2\,dp^3}{-p_0}.
\end{equation}  
To impose that all the Vlasov-matter particles have constant mass 1 means that 
we take $f$ of the form $f(x^\alpha, m, p^b)=\delta(m-1)\tilde{f}(x^\alpha, p^b)$, which results in  
\begin{equation}\label{tab3} 
   T_{\alpha\beta}=c\,|g|^{1/2}\int_{\R^3} p_\alpha p_\beta\,f\,
   \frac{dp^1\,dp^2\,dp^3}{-p_0},
\end{equation}
where we dropped the tilde, so that $f=f(x^\alpha, p^b)$. 
To rewrite the Vlasov equation $\frac{d}{d\tau}\,f(x^\alpha(\tau), p^b(\tau))=0$ 
for fixed mass $m=1$ in terms of the coordinate time $t$, 
note that $c\,\frac{dt}{d\tau}=\frac{dx^0}{d\tau}=p^0>0$ on $P{\cal M}$. Let 
\[ \tilde{x}^\alpha(t)=x^\alpha(\tau(t)),
   \quad\tilde{p}^b(t)=p^b(\tau(t)),
   \quad\tilde{f}(t, \tilde{x}^a, \tilde{p}^b)=f(ct, \tilde{x}^a, \tilde{p}^b). \] 
Then $\tilde{x}^0(t)=x^0(\tau)=ct$ yields 
\begin{eqnarray}\label{adr}  
   \frac{d}{dt}\,\tilde{f}(t, \tilde{x}^a(t), \tilde{p}^b(t)) 
   & = & \frac{d}{dt}\,f(ct, \tilde{x}^a(t), \tilde{p}^b(t))
   =\frac{d}{dt}\,f(\tilde{x}^\alpha(t), \tilde{p}^b(t))
   \nonumber
   \\ & = & \frac{c}{p^0}\,\frac{d}{d\tau}\,f(\tilde{x}^\alpha(t(\tau)), \tilde{p}^b(t(\tau)))
   =\frac{c}{p^0}\,\frac{d}{d\tau}\,f(x^\alpha(\tau), p^b(\tau))=0. 
\end{eqnarray} 
Since 
\begin{equation}\label{chart} 
   \frac{d\tilde{x}^a}{dt}=\frac{dx^a}{d\tau}\,\frac{d\tau}{dt} 
   =\frac{c}{p^0}\,\tilde{p}^a,
   \quad\frac{d\tilde{p}^b}{dt}=\frac{dp^b}{d\tau}\,\frac{d\tau}{dt}
   =-\frac{c}{p^0}\,\Gamma_{\beta\gamma}^b\,\tilde{p}^\beta\tilde{p}^\gamma,
\end{equation} 
relation (\ref{adr}) finally leads to 
\begin{equation}\label{vlas3} 
   c^{-1}\partial_t f+\frac{p^a}{p^0}\,\partial_{x^a}f
   -\frac{1}{p^0}\,\Gamma_{\beta\gamma}^a\,p^\beta p^\gamma\,\partial_{p^a} f=0,
\end{equation}
where all tildes have been dropped.

\medskip
Throughout the paper, what we refer to by the name \textit{Einstein-Vlasov system} 
is the set of equations (\ref{einst}), (\ref{tab3}) and (\ref{vlas3}) 
for the unknowns $g_{\alpha\beta} = g_{\alpha\beta}(t,x^a)$ and $f=f(t,x^a,p^b)$ 
defined globally as functions of $(t,x^a)\in\bbR^4$ and $(t,x^a,p^b)\in\bbR^7$. 
It should be noted, however, that in these variables (and re-introducing 
the tildes for the moment) relation (\ref{tab3}) is in fact to be read as 
\begin{equation}\label{tab3b} 
   \tilde{T}_{\alpha\beta}(t, x^a)
   =c\,|g|^{1/2}\int_{\R^3} p_\alpha p_\beta\,\tilde{f}(t, x^a)\,
   \frac{dp^1\,dp^2\,dp^3}{-p_0}, 
\end{equation}     
i.e., $\tilde{T}_{\alpha\beta}(t, x^a)=T_{\alpha\beta}(ct, x^a)$. This identification 
is also used for other functions throughout the paper.

\smallskip
In order to keep a correct track of relevant physical constants, we let $\lambda>0$ 
be a fixed constant with a dimension of length. Using $\lambda$ and the constants $G$ and $c$, 
we can produce any other dimension relevant to the problem, for example fundamental scales of time, 
mass and momentum given respectively by $c^{-1}\lambda$, $G^{-1}\lambda c^2$ and $G^{-1}\lambda c^3$. 
In particular, the particles' constant mass equal to 1 in the above paragraph should be understood 
as having the value $G^{-1}\lambda c^2$.

\subsection{The quadrupole formula}

If one considers electromagnetic fields $E,B$ (functions of a spacetime argument $(t,x)\in\bbR^4$) 
satisfying the Maxwell equations in Minkowski space, then there is a scalar field $e$ 
denoting the electromagnetic-energy density and, given a fixed event $(t,x)$, 
it makes sense to define the energy content at time $t$ in the ball of radius $r=|x|$ 
and center at the origin by
\begin{equation}\label{pre-poynt} 
    {\cal E}(t,x)=\int_{B_r(0)} e(t, \bar{x})\,d\bar{x}.
\end{equation}   
Using the divergence theorem, the rate of change of this energy is found to equal the surface integral 
\begin{equation}\label{poynt} 
   \frac{d}{dt}\,{\cal E}(t,x)=\frac{c}{4\pi}\int_{\partial B_r(0)} 
   \frac{\bar{x}}{r}\cdot (B\wedge E)(t, \bar{x})\,dS(\bar{x}),
\end{equation}   
which measures the flow of energy over the boundary to infinity. Reference \cite[p.~268/269]{BKRR} 
studies the Vlasov-Maxwell and Vlasov-Nordstr\"om systems and derives a dipole radiation formula 
for the Vlasov-Maxwell case using~\eqref{poynt}.

However, in the context of General Relativity, in order to measure the amount 
of gravitational energy radiated from a faraway matter source with associated 
energy-momentum tensor $T^{\alpha\beta}$, a modification of this approach is needed, 
since the gravitational energy is not localizable; see e.g.~\cite[Section 14]{stephani} 
for further discussion. 
For this purpose, we will need the Landau-Lifschitz energy-momentum pseudo tensor $t^{\alpha\beta}$. 
Its definition is rather involved and given in (\ref{LaLi}), 
with Appendix \ref{sec_en_mom_ps_tensor} containing the necessary background material. 
The upshot in using the energy-momentum pseudo tensor is that it leads 
to a physically meaningful notion of the total energy contained in $B_r(0)$, namely
\[ \mathfrak{p}^0(t,x)
   =\int_{B_r(0)} |g(t, \bar{x})|\,(T^{00}(t, \bar{x})+t^{00}(t, \bar{x})) \,d\bar{x}. \]
The divergence theorem then implies (see (\ref{morgo}))
\begin{equation}\label{ogin} 
   \frac{d}{dt}\,\mathfrak{p}^0 (t,x)
   =-c\,\sum_{i=1}^3\int_{\partial B_r(0)} |g(t, \bar{x})|\,\frac{\bar{x}^i}{r}
   \,(T^{0i}(t, \bar{x})+t^{0i}(t, \bar{x}))\,dS((t, \bar{x})),
\end{equation}
Therefore (\ref{ogin}) should be viewed as the appropriate replacement for (\ref{poynt}) 
when it comes to measuring gravitational radiation.
\smallskip 

The \textit{quadrupole-formula approximation}, loosely speaking, is the statement that, 
for $x$ far enough away from the gravitational source and $t$ close enough to $c^{-1}|x|$, 
the following approximation holds up to corresponding orders in $h$, $T$ and $1/r$ terms:
\begin{equation} \label{quadruformu}
    \frac{d}{dt}\,\mathfrak{p}^0 (t,x) \approx -\frac{G}{5c^5}\bigg[\sum_{j, k=1}^3 (\dddot{D}^{\,jk}(u))^2  
   -\frac{1}{3}\sum_{j=1}^3(\tensor{\dddot{D}}{^{jj}}(u))^2\bigg], 
\end{equation}
where $u = t - c^{-1}|x|$ is the \textit{retarded time} corresponding to the event $(t,x)$, and
\begin{equation}\label{ito} 
   D^{\alpha\beta}(u)
   =\frac{1}{c^2}\int_{\R^3} T_{00}(u, \bar{x})\,\bar{x}^\alpha
   \,\bar{x}^\beta\,d\bar{x}
\end{equation}
is the \textit{quadrupole moment} of the gravitational source. The dots denote time derivatives; 
thus (note that $\partial_{x^0}=c^{-1}\partial_t$)
\[ \dddot{D}^{\alpha\beta}(u) 
   =c\int_{\R^3}\partial^3_{x^0} T_{00}(u, \bar{x})\,\bar{x}^\alpha
   \,\bar{x}^\beta\,d\bar{x} . \] 
Also note that the right-hand side of (\ref{quadruformu}) is negative, so there 
is indeed a loss of gravitational energy.

We assume throughout the paper that global solutions $((g_{\alpha\beta}), f)$ 
to the Einstein-Vlasov system exist satisfying certain technical assumptions, 
and our main result, Theorem~\ref{mainthm}, is an estimate for the error 
in the approximation~\eqref{quadruformu} expressed in terms of powers of $|x|$ 
and of quantities that measure the smallness of $|T^{\alpha\beta}|$ and $|h_{\alpha\beta}|$ 
(for the definition of the latter see~\eqref{grossH} just ahead).

\subsection{\textit{A priori} assumptions on the solutions}
\label{subsec_apriori}

In this subsection we explain what the assumptions on our solutions are going to be.

\subsubsection{Absence of incoming radiation: Blanchet's condition}  
\label{swnir} 

We decompose our metric into the sum of the Minkowski metric and a small perturbation term. 
Letting $(\eta_{\alpha\beta})={\rm diag}\,(-1, 1, 1, 1)$ and $(\eta^{\alpha\beta})
=(\eta_{\alpha\beta})$, the Minkowski metric is given by
\[ -c^2 d\tau^2=ds^2=\eta_{\alpha\beta}\,dx^\alpha\,dx^\beta=-c^2\,dt^2+(dx^1)^2+(dx^2)^2+(dx^3)^2, \] 
where $x^0=ct$ as before, and we define the following coefficients measuring 
the deviation of $g$ from $\eta$:
\begin{equation}\label{grossH} 
   h_{\alpha\beta}=g_{\alpha\beta}-\eta_{\alpha\beta}
   \quad\mbox{as well as}\quad H^{\alpha\beta}=g^{\alpha\beta}-\eta^{\alpha\beta}.
\end{equation} 
We also let
\begin{equation}\label{grossH2} 
   h^{\alpha\beta}=\eta^{\alpha\alpha'}\eta^{\beta\beta'} h_{\alpha'\beta'}
   \quad\mbox{and}\quad h=\tensor{h}{_\alpha^\alpha}=\eta^{\alpha\beta} h_{\alpha\beta},
\end{equation} 
and furthermore 
\begin{equation}\label{barhdef}
   \bar{h}_{\alpha\beta}=h_{\alpha\beta}-\frac{1}{2}\,h\,\eta_{\alpha\beta}, 
\end{equation} 
so that 
\begin{equation}\label{barhh} 
   h_{\alpha\beta}=\bar{h}_{\alpha\beta}-\frac{1}{2}\,\bar{h}\,\eta_{\alpha\beta}
   \quad\mbox{for}\quad\bar{h}=\eta^{\alpha\beta}\,\bar{h}_{\alpha\beta}=-h.
\end{equation} 
Finally, we put
\begin{equation}\label{ingem} 
   \bar{h}^{\alpha\beta}=\eta^{\alpha\alpha'}\eta^{\beta\beta'}\,\bar{h}_{\alpha'\beta'},
\end{equation}  
and hence, cf.~(\ref{grossH2}),  
\begin{equation}\label{grossH3} 
   h^{\alpha\beta}=\bar{h}^{\alpha\beta}-\frac{1}{2}\,\bar{h}\,\eta^{\alpha\beta}
   =\bar{h}^{\alpha\beta}+\frac{1}{2}\,h\,\eta^{\alpha\beta}
\end{equation} 
is verified. 
\medskip 

In order for the quantity approximated by the quadrupole formula to truly represent 
the rate of change of energy radiated from the source matter, it is necessary 
that the spacetime metric satisfy a condition that can be physically interpreted 
as meaning the absence of radiation incoming from spatial infinity. For the solutions 
considered in this paper, we are going to assume the following condition that is inspired 
by (and very close to) the one put forward by Blanchet \cite[equ.~(19)]{B_LRR}: 

\medskip 
\textbf{No-incoming-radiation (NIR) condition:} \textit{We say that a solution 
to the Einstein-Vlasov system is \textbf{isolated from incoming radiation}, 
if there exist parameters ${\cal T} > 0$ and ${\cal S} > 0$ and functions 
$p_{\alpha\beta} = p_{\alpha\beta}(x)$ such that for $\alpha, \beta=0, 1, 2, 3$ 
\begin{equation}\label{Blanch} 
   h_{\alpha\beta}(t, x)=p_{\alpha\beta}(x) \quad \text{and} \quad 
   |\partial^\kappa p_{\alpha\beta}(x)| \le \frac{1}{\lambda^{|\kappa|}} \frac{\delta_h}{|x|^{1+|\kappa|}}.
\end{equation}
for all $t$ with $t\leq -{\cal T}$, all $x$ with $|x|\geq {\cal S}$ 
and all multi-indices $\kappa$ with $|\kappa|\leq 3$. Here $\delta_h$ denotes 
a universal constant depending on the particular solution, but not on $t$, $x$ or $\kappa$, 
which is used to measure ``smallness''.}
\medskip

The relations (\ref{Blanch}) will render several of the terms that appear 
in our estimates technically tractable. A more thorough discussion of Blanchet's condition 
and how it relates to our NIR condition above can be found in Appendix \ref{Blanch_discuss}.

\subsubsection{Support and decay properties}
\label{LT_intro} 

Through the recent papers \cite{FJS} and \cite{LT} it is known that Minkowski space 
is asymptotically stable with respect to small perturbations of the Einstein-Vlasov system, 
where the ``smallness'' has to be specified in technical terms. The solutions that 
we are going to consider are modeled after the work \cite{LT}, which shows global existence 
of the Einstein-Vlasov system posed as a Cauchy problem with compactly supported 
and sufficiently smooth initial data $f_0(x,p) = f(0,x,p)$, with the solutions additionally 
satisfying certain support properties for $f$ and $T^{\alpha\beta}$ as well 
as decay properties for $T^{\alpha\beta}$ and $h_{\alpha\beta}$. In particular, 
as in that work, we will be using the harmonic gauge condition $\Box_g x^\lambda=0$; 
see Appendix \ref{hargau} for more details on this gauge.

\smallskip
The following conditions express the fact that the Vlasov matter propagates 
at a subluminal speed that remains bounded. See Lemma~\ref{suppft} and Corollary~\ref{suppon} 
in Appendix~\ref{sec_app_technical} to see how these conditions can be derived 
from a natural assumption on the support of the initial data $f_0$ for the Einstein-Vlasov system.

\medskip 
\textbf{Compact-support conditions:} \textit{We say that a solution 
to the Einstein-Vlasov system satisfies the \textbf{compact-support conditions} 
when there exist positive parameters $\beta_0$, $K_0$, $K_0'$ 
and $K_\ast$, with $\beta_0\in (0,1)$, such that:}
    
    \begin{itemize}
    
        \item[(a)] \textit{For all $t\in\bbR$,}
        \begin{equation} \label{CS_cond_f}
           \mathrm{supp}(f(t,\cdot,\cdot))\subset\Big\{(x, p): |x|\le \lambda K_0+\beta_0 c|t|, 
           \ |p|\le c(K'_0+1)\Big\} .
        \end{equation}
        
        \item[(b)] \textit{For all $(t,x)\in\bbR^4$ with $|t-c^{-1}|x|| \leq \lambda c^{-1}$ 
        and all $\bar{x}\in\bbR^3$ with $|\bar{x}|\geq \lambda K_\ast$,}
        \begin{equation} \label{CS_cond_T}
            T_{\alpha\beta}(t-c^{-1}|x-\bar{x}|, \ \bar{x}) = 0 \quad\text{and}
            \quad T_{\alpha\beta}(t, \bar{x}) = 0 ,
        \end{equation}
        \textit{the same also being true for any derivatives of $T_{\alpha\beta}$, 
        as well as for $T^{\alpha\beta}$ and its derivatives.}
        
    \end{itemize}
\medskip
We remark that Corollary~\ref{suppon} will show that (b) above follows from (a) if one defines
\begin{equation}\label{R0def} 
   K_\ast=\frac{K_0+\beta_0}{1-\beta_0},
\end{equation} 
whereas (a), for solutions evolving from initial data satisfying appropriate conditions, 
holds for a given $K_0$ and $K_0'$ if $\beta_0$ is defined by the relation
\begin{equation}\label{beta0def}
    1-\beta_0^2 = \frac{1}{4(8K'_0+1)}.
\end{equation}

Next, to describe the final set of conditions, once again following \cite{LT}, we need a further splitting 
of the weak-field coefficients $h_{\alpha\beta}$ into a Schwarzschild-like piece and a remainder piece. Let
\begin{equation} \label{eq_chichi}
    \chi: [0,\infty) \longrightarrow [0,1] \quad \text{such that } \chi(\xi) = \left\{\begin{array}{ll}
    0 &\text{ for } \xi\leq 1/2 , \\
    1 &\text{ for } \xi\geq 3/4 \\
\end{array}\right.
\end{equation}
be a smooth cutoff function. Given a solution to the Einstein-Vlasov system with corresponding 
ADM mass $M$ as well as $h_{\alpha\beta}$ defined in~\eqref{grossH}, let
\begin{equation}\label{def_h0h1}
    h^0_{\alpha\beta}(x) = \chi\left( \frac{|x|}{\lambda+|t|} \right)
    \frac{GM}{c^2|x|}\delta_{\alpha\beta} \ , \quad h^1_{\alpha\beta}(x) 
    = h_{\alpha\beta}(x) - h^0_{\alpha\beta}(x) .
\end{equation}
In Appendix \ref{LTbounds_sect}, Corollary \ref{LT_cor_cG}, we outline the relevant estimates from \cite{LT} 
that motivate the following conditions. For now it suffices to say that they can be derived from 
a set of natural assumptions concerning the smallness of the initial data of the Einstein-Vlasov system.

\medskip 
\textbf{Boundedness and decay conditions:} \textit{We say that a solution 
to the Einstein-Vlasov system satisfies the \textbf{boundedness and decay conditions} 
if there exist parameters $\gamma\in (0,1)$, $\delta\in (0,1)$, $\delta_h > 0$ 
and $\delta_{T,\mu\nu}^{\kappa}$ (the latter for each $\mu,\nu=0, 1, 2, 3$ 
and multi-index $\kappa$ with $|\kappa|\leq 4$) such that, for all $\mu,\nu=0, 1, 2, 3$, 
all $(x^0, x)\in\bbR^4$ (representing harmonic-gauge coordinates) and all multi-indices $\kappa$ 
with $|\kappa|\leq 4$,}
        \begin{equation}\label{fuvor0} 
           |\partial^\kappa T_{\mu\nu}(x^0, x)| 
           \leq\frac{c^4}{G\lambda^{2+|\kappa|}}\,\delta_{T,\mu\nu}^\kappa,
        \end{equation}
        \begin{equation}\label{fuvor} 
           |\partial h_{\mu\nu}(x^0, x)|
           \le\frac{1}{\lambda}\cdot
           \frac{\delta_h}{\big(1+\lambda^{-1}|x^0|\big)^{1-2\delta} \ 
           \big(1+\lambda^{-1}\big||x|-|x^0|\big|\big)},
        \end{equation}      
        \begin{equation}\label{ils} 
           |\partial^\kappa (h^1)_{\mu\nu}(x^0, x)|
           \le\dfrac{1}{\lambda^{|\kappa|}}\cdot
           \dfrac{\delta_h (1+\lambda^{-1}|x^0|)^{\delta_h}}
           {\big( 1+\lambda^{-1}(|x|+|x^0|)\big) \ \big(1+\lambda^{-1}(|x|-|x^0|)_+\big)^\gamma},
        \end{equation} 
        \textit{where $\partial h_{\mu\nu}$ stands for 
        any of the four first-order derivatives of $h_{\mu\nu}$, 
        and $(|x|-|x^0|)_+$ means $|x|-|x^0|$ if $|x| \geq |x^0|$ and zero otherwise.}
        
\medskip
\noindent For a more finely tuned result, one could also assume modified bounds 
on the different $\partial^\kappa h_{\mu\nu}$ by introducing separate constants 
$\delta_{h,\mu\nu}^{(\kappa)}$ for them, but we refrain from doing so. For our purposes here, 
this kind of separation will only be needed for the $\partial^\kappa T_{\mu\nu}$ terms, 
as above. Also note that, for the proof of Theorem \ref{mainthm}, we do not assume 
any \textit{a priori} relation between the constants $\delta_h$ 
and $\delta_{T,\mu\nu}^\kappa$ (but cf.~\cite[p.~330]{PoWi}).

\begin{remark}{\rm Since our intention is to study an isolated collisionless gravitational system 
close to Minkowski space, it makes sense to consider solutions 
to the Einstein-Vlasov system satisfying the conditions (\ref{Blanch}), (\ref{CS_cond_f}), 
(\ref{CS_cond_T}) and \eqref{fuvor0}--\eqref{ils}. 
Let us clearly state that we do not know at present if such solutions do actually exist, 
and in particular that we do not show this in the present paper. 

In a similar vein, it is generally accepted 
that a ``no incoming radiation condition'' has to be imposed 
in order to not obtain unwanted and unphysical contributions. 
For this, we take a formulation of such a condition (set out by Blanchet, as introduced above) 
that still is manageable analytically. It has to be clarified in future work, 
if the NIR condition (\ref{Blanch}) could be replaced by an asymptotic condition as $t\to -\infty$. 

Put together, we currently do not know to which solutions our main result applies, 
but we nevertheless consider this work to be a valid initial step, 
since the quadrupole formula is derived from clear-cut assumptions. 
} 
\end{remark}

\subsection{Statement of the main result}\label{sub_main}

The main result of this paper is the next theorem. 

\begin{theorem}\label{mainthm} 
Let positive constants $C_0$, $\eps_0$, $\mathcal{T}\geq 2c^{-1}\lambda$, 
$\mathcal{S}\geq\max\{2\mathcal{T}c,6\lambda\}$, $K_0$, $K_0'$, $\gamma\in [\frac{3}{4},1)$ 
and $\delta\in(0,\frac{1}{2})$ be given, and define $\beta_0$ as in~\eqref{beta0def} 
and $K_\ast$ as in~\eqref{R0def}. Furthermore, suppose that, for all $\eps\in (0,\eps_0]$, 
there exist constants
\[ \eps_h\in (0,\eps) \quad\text{and}\quad \eps_{T,\mu\nu}^\kappa\in (0,\eps) \]
(for each $\mu,\nu=0, 1, 2, 3$ and each multi-index $\kappa$ such that $|\kappa|\leq 4$) 
and a set of initial data $(g_{\alpha\beta}^{(\eps)}, 
\partial_t g_{\alpha\beta}^{(\eps)}, f_0^{(\eps)})\big|_{t=0}$ for the Einstein-Vlasov system satisfying
\[ \mathrm{supp}(f_0^{(\eps)})\subset\{(x, p): |x|\leq\lambda K_0, \ |p|\leq cK'_0 \} \]
and leading to a globally defined solution $(g_{\alpha\beta}^{(\eps)}, f^{(\eps)})$ 
in harmonic coordinates with the following properties:
\begin{itemize}

    \item[(i)] the NIR condition~\eqref{Blanch} for $h_{\alpha\beta}^{(\eps)} 
    = g_{\alpha\beta}^{(\eps)} - \eta_{\alpha\beta}$ holds with the constant $\delta_h = C_0\eps_h$;

    \item[(ii)] the compact-support conditions~\eqref{CS_cond_f} and \eqref{CS_cond_T} 
    hold for $f^{(\eps)}$ and the corresponding $T_{\alpha\beta}^{(\eps)}$;

    \item[(iii)] the boundedness and decay conditions~\eqref{fuvor}--\eqref{ils} 
    hold for $T_{\alpha\beta}^{(\eps)}$, $h^{(\eps)}_{\mu\nu}$ and $h^{(\eps),1}_{\mu\nu}$ 
    (the latter defined as in~\eqref{def_h0h1}) with the constants 
    $\delta_{T,\mu\nu}^\kappa = C_0\eps_{T,\mu\nu}^\kappa$ and $\delta_h = C_0\eps_h$;

    \item[(iv)] the ADM mass $M^{(\eps)}$ corresponding to the metric $g^{(\eps)}$ satisfies
    \begin{equation}\label{ADM_boundbound}
        \frac{GM^{(\eps)}}{c^2\lambda} < \eps_h .
    \end{equation}
\end{itemize}
Then there are $\eps_1\in (0, \eps_0]$ and a constant $C_1>0$ 
such that, for all $\eps\in (0, \eps_1]$ and events $(t, x)$ so that
\[ |x|\ge\max\{12\lambda, c{\cal T}, 2{\cal S}, 5\lambda K_\ast\} \quad\text{and}
   \quad |u|=|t-c^{-1}|x||\le\lambda c^{-1} , \]
the following quadrupole approximation for gravitational radiation holds 
for the solution $(g_{\alpha\beta}^{(\eps)}, f^{(\eps)})$:
\begin{eqnarray}\label{quadru} 
   \lefteqn{\bigg|\partial_t\,\mathfrak{p}^0(t, x)
   -\bigg(-\frac{G}{5c^5}\sum_{j, k=1}^3 \bigg[ (\dddot{D}^{\,jk}(u))^2  
   -\frac{1}{3}(\tensor{\dddot{D}}{_j^j}(u))^2\bigg]\bigg)\bigg|} 
   \nonumber 
   \\ & \le & C_1\,\frac{c^5}{G}\,
   \Bigg[(\lambda^{-1}|x|)^{1-\gamma_0+C_0\eps_h}\,\eps_h^2
   +\frac{1}{\lambda^{-1}|x|}\,\Big(\sum_{j=0}^3\eps_T^{(j)}\Big)
   \nonumber
   \\ & & \hspace{4em} +\,\Big(\eps_{T,\,00}^{(3), x^0}+\eps_{T,\,00}^{(4), x^0}
   +\sum_{i, j=1}^3\eps_{T,\,ij}^{(4), x^0}
   +\sum_{j=1}^3\eps_{T,\,0j}^{(4), x^0}
   \nonumber   
   \\ & & \hspace{6em} +\,\sum_{i, j=1}^3\eps_{T,\,ij}^{(3), x^0}
   +\sum_{j=1}^3\eps_{T,\,0j}^{(3), x^0}
   +\sum_{i, j=1}^3\eps_{T,\,ij}^{(2), x^0}   
   +\Big(\sum_{j=0}^3\eps_T^{(j)}\Big)\eps_h\Big)\Bigg]
   \nonumber 
   \\ & & \hspace{3em}\times\Bigg[
   (\lambda^{-1}|x|)^{1-\gamma_0+C_0\eps_h}\,\eps_h^2
   +\frac{1}{\lambda^{-1}|x|}\,\Big(\sum_{j=0}^3\eps_T^{(j)}\Big)
   \nonumber
   \\ & & \hspace{5em} +\,\Big(\eps_{T,\,00}^{(4), x^0}
   +\sum_{i, j=1}^3\eps_{T,\,ij}^{(4), x^0}
   +\sum_{j=1}^3\eps_{T,\,0j}^{(4), x^0}   
   \nonumber 
   \\ & & \hspace{7em}
   +\,\sum_{i, j=1}^3\eps_{T,\,ij}^{(3), x^0}
   +\sum_{j=1}^3\eps_{T,\,0j}^{(3), x^0}
   +\sum_{i, j=1}^3\eps_{T,\,ij}^{(2), x^0}   
   +\Big(\sum_{j=0}^3\eps_T^{(j)}\Big)\eps_h\Big)\Bigg]
   \nonumber
   \\ & & +\,C_1\,\frac{c^5}{G}
   \,\frac{1}{(\lambda^{-1}|x|)^{1-3\eps_h}}\,\eps_h^3,
\end{eqnarray}
where we have used the symbols
\[ \eps_T^{(j)} = \max\{\eps_{T,\,\mu\nu}^\kappa: |\kappa|=j, \ \mu, \nu=0, 1, 2, 3\} \]
and
\[ \eps_{T,\,\mu\nu}^{(j), x^0} = \eps_{T,\,\mu\nu}^{(0\ldots 0)}, \] 
i.e., the multi-index $\kappa$ corresponding to the latter has $j$ entries all equal to 0.
\end{theorem}

A word on the proof of this result: Through a sequence of interrelated remainder terms 
(of which we will keep precise track), we will carve out the quadrupole expression 
from $\partial_t\,\mathfrak{p}^0(t, x)$, culminating at the relation~\eqref{knigh}, namely
\begin{equation}\label{brill40} 
   \partial_t\,\mathfrak{p}^0(t, x)
   =-\frac{G}{5c^5}\bigg[\sum_{j, k=1}^3 (\dddot{D}^{\,jk}(u))^2  
   -\frac{1}{3}\sum_{j=1}^3(\tensor{\dddot{D}}{^{jj}}(u))^2\bigg]+({\rm rem}_{50})(t, x) .
\end{equation} 
The term at the end is a remainder term, the remainder number 50 in our estimates. 
As soon as it is bounded in (\ref{maumau50}), the proof of the theorem will be complete. 

\begin{remark}
{\rm
\noindent 
(a) We now explain why the estimate~\eqref{quadru} is meaningful with respect 
to the relative sizes of the objects 
being compared. The left-hand side is the difference between two expressions 
whose orders of magnitude are as follows, in self-explanatory symbolic notation:
\begin{equation} \label{ord_mag_1}
    \partial_t\,\mathfrak{p}^0
   \sim t^{0i}\sim (\partial_{x^\gamma}\bar{h}^{0i})(\partial_{x^\delta}\bar{h}^{\delta\gamma})
   \sim (\partial_{x^\gamma} h_{0i})(\partial_{x^\delta} h_{\delta\gamma})
   \sim\delta_h^2
\end{equation}
and
\begin{equation} \label{ord_mag_2}
    (\dddot{D}^{\,jk})^2\sim (\partial_{x^0}^3 T_{00})^2\sim (\eps_{T,\,00}^{(3), x^0})^2.
\end{equation}
Let us check that the right-hand side of (\ref{quadru}) has a smaller order of magnitude 
in comparison to~\eqref{ord_mag_1} and ~\eqref{ord_mag_2}. The last term in there 
is of order $\sim\frac{1}{|x|^{1-3C_0\eps_h}}\,\delta_h^3$ and thus can be expected 
to be ``small'' as compared to the terms on the left-hand side. As for the 
$[\ldots]\times [\ldots]$ term, there is the factor $|x|^{1-\gamma_0+C_0\eps_h}\eps_h^2$ 
that leads to at least ${\cal O}(\eps^3)$ upon multiplication with another factor. 
However, $1-\gamma_0+C_0\eps_h>0$ for $\eps_h$ small, 
so there is a positive power of $|x|$ to deal with when discussing 
the range of validity of the quadrupole approximation; also recall that we are allowed 
to take $\gamma_0<1$ as close to $1$ as we wish initially. It should be stressed that the positive power 
is {\em not} an artifact of the proof, but actually the best bound that 
one can get under the decay assumptions that we impose on $h$. 
From a mathematics-of-wave-equations perspective, 
the estimate perhaps could be improved to something like $\log(3+|x|)$, 
but this would not make a conceptual difference, since the whole point 
is that the available estimates on $h$ degenerate close to the light cone; see (\ref{fuvor}) 
and (\ref{ils}). Certainly, if there were rigorous results that showed an improved decay for $h$, 
the factor $|x|^{1-\gamma_0+C_0\eps_h}$ would disappear. For the other orders, 
in a physically realistic slow-motion situation, one expects to have
\[ |T_{00}|\gg |T_{0j}|\gg |T_{ij}|, \] 
see \cite[p.~222]{stephani} or \cite[p.~330]{PoWi}. In terms of a ``typical velocity'' $\bar{v}$ 
in the system and under the slow-motion assumption $(\frac{\bar{v}}{c})\ll 1$, it is the case that 
\[ \Big(\frac{\bar{v}}{c}\Big)^2 |T_{00}|\sim\Big(\frac{\bar{v}}{c}\Big)|T_{0j}|\sim |T_{ij}|, \] 
which can be seen from (\ref{tab3}) and the fact that $\sim -p_0 f\sim\sqrt{1+|p|^2} f$ 
appears integrated in the expression for $T_{00}$, $\sim p_j f\sim\sqrt{1+|p|^2}v_j f$ 
is integrated in the expression for $T_{0j}$ 
and $\sim\frac{p_i p_j}{\sqrt{1+|p|^2}} f\sim\sqrt{1+|p|^2}v_i v_j f$ is integrated 
in the expression for $T_{ij}$. Furthermore, 
\[ \frac{\partial}{\partial x^j}\sim\frac{1}{\lambda},\quad 
   \frac{\partial}{\partial x^0}\sim\frac{(\bar{v}/c)}{\lambda}, \] 
for taking derivatives \cite[p.~303]{strau2}. 
Therefore the following orders are obtained: 
\begin{eqnarray*} 
   & & \eps_{T,\,00}^{(3), x^0}\sim \Big(\frac{\bar{v}}{c}\Big)^3 |T_{00}|,
   \quad\eps_{T,\,00}^{(4), x^0}\sim\Big(\frac{\bar{v}}{c}\Big)^4 |T_{00}|,
   \quad\eps_{T,\,ij}^{(4), x^0}\sim\Big(\frac{\bar{v}}{c}\Big)^6 |T_{00}|,
   \quad\eps_{T,\,0j}^{(4), x^0}\sim\Big(\frac{\bar{v}}{c}\Big)^5 |T_{00}|,
   \\ & & \quad\eps_{T,\,ij}^{(3), x^0}\sim\Big(\frac{\bar{v}}{c}\Big)^5 |T_{00}|,
   \quad\eps_{T,\,0j}^{(3), x^0}\sim\Big(\frac{\bar{v}}{c}\Big)^4 |T_{00}|, 
   \quad\eps_{T,\,ij}^{(2), x^0}\sim\Big(\frac{\bar{v}}{c}\Big)^4 |T_{00}|.   
\end{eqnarray*} 
Hence, if we look at the different ${\cal O}(\eps^2)$-products 
that appear on the right-hand side of (\ref{quadru}), the worst order results from  
\[ \eps_{T,\,00}^{(3), x^0}(\eps_{T,\,00}^{(4), x^0}+\eps_{T,\,0j}^{(3), x^0}
   +\eps_{T,\,ij}^{(2), x^0})\sim\Big(\frac{\bar{v}}{c}\Big)^7 |T_{00}|^2, \] 
which is one order better than 
\[ (\dddot{D}^{\,jk})^2\sim\Big(\frac{\bar{v}}{c}\Big)^6 |T_{00}|^2, \]  
and thus indeed ``small''. 
\smallskip 

\noindent 
(b) We will not carry out an explicit post-Newtonian expansion ``$c\to\infty$'' 
in this paper, but it should be remarked that the Newtonian limit 
of the Einstein-Vlasov system is the Vlasov-Poisson system; 
see \cite{Rendall, ReinRendall}, and also \cite[p.~176-178]{stephani} 
and \cite[p.~219/220]{strau2}. If $\rho_{{\rm VP}}(t, x)=\int_{\R^3} f_{{\rm VP}}(t, x, v)\,dv$ 
and $U_{{\rm VP}}$ denote the density and the Newtonian potential 
for this Vlasov-Poisson system, respectively, then one readily sees 
$U_{{\rm VP}}\sim -\frac{c^2}{2}h_{00}$ and $T_{00}\sim c^2\rho_{{\rm VP}}$, 
so that $D^{\alpha\beta}\sim\int_{\R^3}\rho_{{\rm VP}}\,x^\alpha x^\beta\,dx$. 
}
\end{remark}


\section{Expansion of the energy-rate formula}
\label{sec_expansion}
\setcounter{equation}{0} 

In this section we carry out the expansion (up to the needed orders with respect 
to powers of the terms $1/|x|$ and $T^{\alpha\beta}$) of all the basic quantities 
that enter the formula~\eqref{ogin} for $\partial_t\mathfrak{p}^0$. 
Most physics-oriented references perform this expansion by simply discarding terms 
that are deemed small due to being of a high order in those terms, but it should be noted 
that our goal here is to keep precise track of these errors, which will then be treated 
in Section \ref{sec_estimating}. We call these terms \textit{remainder terms} 
and denote them by the symbol $\rem{i}$, and the point is to use only those bounds 
that can be justified mathematically. 

Let a solution to the Einstein-Vlasov system satisfy the NIR condition~\eqref{Blanch}, 
compact-support conditions~\eqref{CS_cond_f},~\eqref{CS_cond_T} 
and decay conditions~\eqref{fuvor0}--\eqref{ils}, and let $(t,x)$ be a fixed spacetime 
event such that $|u|=|t-c^{-1}|x|| \leq \lambda c^{-1}$ and $r= |x|\ge\lambda K_\ast$.

\subsection{Einstein tensor} 
\label{expa_sect}

According to (\ref{Ric_harmo}) in Lemma \ref{Ric_expr}, the Ricci tensor 
in the harmonic gauge is given by 
\[ R_{\alpha\beta}=-\frac{1}{2}\,\Box'_g\,g_{\alpha\beta}
   +\frac{1}{2}\,\tilde{P}(g; \partial_{x^\alpha} g, \partial_{x^\beta} g)
   +\frac{1}{2}\,\tilde{Q}_{\alpha\beta}(g; \partial g, \partial g), \] 
where $\Box'_g=g^{\alpha\beta}\,\partial^2_{x^\alpha x^\beta}$; 
the expressions $\tilde{P}$ and $\tilde{Q}$ 
are defined in (\ref{tildP}) and (\ref{tildQ}). 
Since $\partial_{x^\gamma} g_{\alpha\beta}=\partial_{x^\gamma} h_{\alpha\beta}$, 
we first observe that 
\begin{eqnarray}\label{gezu}  
   R_{\alpha\beta}
   & = & -\frac{1}{2}\,\Box'_g\,h_{\alpha\beta}
   +\frac{1}{2}\,\tilde{P}(g; \partial_{x^\alpha} h, \partial_{x^\beta} h)
   +\frac{1}{2}\,\tilde{Q}_{\alpha\beta}(g; \partial h, \partial h)
   \nonumber
   \\ & = & -\frac{1}{2}\,\Box'_g\,h_{\alpha\beta}
   +\frac{1}{2}\,P(\partial_{x^\alpha} h, \partial_{x^\beta} h)
   +\frac{1}{2}\,Q_{\alpha\beta}(\partial h, \partial h)+{({\rm rem}_1)}_{\alpha\beta}, 
\end{eqnarray} 
where 
\[ P(\partial_{x^\alpha} h, \partial_{x^\beta} h)
   =\tilde{P}(\eta; \partial_{x^\alpha} h, \partial_{x^\beta} h),
   \quad Q_{\alpha\beta}(\partial h, \partial h)
   =\tilde{Q}_{\alpha\beta}(\eta; \partial h, \partial h), \] 
as well as 
\begin{equation}
    {({\rm rem}_1)}_{\alpha\beta}
   =\frac{1}{2}\,\tilde{P}(g; \partial_{x^\alpha} h, \partial_{x^\beta} h)
   -\frac{1}{2}\,P(\partial_{x^\alpha} h, \partial_{x^\beta} h)
   +\frac{1}{2}\,\tilde{Q}_{\alpha\beta}(g; \partial h, \partial h)
   -\frac{1}{2}\,Q_{\alpha\beta}(\partial h, \partial h), \label{eq_rem1}
\end{equation}
which is a remainder term that is (formally) cubic in $h$. The relation (\ref{gezu}) 
corresponds to \cite[Lemma 3.2]{LR}. Explicitly, (\ref{tildP}) and (\ref{tildQ}) yield 
\begin{eqnarray*} 
   P(\partial_{x^\mu} h, \partial_{x^\nu} h) 
   & = & \frac{1}{4}\,\eta^{\alpha\alpha'}\partial_{x^\mu} h_{\alpha\alpha'}
   \,\eta^{\beta\beta'}\partial_{x^\nu} h_{\beta\beta'}
   -\frac{1}{2}\,\eta^{\alpha\alpha'} \eta^{\beta\beta'}
   \,\partial_{x^\mu} h_{\alpha\beta}\,\partial_{x^\nu} h_{\alpha'\beta'}, 
   \\ Q_{\mu\nu}(\partial h, \partial h) 
   & = & \eta^{\alpha\alpha'} \eta^{\beta\beta'}\,\partial_{x^\alpha} h_{\beta\mu}
   \,\partial_{x^{\alpha'}} h_{\beta'\nu}
   \\ & & -\,\eta^{\alpha\alpha'} \eta^{\beta\beta'} 
   \,(\partial_{x^\alpha} h_{\beta\mu}\,\partial_{x^{\beta'}} h_{\alpha'\nu}
   -\partial_{x^{\beta'}} h_{\beta\mu}\,\partial_{x^\alpha} h_{\alpha'\nu})
   \\ & & +\,\eta^{\alpha\alpha'} \eta^{\beta\beta'} 
   \,(\partial_{x^\mu} h_{\alpha'\beta'}\,\partial_{x^\alpha} h_{\beta\nu}
   -\partial_{x^\alpha} h_{\alpha'\beta'}\,\partial_{x^\mu} h_{\beta\nu})
   \\ & & +\,\eta^{\alpha\alpha'} \eta^{\beta\beta'} 
   \,(\partial_{x^\nu} h_{\alpha'\beta'}\,\partial_{x^\alpha} h_{\beta\mu}
   -\partial_{x^\alpha} h_{\alpha'\beta'}\,\partial_{x^\nu} h_{\beta\mu})
   \\ & & +\,\frac{1}{2}\,\eta^{\alpha\alpha'} \eta^{\beta\beta'} 
   \,(\partial_{x^{\beta'}} h_{\alpha\alpha'}\,\partial_{x^\mu} h_{\beta\nu}
   -\partial_{x^\mu} h_{\alpha\alpha'}\,\partial_{x^{\beta'}} h_{\beta\nu})
   \\ & & +\,\frac{1}{2}\,\eta^{\alpha\alpha'} \eta^{\beta\beta'} 
   \,(\partial_{x^{\beta'}} h_{\alpha\alpha'}\,\partial_{x^\nu} h_{\beta\mu}
   -\partial_{x^\nu} h_{\alpha\alpha'}\,\partial_{x^{\beta'}} h_{\beta\mu}). 
\end{eqnarray*}  
If one is interested only in an expansion up to quadratic errors in $h$, 
then using (\ref{grossH}) in (\ref{gezu}) gives 
\begin{eqnarray}\label{tbfw1}  
   R_{\alpha\beta} & = & -\frac{1}{2}\,
   (\eta^{\alpha'\beta'}+H^{\alpha'\beta'})
   \,\partial^2_{x^{\alpha'} x^{\beta'}}\,h_{\alpha\beta}
   +\frac{1}{2}\,P(\partial_{x^\alpha} h, \partial_{x^\beta} h)
   +\frac{1}{2}\,Q_{\alpha\beta}(\partial h, \partial h)+{({\rm rem}_1)}_{\alpha\beta}
   \nonumber
   \\ & = & -\frac{1}{2}\,\Box\,h_{\alpha\beta}+{({\rm rem}_0)}_{\alpha\beta} 
   +{({\rm rem}_1)}_{\alpha\beta}+{({\rm rem}_2)}_{\alpha\beta}, 
\end{eqnarray} 
defining $\Box=\eta^{\alpha\beta}\,\partial^2_{x^\alpha x^\beta}$, and furthermore
\begin{equation}
    {({\rm rem}_0)}_{\alpha\beta}=\frac{1}{2}\,P(\partial_{x^\alpha} h, \partial_{x^\beta} h)
   +\frac{1}{2}\,Q_{\alpha\beta}(\partial h, \partial h),\quad 
   {({\rm rem}_2)}_{\alpha\beta}
   =-\frac{1}{2}\,H^{\alpha'\beta'}
   \,\partial^2_{x^{\alpha'} x^{\beta'}}\,h_{\alpha\beta}. \label{eq_rem0_rem2}
\end{equation}
The term ${({\rm rem}_2)}_{\alpha\beta}$ is also expected to be small, 
since $H^{\alpha'\beta'}=(H^{\alpha'\beta'}+h^{\alpha'\beta'})-h^{\alpha'\beta'}$, 
where the term in parentheses turns out to be quadratic in $h$. For the scalar curvature, this implies 
\begin{eqnarray}\label{tbfw2}  
   R & = & g^{\alpha\beta} R_{\alpha\beta}
   \nonumber
   \\ & = & (\eta^{\alpha\beta}+H^{\alpha\beta})
   \Big(-\frac{1}{2}\,\Box\,h_{\alpha\beta}+{({\rm rem}_0)}_{\alpha\beta} 
   +{({\rm rem}_1)}_{\alpha\beta}+{({\rm rem}_2)}_{\alpha\beta}\Big)
   \nonumber
   \\ & = & -\frac{1}{2}\,\eta^{\alpha\beta}\,\Box\,h_{\alpha\beta}+({\rm rem}_3)
   \nonumber
   \\ & = & -\frac{1}{2}\,\Box\,h+({\rm rem}_3)
\end{eqnarray} 
for 
\begin{eqnarray*} 
   ({\rm rem})_3 & = & \eta^{\alpha\beta}\,({({\rm rem}_0)}_{\alpha\beta}
   +{({\rm rem}_1)}_{\alpha\beta}+{({\rm rem}_2)}_{\alpha\beta})
   \\ & & +\,H^{\alpha\beta}\Big(-\frac{1}{2}\,\Box\,h_{\alpha\beta}+{({\rm rem}_0)}_{\alpha\beta}  
   +{({\rm rem}_1)}_{\alpha\beta}+{({\rm rem}_2)}_{\alpha\beta}\Big). \label{eq_rem3}
\end{eqnarray*} 
Thus it follows from (\ref{tbfw1}) and (\ref{tbfw2}) that the Einstein tensor can be written as
\begin{eqnarray}\label{tfte} 
   G_{\alpha\beta} & = & R_{\alpha\beta}-\frac{1}{2}\,R\,g_{\alpha\beta}
   \nonumber\\ & = & -\frac{1}{2}\,\Box\,\Big(h_{\alpha\beta}-\frac{1}{2}\,h\,\eta_{\alpha\beta}\Big)
   +{({\rm rem}_4)}_{\alpha\beta},  
\end{eqnarray}
where 
\begin{eqnarray}\label{eq_rem4} 
   {({\rm rem}_4)}_{\alpha\beta} 
   & = & {({\rm rem}_0)}_{\alpha\beta}+{({\rm rem}_1)}_{\alpha\beta}+{({\rm rem}_2)}_{\alpha\beta}
   -\frac{1}{2}\,({\rm rem}_3)\,g_{\alpha\beta}
   +\frac{1}{4}\,(\Box\,h)\,h_{\alpha\beta} \nonumber
   \\ & = & \frac{1}{2}\,\tilde{P}(g; \partial_{x^\alpha} h, \partial_{x^\beta} h)
   +\frac{1}{2}\,\tilde{Q}_{\alpha\beta}(g; \partial h, \partial h)
   -\frac{1}{2}\,H^{\alpha'\beta'}
   \,\partial^2_{x^{\alpha'} x^{\beta'}}\,h_{\alpha\beta} \nonumber
   \\ & & -\,\frac{1}{4}\,(\eta^{\alpha'\beta'}+H^{\alpha'\beta'})
   \bigg(\tilde{P}(g; \partial_{x^{\alpha'}} h, \partial_{x^{\beta'}} h)
   +\tilde{Q}_{\alpha'\beta'}(g; \partial h, \partial h)
   -H^{\gamma\delta}\,\partial^2_{x^\gamma x^\delta}\,h_{\alpha'\beta'}\bigg)\,g_{\alpha\beta} \nonumber
   \\ & & +\,\frac{1}{4}\,\Big(H^{\alpha'\beta'}\Box\,h_{\alpha'\beta'}\Big)\,g_{\alpha\beta}
   +\frac{1}{4}\,(\Box\,h)\,h_{\alpha\beta}
\end{eqnarray} 
is quadratic in $h$. 

\subsection{Metric determinant}
\label{metric_det_sect}

Let $\hat{g}=(g_{\alpha\beta})$, $\hat{m}=(\eta_{\alpha\beta})$ 
and $\hat{h}=(h_{\alpha\beta})$ be viewed as matrices of tensor components. Then 
\begin{eqnarray*} 
   g & = & \det\hat{g}=\det(\hat{m}+\hat{h})
   =\det(\hat{m})\det(I+\hat{m}^{-1}\hat{h})
   =-\det(I+\hat{m}^{-1}\hat{h})
   \\ & = & -\det\bigg(I+\|(h_{\alpha\beta})\|\,\hat{m}^{-1}
   \frac{\hat{h}}{\|(h_{\alpha\beta})\|}\bigg)
   =-1-\|(h_{\alpha\beta})\|\,{\rm tr}\,\Big(\hat{m}^{-1}
   \frac{\hat{h}}{\|(h_{\alpha\beta})\|}\Big)+({\rm rem}_5)
   \\ & = & -1-\,{\rm tr}\,(\hat{m}^{-1}\hat{h})+({\rm rem}_5)
   =-1-h-({\rm rem}_5)
\end{eqnarray*} 
for 
\begin{equation}\label{eq_rem5}
    ({\rm rem}_5)=\det(I+\hat{m}^{-1}\hat{h})
   -1-{\rm tr}\,(\hat{m}^{-1}\hat{h}), 
\end{equation}
cf.~also (\ref{grossH2}). It follows that 
\[ |g|=-g=1+h+({\rm rem}_5) \] 
and hence 
\begin{equation}\label{det12} 
   |g|^{1/2}=\sqrt{1+h+({\rm rem}_5)}
   =1+\frac{1}{2}\,h+({\rm rem}_6),
   \quad ({\rm rem}_6)=\sqrt{1+h+({\rm rem}_5)}-1-\frac{1}{2}\,h.
\end{equation} 

\subsection{Energy-momentum pseudo tensor}
\label{en_mom_ps_tensor_sect}

Our restrictions $|t-c^{-1}r| < \lambda c^{-1}$ and $r\geq \lambda K_\ast$ (where $r = |x|$) 
imply through the compact-support condition~\eqref{CS_cond_T} 
that the $T^{0i} = T^{0i}(t, \bar{x})$ term in the surface-integral formula~\eqref{ogin} 
for $\partial_t\mathfrak{p}^0$ vanishes. Thus
\begin{equation}\label{thores} 
   \partial_t\,\mathfrak{p}^0 
   =-c\,\sum_{i=1}^3\int_{\partial B_r(0)} |g|\,\frac{x^i}{r}\,t^{0i}\,dS(x). 
\end{equation}
We remind the reader that the Landau-Lifschitz energy-momentum pseudo tensor 
$t^{\alpha\beta}$ is defined in~\eqref{LaLi}. Its components are given in terms 
of derivatives of the metric components as well as of the quantity 
$\mathfrak{g}^{\alpha\beta}=|g|^{1/2}\,g^{\alpha\beta}$. Using (\ref{mfg}), 
(\ref{det12}), (\ref{grossH}) and (\ref{grossH3}),
\begin{eqnarray*} 
   \mathfrak{g}^{\alpha\beta} 
   & = & |g|^{1/2}\,g^{\alpha\beta}
   \\ & = & \Big(1+\frac{1}{2}\,h+({\rm rem}_6)\Big)(\eta^{\alpha\beta}+H^{\alpha\beta})
   \\ & = & \Big(1+\frac{1}{2}\,h+({\rm rem}_6)\Big)
   \Big(\eta^{\alpha\beta}-h^{\alpha\beta}+({\rm rem}_7)^{\alpha\beta}\Big)
   \\ & = & \eta^{\alpha\beta}-h^{\alpha\beta}
   +\frac{1}{2}\,h\,\eta^{\alpha\beta}+({\rm rem}_8)^{\alpha\beta}
   \\ & = & \eta^{\alpha\beta}-\bar{h}^{\alpha\beta}
   +({\rm rem}_8)^{\alpha\beta}   
\end{eqnarray*} 
for
\begin{align}
    ({\rm rem}_7)^{\alpha\beta} &= h^{\alpha\beta}+H^{\alpha\beta} , \label{eq_rem7} \\
    ({\rm rem}_8)^{\alpha\beta} & = -\frac{1}{2}\,h\,h^{\alpha\beta}
    + ({\rm rem}_7)^{\alpha\beta}\Big(1+\frac{1}{2}\,h\Big)
    + ({\rm rem}_6)\Big(\eta^{\alpha\beta}-h^{\alpha\beta}+({\rm rem}_7)^{\alpha\beta}\Big) . \label{eq_rem8}
\end{align}
As a consequence, 
\begin{equation}\label{hast1} 
   \partial_{x^\gamma}\mathfrak{g}^{\alpha\beta}
   =-\partial_{x^\gamma}\bar{h}^{\alpha\beta}+\tensor{({\rm rem}_9)}{_\gamma^\alpha^\beta},    
\end{equation}  
defining 
\begin{equation}\label{eq_rem9}
    \tensor{({\rm rem}_9)}{_\gamma^\alpha^\beta}
   =\partial_{x^\gamma}({\rm rem}_8)^{\alpha\beta}. 
\end{equation}
Therefore (\ref{LaLi}) in conjunction with (\ref{hast1}) and (\ref{grossH}) yields 
\begin{eqnarray*} 
   \frac{16\pi G}{c^4}\,|g|\,t^{\alpha\beta} 
   & = & \partial_{x^\gamma}\mathfrak{g}^{\alpha\beta}\,\partial_{x^\delta}\mathfrak{g}^{\delta\gamma}
   -\partial_{x^\gamma}\mathfrak{g}^{\alpha\gamma}\,\partial_{x^\delta}\mathfrak{g}^{\beta\delta}
   +\frac{1}{2}\,g^{\alpha\beta}\,g_{\delta\eps}\,\partial_{x^\sigma}\mathfrak{g}^{\delta\gamma}
   \,\partial_{x^\gamma}\mathfrak{g}^{\sigma\eps}
   \\ & & -g_{\eps\gamma}\,\partial_{x^\delta}\mathfrak{g}^{\eps\sigma}\,
   (g^{\alpha\delta}\,\partial_{x^\sigma}\mathfrak{g}^{\beta\gamma}
   +g^{\beta\delta}\,\partial_{x^\sigma}\mathfrak{g}^{\alpha\gamma})
   +g_{\delta\eps}\,g^{\gamma\sigma}\,\partial_{x^\gamma}\mathfrak{g}^{\beta\delta}
   \,\partial_{x^\sigma}\mathfrak{g}^{\alpha\eps}
   \\ & & +\frac{1}{8}\,(2g^{\alpha\delta}\,g^{\beta\gamma}-g^{\alpha\beta}\,g^{\delta\gamma})
   (2g_{\eps\sigma}\,g_{\kappa\nu}-g_{\sigma\kappa}\,g_{\eps\nu})
   \,\partial_{x^\delta}\mathfrak{g}^{\eps\nu}\,\partial_{x^\gamma}\mathfrak{g}^{\sigma\kappa}
   \\ & = & 
   \Big(-\partial_{x^\gamma}\bar{h}^{\alpha\beta}+\tensor{({\rm rem}_9)}{_\gamma^\alpha^\beta}\Big)
   \Big(-\partial_{x^\delta}\bar{h}^{\delta\gamma}+\tensor{({\rm rem}_9)}{_\delta^\delta^\gamma}\Big)   
   \\ & & 
   -\Big(-\partial_{x^\gamma}\bar{h}^{\alpha\gamma}+\tensor{({\rm rem}_9)}{_\gamma^\alpha^\gamma}\Big)
   \Big(-\partial_{x^\delta}\bar{h}^{\beta\delta}+\tensor{({\rm rem}_9)}{_\delta^\beta^\delta}\Big)
   \\ & & +\frac{1}{2}\,(\eta^{\alpha\beta}+H^{\alpha\beta})
   \,(\eta_{\delta\eps}+h_{\delta\eps})
   \Big(-\partial_{x^\sigma}\bar{h}^{\delta\gamma}+\tensor{({\rm rem}_9)}{_\sigma^\delta^\gamma}\Big)   
   \Big(-\partial_{x^\gamma}\bar{h}^{\sigma\eps}+\tensor{({\rm rem}_9)}{_\gamma^\sigma^\eps}\Big)
   \\ & & -(\eta_{\eps\gamma}+h_{\eps\gamma})
   \,(\eta^{\alpha\delta}+H^{\alpha\delta})
   \Big(-\partial_{x^\delta}\bar{h}^{\eps\sigma}+\tensor{({\rm rem}_9)}{_\delta^\eps^\sigma}\Big)
   \Big(-\partial_{x^\sigma}\bar{h}^{\beta\gamma}+\tensor{({\rm rem}_9)}{_\sigma^\beta^\gamma}\Big)
   \\ & & -(\eta_{\eps\gamma}+h_{\eps\gamma})
   \,(\eta^{\beta\delta}+H^{\beta\delta})
   \Big(-\partial_{x^\delta}\bar{h}^{\eps\sigma}+\tensor{({\rm rem}_9)}{_\delta^\eps^\sigma}\Big)   
   \Big(-\partial_{x^\sigma}\bar{h}^{\alpha\gamma}+\tensor{({\rm rem}_9)}{_\sigma^\alpha^\gamma}\Big)
   \\ & & +(\eta_{\delta\eps}+h_{\delta\eps})
   \,(\eta^{\gamma\sigma}+H^{\gamma\sigma})
   \Big(-\partial_{x^\gamma}\bar{h}^{\beta\delta}+\tensor{({\rm rem}_9)}{_\gamma^\beta^\delta}\Big)   
   \Big(-\partial_{x^\sigma}\bar{h}^{\alpha\eps}+\tensor{({\rm rem}_9)}{_\sigma^\alpha^\eps}\Big)
   \\ & & +\frac{1}{8}\,\Big(2(\eta^{\alpha\delta}+H^{\alpha\delta})
   \,(\eta^{\beta\gamma}+H^{\beta\gamma})
   -(\eta^{\alpha\beta}+H^{\alpha\beta})\,(\eta^{\delta\gamma}+H^{\delta\gamma})\Big)
   \\ & & \hspace{3em}\times
   \Big(2(\eta_{\eps\sigma}+h_{\eps\sigma})\,(\eta_{\kappa\nu}+h_{\kappa\nu})
   -(\eta_{\sigma\kappa}+h_{\sigma\kappa})\,(\eta_{\eps\nu}+h_{\eps\nu})\Big)
   \\ & & \hspace{3em}\times
   \Big(-\partial_{x^\delta}\bar{h}^{\eps\nu}+\tensor{({\rm rem}_9)}{_\delta^\eps^\nu}\Big)
   \Big(-\partial_{x^\gamma}\bar{h}^{\sigma\kappa}+\tensor{({\rm rem}_9)}{_\gamma^\sigma^\kappa}\Big)
   \\ & = & \partial_{x^\gamma}\bar{h}^{\alpha\beta}\,\partial_{x^\delta}\bar{h}^{\delta\gamma}
   -\partial_{x^\gamma}\bar{h}^{\alpha\gamma}\,\partial_{x^\delta}\bar{h}^{\beta\delta}
   +\frac{1}{2}\,\eta^{\alpha\beta}\,\eta_{\delta\eps}\,\partial_{x^\sigma}\bar{h}^{\delta\gamma}
   \,\partial_{x^\gamma}\bar{h}^{\sigma\eps}
   \\ & & -\,\eta_{\eps\gamma}\,\partial_{x^\delta}\bar{h}^{\eps\sigma}\,
   (\eta^{\alpha\delta}\,\partial_{x^\sigma}\bar{h}^{\beta\gamma}
   +\eta^{\beta\delta}\,\partial_{x^\sigma}\bar{h}^{\alpha\gamma})
   +\eta_{\delta\eps}\,\eta^{\gamma\sigma}\,\partial_{x^\gamma}\bar{h}^{\beta\delta}
   \,\partial_{x^\sigma}\bar{h}^{\alpha\eps}
   \\ & & +\,\frac{1}{8}\,(2\eta^{\alpha\delta}\,\eta^{\beta\gamma}
   -\eta^{\alpha\beta}\,\eta^{\delta\gamma})
   (2\eta_{\eps\sigma}\,\eta_{\kappa\nu}-\eta_{\sigma\kappa}\,\eta_{\eps\nu})
   \,\partial_{x^\delta}\bar{h}^{\eps\nu}\,\partial_{x^\gamma}\bar{h}^{\sigma\kappa}
   \\ & & +\,({\rm rem}_{10})^{\alpha\beta}, 
\end{eqnarray*}  
where 
\begin{eqnarray} 
   \lefteqn{({\rm rem}_{10})^{\alpha\beta}} \nonumber 
   \\ & = & -\partial_{x^\gamma}\bar{h}^{\alpha\beta}\tensor{({\rm rem}_9)}{_\delta^\delta^\gamma}
   -\partial_{x^\delta}\bar{h}^{\delta\gamma}\tensor{({\rm rem}_9)}{_\gamma^\alpha^\beta}
   +\tensor{({\rm rem}_9)}{_\gamma^\alpha^\beta}\tensor{({\rm rem}_9)}{_\delta^\delta^\gamma} \nonumber
   \\ & & +\partial_{x^\gamma}\bar{h}^{\alpha\gamma}\tensor{({\rm rem}_9)}{_\delta^\beta^\delta}
   +\partial_{x^\delta}\bar{h}^{\beta\delta}\tensor{({\rm rem}_9)}{_\gamma^\alpha^\gamma}
   -\tensor{({\rm rem}_9)}{_\gamma^\alpha^\gamma}\tensor{({\rm rem}_9)}{_\delta^\beta^\delta} \nonumber
   \\ & & +\frac{1}{2}\,\eta^{\alpha\beta}h_{\delta\eps}
   \partial_{x^\sigma}\bar{h}^{\delta\gamma}\partial_{x^\gamma}\bar{h}^{\sigma\eps}  \nonumber
   \\ & & +\frac{1}{2}\,g^{\alpha\beta} g_{\delta\eps}
   \Big[-\partial_{x^\sigma}\bar{h}^{\delta\gamma}\tensor{({\rm rem}_9)}{_\gamma^\sigma^\eps}
   -\partial_{x^\gamma}\bar{h}^{\sigma\eps}\tensor{({\rm rem}_9)}{_\sigma^\delta^\gamma}
   +\tensor{({\rm rem}_9)}{_\sigma^\delta^\gamma}\tensor{({\rm rem}_9)}{_\gamma^\sigma^\eps}\Big] \nonumber
   \\ & & +\frac{1}{2}\,H^{\alpha\beta}g_{\delta\eps}
   \,\partial_{x^\sigma}\bar{h}^{\delta\gamma}\partial_{x^\gamma}\bar{h}^{\sigma\eps} \nonumber
   \\ & & -h_{\eps\gamma}
   \,(\eta^{\alpha\delta}+H^{\alpha\delta})
   \Big(-\partial_{x^\delta}\bar{h}^{\eps\sigma}+\tensor{({\rm rem}_9)}{_\delta^\eps^\sigma}\Big)
   \Big(-\partial_{x^\sigma}\bar{h}^{\beta\gamma}+\tensor{({\rm rem}_9)}{_\sigma^\beta^\gamma}\Big) \nonumber
   \\ & & -\eta_{\eps\gamma}\,H^{\alpha\delta}
   \Big(-\partial_{x^\delta}\bar{h}^{\eps\sigma}+\tensor{({\rm rem}_9)}{_\delta^\eps^\sigma}\Big)
   \Big(-\partial_{x^\sigma}\bar{h}^{\beta\gamma}+\tensor{({\rm rem}_9)}{_\sigma^\beta^\gamma}\Big) \nonumber
   \\ & & -\eta_{\eps\gamma}\,\eta^{\alpha\delta}
   \Big[-\partial_{x^\delta}\bar{h}^{\eps\sigma}\tensor{({\rm rem}_9)}{_\sigma^\beta^\gamma}
   -\partial_{x^\sigma}\bar{h}^{\beta\gamma}\tensor{({\rm rem}_9)}{_\delta^\eps^\sigma}
   +\tensor{({\rm rem}_9)}{_\delta^\eps^\sigma}\tensor{({\rm rem}_9)}{_\sigma^\beta^\gamma}\Big] \nonumber
   \\ & & -h_{\eps\gamma}
   \,(\eta^{\beta\delta}+H^{\beta\delta})
   \Big(-\partial_{x^\delta}\bar{h}^{\eps\sigma}+\tensor{({\rm rem}_9)}{_\delta^\eps^\sigma}\Big)   
   \Big(-\partial_{x^\sigma}\bar{h}^{\alpha\gamma}+\tensor{({\rm rem}_9)}
   {_\sigma^\alpha^\gamma}\Big)  \nonumber
   \\ & & -\eta_{\eps\gamma} H^{\beta\delta}
   \Big(-\partial_{x^\delta}\bar{h}^{\eps\sigma}+\tensor{({\rm rem}_9)}{_\delta^\eps^\sigma}\Big)   
   \Big(-\partial_{x^\sigma}\bar{h}^{\alpha\gamma}+\tensor{({\rm rem}_9)}
   {_\sigma^\alpha^\gamma}\Big) \nonumber
   \\ & & -\eta_{\eps\gamma}\eta^{\beta\delta}
   \Big[-\partial_{x^\delta}\bar{h}^{\eps\sigma}\tensor{({\rm rem}_9)}{_\sigma^\alpha^\gamma}   
   -\partial_{x^\sigma}\bar{h}^{\alpha\gamma}\tensor{({\rm rem}_9)}{_\delta^\eps^\sigma}
   +\tensor{({\rm rem}_9)}{_\delta^\eps^\sigma}\tensor{({\rm rem}_9)}{_\sigma^\alpha^\gamma}\Big] \nonumber
   \\ & & +h_{\delta\eps}
   \,(\eta^{\gamma\sigma}+H^{\gamma\sigma})
   \Big(-\partial_{x^\gamma}\bar{h}^{\beta\delta}+\tensor{({\rm rem}_9)}{_\gamma^\beta^\delta}\Big)   
   \Big(-\partial_{x^\sigma}\bar{h}^{\alpha\eps}+\tensor{({\rm rem}_9)}{_\sigma^\alpha^\eps}\Big) \nonumber
   \\ & & +\eta_{\delta\eps} H^{\gamma\sigma}
   \Big(-\partial_{x^\gamma}\bar{h}^{\beta\delta}+\tensor{({\rm rem}_9)}{_\gamma^\beta^\delta}\Big)   
   \Big(-\partial_{x^\sigma}\bar{h}^{\alpha\eps}+\tensor{({\rm rem}_9)}{_\sigma^\alpha^\eps}\Big) \nonumber
   \\ & & +\eta_{\delta\eps}\eta^{\gamma\sigma}
   \Big[-\partial_{x^\gamma}\bar{h}^{\beta\delta}\tensor{({\rm rem}_9)}{_\sigma^\alpha^\eps}
   -\tensor{({\rm rem}_9)}{_\gamma^\beta^\delta}\partial_{x^\sigma}\bar{h}^{\alpha\eps}
   +\tensor{({\rm rem}_9)}{_\gamma^\beta^\delta}\tensor{({\rm rem}_9)}{_\sigma^\alpha^\eps}\Big] \nonumber
   \\ & & +\frac{1}{8}\,\Big(2H^{\alpha\delta}H^{\beta\gamma}
   -H^{\alpha\beta}H^{\delta\gamma}\Big)
   (2g_{\eps\sigma} g_{\kappa\nu}-g_{\sigma\kappa} g_{\eps\nu}) \nonumber
   \\ & & \hspace{3em}\times
   \Big(-\partial_{x^\delta}\bar{h}^{\eps\nu}+\tensor{({\rm rem}_9)}{_\delta^\eps^\nu}\Big)
   \Big(-\partial_{x^\gamma}\bar{h}^{\sigma\kappa}+\tensor{({\rm rem}_9)}
   {_\gamma^\sigma^\kappa}\Big) \nonumber
   \\ & & +\frac{1}{8}\,\Big(
   2\eta^{\alpha\delta}\eta^{\beta\gamma}
   -\eta^{\alpha\beta}\eta^{\delta\gamma}
   +2\eta^{\alpha\delta}H^{\beta\gamma}
   +2\eta^{\beta\gamma}H^{\alpha\delta}
   -\eta^{\alpha\beta}H^{\delta\gamma}
   -\eta^{\delta\gamma}H^{\alpha\beta}\Big) \nonumber
   \\ & & \hspace{3em}\times
   \Big(2h_{\eps\sigma}h_{\kappa\nu}-h_{\sigma\kappa}h_{\eps\nu}\Big)
   \Big(-\partial_{x^\delta}\bar{h}^{\eps\nu}+\tensor{({\rm rem}_9)}{_\delta^\eps^\nu}\Big)
   \Big(-\partial_{x^\gamma}\bar{h}^{\sigma\kappa}+\tensor{({\rm rem}_9)}
   {_\gamma^\sigma^\kappa}\Big) \nonumber
   \\ & & +\frac{1}{8}\,\Big(
   \Big[2\eta^{\alpha\delta}\eta^{\beta\gamma}-\eta^{\alpha\beta}\eta^{\delta\gamma}\Big]   
   \Big[2\eta_{\eps\sigma}h_{\kappa\nu}+2\eta_{\kappa\nu}h_{\eps\sigma}
   -\eta_{\sigma\kappa}h_{\eps\nu}-\eta_{\eps\nu} h_{\sigma\kappa}\Big] \nonumber
   \\ & & \hspace{3em} +\Big[2\eta^{\alpha\delta}H^{\beta\gamma}+2\eta^{\beta\gamma}H^{\alpha\delta}
   -\eta^{\alpha\beta}H^{\delta\gamma}-\eta^{\delta\gamma}H^{\alpha\beta}\Big]
   \Big[2\eta_{\eps\sigma}\eta_{\kappa\nu}-\eta_{\sigma\kappa}\eta_{\eps\nu}\Big] \nonumber
   \\ & & \hspace{3em} +\Big[2\eta^{\alpha\delta}H^{\beta\gamma}+2\eta^{\beta\gamma}H^{\alpha\delta}
   -\eta^{\alpha\beta}H^{\delta\gamma}-\eta^{\delta\gamma}H^{\alpha\beta}\Big]
   \Big[2\eta_{\eps\sigma}h_{\kappa\nu}+2\eta_{\kappa\nu}h_{\eps\sigma}
   -\eta_{\sigma\kappa}h_{\eps\nu}-\eta_{\eps\nu} h_{\sigma\kappa}\Big]\Big) \nonumber
   \\ & & \hspace{1em}\times
   \Big(-\partial_{x^\delta}\bar{h}^{\eps\nu}+\tensor{({\rm rem}_9)}{_\delta^\eps^\nu}\Big)
   \Big(-\partial_{x^\gamma}\bar{h}^{\sigma\kappa}+\tensor{({\rm rem}_9)}
   {_\gamma^\sigma^\kappa}\Big) \nonumber
   \\ & & +\frac{1}{8}
   \,(2\eta^{\alpha\delta}\eta^{\beta\gamma}-\eta^{\alpha\beta}\eta^{\delta\gamma})
   (2\eta_{\eps\sigma}\eta_{\kappa\nu}-\eta_{\sigma\kappa}\eta_{\eps\nu}) \nonumber
   \\ & & \hspace{1em}\times
   \Big[-\partial_{x^\delta}\bar{h}^{\eps\nu}\tensor{({\rm rem}_9)}{_\gamma^\sigma^\kappa}   
   -\partial_{x^\gamma}\bar{h}^{\sigma\kappa}\tensor{({\rm rem}_9)}{_\delta^\eps^\nu}
   +\tensor{({\rm rem}_9)}{_\delta^\eps^\nu}\tensor{({\rm rem}_9)}{_\gamma^\sigma^\kappa}\Big].    
   \label{eq_rem10}
\end{eqnarray} 
For mixed time/space indices this can be simplified as

\begin{eqnarray}\label{mtsi} 
   \frac{16\pi G}{c^4}\,|g|\,t^{0i} 
   & = & \frac{1}{2}\,\partial_{x^i}\bar{h}^{00}\,\partial_{x^0}\bar{h}^{00} 
   +\partial_{x^j}\bar{h}^{0i}\,\partial_{x^k}\bar{h}^{kj}
   -\partial_{x^0}\bar{h}^{00}\,\partial_{x^k}\bar{h}^{ik}
   -\partial_{x^j}\bar{h}^{0j}\,\partial_{x^k}\bar{h}^{ik}
   +\partial_{x^i}\bar{h}^{0j}\,\partial_{x^j}\bar{h}^{00}
   \nonumber 
   \\ & & +\sum_{j=1}^3\partial_{x^0}\bar{h}^{jk}\,\partial_{x^k}\bar{h}^{ij}
   -\sum_{j=1}^3\partial_{x^i}\bar{h}^{jk}\,\partial_{x^k}\bar{h}^{0j}
   -\sum_{j=1}^3\partial_{x^j}\bar{h}^{i0}\,\partial_{x^j}\bar{h}^{00}
   +\sum_{j, k=1}^3\partial_{x^k}\bar{h}^{ij}\,\partial_{x^k}\bar{h}^{0j}
   \nonumber 
   \\ & &    
   -\frac{1}{2}\sum_{j, k=1}^3\partial_{x^0}\bar{h}^{jk}\,\partial_{x^i}\bar{h}^{jk}
   +\frac{1}{4}\,\partial_{x^0}\bar{h}\,\partial_{x^i}\bar{h}
   +({\rm rem}_{10})^{0i}.   
\end{eqnarray}

\subsection{Moments} 
\label{moments_sect}

We define, following \cite[p.~223]{stephani}, the auxiliary functions 
\begin{eqnarray}
   m(t) & = & \int_{\R^3} T_{00}(t, \bar{x})\,d\bar{x},
   \quad d^\alpha(t)=\int_{\R^3} T_{00}(t, \bar{x})\,\bar{x}^\alpha\,d\bar{x},
   \quad d^{\alpha\beta}(t)=\int_{\R^3} T_{00}(t, \bar{x})
   \,\bar{x}^\alpha\bar{x}^\beta\,d\bar{x},
   \qquad\label{auxif1}  
   \\ -p_\nu(t) & = & \int_{\R^3} T_{0\nu}(t, \bar{x})\,d\bar{x},\quad
   \tensor{b}{_\nu^\alpha}(t)=\int_{\R^3} T_{0\nu}(t, \bar{x})\,\bar{x}^\alpha
   \,d\bar{x},\quad a_{\alpha\beta}(t)=\int_{\R^3} T_{\alpha\beta}(t, \bar{x})
   \,d\bar{x},
   \qquad\label{auxif2}   
\end{eqnarray}
as well as
\[ \tilde{d}^{\,\alpha\beta}(t)=\int_{\R^3} T^{00}(t, \bar{x})
   \,\bar{x}^\alpha\bar{x}^\beta\,d\bar{x},
   \quad\tilde{a}^{\,\alpha\beta}(t)=\int_{\R^3} T^{\alpha\beta}(t, \bar{x})\,d\bar{x}. \] 
The contracted Bianchi identities together with the Einstein equations 
imply $\nabla_\nu T^{\mu\nu}=0$, where $\nabla$ denotes the covariant derivative. From this identity follows
\[ 0=\nabla_\nu T^{\mu\nu}=\partial_{\nu} T^{\mu\nu}
   +\tensor{\Gamma}{^\mu_\nu_\lambda}\,T^{\lambda\nu}
   +\tensor{\Gamma}{^\nu_\nu_\lambda}\,T^{\mu\lambda}, \] 
and hence 
\begin{equation}\label{knu} 
   c^{-1}\partial_t T^{0\mu}=\partial_{x^0} T^{0\mu}
   =-\partial_{x^m} T^{m\mu}+({\rm rem}_{11})^\mu
\end{equation}          	
for 
\begin{equation} \label{eq_rem11}
    ({\rm rem}_{11})^\mu=-\tensor{\Gamma}{^\mu_\nu_\lambda}\,T^{\lambda\nu}
   -\tensor{\Gamma}{^\nu_\nu_\lambda}\,T^{\mu\lambda}.
\end{equation}
Now integration by parts gives
\begin{eqnarray}\label{wrpd}  
   c^{-1}\partial_t\int_{\R^3} T^{0i}\,{\bar x}^j\,d\bar{x}
   & = & \int_{\R^3}(\partial_{\bar{x}^0} T^{0i})\,\bar{x}^j\,d\bar{x}
   =\int_{\R^3}(-\partial_{\bar{x}^m} T^{mi}+({\rm rem}_{11})^i)\,\bar{x}^j\,d\bar{x}
   \nonumber
   \\ & = & \int_{\R^3} T^{ij}\,d\bar{x}+({\rm rem}_{12})^{ij}, 
\end{eqnarray}  
where 
\begin{equation} \label{eq_rem12}
    ({\rm rem}_{12})^{ij}=\int_{\R^3} ({\rm rem}_{11})^i\,\bar{x}^j\,d\bar{x}.
\end{equation}
With another application of (\ref{knu}), integration by parts 
and (\ref{wrpd}), we are led to
\begin{eqnarray}\label{grec} 
   \partial_t^2\tilde{d}^{\,ij}
   & = & c\partial_t\int_{\R^3} \partial_{\bar{x}^0} T^{00}
   \,\bar{x}^i\bar{x}^j\,d\bar{x}
   =c\partial_t\int_{\R^3} (-\partial_{\bar{x}^m} T^{m0})
   \,\bar{x}^i\bar{x}^j\,d\bar{x}+({\rm rem}_{13})^{ij}
   \nonumber
   \\ & = & c\partial_t\int_{\R^3} (T^{j0}\bar{x}^i+T^{i0}\bar{x}^j)\,d\bar{x}
   +({\rm rem}_{13})^{ij}
   =2c^2\tilde{a}^{\,ij}+({\rm rem}_{14})^{ij}, 
\end{eqnarray} 
denoting
\begin{equation} \label{eq_rem13}
   ({\rm rem}_{13})^{ij}
   =c\int_{\R^3} \partial_t ({\rm rem}_{11})^0\,\bar{x}^i\bar{x}^j\,d\bar{x},
   \quad ({\rm rem}_{14})^{ij}=c^2(({\rm rem}_{12})^{ij}+({\rm rem}_{12})^{ji})
   +({\rm rem}_{13})^{ij}.
\end{equation}
Hence 
\begin{eqnarray}\label{heli} 
   a_{ij} & = & \int_{\R^3} T_{ij}\,d\bar{x}
   =\int_{\R^3} g_{i\alpha} g_{j\beta} T^{\alpha\beta}\,d\bar{x}
   =\int_{\R^3} (\eta_{i\alpha}+h_{i\alpha})\,
   (\eta_{j\beta}+h_{j\beta})\,T^{\alpha\beta}\,d\bar{x}
   \nonumber
   \\ & = & \int_{\R^3} T^{ij}\,d\bar{x}+({\rm rem}_{15})_{ij}
   =\tilde{a}^{\,ij}+({\rm rem}_{15})_{ij}
   =\frac{1}{2c^2}\,\partial_t^2\tilde{d}^{\,ij}+({\rm rem}_{16})_{ij}
\end{eqnarray} 
for 
\begin{eqnarray} 
   ({\rm rem}_{15})_{ij}
   & = & \int_{\R^3} (\eta_{i\alpha} h_{j\beta}
   +h_{i\alpha}\eta_{j\beta}
   +h_{i\alpha} h_{j\beta})\,T^{\alpha\beta}\,d\bar{x}, 
   \label{eq_rem15}
   \\ ({\rm rem}_{16})_{ij} & = & -\frac{1}{2c^2}\,({\rm rem}_{14})^{ij}+({\rm rem}_{15})_{ij}.    
   \label{eq_rem16}
\end{eqnarray} 
Now 
\begin{eqnarray}\label{won} 
   \tilde{d}^{\,ij} & = & \int_{\R^3} T^{00}\bar{x}^i\bar{x}^j\,d\bar{x}
   =\int_{\R^3} g^{0\alpha} g^{0\beta} T_{\alpha\beta}\,\bar{x}^i\bar{x}^j\,d\bar{x}
   =\int_{\R^3} (\eta^{0\alpha}+H^{0\alpha})
   (\eta^{0\beta}+H^{0\beta})\,T_{\alpha\beta}\,\bar{x}^i\bar{x}^j\,d\bar{x}
   \nonumber
   \\ & = & d^{\,ij}+({\rm rem}_{17})^{ij}, 
\end{eqnarray} 
where 
\begin{equation}\label{eq_rem17}
    ({\rm rem}_{17})^{ij}
   =\int_{\R^3} (\eta^{0\alpha}H^{0\beta}+H^{0\alpha}\eta^{0\beta}
   +H^{0\alpha}H^{0\beta})\,T_{\alpha\beta}\,\bar{x}^i\bar{x}^j\,d\bar{x}.
\end{equation}
Going back to (\ref{heli}), we obtain 
\begin{equation}\label{bwh1} 
   a_{ij}=\frac{1}{2c^2}\,\ddot{d}^{\,ij}+({\rm rem}_{18})_{ij}
   \quad\mbox{for}\quad    
   ({\rm rem}_{18})_{ij}=\frac{1}{2c^2}\,\partial_t^2 ({\rm rem}_{17})^{ij}
   +({\rm rem}_{16})_{ij}.
\end{equation}  
Also by (\ref{knu}) and Gauss's Theorem
\begin{equation}\label{fas} 
   \partial_t\int_{\R^3} T^{0\nu}\,d\bar{x}
   =c\int_{\R^3}\partial_{\bar{x}^0} T^{0\nu}\,d\bar{x}
   =c\int_{\R^3} (-\partial_{x^m} T^{m\nu}+({\rm rem}_{11})^\nu)\,d\bar{x}
   =({\rm rem}_{19})^\nu,
\end{equation}  
defining 
\begin{equation}\label{eq_rem19}
    ({\rm rem}_{19})^\nu=c\int_{\R^3} ({\rm rem}_{11})^{\nu}\,d\bar{x}
\end{equation}
According to (\ref{heli}) we have 
\[ \int_{\R^3} T_{0\nu}\,d\bar{x}=a_{0\nu}=\tilde{a}^{\,0\nu}+({\rm rem}_{15})_{0\nu}
   =\int_{\R^3} T^{0\nu}\,d\bar{x}+({\rm rem}_{15})_{0\nu}, \] 
so that in particular
\begin{equation}\label{bwh2} 
   \dot{m}=\partial_t\int_{\R^3} T_{00}\,d\bar{x}={\rm rem}_{20}
\end{equation}  
for 
\begin{equation}\label{eq_rem20}
    {\rm rem}_{20}=({\rm rem}_{19})^0+\partial_t ({\rm rem}_{15})_{00}, 
\end{equation}
and in addition 
\begin{equation}\label{bwh3}
   \dot{p}_j=-\partial_t \int_{\R^3} T_{0j}\,d\bar{x}=({\rm rem}_{21})^j
\end{equation} 
for 
\begin{equation}\label{eq_rem21}
    ({\rm rem}_{21})^j=-({\rm rem}_{19})^j-\partial_t ({\rm rem}_{15})_{0j}. 
\end{equation}
The next terms to consider are the $d^{\,i}$. 
First we deduce from (\ref{wrpd}) and (\ref{fas}) that 
\[ \partial_t^2\int_{\R^3} T^{00}\,{\bar x}^i\,d\bar{x}
   =c\partial_t\int_{\R^3} T^{0i}\,d\bar{x}+c\partial_t ({\rm rem}_{12})^{0i}
   =({\rm rem}_{22})^i, \]
where 
\begin{equation} \label{eq_rem22}
    ({\rm rem}_{22})^i=c({\rm rem}_{19})^i+c\partial_t ({\rm rem}_{12})^{0i}. 
\end{equation}
Also 
\begin{eqnarray*} 
   d^{\,i} & = & \int_{\R^3} T_{00}\,{\bar x}^i\,d\bar{x}
   =\int_{\R^3} g_{0\alpha} g_{0\beta}\,T^{\alpha\beta}\,{\bar x}^i\,d\bar{x}
   =\int_{\R^3} (\eta_{0\alpha}+h_{0\alpha})
   (\eta_{0\beta}+h_{0\beta})\,T^{\alpha\beta}\,{\bar x}^i\,d\bar{x} 
   \\ & = & \int_{\R^3} T^{00}\,{\bar x}^i\,d\bar{x} 
   +({\rm rem}_{23})^i
\end{eqnarray*}    
for
\begin{equation}\label{eq_rem23}
    ({\rm rem}_{23})^i=\int_{\R^3} (\eta_{0\alpha} h_{0\beta}
    +h_{0\alpha}\eta_{0\beta}+h_{0\alpha}h_{0\beta})
   \,T^{\alpha\beta}\,{\bar x}^i\,d\bar{x},
\end{equation}
and accordingly 
\begin{equation}\label{bwh4}
   \ddot{d}^{\,i}=({\rm rem}_{24})^i, 
\end{equation} 
where 
\begin{equation}\label{eq_rem24}
    ({\rm rem}_{24})^i=({\rm rem}_{22})^i+\partial_t^2 ({\rm rem}_{23})^i.     
\end{equation}
Finally, from (\ref{wrpd}) we infer that 
\begin{equation}\label{blom} 
   \partial_t\int_{\R^3} (T^{0m}\,{\bar x}^j-T^{0j}\,{\bar x}^m)\,d\bar{x}   
   =c({\rm rem}_{12})^{mj}-c({\rm rem}_{12})^{jm}.  
\end{equation} 
As in (\ref{grec}),  
\[ c^{-1}\partial_t\,\tilde{d}^{\,jm}
   =c^{-1}\partial_t\int_{\R^3} T^{00}\,\bar{x}^j\,\bar{x}^m\,d\bar{x}
   =\int_{\R^3} (T^{m0}\,\bar{x}^j+T^{j0}\,\bar{x}^m)\,d\bar{x}
   +\int_{\R^3} ({\rm rem}_{11})^0\,\bar{x}^j\,\bar{x}^m\,d\bar{x}, \] 
and this together with (\ref{won}) leads to 
\begin{eqnarray*}
   c^{-1}\dot{d}^{\,jm}=c^{-1}\partial_t\,(\tilde{d}^{\,jm}-({\rm rem}_{17})^{jm})
   =\int_{\R^3} (T^{m0}\,\bar{x}^j+T^{j0}\,\bar{x}^m)\,d\bar{x}
   +({\rm rem}_{25})^{jm}, 
\end{eqnarray*}  
for 
\begin{equation}\label{eq_rem25}
    ({\rm rem}_{25})^{jm}
   =\int_{\R^3} ({\rm rem}_{11})^0\,\bar{x}^j\,\bar{x}^m\,d\bar{x}
   -c^{-1}\partial_t\,({\rm rem}_{17})^{jm}.
\end{equation}
Hence (\ref{blom}) implies 
\begin{eqnarray}\label{alkim} 
   c^{-1}\ddot{d}^{\,jm}
   & = & \partial_t\int_{\R^3} (T^{m0}\,\bar{x}^j+T^{j0}\,\bar{x}^m)\,d\bar{x}
   +\partial_t ({\rm rem}_{25})^{jm}
   \nonumber 
   \\ & = & 2\partial_t\int_{\R^3} T^{0j}\,{\bar x}^m\,d\bar{x}  
   +({\rm rem}_{26})^{jm}, 
\end{eqnarray} 
defining 
\begin{equation}\label{eq_rem26}
    ({\rm rem}_{26})^{jm}=c({\rm rem}_{12})^{mj}-c({\rm rem}_{12})^{jm}
   +\partial_t ({\rm rem}_{25})^{jm}.
\end{equation}
Now 
\begin{eqnarray*} 
   \tensor{b}{_j^m}
   & = & \int_{\R^3} T_{0j}\,\bar{x}^m\,d\bar{x}
   =\int_{\R^3} g_{\alpha 0}\,g_{\beta j}\,T^{\alpha\beta}
   \,\bar{x}^m\,d\bar{x}
   =\int_{\R^3} (\eta_{\alpha 0}+h_{\alpha 0})\,(\eta_{\beta j}+h_{\beta j})\,T^{\alpha\beta}
   \,\bar{x}^m\,d\bar{x}
   \\ & = & -\int_{\R^3} T^{0j}\,\bar{x}^m\,d\bar{x}
   +\tensor{({\rm rem}_{27})}{_j^m}, 
\end{eqnarray*} 
where 
\begin{equation}\label{eq_rem27}
   \tensor{({\rm rem}_{27})}{_j^m}=\int_{\R^3} (\eta_{\alpha 0} h_{\beta j}
   +h_{\alpha 0}\eta_{\beta j}+h_{\alpha 0} h_{\beta j})
   \,T^{\alpha\beta}\,\bar{x}^m\,d\bar{x}.
\end{equation}
Thus we get from (\ref{alkim}) that 
\[ \frac{1}{2c}\,\ddot{d}^{\,jm}
   =-\tensor{\dot{b}}{_j^m}
   +\partial_t\tensor{({\rm rem}_{27})}{_j^m}  
   +\frac{1}{2}\,({\rm rem}_{26})^{jm}, \] 
and in particular 
\begin{equation}\label{bwh5} 
   \frac{1}{2c}\,\dddot{d}^{\,jm}
   =-\tensor{\ddot{b}}{_j^m}
   +({\rm rem}_{28})^{jm}
\end{equation}  
for 
\begin{equation}\label{eq_rem28} 
    ({\rm rem}_{28})^{jm}=\partial_t^2\tensor{({\rm rem}_{27})}{_j^m}  
   +\frac{1}{2}\,\partial_t ({\rm rem}_{26})^{jm}. 
\end{equation}

\subsection{Derivatives of weak-field coefficients} 
\label{eotm_sect}

In order to evaluate the expression (\ref{mtsi}) we will need the derivatives 
$\partial_{x^\gamma}\bar{h}^{\alpha\beta}$. Hence our next step is 
to solve for $\bar{h}_{\alpha\beta}$ from the Einstein equations (\ref{einst}), 
which, owing to (\ref{tfte}) and (\ref{barhdef}), are written 
in the Gaussian coordinates ${(x^\mu)}_{\mu=0, 1, 2, 3}$ as
\begin{equation}\label{jret} 
   -\frac{1}{2}\,\Box\,\bar{h}_{\alpha\beta}
   +{({\rm rem}_4)}_{\alpha\beta}=G_{\alpha\beta}=\frac{8\pi G}{c^4}\,T_{\alpha\beta}
\end{equation}
for $\Box=\eta^{\alpha\beta}\,\partial_{x^\alpha x^\beta}^2$. Identifying 
$T_{\alpha\beta}(x^\mu)$ with $T_{\alpha\beta}(t, x^a)$ [cf.~the remark concerning 
(\ref{tab3b})] and $\bar{h}_{\alpha\beta}(x^\mu)$ with $\bar{h}_{\alpha\beta}(t, x^a)$, they turn into
\begin{equation}\label{jret2} 
   -\frac{1}{2}\,\Big(c^{-2}\partial_t^2+\sum_{a=1}^3\partial_{x^a}^2\Big)
   \bar{h}_{\alpha\beta}(t, x)
   +{({\rm rem}_4)}_{\alpha\beta}(t, x)
   =\frac{8\pi G}{c^4}\,T_{\alpha\beta}(t, x). 
\end{equation}
After moving the $\rem{4}$ term to the right, the solution to the wave equation~\eqref{jret2} 
that reflects a situation where no radiation comes in from spatial infinity is found 
by a spatial integral involving the source evaluated at the \textit{retarded time}, 
that is, 
\begin{eqnarray}\label{eq_noinc_radiation}
   \bar{h}_{\alpha\beta}(t, x)
   & = & \frac{4G}{c^4}\int_{\R^3}\frac{1}{|x-\bar{x}|}
   \,T_{\alpha\beta}(t-c^{-1}|x-\bar{x}|, \bar{x})\,d\bar{x}
   \nonumber
   \\ & & -\frac{1}{2\pi}\,\int_{\R^3}\frac{1}{|x-\bar{x}|}
   {({\rm rem}_4)}_{\alpha\beta}(t-c^{-1}|x-\bar{x}|, \bar{x})\,d\bar{x}. 
\end{eqnarray}
The integral containing the matter terms $T_{\alpha\beta}$ in (\ref{eq_noinc_radiation}) 
can be reduced to the bounded domain $\{x\in\bbR^3: \ |x|\leq\lambda K_\ast\}$ 
due to the compact-support condition~\eqref{CS_cond_T} on $T_{\alpha\beta}$. 
Similarly, other $d\bar{x}$-integrals of matter terms that appear in what follows 
only need to be performed over a bounded ball. However, the finiteness 
of the second integral in (\ref{eq_noinc_radiation}), namely
\begin{equation}\label{eq_rem29}
    {({\rm rem}_{29})}_{\alpha\beta}(t, x)   
   =-\frac{1}{2\pi}\,\int_{\R^3}\frac{1}{|x-\bar{x}|}
   {({\rm rem}_4)}_{\alpha\beta}(t-c^{-1}|x-\bar{x}|, \bar{x})\,d\bar{x},
\end{equation}
depends crucially on the NIR condition (\ref{Blanch}), reminiscent of Blanchet's 
no-incoming-radiation condition. In general~\eqref{eq_rem29} will be infinite 
(cf.~\cite[p.~4814]{WiWi} for instance, and also Section \ref{Blanch_discuss}), 
and several ways to remedy this problem 
have been suggested, like the tail-integral approach or the Will-Wiseman 
formalism \cite{WiWi,PoWi}. However, we found condition (\ref{Blanch}) 
to be the most analytically tractable. One may consult Appendix~\ref{Blanch_discuss} 
for a more in-depth discussion, and also Subsection~\ref{subsubsec_rem29}, 
where we will in particular see that $\rem{29}$ is finite.

It follows that
\begin{equation} \label{h_T_rem29}
    \bar{h}_{\alpha\beta}(t, x)
   =\frac{4G}{c^4}\int_{|\bar{x}|\le\lambda K_\ast}\frac{1}{|x-\bar{x}|}
   \,T_{\alpha\beta}(t-c^{-1}|x-\bar{x}|, \bar{x})\,d\bar{x}
   +{({\rm rem}_{29})}_{\alpha\beta}(t, x).
\end{equation}
We now find the derivatives of $\bar{h}^{\alpha\beta}$. To begin with, from (\ref{ingem}) we obtain 
\begin{eqnarray*} 
   \bar{h}^{\alpha\beta}(t, x)
   & = & \eta^{\alpha\alpha'}\eta^{\beta\beta'}\,\bar{h}_{\alpha'\beta'}(t, x)
   \\ & = & \frac{4G}{c^4}\int_{|\bar{x}|\le\lambda K_\ast}\frac{1}{|x-\bar{x}|}
   \,\hat{T}^{\alpha\beta}(t-c^{-1}|x-\bar{x}|, \bar{x})\,d\bar{x}
   +{({\rm rem}_{30})}^{\alpha\beta}(t, x)
\end{eqnarray*}  
for
\begin{equation}\label{That}
    \hat{T}^{\alpha\beta}=\eta^{\alpha\alpha'}\eta^{\beta\beta'}\,T_{\alpha'\beta'}
   \quad\mbox{and}\quad {({\rm rem}_{30})}^{\alpha\beta}
   =\eta^{\alpha\alpha'}\eta^{\beta\beta'} {({\rm rem}_{29})}_{\alpha'\beta'}.
\end{equation}
Therefore 
\begin{eqnarray} 
   \partial_{x^0}\bar{h}^{\alpha\beta}(t, x)
   & = & c^{-1}\partial_t\bar{h}^{\alpha\beta}(t, x)
   \nonumber
   \\ & = & \frac{4G}{c^5}\int_{|\bar{x}|\le\lambda K_\ast}\frac{1}{|x-\bar{x}|}
   \,\partial_t\hat{T}^{\alpha\beta}(t-c^{-1}|x-\bar{x}|, \bar{x})\,d\bar{x}
   +{({\rm rem}_{31})}^{\alpha\beta}(t, x),
   \qquad\label{noett1}
   \\ \partial_{x^k}\bar{h}^{\alpha\beta}(t, x)
   & = & -\frac{4G}{c^5}\int_{|\bar{x}|\le\lambda K_\ast}\frac{x^k}{|x-\bar{x}|^2}
   \,\partial_t\hat{T}^{\alpha\beta}(t-c^{-1}|x-\bar{x}|, \bar{x})\,d\bar{x}
   +\tensor{({\rm rem}_{32})}{^\alpha^\beta^k}(t, x),
   \qquad\label{noett2} 
\end{eqnarray} 
defining 
\begin{eqnarray} 
   {({\rm rem}_{31})}^{\alpha\beta}(t, x) 
   & = & c^{-1}\partial_t {({\rm rem}_{30})}^{\alpha\beta}(t, x), \label{eq_rem31}
   \\ \tensor{({\rm rem}_{32})}{^\alpha^\beta^k}(t, x) & = &
   -\frac{4G}{c^4}\int_{|\bar{x}|\le\lambda K_\ast}
   \frac{x^k}{|x-\bar{x}|^3}
   \,\hat{T}^{\alpha\beta}(t-c^{-1}|x-\bar{x}|, \bar{x})\,d\bar{x}
   +\partial_{x^k}{({\rm rem}_{30})}^{\alpha\beta}(t, x).\qquad \label{eq_rem32}
\end{eqnarray} 
Next we are going to use Lemma \ref{mupol}(b) and (c) 
from Appendix~\ref{sec_app_technical} to expand the right-hand sides 
of (\ref{noett1}) and (\ref{noett2}), letting $r_\ast=\lambda K_\ast$ in the notation 
of that Lemma. It follows that, for $|u|\le\lambda c^{-1}$ and $|x|\ge 5\lambda K_\ast$,
\begin{eqnarray*} 
   \partial_{x^0}\bar{h}^{\alpha\beta}(t, x)
   & = & \frac{4G}{c^5}\int_{|\bar{x}|\le\lambda K_\ast}
   \,\bigg[\frac{1}{r}\,\partial_t\hat{T}^{\alpha\beta}(t-c^{-1}r, \bar{x})
   +\frac{(x\cdot\bar{x})}{cr^2}\,\partial^2_t\hat{T}^{\alpha\beta}(t-c^{-1}r, \bar{x})
   \\ & & \hspace{6em} 
   +\frac{(x\cdot\bar{x})^2}{2c^2r^3}\,\partial^3_t\hat{T}^{\alpha\beta}(t-c^{-1}r, \bar{x})
   \bigg]\,d\bar{x}+{({\rm rem}_{33})}^{\alpha\beta}(t, x)
\end{eqnarray*} 
for 
\begin{equation}\label{eq_rem33}
    {({\rm rem}_{33})}^{\alpha\beta}(t, x)
   ={({\rm rem}_{31})}^{\alpha\beta}(t, x)
   +{\cal O}\Big(c^{-5}r^{-2}\sum_{j=0}^2 c^{-j}{\|\partial_t^{j+1}\hat{T}^{\alpha\beta}\|}_\infty
   +c^{-8}r^{-1} {\|\partial_t^4\hat{T}^{\alpha\beta}\|}_\infty\Big) 
\end{equation}
and $r=|x|$. Similarly, 
\begin{eqnarray*} 
   \partial_{x^k}\bar{h}^{\alpha\beta}(t, x)
   & = & -\frac{4G}{c^5}\int_{|\bar{x}|\le\lambda K_\ast}
   \bigg[\frac{x^k}{r^2}\,\partial_t\hat{T}^{\alpha\beta}(t-c^{-1}r, \bar{x})
   +x^k\,\frac{(x\cdot\bar{x})}{cr^3}
   \,\partial^2_t\hat{T}^{\alpha\beta}(t-c^{-1}r, \bar{x})
   \\ & & \hspace{7em} +\,x^k\,\frac{(x\cdot\bar{x})^2}{2c^2r^4}
   \,\partial^3_t\hat{T}^{\alpha\beta}(t-c^{-1}r, \bar{x})
   \bigg]\,d\bar{x}
   +\tensor{({\rm rem}_{34})}{^\alpha^\beta^k}(t, x)
\end{eqnarray*} 
for 
\begin{equation}\label{eq_rem34}
    \tensor{({\rm rem}_{34})}{^\alpha^\beta^k}(t, x)
   =\tensor{({\rm rem}_{32})}{^\alpha^\beta^k}(t, x)
   +{\cal O}\Big(c^{-5}r^{-2}\sum_{j=0}^2 c^{-j}{\|\partial_t^{j+1}\hat{T}^{\alpha\beta}\|}_\infty
   +c^{-8}r^{-1} {\|\partial_t^4\hat{T}^{\alpha\beta}\|}_\infty\Big). 
\end{equation}
To further simplify the right-hand sides of the $\partial_{x^\gamma}\bar{h}^{\alpha\beta}$, 
recall the auxiliary functions that have been defined in (\ref{auxif1}) and (\ref{auxif2}). 
Since $\hat{T}^{00}=T_{00}$, it follows that  
\begin{equation}\label{djur1} 
   \partial_{x^0}\bar{h}^{00}(t, x)
   =\frac{4G}{c^5}\bigg[\frac{1}{r}\,\dot{m}(u)
   +\frac{1}{cr^2}\,\sum_{i=1}^3 x^i \ddot{d}^{\,i}(u)
   +\frac{1}{2c^2r^3}\,\sum_{i, j=1}^3 x^i x^j\dddot{d}^{ij}(u)\bigg] 
   +{({\rm rem}_{33})}^{00}(t, x),
\end{equation}   
where $u=t-c^{-1}r$ as before. Also $\hat{T}^{ij}=T_{ij}$ implies that
\begin{equation}\label{djur2}  
   \partial_{x^0}\bar{h}^{ij}(t, x)
   =\frac{4G}{c^5}\,\frac{1}{r}\,\dot{a}_{ij}(u)
   +{({\rm rem}_{35})}^{ij}(t, x)
\end{equation} 
for 
\begin{equation}\label{eq_rem35}
   {({\rm rem}_{35})}^{ij}(t, x)
   ={({\rm rem}_{33})}^{ij}(t, x)
   +\frac{4G}{c^5}\int_{|\bar{x}|\le\lambda K_\ast}
   \,\bigg[\frac{(x\cdot\bar{x})}{cr^2}\,\partial^2_t T_{ij}(u, \bar{x})
   +\frac{(x\cdot\bar{x})^2}{2c^2r^3}\,\partial^3_t T_{ij}(u, \bar{x})
   \bigg]\,d\bar{x}. 
\end{equation}
Furthermore, 
\begin{equation}\label{djur3} 
   \partial_{x^k}\bar{h}^{00}(t, x)
   =-\frac{4G}{c^5}\bigg[\frac{x^k}{r^2}\,\dot{m}(u)
   +\frac{x^k}{cr^3}\,\sum_{i=1}^3 x^i\ddot{d}^{\,i}(u)
   +\frac{x^k}{2c^2r^4}\,\sum_{i, j=1}^3 x^i x^j\dddot{d}^{ij}(u)\bigg]
   +\tensor{({\rm rem}_{34})}{^0^0^k}(t, x).
\end{equation}  
Noting $\hat{T}^{0j}=-T_{0j}$, we moreover have 
\begin{equation}\label{djur4} 
   \partial_{x^k}\bar{h}^{0j}(t, x)
   =\frac{4G}{c^5}\,\bigg[-\frac{x^k}{r^2}\,\dot{p}_j(u)
   +\frac{x^k}{cr^3}\,\sum_{m=1}^3 x^m\tensor{\ddot{b}}{_j^m}(u)\bigg]	
   +\tensor{({\rm rem}_{36})}{^0^j^k}(t, x)
\end{equation}  
for 
\begin{equation}\label{eq_rem36}
    \tensor{({\rm rem}_{36})}{^0^j^k}(t, x)
   =\tensor{({\rm rem}_{34})}{^0^j^k}(t, x)	
   +\frac{4G}{c^5}\int_{|\bar{x}|\le\lambda K_\ast}
   x^k\,\frac{(x\cdot\bar{x})^2}{2c^2r^4}
   \,\partial^3_t T_{0j}(u, \bar{x})\,d\bar{x}. 
\end{equation}
Lastly, 
\begin{equation}\label{djur5} 
   \partial_{x^k}\bar{h}^{ij}(t, x)
   =-\frac{4G}{c^5}\,\frac{x^k}{r^2}\,\dot{a}_{ij}(u) 
   +\tensor{({\rm rem}_{37})}{^i^j^k}(t, x),
\end{equation}  
defining 
\begin{equation}\label{eq_rem37}
    \tensor{({\rm rem}_{37})}{^i^j^k}(t, x)
   =\tensor{({\rm rem}_{34})}{^i^j^k}(t, x)
   -\frac{4G}{c^5}\int_{|\bar{x}|\le\lambda K_\ast}
   \bigg[x^k\,\frac{(x\cdot\bar{x})}{cr^3}
   \,\partial^2_t T_{ij}(u, \bar{x})
   +\,x^k\,\frac{(x\cdot\bar{x})^2}{2c^2r^4}
   \,\partial^3_t T_{ij}(u, \bar{x})
   \bigg]\,d\bar{x}. 
\end{equation}
If we also invoke (\ref{bwh1}), (\ref{bwh2}), (\ref{bwh3}), (\ref{bwh4}) 
and (\ref{bwh5}), then we can summarize (\ref{djur1})--(\ref{djur5}) as 
\begin{eqnarray}
   \partial_{x^0}\bar{h}^{00}(t, x)
   & = & \frac{4G}{c^5}\,\frac{1}{2c^2r^3}\,\sum_{i, j=1}^3 x^i x^j\dddot{d}^{\,ij}(u) 
   +{({\rm rem}_{38})}^{00}(t, x), 
   \label{mcla1} 
   \\[1ex] \partial_{x^0}\bar{h}^{ij}(t, x)
   & = & \frac{4G}{c^5}\,\frac{1}{2c^2r}\,\dddot{d}^{\,ij}(u)
   +{({\rm rem}_{39})}^{ij}(t, x),  
   \label{mcla2} 
   \\[1ex] \partial_{x^k}\bar{h}^{00}(t, x)
   & = & -\frac{4G}{c^5}\,\,\frac{x^k}{2c^2r^4}\,\sum_{i, j=1}^3 x^i x^j\dddot{d}^{\,ij}(u)
   +\tensor{({\rm rem}_{40})}{^0^0^k}(t, x), 
   \label{mcla3} 
   \\[1ex] 
   \partial_{x^k}\bar{h}^{0j}(t, x)
   & = & -\frac{4G}{c^5}\,\,\frac{x^k}{2c^2r^3}\,\sum_{m=1}^3 x^m\dddot{d}^{\,jm}(u)	
   +\tensor{({\rm rem}_{41})}{^0^j^k}(t, x), 
   \label{mcla4} 
   \\[1ex] 
   \partial_{x^k}\bar{h}^{ij}(t, x)
   & = & -\frac{4G}{c^5}\,\,\frac{x^k}{2c^2r^2}\,\dddot{d}^{\,ij}(u) 
   +\tensor{({\rm rem}_{42})}{^i^j^k}(t, x), 
   \label{mcla5} 
\end{eqnarray} 
for 
\begin{eqnarray}
   {({\rm rem}_{38})}^{00}(t, x)
   & = & \frac{4G}{c^5}\,\bigg[\frac{1}{r}\,{\rm rem}_{20}(u)
   +\frac{1}{cr^2}\,\sum_{i=1}^3 x^i ({\rm rem}_{24})^i(u)\bigg] 
   +{({\rm rem}_{33})}^{00}(t, x),  \label{eq_rem38}
   \\ {({\rm rem}_{39})}^{ij}(t, x)
   & = & \frac{4G}{c^5}\,\frac{1}{r}\,\partial_t ({\rm rem}_{18})_{ij}(u)
   +{({\rm rem}_{35})}^{ij}(t, x), \label{eq_rem39}
   \\ \tensor{({\rm rem}_{40})}{^0^0^k}(t, x)
   & = & -\frac{4G}{c^5}\,\bigg[\frac{x^k}{r^2}\,{\rm rem}_{20}(u)
   +\frac{x^k}{cr^3}\,\sum_{i=1}^3 x^i({\rm rem}_{24})^i(u)\bigg]
   +\tensor{({\rm rem}_{34})}{^0^0^k}(t, x), \label{eq_rem40}
   \\ \tensor{({\rm rem}_{41})}{^0^j^k}(t, x)
   & = & \frac{4G}{c^5}\,\bigg[-\frac{x^k}{r^2}\,({\rm rem}_{21})^j(u)
   +\frac{x^k}{cr^3}\,\sum_{m=1}^3 x^m ({\rm rem}_{28})^{jm}(u)\bigg]	
   +\tensor{({\rm rem}_{36})}{^0^j^k}(t, x),\qquad\label{eq_rem41}
   \\ \tensor{({\rm rem}_{42})}{^i^j^k}(t, x)
   & = & -\frac{4G}{c^5}\,\frac{x^k}{r^2}\,\partial_t ({\rm rem}_{18})_{ij}(u)
   +\tensor{({\rm rem}_{37})}{^i^j^k}(t, x), \label{eq_rem42}
\end{eqnarray} 
where $r=|x|$ and $u=t-c^{-1}r$. In terms of  
\[ D^{\alpha\beta}(u)
   =\frac{1}{c^2}\int_{\R^3} T_{00}(u, \bar{x})\,\bar{x}^\alpha
   \,\bar{x}^\beta\,d\bar{x}
   =\frac{1}{c^2}\,d^{\alpha\beta}(u) \] 
from (\ref{ito}), the expansions (\ref{mcla1})--(\ref{mcla5}) read as 
\begin{eqnarray}
   \partial_{x^0}\bar{h}^{00}(t, x)
   & = & \frac{2G}{c^5 r^3}\,\sum_{i, j=1}^3 x^i x^j\dddot{D}^{\,ij}(u) 
   +{({\rm rem}_{38})}^{00}(t, x), 
   \label{haste1} 
   \\[1ex] \partial_{x^0}\bar{h}^{ij}(t, x)
   & = & \frac{2G}{c^5 r}\,\dddot{D}^{\,ij}(u)
   +{({\rm rem}_{39})}^{ij}(t, x), 
   \label{haste2} 
   \\[1ex] \partial_{x^k}\bar{h}^{00}(t, x)
   & = & -\frac{2Gx^k}{c^5 r^4}\,\sum_{i, j=1}^3 x^i x^j\dddot{D}^{\,ij}(u)
   +\tensor{({\rm rem}_{40})}{^0^0^k}(t, x), 
   \label{haste3} 
   \\[1ex] 
   \partial_{x^k}\bar{h}^{0j}(t, x)
   & = & -\frac{2Gx^k}{c^5 r^3}\,\sum_{m=1}^3 x^m\dddot{D}^{\,jm}(u)	
   +\tensor{({\rm rem}_{41})}{^0^j^k}(t, x), 
   \label{haste4} 
   \\[1ex] 
   \partial_{x^k}\bar{h}^{ij}(t, x)
   & = & -\frac{2Gx^k}{c^5 r^2}\,\dddot{D}^{\,ij}(u) 
   +\tensor{({\rm rem}_{42})}{^i^j^k}(t, x). 
   \label{haste5} 
\end{eqnarray}


\subsection{Energy-rate formula}
\label{expansion_of_energy_flux_sect}

By (\ref{barhh}) and (\ref{ingem}), 
\[ \bar{h}=\eta^{\alpha\beta}\bar{h}_{\alpha\beta}
   =-\bar{h}_{00}+\bar{h}_{11}+\bar{h}_{22}+\bar{h}_{33}
   =-\bar{h}^{00}+\bar{h}^{11}+\bar{h}^{22}+\bar{h}^{33}. \] 
In particular, denoting 
\[ \tensor{\dddot{D}}{_j^j}(u)=\sum_{j=1}^3\dddot{D}^{\,jj}(u), \] 
we get from (\ref{haste1})--(\ref{haste5}) 
\[ \partial_{x^0}\bar{h}
   =-\frac{2G}{c^5 r^3}\,\sum_{i, j=1}^3 x^i x^j\dddot{D}^{\,ij}
   +\frac{2G}{c^5 r}\,\tensor{\dddot{D}}{_j^j}
   +({\rm rem}_{43}), \] 
where
\begin{equation}\label{eq_rem43}
    ({\rm rem}_{43})=-{({\rm rem}_{38})}^{00}
   +\sum_{j=1}^3 {({\rm rem}_{39})}^{jj} \ ,
\end{equation}
and in addition 
\[ \partial_{x^i}\bar{h}
   =\frac{2Gx^i}{c^5 r^4}\,\sum_{m, n=1}^3 x^m x^n\dddot{D}^{\,mn}
   -\frac{2Gx^i}{c^5 r^2}\,\tensor{\dddot{D}}{_j^j}
   +({\rm rem}_{44})^i \]  
for
\begin{equation}\label{eq_rem44}
    ({\rm rem}_{44})^i=-\tensor{({\rm rem}_{40})}{^0^0^i} 
   +\sum_{j=1}^3\tensor{({\rm rem}_{42})}{^j^j^i}.
\end{equation}
Hence also 
\[ \sum_{i=1}^3\frac{x^i}{r}\,\partial_{x^i}\bar{h}
   =\frac{2G}{c^5 r^3}\,\sum_{m, n=1}^3 x^m x^n\dddot{D}^{\,mn}
   -\frac{2G}{c^5 r}\,\tensor{\dddot{D}}{_j^j}
   +({\rm rem}_{45}), \]    
defining
\begin{equation}\label{eq_rem45}
    ({\rm rem}_{45})=\sum_{i=1}^3 \frac{x_i}{r}\,({\rm rem}_{44})^i.
\end{equation}
It follows that 
\begin{eqnarray}\label{barbarhh} 
   \partial_{x^0}\bar{h}\,\sum_{i=1}^3\frac{x^i}{r}\,\partial_{x^i}\bar{h}
   & = & \Big(-\frac{2G}{c^5 r^3}\,\sum_{i, j=1}^3 x^i x^j\dddot{D}^{\,ij}
   +\frac{2G}{c^5 r}\,\tensor{\dddot{D}}{_j^j}
   +({\rm rem}_{43})\Big)
   \nonumber
   \\ & & \times\Big(\frac{2G}{c^5 r^3}\,\sum_{m, n=1}^3 x^m x^n\dddot{D}^{\,mn}
   -\frac{2G}{c^5 r}\,\tensor{\dddot{D}}{_j^j}
   +({\rm rem}_{45})\Big)
   \nonumber
   \\ & = &  -\frac{4G^2}{c^{10} r^6}\,\sum_{i, j, m, n=1}^3 x^i x^j x^m x^n
   \,\dddot{D}^{\,ij}\,\dddot{D}^{\,mn}
   +\frac{8G^2}{c^{10} r^4}\,\tensor{\dddot{D}}{_k^k}\,\sum_{i, j=1}^3 x^i x^j\dddot{D}^{\,ij}
   -\frac{4G^2}{c^{10} r^2}\,(\tensor{\dddot{D}}{_j^j})^2
   \nonumber
   \\ & & +({\rm rem}_{46}), 
\end{eqnarray} 
for 
\begin{eqnarray}\label{eq_rem46} 
   ({\rm rem}_{46}) 
   & = & -\frac{2G}{c^5 r^3}\,\sum_{i, j=1}^3 x^i x^j\dddot{D}^{\,ij}({\rm rem}_{45})
   +\frac{2G}{c^5 r}\,\tensor{\dddot{D}}{_j^j}({\rm rem}_{45}) \nonumber
   \\ & & +\,({\rm rem}_{43})\Big(\frac{2G}{c^5 r^3}\,\sum_{m, n=1}^3 x^m x^n\dddot{D}^{\,mn}
   -\frac{2G}{c^5 r}\,\tensor{\dddot{D}}{_j^j}
   +({\rm rem}_{45})\Big). 
\end{eqnarray} 
From (\ref{mtsi}) we recall that 
\begin{eqnarray*} 
   \frac{16\pi G}{c^4}\,|g|\,t^{0i} 
   & = & \frac{1}{2}\,\partial_{x^i}\bar{h}^{00}\,\partial_{x^0}\bar{h}^{00} 
   +\partial_{x^j}\bar{h}^{0i}\,\partial_{x^k}\bar{h}^{kj}
   -\partial_{x^0}\bar{h}^{00}\,\partial_{x^k}\bar{h}^{ik}
   -\partial_{x^j}\bar{h}^{0j}\,\partial_{x^k}\bar{h}^{ik}
   +\partial_{x^i}\bar{h}^{0j}\,\partial_{x^j}\bar{h}^{00}
   \\ & & +\sum_{j=1}^3\partial_{x^0}\bar{h}^{jk}\,\partial_{x^k}\bar{h}^{ij}
   -\sum_{j=1}^3\partial_{x^i}\bar{h}^{jk}\,\partial_{x^k}\bar{h}^{0j}
   -\sum_{j=1}^3\partial_{x^j}\bar{h}^{i0}\,\partial_{x^j}\bar{h}^{00}
   +\sum_{j, k=1}^3\partial_{x^k}\bar{h}^{ij}\,\partial_{x^k}\bar{h}^{0j}
   \\ & & -\frac{1}{2}\sum_{j, k=1}^3\partial_{x^0}\bar{h}^{jk}\,\partial_{x^i}\bar{h}^{jk}
   +\frac{1}{4}\,\partial_{x^0}\bar{h}\,\partial_{x^i}\bar{h}
   +({\rm rem}_{10})^{0i},   
\end{eqnarray*}  
and therefore (\ref{haste1})--(\ref{haste5}) leads to  
\begin{eqnarray*} 
   |g|\,t^{0i} 
   & = & \frac{c^4}{16\pi G}\bigg[
   \frac{1}{2}\,\Big(-\frac{2Gx^i}{c^5 r^4}\,\sum_{m, n=1}^3 x^m x^n\dddot{D}^{\,mn}
   +\tensor{({\rm rem}_{40})}{_i^0^0}\Big)
   \Big(\frac{2G}{c^5 r^3}\,\sum_{m, n=1}^3 x^m x^n\dddot{D}^{\,mn} 
   +{({\rm rem}_{38})}^{00}\Big) 
   \\ & & +\sum_{j, k=1}^3\Big(-\frac{2Gx^j}{c^5 r^3}\,\sum_{m=1}^3 x^m\dddot{D}^{\,im}
   +\tensor{({\rm rem}_{41})}{_j^0^i}\Big)
   \,\Big(-\frac{2Gx^k}{c^5 r^2}\,\dddot{D}^{\,jk}
   +\tensor{({\rm rem}_{42})}{_k^j^k}\Big)
   \\ & & -\Big(\frac{2G}{c^5 r^3}\,\sum_{m, n=1}^3 x^m x^n\dddot{D}^{\,mn}
   +{({\rm rem}_{38})}^{00}\Big)\,
   \sum_{k=1}^3\Big(-\frac{2G x^k}{c^5 r^2}\,\dddot{D}^{\,ik}
   +\tensor{({\rm rem}_{42})}{_k^i^k}\Big)
   \\ & & -\sum_{j=1}^3\Big(-\frac{2G x^j}{c^5 r^3}\,\sum_{m=1}^3 x^m\dddot{D}^{\,jm}
   +\tensor{({\rm rem}_{41})}{_j^0^j}\Big)\,
   \sum_{k=1}^3\Big(-\frac{2G x^k}{c^5 r^2}\,\dddot{D}^{\,ik}
   +\tensor{({\rm rem}_{42})}{_k^i^k}\Big)
   \\ & & +\sum_{j=1}^3\Big(-\frac{2G x^i}{c^5 r^3}\,\sum_{m=1}^3 x^m\dddot{D}^{\,jm}
   +\tensor{({\rm rem}_{41})}{_i^0^j}\Big)
   \,\Big(-\frac{2G x^j}{c^5 r^4}\,\sum_{m, n=1}^3 x^m x^n\dddot{D}^{\,mn}
   +\tensor{({\rm rem}_{40})}{_j^0^0}\Big)
   \\ & & +\sum_{j, k=1}^3\Big(\frac{2G}{c^5 r}\,\dddot{D}^{\,jk}
   +{({\rm rem}_{39})}^{jk}\Big)\,
   \Big(-\frac{2G x^k}{c^5 r^2}\,\dddot{D}^{\,ij}
   +\tensor{({\rm rem}_{42})}{_k^i^j}\Big)
   \\ & & -\sum_{j, k=1}^3
   \Big(-\frac{2G x^i}{c^5 r^2}\,\dddot{D}^{\,jk}
   +\tensor{({\rm rem}_{42})}{_i^j^k}\Big)\,
   \Big(-\frac{2G x^k}{c^5 r^3}\,\sum_{m=1}^3 x^m\dddot{D}^{\,jm}
   +\tensor{({\rm rem}_{41})}{_k^0^j}\Big)
   \\ & & -\sum_{j=1}^3\Big(-\frac{2G x^j}{c^5 r^3}\,\sum_{m=1}^3 x^m\dddot{D}^{\,im}
   +\tensor{({\rm rem}_{41})}{_j^0^i}\Big)
   \,\Big(-\frac{2Gx^j}{c^5 r^4}\,\sum_{m, n=1}^3 x^m x^n\dddot{D}^{\,mn}
   +\tensor{({\rm rem}_{40})}{_j^0^0}\Big)
   \\ & & +\sum_{j, k=1}^3\Big(-\frac{2G x^k}{c^5 r^2}\,\dddot{D}^{\,ij}
   +\tensor{({\rm rem}_{42})}{_k^i^j}\Big)\,
   \Big(-\frac{2G x^k}{c^5 r^3}\,\sum_{m=1}^3 x^m\dddot{D}^{\,jm}
   +\tensor{({\rm rem}_{41})}{_k^0^j}\Big)
   \\ & & -\frac{1}{2}\sum_{j, k=1}^3
   \Big(\frac{2G}{c^5 r}\,\dddot{D}^{\,jk}
   +{({\rm rem}_{39})}^{jk}\Big)\,
   \Big(-\frac{2G x^i}{c^5 r^2}\,\dddot{D}^{\,jk}
   +\tensor{({\rm rem}_{42})}{_i^j^k}\Big)\bigg]
   \\ & & +\frac{c^4}{64\pi G}\,\partial_{x^0}\bar{h}\,\partial_{x^i}\bar{h}
   +\frac{c^4}{16\pi G}\,({\rm rem}_{10})^{0i}. 
\end{eqnarray*} 
Now we set aside the terms in this expression which are small compared to the others, 
collecting them into a new remainder:
\begin{eqnarray*} 
   |g|\,t^{0i} 
   & = & \frac{c^4}{16\pi G}\bigg[
   -\frac{2G^2 x^i}{c^{10} r^7}\,\sum_{k, l, m, n=1}^3 x^k x^l x^m x^n\,\dddot{D}^{\,kl}\dddot{D}^{\,mn}
   +\sum_{j, k, m=1}^3\frac{4G^2 x^j x^k x^m}{c^{10} r^5}\,\dddot{D}^{\,im}\,\dddot{D}^{\,jk}
   \\ & & \hspace{4em} +\sum_{k, m, n=1}^3\frac{4G^2 x^k x^m x^n}{c^{10} r^5}\,
   \dddot{D}^{\,mn}\,\dddot{D}^{\,ik}
   -\sum_{j, k, m=1}^3\frac{4G^2 x^j x^k x^m}{c^{10} r^5}\,\dddot{D}^{\,jm}\,\dddot{D}^{\,ik}
   \\ & & \hspace{4em} +\sum_{j, k, m, n=1}^3\frac{4G^2 x^i x^j x^k x^m x^n}{c^{10} r^7}
   \,\dddot{D}^{\,jk}\,\dddot{D}^{\,mn}
   -\sum_{j, k=1}^3\frac{4G^2 x^k}{c^{10} r^3}\,\dddot{D}^{\,jk}\,\dddot{D}^{\,ij}
   \\ & & \hspace{4em} -\sum_{j, k, m=1}^3\frac{4G^2 x^i x^k x^m}{c^{10} r^5}
   \,\dddot{D}^{\,jk}\,\dddot{D}^{\,jm}
   -\sum_{j, k, m, n=1}^3\frac{4G^2 (x^j)^2 x^k x^m x^n}{c^{10} r^7}\,\dddot{D}^{\,ik}\dddot{D}^{\,mn}
   \\ & & \hspace{4em} +\sum_{j, k, m=1}^3\frac{4G^2 (x^k)^2 x^m}{c^{10} r^5}
   \,\dddot{D}^{\,ij}\,\dddot{D}^{\,jm}
   +\frac{2G^2 x^i}{c^{10} r^3}\,\sum_{j, k=1}^3 (\dddot{D}^{\,jk})^2\bigg]   
   \\ & & +\frac{c^4}{64\pi G}\,\partial_{x^0}\bar{h}\,\partial_{x^i}\bar{h}+({\rm rem}_{47})^i, 
\end{eqnarray*}  
where
\begin{eqnarray}\label{bigrem1}
   ({\rm rem}_{47})^i
   & = & \frac{c^4}{16\pi G}\bigg[\Big(-{({\rm rem}_{38})}^{00}\frac{x^i}{c^3 r^4}
   +\tensor{({\rm rem}_{40})}{^0^0^i}\frac{1}{c^3 r^3}
   -\sum_{k=1}^3\frac{2}{c^3 r^3}\,\tensor{({\rm rem}_{42})}{^i^k^k}\Big)
   \frac{G}{c^2}\sum_{m, n=1}^3 x^m x^n\dddot{D}^{\,mn} \nonumber
   \\ & & \hspace{3em}
   +\sum_{j, k, m=1}^3\frac{2G x^j x^m}{c^5 r^3}\,\Big(\dddot{D}^{\,jm}\tensor{({\rm rem}_{42})}{^i^k^k}   
   -\dddot{D}^{\,im}\tensor{({\rm rem}_{42})}{^j^k^k}\Big)  \nonumber
   \\ & & \hspace{3em}
   +\,{({\rm rem}_{38})}^{00}\sum_{k=1}^3\frac{2G x^k}{c^5 r^2}\,\dddot{D}^{\,ik}
   +\sum_{j, k=1}^3\frac{2G x^k}{c^5 r^2}\,\Big(\dddot{D}^{\,ik}\tensor{({\rm rem}_{41})}{^0^j^j}
   -\dddot{D}^{\,jk}\tensor{({\rm rem}_{41})}{^0^i^j}\Big)  \nonumber
   \\ & & \hspace{3em}
   -\frac{2G x^i}{c^5 r^3}\,\sum_{j, k=1}^3 x^k\dddot{D}^{\,jk}\tensor{({\rm rem}_{40})}{^0^0^j}
   -\frac{2G}{c^5 r^4}\,\sum_{j, m, n=1}^3 x^j x^m x^n\dddot{D}^{\,mn}
   \tensor{({\rm rem}_{41})}{^0^j^i}   \nonumber
   \\ & & \hspace{3em}
   +\sum_{j, k=1}^3\frac{2G}{c^5 r}\,\dddot{D}^{\,jk}\tensor{({\rm rem}_{42})}{^i^j^k}
   -\sum_{j, k=1}^3 \frac{2G x^k}{c^5 r^2}\,\dddot{D}^{\,ij} {({\rm rem}_{39})}^{jk} \nonumber
   \\ & & \hspace{3em}
   +\sum_{j, k, m=1}^3 \frac{2G x^k x^m}{c^5 r^3}\,\dddot{D}^{\,jm}\,
   \Big(\tensor{({\rm rem}_{42})}{^j^k^i}-\tensor{({\rm rem}_{42})}{^i^j^k}\Big)  \nonumber
   \\ & & \hspace{3em}
   +\sum_{j, k=1}^3\frac{2G x^j x^k}{c^5 r^3}\,\dddot{D}^{\,ik}\tensor{({\rm rem}_{40})}{^0^0^j}
   +\sum_{j, m, n=1}^3\frac{2G x^j x^m x^n}{c^5 r^4}\,\dddot{D}^{\,mn}\tensor{({\rm rem}_{41})}{^0^i^j}  \nonumber
   \\ & & \hspace{3em}
   +\frac{2G x^i}{c^5 r^2}\sum_{j, k=1}^3\,\dddot{D}^{\,jk}\tensor{({\rm rem}_{41})}{^0^j^k}
   -\sum_{j, k=1}^3\frac{2Gx^k}{c^5 r^2}\,\dddot{D}^{\,ij}\tensor{({\rm rem}_{41})}{^0^j^k}  \nonumber
   \\ & & \hspace{3em}
   +\frac{G x^i}{c^5 r^2}\,\sum_{j, k=1}^3\dddot{D}^{\,jk}\,{({\rm rem}_{39})}^{jk} 
   -\frac{G}{c^5 r}\,\sum_{j, k=1}^3\dddot{D}^{\,jk}\tensor{({\rm rem}_{42})}{^j^k^i}  \nonumber
   \\ & & \hspace{3em}
   +\sum_{j, k=1}^3\tensor{({\rm rem}_{41})}{^0^i^j}\tensor{({\rm rem}_{42})}{^j^k^k}
   -{({\rm rem}_{38})}^{00}\sum_{k=1}^3\tensor{({\rm rem}_{42})}{^i^k^k}
   +\frac{1}{2}\,{({\rm rem}_{38})}^{00}\tensor{({\rm rem}_{40})}{^0^0^i}  \nonumber
   \\ & & \hspace{3em}
   -\sum_{j, k=1}^3\tensor{({\rm rem}_{41})}{^0^j^j}\tensor{({\rm rem}_{42})}{^i^k^k}
   +\sum_{j=1}^3\tensor{({\rm rem}_{41})}{^0^j^i}\tensor{({\rm rem}_{40})}{^0^0^j}
   +\sum_{j, k=1}^3{({\rm rem}_{39})}^{jk}\tensor{({\rm rem}_{42})}{^i^j^k}  \nonumber
   \\ & & \hspace{3em}
   -\sum_{j, k=1}^3\tensor{({\rm rem}_{42})}{^j^k^i}\tensor{({\rm rem}_{41})}{^0^j^k}
   -\sum_{j=1}^3\tensor{({\rm rem}_{41})}{^0^i^j}\tensor{({\rm rem}_{40})}{^0^0^j}
   +\sum_{j, k=1}^3\tensor{({\rm rem}_{42})}{^i^j^k}\tensor{({\rm rem}_{41})}{^0^j^k}  \nonumber
   \\ & & \hspace{3em}
   -\frac{1}{2}\,\sum_{j, k=1}^3{({\rm rem}_{39})}^{jk}\tensor{({\rm rem}_{42})}{^j^k^i}\bigg]
   +\frac{c^4}{16\pi G}\,({\rm rem}_{10})^{0i}.  
\end{eqnarray} 
Thus, for $|x|=r$,  
\begin{eqnarray}\label{bts} 
   \sum_{i=1}^3 |g|\,\frac{x^i}{r}\,t^{0i} 
   & = & \frac{c^4}{16\pi G}\bigg[
   -\frac{2G^2}{c^{10} r^6}\,\sum_{k, l, m, n=1}^3 x^k x^l x^m x^n\,\dddot{D}^{\,kl}\dddot{D}^{\,mn}
   +\sum_{i, j, k, m=1}^3\frac{4G^2 x^i x^j x^k x^m}{c^{10} r^6}\,\dddot{D}^{\,im}\,\dddot{D}^{\,jk}
   \nonumber
   \\ & & \hspace{4em} +\sum_{i, k, m, n=1}^3\frac{4G^2 x^i x^k x^m x^n}{c^{10} r^6}
   \,\dddot{D}^{\,mn}\,\dddot{D}^{\,ik}
   -\sum_{i, j, k, m=1}^3\frac{4G^2 x^i x^j x^k x^m}{c^{10} r^6}\,\dddot{D}^{\,jm}\,\dddot{D}^{\,ik}
   \nonumber
   \\ & & \hspace{4em} +\sum_{j, k, m, n=1}^3\frac{4G^2 x^j x^k x^m x^n}{c^{10} r^6}
   \,\dddot{D}^{\,jk}\,\dddot{D}^{\,mn}
   -\sum_{i, j, k=1}^3\frac{4G^2 x^i x^k}{c^{10} r^4}\,\dddot{D}^{\,jk}\,\dddot{D}^{\,ij}
   \nonumber
   \\ & & \hspace{4em} -\sum_{j, k, m=1}^3\frac{4G^2 x^k x^m}{c^{10} r^4}\,\dddot{D}^{\,jk}\,\dddot{D}^{\,jm}
   -\sum_{i, k, m, n=1}^3\frac{4G^2 x^i x^k x^m x^n}{c^{10} r^6}\,\dddot{D}^{\,ik}\dddot{D}^{\,mn}
   \nonumber
   \\ & & \hspace{4em} +\sum_{i, j, m=1}^3\frac{4G^2 x^i x^m}{c^{10} r^4}\,\dddot{D}^{\,ij}\,\dddot{D}^{\,jm}
   +\frac{2G^2}{c^{10} r^2}\,\sum_{j, k=1}^3 (\dddot{D}^{\,jk})^2\bigg]   
   \nonumber
   \\ & & +\frac{c^4}{64\pi G}\,\partial_{x^0}\bar{h}\,\sum_{i=1}^3\frac{x^i}{r}\,\partial_{x^i}\bar{h}
   +({\rm rem}_{48})
   \nonumber 
   \\ & = & \frac{c^4}{16\pi G}\bigg[
   \frac{2G^2}{c^{10} r^6}\,\sum_{k, l, m, n=1}^3 x^k x^l x^m x^n\,\dddot{D}^{\,kl}\dddot{D}^{\,mn}
   -\sum_{i, j, k=1}^3\frac{4G^2 x^i x^k}{c^{10} r^4}\,\dddot{D}^{\,jk}\,\dddot{D}^{\,ij}
   \nonumber
   \\ & & \hspace{4em} +\frac{2G^2}{c^{10} r^2}\,\sum_{j, k=1}^3 (\dddot{D}^{\,jk})^2\bigg]   
   +\frac{c^4}{64\pi G}\,\partial_{x^0}\bar{h}\,\sum_{i=1}^3\frac{x^i}{r}\,\partial_{x^i}\bar{h}
   +({\rm rem}_{48}),  
\end{eqnarray} 
defining
\begin{equation} \label{eq_rem48}
    ({\rm rem}_{48})=\sum_{i=1}^3 \frac{x^i}{r} ({\rm rem}_{47})^i.
\end{equation}
Next we substitute (\ref{barbarhh}) into (\ref{bts}) to find, for $|x|=r$,  
\begin{eqnarray}\label{bts2}  
   \sum_{i=1}^3 |g|\,\frac{x^i}{r}\,t^{0i} 
   & = & \frac{c^4}{16\pi G}\bigg[
   \frac{2G^2}{c^{10} r^6}\,\sum_{k, l, m, n=1}^3 x^k x^l x^m x^n\,\dddot{D}^{\,kl}\dddot{D}^{\,mn}
   -\sum_{i, j, k=1}^3\frac{4G^2 x^i x^k}{c^{10} r^4}\,\dddot{D}^{\,jk}\,\dddot{D}^{\,ij}
   \nonumber
   \\ & & \hspace{4em} +\,\frac{2G^2}{c^{10} r^2}\,\sum_{j, k=1}^3 (\dddot{D}^{\,jk})^2\bigg]   
   \nonumber
   \\ & & +\frac{c^4}{16\pi G}\,\bigg[-\frac{G^2}{c^{10} r^6}\,\sum_{i, j, m, n=1}^3 x^i x^j x^m x^n
   \,\dddot{D}^{\,ij}\,\dddot{D}^{\,mn}
   +\frac{2G^2}{c^{10} r^4}\,\tensor{\dddot{D}}{_k^k}\,\sum_{i, j=1}^3 x^i x^j\dddot{D}^{\,ij}
   -\frac{G^2}{c^{10} r^2}\,(\tensor{\dddot{D}}{_j^j})^2\bigg]
   \nonumber
   \\ & & +\,({\rm rem}_{49})
   \nonumber
   \\ & = & \frac{c^4}{16\pi G}\bigg[
   \frac{G^2}{c^{10} r^6}\,\sum_{k, l, m, n=1}^3 x^k x^l x^m x^n\,\dddot{D}^{\,kl}\dddot{D}^{\,mn}
   -\sum_{i, j, k=1}^3\frac{4G^2 x^i x^k}{c^{10} r^4}\,\dddot{D}^{\,jk}\,\dddot{D}^{\,ij}
   \nonumber
   \\ & & \hspace{4em} +\,\frac{2G^2}{c^{10} r^2}\,\sum_{j, k=1}^3 (\dddot{D}^{\,jk})^2  
   +\frac{2G^2}{c^{10} r^4}\,\tensor{\dddot{D}}{_k^k}\,\sum_{i, j=1}^3 x^i x^j\dddot{D}^{\,ij}
   -\frac{G^2}{c^{10} r^2}\,(\tensor{\dddot{D}}{_j^j})^2\bigg]
   \nonumber
   \\ & & +\,({\rm rem}_{49})
\end{eqnarray} 
for
\begin{equation}\label{eq_rem49}
    ({\rm rem}_{49})=\frac{c^4}{64\pi G}({\rm rem}_{46})+({\rm rem}_{48}).
\end{equation}
Using (\ref{bts2}) in (\ref{thores}) and applying Lemma \ref{pients}, 
we conclude that for $r\ge\lambda K_\ast$ and $|u|\le\lambda c^{-1}$: 
\begin{eqnarray}\label{knigh} 
   \partial_t\,\mathfrak{p}^0 
   & = & -c\,\sum_{i=1}^3\int_{\partial B_r(0)} |g|\,\frac{x^i}{r}\,t^{0i}\,dS(x)
   \nonumber
   \\ & = & - \frac{c^5}{16\pi G}\int_{\partial B_r(0)} 
  \bigg[\frac{G^2}{c^{10} r^6}\,\sum_{k, l, m, n=1}^3 x^k x^l x^m x^n\,\dddot{D}^{\,kl}\dddot{D}^{\,mn}
   -\sum_{i, j, k=1}^3\frac{4G^2 x^i x^k}{c^{10} r^4}\,\dddot{D}^{\,jk}\,\dddot{D}^{\,ij}
   \nonumber
   \\ & & \hspace{7em} +\frac{2G^2}{c^{10} r^2}\,\sum_{j, k=1}^3 (\dddot{D}^{\,jk})^2  
   +\frac{2G^2}{c^{10} r^4}\,\tensor{\dddot{D}}{_k^k}\,\sum_{i, j=1}^3 x^i x^j\dddot{D}^{\,ij}
   -\frac{G^2}{c^{10} r^2}\,(\tensor{\dddot{D}}{_j^j})^2\bigg]\,dS(x)
   \nonumber
   \\ & & +\,({\rm rem}_{50})
   \nonumber
   \\ & = & - \frac{c^5}{16 G}
   \bigg[\frac{4G^2}{15 c^{10}}\,\sum_{k, j, m, n=1}^3 
   \,(\delta_{jk}\delta_{nm}+\delta_{jn}\delta_{km}+\delta_{jm}\delta_{kn})\,\dddot{D}^{\,kj}\dddot{D}^{\,mn}
   -\frac{16G^2}{3 c^{10}}\sum_{j, k=1}^3\,(\dddot{D}^{\,jk})^2
   \nonumber
   \\ & & \hspace{7em} +\frac{8G^2}{c^{10}}\sum_{j, k=1}^3 (\dddot{D}^{\,jk})^2  
   +\frac{8G^2}{3 c^{10}}\,(\tensor{\dddot{D}}{_j^j})^2
   -\frac{4G^2}{c^{10}}\,(\tensor{\dddot{D}}{_j^j})^2\bigg]
   \nonumber
   \\ & & +\,({\rm rem}_{50})
   \nonumber
   \\ & = & - \frac{G}{5c^5}\bigg[\sum_{j, k=1}^3 (\dddot{D}^{\,jk})^2  
   -\frac{1}{3}(\tensor{\dddot{D}}{_j^j})^2\bigg]+({\rm rem}_{50}), 
\end{eqnarray} 
where
\begin{equation}\label{eq_rem50}
    ({\rm rem}_{50})=-c\int_{\partial B_r(0)}({\rm rem}_{49})\,dS(x). 
\end{equation}
Note that we were justified in pulling the $\dddot{D}^{\,jk}=\dddot{D}^{\,jk}(t-c^{-1}r)$ 
terms from inside the integrals, since they only depend on $x$ through $r$. 
Also observe the following simplification:
\begin{eqnarray*} 
   \lefteqn{\Big(\dddot{D}_{jk}-\frac{1}{3}\,\eta_{jk}\tensor{\dddot{D}}{_m^m}\Big)
   \Big(\dddot{D}^{\,jk}-\frac{1}{3}\,\eta^{jk}\tensor{\dddot{D}}{_n^n}\Big)}
   \\ & = & \dddot{D}_{jk}\dddot{D}^{\,jk}
   -\frac{2}{3}\dddot{D}^{\,jk}\eta^{jk}\tensor{\dddot{D}}{_n^n}
   +\frac{1}{9}\,\eta_{jk}\,\eta^{jk}\,(\tensor{\dddot{D}}{_n^n})^2
   \\ & = & \sum_{j, k=1}^3 (\dddot{D}^{\,jk})^2
   -\frac{1}{3}\,(\tensor{\dddot{D}}{_j^j})^2, 
\end{eqnarray*}  
if it is understood that the indices are raised and lowered using $\eta$. 
By (\ref{knigh}), re-inserting the correct function arguments, we finally arrive at
\begin{equation} \label{absvalue_final}
    \bigg|\partial_t\,\mathfrak{p}^0(t, x)
   -\bigg(-\frac{G}{5c^5}\bigg[\sum_{j, k=1}^3 (\dddot{D}^{\,jk}(u))^2  
   -\frac{1}{3}(\tensor{\dddot{D}}{_j^j}(u))^2\bigg]\bigg)\bigg|=|({\rm rem}_{50})(t, x)|.    
\end{equation}
To finish the proof of Theorem \ref{mainthm}, we now need to establish that $|\rem{50}(t,x)|$ 
can be bounded by the expression on the right-hand side of~\eqref{quadru}. 
This will be achieved in (\ref{maumau50}) ahead.
 
\section{Estimating the remainders}
\setcounter{equation}{0}
\label{sec_estimating}

In this section we find estimates for all the remainder terms obtained in the above derivation 
of the quadrupole formula. For clarity, the calculations are divided into subsections based 
on the direct dependence between the corresponding remainder terms. Throughout our estimates, 
we will denote by $C>0$ universal constants that may change from line to line.

It should be mentioned that remainder 29 from (\ref{eq_rem29}) 
is the most complicated one to control (which is no surprise, 
see the discussion in Appendix \ref{Blanch_discuss}). 
In fact at this point the precise decay estimates 
on $h$ and its derivatives are required, whereas for all of the other terms 
simpler pointwise bounds would be sufficient. 

\subsection{Remainders 5 and 6: (\ref{eq_rem5}) and (\ref{det12})}

Recall that $\hat{h}=(h_{\alpha\beta})$ and $h=\eta^{\alpha\beta} h_{\alpha\beta}$. 
We can apply (\ref{2behc}) from Lemma \ref{tele}(a)  
with $n=4$ and the matrix $\hat{m}^{-1}\hat{h}$ to obtain the pointwise bound
\begin{equation}\label{calmy} 
   |({\rm rem}_5)|\le C |\hat{h}|^2,
\end{equation}  
where $|\hat{h}|$ means $\max_{\alpha,\beta=0, 1, 2, 3}|\hat{h}_{\alpha\beta}|$ 
pointwise in $(t, x)$, with the arguments $(t,x)$ of $\rem{5}$ and $\hat{h}$ 
having been suppressed for brevity; such abbreviations will be used throughout this entire section. 
Therefore Lemma \ref{tele}(b) with $x=h+({\rm rem}_5)$ yields 
\begin{equation}\label{parac} 
   |({\rm rem}_6)|\le C(h+({\rm rem}_5))^2\le C(h^2+|\hat{h}|^4)
   \le C(|\hat{h}|^2+|\hat{h}|^4)\le C|\hat{h}|^2.
\end{equation}  
To bound the derivatives of these remainders, we use the Jacobi formula 
\[ \partial_{x^\gamma}(\det A)=(\det A)\,{\rm tr}\,(A^{-1}\partial_{x^\gamma}A), \] 
valid for any invertible-matrix-valued function $A$. Using the notation from Section \ref{boH} and the function $A=I+\hat{m}^{-1}\hat{h}$, we have $A=\hat{m}^{-1}(\hat{m}+\hat{h})
=\hat{m}^{-1}\hat{g}$, and hence $A^{-1}=\hat{g}^{-1}\hat{m}=(\hat{m}^{-1}+\hat{H})\hat{m}
=I+\hat{H}\hat{m}$, where $\hat{H}=(H^{\alpha\beta})$. It follows that 
\begin{eqnarray*} 
   \partial_{x^\gamma} ({\rm rem}_5)
   & = & \partial_{x^\gamma}\Big(\det(I+\hat{m}^{-1}\hat{h}) - 1 - {\rm tr}\,(\hat{m}^{-1}\hat{h})\Big)
   \\ & = & \det(I+\hat{m}^{-1}\hat{h})\,{\rm tr}\Big((I+\hat{H}\hat{m})\hat{m}^{-1}\partial_{x^\gamma}\hat{h}\Big)
   -{\rm tr}\,(\hat{m}^{-1}\partial_{x^\gamma}\hat{h})
   \\ & = & (\det(I+\hat{m}^{-1}\hat{h})-1)\,{\rm tr}\Big((\hat{m}^{-1}+\hat{H})\partial_{x^\gamma}\hat{h}\Big)
   +{\rm tr}\,(\hat{H}\partial_{x^\gamma}\hat{h}), 
\end{eqnarray*} 
and thus from (\ref{2behc}) in Lemma \ref{tele}(a) we find 
\begin{equation}\label{calmyd} 
   |\partial_{x^\gamma} ({\rm rem}_5)| 
   \le C(|\hat{h}|+|\hat{H}|)\,|\partial_{x^\gamma}\hat{h}|, 
\end{equation}  
Next we calculate 
\[ \partial_{x^\gamma} ({\rm rem}_6)
   =\partial_{x^\gamma}\,\Big(\sqrt{1+h+({\rm rem}_5)}-1-\frac{1}{2}\,h\Big)
   =\frac{1}{2}\,\frac{\partial_{x^\gamma}h+\partial_{x^\gamma}({\rm rem}_5)}
   {\sqrt{1+h+({\rm rem}_5)}}-\frac{1}{2}\,\partial_{x^\gamma}h. \]  
Due to (\ref{calmyd}), Lemma \ref{tele}(b) and (\ref{calmy}), we find the bound
\begin{eqnarray}\label{paracd}  
   |\partial_{x^\gamma} ({\rm rem}_6)|
   & \le & C|\partial_{x^\gamma}({\rm rem}_5)|
   +C|\partial_{x^\gamma}h|\,|\sqrt{1+h+({\rm rem}_5)}-1|
   \nonumber
   \\ & \le & C(|\hat{h}|+|\hat{H}|)\,|\partial_{x^\gamma}\hat{h}|
   +C(|h|+|\hat{h}|^2)\,|\partial_{x^\gamma}h|
   \nonumber
   \\ & \le & C(|\hat{h}|+|\hat{H}|)
   \,|\partial_{x^\gamma}\hat{h}|.     
\end{eqnarray} 

\subsection{Remainders 7, 8 and 9: (\ref{eq_rem7}), (\ref{eq_rem8}) and (\ref{eq_rem9})}

Using the notation of equation (\ref{eq_names_gh}), 
we have $\hat{g}\hat{H}\hat{m}=\hat{g}(\hat{g}^{-1}-\hat{m}^{-1})\hat{m}=-\hat{h}$, 
and hence $\hat{H}=-\hat{g}^{-1}\hat{h}\hat{m}^{-1}$. Therefore 
\begin{eqnarray*} 
   \rem{7}^{\alpha\beta} 
   & = & h^{\alpha\beta} + H^{\alpha\beta} 
   = \eta^{\alpha\mu} h_{\mu\nu} \eta^{\nu\beta} + H^{\alpha\beta} 
   = (\hat{m}^{-1}\hat{h}\,\hat{m}^{-1})^{\alpha\beta} + \hat{H}^{\alpha\beta} 
   \\ & = & \Big((\hat{m}^{-1}-\hat{g}^{-1})\hat{h}\,\hat{m}^{-1}\Big)^{\alpha\beta} 
   = -(\hat{H}\hat{h}\hat{m}^{-1})^{\alpha\beta} \ ,
\end{eqnarray*} 
that is,
\[ \rem{7}^{\alpha\beta} = -\eta^{\nu\beta}h_{\mu\nu}H^{\alpha\mu}. \]
This yields the bound 
\begin{equation}\label{alfers} 
   |\rem{7}|\le C|\hat{h}||\hat{H}|.
\end{equation}  
Next, taking a derivative, 
\begin{equation}\label{alfersd} 
   |\partial_{x^\gamma}\rem{7}|
   \le C(|\hat{h}||\partial_{x^\gamma}\hat{H}|+|\hat{H}||\partial_{x^\gamma}\hat{h}|).
\end{equation}  
In our derivation of the quadrupole formula in Section~\ref{sec_expansion}, we don't need any bounds for $\rem{8}$ itself, but rather only for its derivative, which is the definition of $\rem{9}$:
\begin{eqnarray*} 
    \tensor{\rem{9}}{_\gamma^{\alpha\beta}} 
    & = & \partial_{x^\gamma}\rem{8}^{\alpha\beta} 
    \\ & = &  -\frac{1}{2}\,(\partial_{x^\gamma} h)h^{\alpha\beta} 
    - \frac{1}{2}h(\partial_{x^\gamma} h^{\alpha\beta}) 
    + \big(\partial_{x^\gamma}\rem{7}^{\alpha\beta}\big)\left(1 + \frac{h}{2}\right) 
    +\frac{1}{2}\,\rem{7}^{\alpha\beta}\,(\partial_{x^\gamma} h) 
    \\ & & +\,\big(\partial_{x^\gamma}\rem{6}\big)(\eta^{\alpha\beta}
    -h^{\alpha\beta}+\rem{7}^{\alpha\beta}) + \rem{6}(-\partial_{x^\gamma} h^{\alpha\beta}
    +\partial_{x^\gamma}\rem{7}^{\alpha\beta}) \ .
\end{eqnarray*} 
Thus it follows from 
$h^{\alpha\beta}=\eta^{\alpha\alpha'}\eta^{\beta\beta'} h_{\alpha'\beta'}$ 
and (\ref{alfersd}), (\ref{alfers}), (\ref{paracd}), (\ref{parac}) that
\begin{eqnarray}\label{feta}  
    |\tensor{\rem{9}}{_\gamma^{\alpha\beta}}| 
    & \le & C\,|\hat{h}| |\partial_{x^\gamma}\hat{h}|  
    +C\,(|\hat{h}||\partial_{x^\gamma}\hat{H}|+|\hat{H}||\partial_{x^\gamma}\hat{h}|) 
    +C\,|\hat{h}||\hat{H}|\,|\partial_{x^\gamma}\hat{h}|
    \nonumber 
    \\ & & +\,C(|\hat{h}|+|\hat{H}|)\,|\partial_{x^\gamma}\hat{h}|
    +C|\hat{h}|^2 \Big(|\partial_{x^\gamma}\hat{h}|
    +|\hat{h}||\partial_{x^\gamma}\hat{H}|+|\hat{H}||\partial_{x^\gamma}\hat{h}|\Big)
    \nonumber
    \\ & \le &  C\,\Big(|\hat{h}| |\partial_{x^\gamma}\hat{h}|  
    +|\hat{h}||\partial_{x^\gamma}\hat{H}|+|\hat{H}||\partial_{x^\gamma}\hat{h}|\Big). 
\end{eqnarray} 

\subsection{Remainder 10: (\ref{eq_rem10})}

To bound the lengthy remainder 10, we can use $\bar{h}^{\alpha\beta}=-h^{\alpha\beta}$, 
(\ref{ginfty3}), (\ref{gm1infty}) and the estimate (\ref{feta}) for $\rem{9}$ just above. We obtain 
\begin{eqnarray}\label{maumau10} 
   |({\rm rem}_{10})^{\alpha\beta}| 
   & \le & C|\partial_{x^\gamma}\hat{h}|\,|\tensor{({\rm rem}_9)}{_\delta^\delta^\gamma}|
   +C|\partial_{x^\gamma}\hat{h}|\,|\tensor{({\rm rem}_9)}{_\gamma^\alpha^\beta}|
   +C|\tensor{({\rm rem}_9)}{_\gamma^\alpha^\beta}|\,|\tensor{({\rm rem}_9)}{_\delta^\delta^\gamma}|
   \nonumber
   \\ & & +C|\partial_{x^\gamma}\hat{h}|\,|\tensor{({\rm rem}_9)}{_\delta^\beta^\delta}|
   +C|\partial_{x^\gamma}\hat{h}|\,|\tensor{({\rm rem}_9)}{_\gamma^\alpha^\gamma}|
   +C|\tensor{({\rm rem}_9)}{_\gamma^\alpha^\gamma}|\,|\tensor{({\rm rem}_9)}{_\delta^\beta^\delta}|
   \nonumber
   \\ & & +C|\hat{h}|\,|\partial_{x^\gamma}\hat{h}|^2
   \nonumber
   \\ & & +C\Big[|\partial_{x^\gamma}\hat{h}|\,|\tensor{({\rm rem}_9)}{_\gamma^\sigma^\eps}|
   +|\partial_{x^\gamma}\hat{h}|\,|\tensor{({\rm rem}_9)}{_\sigma^\delta^\gamma}|
   +|\tensor{({\rm rem}_9)}{_\sigma^\delta^\gamma}|\,|\tensor{({\rm rem}_9)}{_\gamma^\sigma^\eps}|\Big]
   \nonumber
   \\ & & +C|\hat{H}|\,|\partial_{x^\gamma}\hat{h}|^2
   \nonumber
   \\ & & +C|\hat{h}|\,
   \Big(|\partial_{x^\gamma}\hat{h}|+|\tensor{({\rm rem}_9)}{_\delta^\eps^\sigma}|\Big)
   \Big(|\partial_{x^\sigma}\hat{h}|+|\tensor{({\rm rem}_9)}{_\sigma^\beta^\gamma}|\Big)
   \nonumber
   \\ & & +C|\hat{H}|\,
   \Big(|\partial_{x^\gamma}\hat{h}|+|\tensor{({\rm rem}_9)}{_\delta^\eps^\sigma}|\Big)
   \Big(|\partial_{x^\gamma}\hat{h}|+|\tensor{({\rm rem}_9)}{_\sigma^\beta^\gamma}|\Big)
   \nonumber
   \\ & & +C\Big[|\partial_{x^\gamma}\hat{h}|\,|\tensor{({\rm rem}_9)}{_\sigma^\beta^\gamma}|
   +|\partial_{x^\gamma}\hat{h}|\,|\tensor{({\rm rem}_9)}{_\delta^\eps^\sigma}|
   +|\tensor{({\rm rem}_9)}{_\delta^\eps^\sigma}|\,|\tensor{({\rm rem}_9)}{_\sigma^\beta^\gamma}|\Big]
   \nonumber
   \\ & & +C|\hat{h}|
   \Big(|\partial_{x^\gamma}\hat{h}|+|\tensor{({\rm rem}_9)}{_\delta^\eps^\sigma}|\Big)   
   \Big(|\partial_{x^\gamma}\hat{h}|+|\tensor{({\rm rem}_9)}{_\sigma^\alpha^\gamma}|\Big)
   \nonumber
   \\ & & +C|\hat{H}|
   \Big(|\partial_{x^\gamma}\hat{h}|+|\tensor{({\rm rem}_9)}{_\delta^\eps^\sigma}|\Big)   
   \Big(|\partial_{x^\gamma}\hat{h}|+|\tensor{({\rm rem}_9)}{_\sigma^\alpha^\gamma}|\Big)
   \nonumber
   \\ & & +C\Big[|\partial_{x^\gamma}\hat{h}|\,|\tensor{({\rm rem}_9)}{_\sigma^\alpha^\gamma}|   
   +|\partial_{x^\gamma}\hat{h}|\,|\tensor{({\rm rem}_9)}{_\delta^\eps^\sigma}|
   +|\tensor{({\rm rem}_9)}{_\delta^\eps^\sigma}|\,|\tensor{({\rm rem}_9)}{_\sigma^\alpha^\gamma}|\Big]
   \nonumber
   \\ & & +C(|\hat{h}|+|\hat{H}|)
   \Big(|\partial_{x^\gamma}\hat{h}|+|\tensor{({\rm rem}_9)}{_\gamma^\beta^\delta}|\Big)   
   \Big(|\partial_{x^\gamma}\hat{h}|+|\tensor{({\rm rem}_9)}{_\sigma^\alpha^\eps}|\Big)
   \nonumber
   \\ & & +C\Big[|\partial_{x^\gamma}\hat{h}|\,|\tensor{({\rm rem}_9)}{_\sigma^\alpha^\eps}|
   +|\partial_{x^\gamma}\hat{h}|\,|\tensor{({\rm rem}_9)}{_\gamma^\beta^\delta}|
   +|\tensor{({\rm rem}_9)}{_\gamma^\beta^\delta}|\,|\tensor{({\rm rem}_9)}{_\sigma^\alpha^\eps}|\Big]
   \nonumber
   \\ & & +C(|\hat{H}|^2+|\hat{h}|^2)\Big(|\partial_{x^\gamma}\hat{h}|+|\tensor{({\rm rem}_9)}{_\delta^\eps^\nu}|\Big)
   \Big(|\partial_{x^\gamma}\hat{h}|+|\tensor{({\rm rem}_9)}{_\gamma^\sigma^\kappa}|\Big)
   \nonumber
   \\ & & +C\Big(|\hat{h}|+|\hat{H}|+|\hat{H}||\hat{h}|\Big)
   \Big(|\partial_{x^\gamma}\hat{h}|+|\tensor{({\rm rem}_9)}{_\delta^\eps^\nu}|\Big)
   \Big(|\partial_{x^\gamma}\hat{h}|+|\tensor{({\rm rem}_9)}{_\gamma^\sigma^\kappa}|\Big)
   \nonumber
   \\ & & +C\Big[|\partial_{x^\gamma}\hat{h}|\,|\tensor{({\rm rem}_9)}{_\gamma^\sigma^\kappa}|   
   +|\partial_{x^\gamma}\hat{h}|\,|\tensor{({\rm rem}_9)}{_\delta^\eps^\nu}|
   +|\tensor{({\rm rem}_9)}{_\delta^\eps^\nu}|\,|\tensor{({\rm rem}_9)}{_\gamma^\sigma^\kappa}|\Big]
   \nonumber
   \\ & \le & C|\partial_{x^\gamma}\hat{h}|\,|\tensor{({\rm rem}_9)}{_\delta^\delta^\gamma}|
   +C|\tensor{({\rm rem}_9)}{_\gamma^\alpha^\beta}|\,|\tensor{({\rm rem}_9)}{_\delta^\delta^\gamma}|
   \nonumber
   \\ & & +C(|\hat{h}|+|\hat{H}|+|\hat{h}|^2+|\hat{H}|^2+|\hat{h}||\hat{H}|)\,
   \Big(|\partial_{x^\gamma}\hat{h}|+|\tensor{({\rm rem}_9)}{_\delta^\eps^\sigma}|\Big)^2
   \nonumber   
   \\ & \le & C(|\hat{h}|+|\hat{H}|)\,
   \Big(|\partial_{x^\gamma}\hat{h}|^2+|\partial_{x^\gamma}\hat{H}|^2\Big) 
\end{eqnarray} 
for $\alpha, \beta=0, 1, 2, 3$.

\subsection{Remainders 18, 20, 21, 24, 28}

Subsection~\ref{moments_sect} defined remainders 11 through 28, and those appearing 
in the title of the present part, defined in equations~\eqref{bwh1}, \eqref{eq_rem20}, \eqref{eq_rem21}, 
\eqref{eq_rem24} and~\eqref{eq_rem28}, are the ones that will figure in the next part. Remainder 11 
was defined as an expression involving $T^{\mu\nu}$ and $\Gamma^\lambda_{\mu\nu}$ terms, 
while remainders 15, 17, 23 and 27 were defined in~\eqref{eq_rem15}, \eqref{eq_rem17}, \eqref{eq_rem23} 
and~\eqref{eq_rem27} as integrals of expressions involving these terms as well as $h_{\mu\nu}$ terms. 
All the other remainders between 11 and 28 were algebraic and/or derivative expressions of these. 
Thus simple pointwise bounds depending on $\delta_T$ and $\delta_h$ terms can be obtained 
for all these remainders one by one, starting with $\rem{11}$ and working our way upwards.

From (\ref{wasa}) we recall that 
\[ \Gamma^{\sigma}_{\mu\nu}
   =\frac{1}{2}\,g^{\sigma\kappa}\,(\partial_{x^\mu} h_{\nu\kappa}
   +\partial_{x^\nu} h_{\mu\kappa}-\partial_{x^\kappa} h_{\mu\nu}), \]   
which together with (\ref{gm1infty}) leads to 
\[ |\Gamma^{\sigma}_{\mu\nu}|\le C|\partial_{x^\gamma}\hat{h}|. \] 
For the derivatives, by (\ref{eq_names_gh}), 
\[ \partial_{x^\gamma}\Gamma^{\sigma}_{\mu\nu}
   =\frac{1}{2}\,\partial_{x^\gamma} H^{\sigma\kappa}\,(\partial_{x^\mu} h_{\nu\kappa}
   +\partial_{x^\nu} h_{\mu\kappa}-\partial_{x^\kappa} h_{\mu\nu})
   +\frac{1}{2}\,g^{\sigma\kappa}\,(\partial^2_{x^\gamma x^\mu} h_{\nu\kappa}
   +\partial^2_{x^\gamma x^\nu} h_{\mu\kappa}-\partial^2_{x^\gamma x^\kappa} h_{\mu\nu}) \] 
and hence 
\[ |\partial_{x^\gamma}\Gamma^{\sigma}_{\mu\nu}|
   \le C(|\partial_{x^\gamma}\hat{H}||\partial_{x^\gamma}\hat{h}|
   +|\partial^2_{x^\beta x^\gamma}\hat{h}|), \]
and similarly 
\[ |\partial^2_{x^\alpha x^\gamma}\Gamma^{\sigma}_{\mu\nu}|
   \le C\Big(|\partial^2_{x^\beta x^\gamma}\hat{H}||\partial_{x^\gamma}\hat{h}|
   +|\partial_{x^\gamma}\hat{H}||\partial^2_{x^\beta x^\gamma}\hat{h}|
   +|\partial^3_{x^\alpha x^\beta x^\gamma}\hat{h}|\Big). \] 
Using (\ref{Tupper}), we deduce that $\rem{11}$ as defined in~\eqref{eq_rem11} can be bounded as 
\begin{equation}\label{pana}
   |({\rm rem}_{11})^\mu|\le C\,\frac{c^4}{G\lambda^2}
   \,\delta_T^{(0)}\,|\partial_{x^\gamma}\hat{h}|
\end{equation}  
and from 
\[ \partial_{x^\gamma}({\rm rem}_{11})^\mu
   =-(\partial_{x^\gamma}\tensor{\Gamma}{^\mu_\nu_\lambda})\,T^{\lambda\nu}
   -\tensor{\Gamma}{^\mu_\nu_\lambda}\,(\partial_{x^\gamma} T^{\lambda\nu})
   -(\partial_{x^\gamma}\tensor{\Gamma}{^\nu_\nu_\lambda})\,T^{\mu\lambda}
   -\tensor{\Gamma}{^\nu_\nu_\lambda}\,(\partial_{x^\gamma} T^{\mu\lambda}) \] 
also 
\begin{eqnarray}\label{panad}  
   |\partial_{x^\gamma}({\rm rem}_{11})^\mu|
   & \le & C(|\partial_{x^\gamma}\hat{H}||\partial_{x^\gamma}\hat{h}|
   +|\partial^2_{x^\beta x^\gamma}\hat{h}|)\,\frac{c^4}{G\lambda^2}\,\delta_T^{(0)}
   +C|\partial_{x^\gamma}\hat{h}|\,\frac{c^4}{G\lambda^3}\,(\delta_T^{(0)}+\delta_T^{(1)})
   \nonumber
   \\ & \le & C\,\frac{c^4}{G\lambda^4}\,(\delta_T^{(0)}+\delta_T^{(1)})
   \,(\lambda |\partial_{x^\gamma}\hat{h}|+\lambda^2 |\partial^2_{x^\beta x^\gamma}\hat{h}|), 
\end{eqnarray} 
and similarly 
\begin{equation}\label{panadd} 
   |\partial^2_{x^\alpha x^\gamma}({\rm rem}_{11})^\mu|
   \le C\,\frac{c^4}{G\lambda^5}\,(\delta_T^{(0)}+\delta_T^{(1)}+\delta_T^{(2)})\,\Big(
   \lambda |\partial_{x^\gamma}\hat{h}|+\lambda^2 |\partial^2_{x^\beta x^\gamma}\hat{h}|
   +\lambda^3 |\partial^3_{x^\alpha x^\beta x^\gamma}\hat{h}|\Big).
\end{equation}  

The compact-support condition~\eqref{CS_cond_T} for $T^{\alpha\beta}$ implies that the integral~\eqref{eq_rem12} 
in the definition of $\rem{12}$ can be reduced to a ball of radius proportional to $\lambda$ 
(the proportionality constant, $K_\ast$, will be incorporated into $C$). In particular 
also the term $|\bar{x}^j|$ inside that integral is bounded in this way, while (\ref{pana}) 
gives an estimate for the rest of the integrand. Thus we find
\begin{equation}\label{fralang} 
   |({\rm rem}_{12})^{ij}|\le C\,\frac{c^4\lambda^2}{G}
   \,\delta_T^{(0)}\,{\|\partial_{x^\gamma}\hat{h}\|}_{L^\infty_x}
\end{equation}  
Similarly, differentiating under the integral and owing to (\ref{panad}),
\begin{equation}\label{fralangd} 
   |\partial_{x^\gamma}({\rm rem}_{12})^{ij}|\le C\,\frac{c^4}{G}\,
   (\delta_T^{(0)}+\delta_T^{(1)})
   \,(\lambda {\|\partial_{x^\gamma}\hat{h}\|}_{L^\infty_x}
   +\lambda^2 {\|\partial^2_{x^\beta x^\gamma}\hat{h}\|}_{L^\infty_x}).
\end{equation}
Analogue reasoning leads to the following for $\rem{13}$ as defined in~\eqref{eq_rem13}
\begin{equation}\label{frasieg} 
   |({\rm rem}_{13})^{ij}|\le C\,\frac{c^6\lambda}{G}
   \,(\delta_T^{(0)}+\delta_T^{(1)})\,(\lambda {\|\partial_{x^\gamma}\hat{h}\|}_{L^\infty_x}
   +\lambda^2 {\|\partial^2_{x^\beta x^\gamma}\hat{h}\|}_{L^\infty_x}),
\end{equation}  
\begin{equation}\label{frasiegd}  
   |\partial_t ({\rm rem}_{13})^{ij}|\le C\,\frac{c^7}{G}
   \,(\delta_T^{(0)}+\delta_T^{(1)}+\delta_T^{(2)})\,\Big(
   \lambda {\|\partial_{x^\gamma}\hat{h}\|}_{L^\infty_x}
   +\lambda^2 {\|\partial^2_{x^\beta x^\gamma}\hat{h}\|}_{L^\infty_x}
   +\lambda^3 {\|\partial^3_{x^\alpha x^\beta x^\gamma}\hat{h}\|}_{L^\infty_x}\Big),
\end{equation}
and with these one immediately finds for $\rem{14}$ as defined in~\eqref{eq_rem13}
\begin{equation}\label{anbu} 
   |({\rm rem}_{14})^{ij}|\le C\,\frac{c^6\lambda}{G}\,
   (\delta_T^{(0)}+\delta_T^{(1)})\,
   (\lambda {\|\partial_{x^\gamma}\hat{h}\|}_{L^\infty_x}
   +\lambda^2 {\|\partial^2_{x^\beta x^\gamma}\hat{h}\|}_{L^\infty_x}),
\end{equation}  
\begin{equation}\label{anbud} 
   |\partial_t ({\rm rem}_{14})^{ij}|
   \le C\,\frac{c^7}{G}\,(\delta_T^{(0)}+\delta_T^{(1)}+\delta_T^{(2)})\,
   \Big(\lambda {\|\partial_{x^\gamma}\hat{h}\|}_{L^\infty_x}
   +\lambda^2 {\|\partial^2_{x^\beta x^\gamma}\hat{h}\|}_{L^\infty_x}
   +\lambda^3 {\|\partial^3_{x^\alpha x^\beta x^\gamma}\hat{h}\|}_{L^\infty_x}\Big).
\end{equation}

For the rest of the pertinent remainders here, we simply state the relevant results of the computation:
\begin{equation}\label{sibu} 
   |({\rm rem}_{15})_{ij}|
   \le C\,\frac{c^4\lambda}{G}\,\delta_T^{(0)}\,{\|\hat{h}\|}_{L^\infty_x},
\end{equation}  
\begin{equation}\label{sibud} 
   |\partial_t ({\rm rem}_{15})_{ij}|\le C\,\frac{c^5}{G}\,
   (\delta_T^{(0)}+\delta_T^{(1)})\,
   ({\|\hat{h}\|}_{L^\infty_x}+\lambda {\|\partial_{x^\gamma}\hat{h}\|}_{L^\infty_x}),
\end{equation}   
\begin{equation}\label{reicht} 
   |({\rm rem}_{16})_{ij}|\le C\,\frac{c^4\lambda}{G}\,
   (\delta_T^{(0)}+\delta_T^{(1)})\,
   \Big({\|\hat{h}\|}_{L^\infty_x}
   +\lambda {\|\partial_{x^\gamma}\hat{h}\|}_{L^\infty_x}
   +\lambda^2 {\|\partial^2_{x^\beta x^\gamma}\hat{h}\|}_{L^\infty_x}\Big) \ ,
\end{equation}
\begin{equation}\label{reichtd} 
   |\partial_t ({\rm rem}_{16})_{ij}| 
   \le C\,\frac{c^5}{G}\,(\delta_T^{(0)}+\delta_T^{(1)}+\delta_T^{(2)})
   \,\Big({\|\hat{h}\|}_{L^\infty_x}
   +\lambda {\|\partial_{x^\gamma}\hat{h}\|}_{L^\infty_x}
   +\lambda^2 {\|\partial^2_{x^\beta x^\gamma}\hat{h}\|}_{L^\infty_x}
   +\lambda^3 {\|\partial^3_{x^\alpha x^\beta x^\gamma}\hat{h}\|}_{L^\infty_x}\Big),
\end{equation}   
\begin{equation}
    |({\rm rem}_{17})^{ij}|\le C\,\frac{c^4\lambda^3}{G}
   \,\delta_T^{(0)}\,{\|\hat{H}\|}_{L^\infty_x},
\end{equation}
\begin{equation}\label{kohue} 
   |\partial_t ({\rm rem}_{17})^{ij}|
   \le C\,\frac{c^5\lambda^2}{G}\,(\delta_T^{(0)}+\delta_T^{(1)})\,
   ({\|\hat{H}\|}_{L^\infty_x}+\lambda {\|\partial_{x^\gamma}\hat{H}\|}_{L^\infty_x}),
\end{equation}  
\begin{equation}\label{awo} 
   |\partial_t^2 ({\rm rem}_{17})^{ij}|
   \le C\,\frac{c^6\lambda}{G}\,(\delta_T^{(0)}+\delta_T^{(1)}+\delta_T^{(2)})\,
   \Big({\|\hat{H}\|}_{L^\infty_x}+\lambda {\|\partial_{x^\gamma}\hat{H}\|}_{L^\infty_x}
   +\lambda^2 {\|\partial^2_{x^\beta x^\gamma}\hat{H}\|}_{L^\infty_x}\Big),
\end{equation}  
\begin{equation}\label{awod}  
   |\partial_t^3 ({\rm rem}_{17})^{ij}|
   \le C\,\frac{c^7}{G}\,\Big(\sum_{j=0}^3 \delta_T^{(j)}\Big)\,
   \Big({\|\hat{H}\|}_{L^\infty_x}+\lambda {\|\partial_{x^\gamma}\hat{H}\|}_{L^\infty_x}
   +\lambda^2 {\|\partial^2_{x^\beta x^\gamma}\hat{H}\|}_{L^\infty_x}
   +\lambda^3 {\|\partial^3_{x^\alpha x^\beta x^\gamma}\hat{H}\|}_{L^\infty_x}\Big),
\end{equation} 
\begin{equation}
    |({\rm rem}_{18})_{ij}|
   \le C\,\frac{c^4\lambda}{G}\,(\delta_T^{(0)}+\delta_T^{(1)}+\delta_T^{(2)})\,
   \Big(\sum_{|\alpha|\le 2}\lambda^{|\alpha|}{\|\partial^\alpha\hat{h}\|}_{L^\infty_x}
   +\sum_{|\alpha|\le 2}\lambda^{|\alpha|}{\|\partial^\alpha\hat{H}\|}_{L^\infty_x}\Big),
\end{equation}
\begin{equation}\label{maumau18} 
   |\partial_t ({\rm rem}_{18})_{ij}|
   \le C\,\frac{c^5}{G}\,\Big(\sum_{j=0}^3\delta_T^{(j)}\Big)\,
   \Big(\sum_{|\alpha|\le 3}\lambda^{|\alpha|}{\|\partial^\alpha\hat{h}\|}_{L^\infty_x}
   +\sum_{|\alpha|\le 3}\lambda^{|\alpha|}{\|\partial^\alpha\hat{H}\|}_{L^\infty_x}\Big),
\end{equation}  
\begin{equation}\label{feinschm} 
   |({\rm rem}_{19})^\nu|
   \le C\,\frac{c^5\lambda}{G}\,\delta_T^{(0)}
   \,{\|\partial_{x^\gamma}\hat{h}\|}_{L^\infty_x},
\end{equation}
\begin{equation}\label{maumau20}  
   |{\rm rem}_{20}|
   \le C\,\frac{c^5}{G}\,(\delta_T^{(0)}+\delta_T^{(1)})
   \,({\|\hat{h}\|}_{L^\infty_x}+\lambda {\|\partial_{x^\gamma}\hat{h}\|}_{L^\infty_x}),
\end{equation}  
\begin{equation}\label{eq_rem21c}
    |({\rm rem}_{21})^j|
    \le C\,\frac{c^5}{G}\,(\delta_T^{(0)}+\delta_T^{(1)})\,
    ({\|\hat{h}\|}_{L^\infty_x}+\lambda {\|\partial_{x^\gamma}\hat{h}\|}_{L^\infty_x}).,
\end{equation}
\begin{equation}\label{johe} 
   |({\rm rem}_{22})^i|
   \le C\,\frac{c^6}{G}\,(\delta_T^{(0)}+\delta_T^{(1)})
   \,(\lambda {\|\partial_{x^\gamma}\hat{h}\|}_{L^\infty_x}
   +\lambda^2 {\|\partial^2_{x^\beta x^\gamma}\hat{h}\|}_{L^\infty_x}),
\end{equation}  
\begin{equation}
   |({\rm rem}_{23})^i|
   \le C\,\frac{c^4\lambda^2}{G}\,\delta_T^{(0)}\,
   {\|\hat{h}\|}_{L^\infty_x},
\end{equation}
\begin{equation}
    |\partial_t ({\rm rem}_{23})^i|
   \le C\,\frac{c^5\lambda}{G}\,(\delta_T^{(0)}+\delta_T^{(1)})\,
   (\lambda {\|\hat{h}\|}_{L^\infty_x}+\lambda^2 {\|\partial_{x^\gamma}\hat{h}\|}_{L^\infty_x}),
\end{equation}
\begin{equation}\label{lihe} 
   |\partial^2_t ({\rm rem}_{23})^i|
   \le C\,\frac{c^6}{G}\,(\delta_T^{(0)}+\delta_T^{(1)}+\delta_T^{(2)})\,
   \Big({\|\hat{h}\|}_{L^\infty_x}+\lambda {\|\partial_{x^\gamma}\hat{h}\|}_{L^\infty_x}
   +\lambda^2 {\|\partial^2_{x^\beta x^\gamma}\hat{h}\|}_{L^\infty_x}\Big),
\end{equation} 
\begin{equation}\label{maumau24} 
   |({\rm rem}_{24})^i|
   \le C\,\frac{c^6}{G}\,(\delta_T^{(0)}+\delta_T^{(1)}+\delta_T^{(2)})\,
   \Big({\|\hat{h}\|}_{L^\infty_x}+\lambda {\|\partial_{x^\gamma}\hat{h}\|}_{L^\infty_x}
   +\lambda^2 {\|\partial^2_{x^\beta x^\gamma}\hat{h}\|}_{L^\infty_x}\Big),
\end{equation}  
\begin{equation}
    |({\rm rem}_{25})^{jm}|
   \le C\,\frac{c^4\lambda^2}{G}\,(\delta_T^{(0)}+\delta_T^{(1)})
   \,\Big(\lambda {\|\partial_{x^\gamma}\hat{h}\|}_{L^\infty_x} 
   +{\|\hat{H}\|}_{L^\infty_x}+\lambda {\|\partial_{x^\gamma}\hat{H}\|}_{L^\infty_x}\Big),
\end{equation}
\begin{equation}\label{waysho} 
   |\partial_t ({\rm rem}_{25})^{jm}|
   \le C\,\frac{c^5\lambda}{G}\,(\delta_T^{(0)}+\delta_T^{(1)}+\delta_T^{(2)})\,
   \Big(\sum_{1\le |\alpha|\le 2}\lambda^{|\alpha|}{\|\partial^\alpha\hat{h}\|}_{L^\infty_x}
   +\sum_{|\alpha|\le 2}\lambda^{|\alpha|}{\|\partial^\alpha\hat{H}\|}_{L^\infty_x}\Big),
\end{equation}   
\begin{equation}\label{wayshod} 
   |\partial^2_t ({\rm rem}_{25})^{jm}|
   \le C\,\frac{c^6}{G}\,\Big(\sum_{j=0}^3\delta_T^{(j)}\Big)\,
   \Big(\sum_{1\le |\alpha|\le 3}\lambda^{|\alpha|}{\|\partial^\alpha\hat{h}\|}_{L^\infty_x}
   +\sum_{|\alpha|\le 3}\lambda^{|\alpha|}{\|\partial^\alpha\hat{H}\|}_{L^\infty_x}\Big),
\end{equation}
\begin{equation}
    |({\rm rem}_{26})^{jm}|
   \le C\,\frac{c^5\lambda}{G}\,(\delta_T^{(0)}+\delta_T^{(1)}+\delta_T^{(2)})
   \,\Big(\sum_{1\le |\alpha|\le 2}\lambda^{|\alpha|}{\|\partial^\alpha\hat{h}\|}_{L^\infty_x}
   +\sum_{|\alpha|\le 2}\lambda^{|\alpha|}{\|\partial^\alpha\hat{H}\|}_{L^\infty_x}\Big),
\end{equation}
\begin{equation}\label{hetask} 
   |\partial_t ({\rm rem}_{26})^{jm}|
   \le C\,\frac{c^6}{G}\,\Big(\sum_{j=0}^3\delta_T^{(j)}\Big)\,
   \Big(\sum_{1\le |\alpha|\le 3}\lambda^{|\alpha|}{\|\partial^\alpha\hat{h}\|}_{L^\infty_x}
   +\sum_{|\alpha|\le 3}\lambda^{|\alpha|}{\|\partial^\alpha\hat{H}\|}_{L^\infty_x}\Big),
\end{equation}  
\begin{equation}
    |\tensor{({\rm rem}_{27})}{_j^m}|
   \le C\,\frac{c^4\lambda^2}{G}\,\delta_T^{(0)}\,
   {\|\hat{h}\|}_{L^\infty_x},
\end{equation}
\begin{equation}
    |\partial_t\tensor{({\rm rem}_{27})}{_j^m}|
   \le C\,\frac{c^5\lambda}{G}\,(\delta_T^{(0)}+\delta_T^{(1)})\,
   ({\|\hat{h}\|}_{L^\infty_x}+\lambda {\|\partial_{x^\gamma}\hat{h}\|}_{L^\infty_x}),
\end{equation}
\begin{equation}\label{notasw} 
   |\partial_t^2\tensor{({\rm rem}_{27})}{_j^m}|
   \le C\,\frac{c^6}{G}\,(\delta_T^{(0)}+\delta_T^{(1)}+\delta_T^{(2)})\,
   \Big({\|\hat{h}\|}_{L^\infty_x}+\lambda {\|\partial_{x^\gamma}\hat{h}\|}_{L^\infty_x}
   +\lambda^2 {\|\partial^2_{x^\beta x^\gamma}\hat{h}\|}_{L^\infty_x}\Big),
\end{equation}  
\begin{equation}\label{maumau28} 
   |({\rm rem}_{28})^{jm}|
   \le C\,\frac{c^6}{G}\,\Big(\sum_{j=0}^3\delta_T^{(j)}\Big)\,
   \Big(\sum_{|\alpha|\le 3}\lambda^{|\alpha|}{\|\partial^\alpha\hat{h}\|}_{L^\infty_x}
   +\sum_{|\alpha|\le 3}\lambda^{|\alpha|}{\|\partial^\alpha\hat{H}\|}_{L^\infty_x}\Big).
\end{equation} 

\subsection{Remainders 4 and 29: (\ref{eq_rem4}) and (\ref{eq_rem29})}
\label{subsubsec_rem29}

From (\ref{eq_rem4}) we recall that 
\begin{eqnarray*} 
   {({\rm rem}_4)}_{\alpha\beta} 
   & = & \frac{1}{2}\,\tilde{P}(g; \partial_{x^\alpha} h, \partial_{x^\beta} h)
   +\frac{1}{2}\,\tilde{Q}_{\alpha\beta}(g; \partial h, \partial h)
   -\frac{1}{2}\,H^{\alpha'\beta'}
   \,\partial^2_{x^{\alpha'} x^{\beta'}}\,h_{\alpha\beta}
   \\ & & -\,\frac{1}{4}\,(\eta^{\alpha'\beta'}+H^{\alpha'\beta'})
   \bigg(\tilde{P}(g; \partial_{x^{\alpha'}} h, \partial_{x^{\beta'}} h)
   +\tilde{Q}_{\alpha'\beta'}(g; \partial h, \partial h)
   -H^{\gamma\delta}\,\partial^2_{x^\gamma x^\delta}\,h_{\alpha'\beta'}\bigg)\,g_{\alpha\beta}
   \\ & & +\,\frac{1}{4}\,\Big(H^{\alpha'\beta'}\Box\,h_{\alpha'\beta'}\Big)\,g_{\alpha\beta}
   +\frac{1}{4}\,(\Box\,h)\,h_{\alpha\beta}
\end{eqnarray*} 
for $\tilde{P}$ and $\tilde{Q}$ given by (\ref{tildP}) and (\ref{tildQ}), respectively. Since $h=\eta^{\alpha\beta} h_{\alpha\beta}$ and $\Box=\eta^{\alpha\beta}\,\partial^2_{x^\alpha x^\beta}$, we see that ${({\rm rem}_4)}_{\alpha\beta}$ is a sum of terms of the following form: 
\[ \phi(\partial_{x^\mu} h_{\alpha\beta})\,(\partial_{x^\nu} h_{\alpha'\beta'}),
   \,\,\phi\,H^{\alpha\beta}\,(\partial^2_{{x^\mu}{x^\nu}} h_{\alpha'\beta'}),
   \,\,\phi\,h_{\alpha\beta}\,(\partial^2_{{x^\mu}{x^\nu}} h^{\alpha'\beta'}), \] 
where $\phi = \phi(t,x)$ is a function such that $\sum_{|\alpha|\le 1}\lambda^{|\alpha|}|\partial^\alpha\phi|\le C$. Then (\ref{eq_rem29}) defines
\[ {({\rm rem}_{29})}_{\alpha\beta}(t, x)   
   =-\frac{1}{2\pi}\,\int_{\R^3}\frac{1}{|x-y|}
   \,{({\rm rem}_4)}_{\alpha\beta}(t-|x-y|, y)\,dy . \] 
It is here that we need the NIR condition~\eqref{Blanch}. Assume, as in the statement of Theorem~\eqref{mainthm}, that
\begin{equation}\label{param} 
   |t-c^{-1}|x||\le\lambda c^{-1},\quad |x|\ge\max\{12\lambda, c{\cal T}, 2{\cal S}\},
   \quad {\cal T}\ge 2\lambda c^{-1} ,\quad {\cal S}\ge\max\{2c{\cal T}, 6\lambda \}.
\end{equation}  
In addition, due to (\ref{Blanch}) and $\delta_h=C_0\eps_h$ we may assume that 
\begin{equation}\label{Blanch_2} 
   \hat{h}(t, x)=p(x)\quad\mbox{for}\quad t\le -{\cal T},
   \quad |\partial^\kappa p(x)|\le\frac{1}{\lambda^{|\alpha|}}
   \,\frac{C\delta_h}{|x|^{1+|\kappa|}},
   \quad |x|\ge {\cal S},\,\,|\kappa|\le 3; 
\end{equation} 
recall that $\hat{h}=(h_{\alpha\beta})$ as a matrix. We will now establish the following result.
\medskip 

\begin{lemma}\label{hinteg} 
Let $\gamma_0\in [\frac{3}{4}, 1)$ and $\delta_h>0$ 
be such that $\gamma_0+\delta_h\le 1$ and let (\ref{param}) 
and (\ref{Blanch_2}) be satisfied. Then 
\[ \lambda\,\bigg|\,\partial_{x^\alpha}\int_{\R^3}
   \frac{1}{|x-y|}\,\zeta(t-c^{-1}|x-y|, y)\,dy\bigg|
   \le\frac{C\delta_h^2}{(\lambda^{-1}|x|)^{\gamma_0-\delta_h}} \] 
for each $\alpha=0, 1, 2, 3$, where $\zeta$ can stand for any of the following expressions:
\begin{align*}
    \zeta &= \phi(\partial_{x^\mu} h_{\alpha\beta})
   \,(\partial_{x^\nu} h_{\alpha'\beta'}) , \\
   \zeta &= \phi\,H^{\alpha\beta}\,(\partial^2_{{x^\mu}{x^\nu}} h_{\alpha'\beta'}) , \ \text{or} \\
   \zeta &= \phi\,h_{\alpha\beta}
   \,(\partial^2_{{x^\mu}{x^\nu}} h^{\alpha'\beta'}) ,
\end{align*}
with $\mu, \nu, \alpha, \beta, \alpha', \beta'=0, 1, 2, 3$ 
and $\phi$ being a function such that $\sum_{|\alpha|\le 1}\lambda^{|\alpha|} |\partial^\alpha\phi|\le C$. In particular,
\begin{equation}\label{mdue} 
   \lambda\,|\partial_{x^\gamma} {({\rm rem}_{29})}_{\alpha\beta}(t, x)|
   \le\frac{C\delta_h^2}{(\lambda^{-1}|x|)^{\gamma_0-\delta_h}},
   \quad\alpha, \beta, \gamma=0, 1, 2, 3.
\end{equation}         
\end{lemma} 
{\bf Proof\,:} Throughout this proof we will simply write 
$h$ for any $h_{\alpha\beta}$ and similarly $H$ for any $H^{\alpha\beta}$. 
Also for simplicity we will let $G=1$, $\lambda=1$ and $c=1$ for the argument, 
so that in particular $x^0=ct=t$. We are going to apply Corollary \ref{I3sum}, 
(\ref{I2sum}) and (\ref{I1sum}) to appropriate 
combinations of terms, where we take $\gamma=\gamma_0$ and $\rho=\delta=\delta_h$. 
\smallskip 

\noindent 
\underline{Case 1:} $\zeta=(\partial_{x^\mu} h)(\partial_{x^\nu} h)$. 
Let $\Psi=\partial_{x^\mu} h$ and $\tilde{\Psi}=h$, 
so that $\zeta=\Psi\partial_{x^\nu}\tilde{\Psi}$. 
(i) assumptions for $I_1$: it follows from (\ref{hinfty3}) that (\ref{I1assum}) holds. 
(ii) assumptions for $I_2$: from (\ref{Blanch_2}) we deduce that 
$\Psi(t, x)=\partial_{x^\mu} p(x)$ for $t\le -{\cal T}$ and furthermore 
$|\partial^\alpha\partial_{x^\mu} p(x)|\le C\delta_h |x|^{-(2+|\alpha|)}$ 
for $|x|\ge {\cal S}$ and $|\alpha|\le 2$. 
Similarly, $\tilde{\Psi}(x)=p(x)$ for $t\le -{\cal T}$ and in addition 
$|\partial^\alpha p(x)|\le C\delta_h |x|^{-(1+|\alpha|)}$ for $|x|\ge {\cal S}$ and $|\alpha|\le 2$. 
Therefore the conditions (\ref{I2psi}) and (\ref{I2psitil}) are verified. 
(iii) assumptions for $I_3$: this requires a more careful analysis. Writing $h=h^0+h^1$, 
we decompose 
\begin{equation}\label{u1dec} 
   \zeta=\psi^0\partial_{x^\nu}\tilde{\psi}^0+\psi^0\partial_{x^\nu}\tilde{\psi}^1
   +\psi^1\partial_{x^\nu}\tilde{\psi}^0+\psi^1\partial_{x^\nu}\tilde{\psi}^1,
\end{equation}  
where 
\[ \psi^0=\partial_{x^\mu} h^0,\,\,\psi^1=\partial_{x^\mu} h^1,\,\,
   \tilde{\psi}^0=h^0,\,\,\tilde{\psi}^1=h^1. \] 
Now we consider each term in (\ref{u1dec}) separately 
and check the assumptions that allow us to bound $I_3$. 
(a) For $\psi^0\partial_{x^\nu}\tilde{\psi}^0$ we will show that 
(\ref{00b})-(\ref{00c}) holds. In fact this is a direct consequence 
of (\ref{00bth}) and (\ref{00cth}). 
(b) For $\psi^0\partial_{x^\nu}\tilde{\psi}^1$
we note that (\ref{0110b}) is satisfied, 
by (\ref{0110bkb}). 
(c) For $\psi^1\partial_{x^\nu}\tilde{\psi}^0$ 
we remark that (\ref{0110bsb}) yields (\ref{0110b}). 
(d) For $\psi^1\partial_{x^\nu}\tilde{\psi}^1$ we have (\ref{11a}) 
due to (\ref{11acpt}). 
\smallskip 

\noindent 
\underline{Case 2:} $\zeta=h(\partial^2_{x^\mu x^\nu} h)$. 
We set $\Psi=h$ and $\tilde{\Psi}=\partial_{x^\mu} h$ 
to get $\zeta=\Psi\partial_{x^\nu}\tilde{\Psi}$.  
(i) assumptions for $I_1$: it follows from (\ref{hinfty3}) that (\ref{I1assum}) holds. 
(ii) assumptions for $I_2$: from (\ref{Blanch_2}) we deduce that 
$\Psi(t, x)=p(x)$ for $t\le -{\cal T}$ and furthermore 
$|\partial^\alpha p(x)|\le C\delta_h |x|^{-(1+|\alpha|)}$ 
for $|x|\ge {\cal S}$ and $|\alpha|\le 2$. 
Similarly, $\tilde{\Psi}(x)=\partial_{x^\mu} p(x)$ for $t\le -{\cal T}$ and in addition 
$|\partial^\alpha\partial_{x^\mu} p(x)|\le C\delta_h |x|^{-(2+|\alpha|)}$ 
for $|x|\ge {\cal S}$ and $|\alpha|\le 2$. 
Therefore the conditions (\ref{I2psi}) and (\ref{I2psitil}) are verified. 
(iii) assumptions for $I_3$: once again this requires a more careful analysis. 
Here we split up 
\[ \zeta=\psi^0\partial_{x^\nu}\tilde{\psi}^0+\psi^0\partial_{x^\nu}\tilde{\psi}^1
   +\psi^1\partial_{x^\nu}\tilde{\psi}^0+\psi^1\partial_{x^\nu}\tilde{\psi}^1, \]   
where 
\[ \psi^0=h^0,\,\,\psi^1=h^1,\,\,
   \tilde{\psi}^0=\partial_{x^\mu} h^0,\,\,\tilde{\psi}^1=\partial_{x^\mu} h^1. \] 
(a) For $\psi^0\partial_{x^\nu}\tilde{\psi}^0$ we need to show that 
(\ref{00b})-(\ref{00c}) is verified. This is in fact due to (\ref{00bco}) and (\ref{00cco}). 
(b) For $\psi^0\partial_{x^\nu}\tilde{\psi}^1$ we can use (\ref{0110brb}) 
to get (\ref{0110b}). 
(c) For $\psi^1\partial_{x^\nu}\tilde{\psi}^0$ we obtain (\ref{0110b}) from (\ref{0110blb}). 
(d) For $\psi^1\partial_{x^\nu}\tilde{\psi}^1$ we note that (\ref{11a}) 
is a consequence of (\ref{11ast}). 
\smallskip 

\noindent 
\underline{Case 3:} $\zeta=H(\partial^2_{x^\mu x^\nu} h)$. 
First recall that $h=h_{\alpha\beta}$ and $H=H^{\gamma\delta}$ denote 
one of the components of $h$ and $H$, respectively. 
For the bounds on $H$ we can rely on Section \ref{boH}, 
where we wrote $Z=\hat{H}\hat{m}$, $X=\hat{m}^{-1}\hat{h}$ 
and $Y=(I+X)^{-1}=(I+\hat{m}^{-1}\hat{h})^{-1}$ 
for the corresponding matrices $\hat{H}$ for $H$, $\hat{m}$ 
for the Minkowski metric and $\hat{h}$ for $h$. Then $Z=-YX$ yields  
\begin{equation}\label{ostl} 
   \hat{H}=-YX\hat{m}^{-1}=-Y\hat{m}^{-1}\hat{h}\hat{m}^{-1},
\end{equation}  
For any derivative therefore 
\begin{equation}\label{hoerma} 
   (\partial_{x^\lambda}\hat{H})
   =-(\partial_{x^\lambda} Y)\hat{m}^{-1}\hat{h}\hat{m}^{-1}
   -Y\hat{m}^{-1}(\partial_{x^\lambda}\hat{h})\hat{m}^{-1}.
\end{equation}  
From (\ref{ostl}) and (\ref{hoerma}) together with Section \ref{boH} 
we deduce in particular that, symbolically, 
\begin{equation}\label{romit} 
   |H|\le C|h|\quad\mbox{and}\quad |\partial_{x^\lambda}H|\le C(|h|+|\partial_{x^\lambda}h|) 
\end{equation} 
holds. Now we set $\Psi=H$ and $\tilde{\Psi}=\partial_{x^\mu} h$ 
to get $\zeta=\Psi\partial_{x^\nu}\tilde{\Psi}$. 
(i) assumptions for $I_1$: from (\ref{romit}) and (\ref{hinfty3}) 
we find $|H|+|\partial_{x^\lambda}H|\le C$. 
Hence (\ref{I1assum}) follows from (\ref{hinfty3}). 
(ii) assumptions for $I_2$: for $\tilde{\Psi}$ we have (\ref{I2psitil}) with $m=1$ by (\ref{Blanch_2}). 
Concerning (\ref{I2psi}), since $Y$ is only depending on $\hat{h}$, (\ref{ostl}) shows 
that the Blanchet condition (\ref{Blanch_2}) for $h$ implies the Blanchet condition also for $H$; 
symbolically, we may write $H=\frac{h}{1+h}$. Thus (\ref{I2psi}) is verified for $l=0$. 
(iii) assumptions for $I_3$: we decompose 
\[ \zeta=\psi^0\partial_{x^\nu}\tilde{\psi}^0+\psi^0\partial_{x^\nu}\tilde{\psi}^1
   +\psi^1\partial_{x^\nu}\tilde{\psi}^0+\psi^1\partial_{x^\nu}\tilde{\psi}^1, \]   
where 
\[ \psi^0=H^0,\,\,\psi^1=H^1,\,\,
   \tilde{\psi}^0=\partial_{x^\mu} h^0,\,\,\tilde{\psi}^1=\partial_{x^\mu} h^1, \] 
and we let $H^0$ and $H^1$ be defined via 
\[ \hat{H}^0=-Y\hat{m}^{-1}\hat{h}^0\hat{m}^{-1}
   \quad\mbox{and}\quad\hat{H}^1=-Y\hat{m}^{-1}\hat{h}^1\hat{m}^{-1}; \] 
note that $Y$ still depends on $h=h^0+h^1$. Then 
\begin{eqnarray*} 
   \partial_{x^\alpha}\hat{H}^0 
   & = & -Y\hat{m}^{-1}(\partial_{x^\alpha}\hat{h}^0)\hat{m}^{-1}
   -(\partial_{x^\alpha}Y)\hat{m}^{-1}\hat{h}^0\hat{m}^{-1}, 
   \\ \partial_{x^\alpha}\hat{H}^1 
   & = & -Y\hat{m}^{-1}(\partial_{x^\alpha}\hat{h}^1)\hat{m}^{-1}
   -(\partial_{x^\alpha}Y)\hat{m}^{-1}\hat{h}^1\hat{m}^{-1}.   
\end{eqnarray*} 
(a) For $\psi^0\partial_{x^\nu}\tilde{\psi}^0$ we need to show that 
(\ref{00b})--(\ref{00c}) is verified. Firstly, 
\[ |\psi^0\partial_{x^\nu}\tilde{\psi}^0|=|H^0\partial^2_{x^\mu x^\nu} h^0|
   \le |h^0\partial^2_{x^\mu x^\nu} h^0|
   \le C\delta_h^2\,{\bf 1}_{\{r\ge\frac{1}{2}(1+|t|)\}}\,\frac{1}{|x|^4}, \] 
by Section \ref{boH} and Section \ref{faten}. Secondly, 
\begin{eqnarray*}
   |\partial_{x^\alpha}(\psi^0\partial_{x^\nu}\tilde{\psi}^0)|
   & = & |\partial_{x^\alpha}(H^0\partial^2_{x^\mu x^\nu} h^0)|
   \\ & \le & |H^0\partial^3_{x^\alpha x^\mu x^\nu} h^0|
   +|(\partial_{x^\alpha}H^0)\partial^2_{x^\mu x^\nu} h^0|
   \\ & \le & C|h^0||\partial^3_{x^\alpha x^\mu x^\nu} h^0|
   +C(|h^0|+|\partial_{x^\alpha}h^0|)|\partial^2_{x^\mu x^\nu} h^0|
   \\ & \le & C\delta_h^2\,{\bf 1}_{\{r\ge\frac{1}{2}(1+|t|)\}}\,\frac{1}{|x|^4}, 
\end{eqnarray*} 
again by Section \ref{boH} and Section \ref{faten}. 
(b) For $\psi^0\partial_{x^\nu}\tilde{\psi}^1$ we estimate 
\begin{eqnarray*} 
   |\psi^0\partial_{x^\nu}\tilde{\psi}^1|
   +|\partial_{x^\alpha}(\psi^0\partial_{x^\nu}\tilde{\psi}^1)| 
   & \le & |H^0\partial^2_{x^\nu x^\mu} h^1|
   +|H^0\partial^3_{x^\alpha x^\nu x^\mu} h^1|
   +|(\partial_{x^\alpha}H^0)\partial^2_{x^\nu x^\mu} h^1| 
   \\ & \le & C|h^0||\partial^2_{x^\nu x^\mu} h^1|
   +C|h^0||\partial^3_{x^\alpha x^\nu x^\mu} h^1|
   \\ & & +\,C(|h^0|+|\partial_{x^\alpha}h^0|)|\partial^2_{x^\nu x^\mu} h^1| 
   \\ & \le & C\delta_h^2\,{\bf 1}_{\{r\ge\frac{1}{2}(1+|t|)\}}
   \,\frac{(1+|t|)^{\delta_h}}{(1+|t|+|x|)(1+(|x|-|t|)_+)^{\gamma_0}}
   \,\frac{1}{|x|},
\end{eqnarray*} 
by means of the bounds on $h^0$ and $h^1$. Therefore (\ref{0110b}) is verified. 
(c) Similarly, for $\psi^1\partial_{x^\nu}\tilde{\psi}^0$ we find 
\begin{eqnarray*} 
   |\psi^1\partial_{x^\nu}\tilde{\psi}^0|
   +|\partial_{x^\alpha}(\psi^1\partial_{x^\nu}\tilde{\psi}^0)| 
   & \le & |H^1\partial^2_{x^\nu x^\mu}h^0|
   +|H^1\partial^3_{x^\alpha x^\nu x^\mu}h^0| 
   +|(\partial_{x^\alpha}H^1)\partial^2_{x^\nu x^\mu}h^0| 
   \\ & \le & C|h^1||\partial^2_{x^\nu x^\mu}h^0|
   +C|h^1||\partial^3_{x^\alpha x^\nu x^\mu}h^0| 
   \\ & & +\,C(|h^1|+|\partial_{x^\alpha} h^1|)|\partial^2_{x^\nu x^\mu}h^0| 
   \\ & \le & C\delta_h^2\,{\bf 1}_{\{r\ge\frac{1}{2}(1+|t|)\}}
   \,\frac{(1+|t|)^{\delta_h}}{(1+|t|+|x|)(1+(|x|-|t|)_+)^{\gamma_0}}
   \,\frac{1}{|x|^3},
\end{eqnarray*} 
and in particular (\ref{0110b}) holds. 
(d) For $\psi^1\partial_{x^\nu}\tilde{\psi}^1$ we deduce, from (\ref{LT116}), 
\begin{eqnarray*} 
   |\psi^1\partial_{x^\lambda}\tilde{\psi}^1|
   +|\partial_{x^\alpha}(\psi^1\partial_{x^\lambda}\tilde{\psi}^1)|
   & \le & |H^1\partial^2_{x^\lambda x^\mu} h^1|
   +|H^1\partial^3_{x^\alpha x^\lambda x^\mu} h^1|
   +|(\partial_{x^\alpha} H^1)\partial^2_{x^\lambda x^\mu} h^1|
   \\ & \le & C|h^1||\partial^2_{x^\lambda x^\mu} h^1|
   +C|h^1||\partial^3_{x^\alpha x^\lambda x^\mu} h^1|
   \\ & & +C(|h^1|+|\partial_{x^\alpha} h^1|)|\partial^2_{x^\lambda x^\mu} h^1|
   \\ & \le & C\delta_h^2\,\frac{(1+|t|)^{2\delta_h}}{(1+|t|+|x|)^2(1+(|x|-|t|)_+)^{2\gamma_0}}, 
\end{eqnarray*} 
which is (\ref{11a}). This completes the proof of Lemma \ref{hinteg}. 
{\hfill$\Box$}\bigskip 

\begin{cor}\label{H49cor} We have 
\begin{equation}\label{H49} 
   |H^{\mu\nu}(x^0, x)|+\lambda |\partial_{x^\alpha} H^{\mu\nu}(x^0, x)|
   \le C\delta_h\,\frac{(1+\lambda^{-1}|x^0|)^{\delta_h}}{1+\lambda^{-1}|x^0|+\lambda^{-1}|x|}. 
\end{equation} 
\end{cor} 
{\bf Proof\,:} This follows from (\ref{romit}) and (\ref{h49}) 
by re-introducing the constants.  
{\hfill$\Box$}\bigskip 

\subsection{Remainders 30 through 37}

The bound (\ref{mdue}) just obtained on $\rem{29}$ immediately implies for $\rem{30}$ defined in~\eqref{That}
\begin{equation}\label{mdue30} 
   \lambda\,|\partial_{x^\gamma} {({\rm rem}_{30})}^{\alpha\beta}(t, x)|
   \le\frac{C\delta_h^2}{(\lambda^{-1}|x|)^{\gamma_0-\delta_h}},\quad\alpha, \beta, \gamma=0, 1, 2, 3,
\end{equation}   
which in turn gives for $\rem{31}$ defined in (\ref{eq_rem31}) (using $c^{-1}\partial_t=\partial_{x^0}$)
\begin{equation}\label{mdue31} 
   |{({\rm rem}_{31})}^{\alpha\beta}(t, x)|
   \le\frac{1}{\lambda}\,\frac{C\delta_h^2}{(\lambda^{-1}|x|)^{\gamma_0-\delta_h}},
   \quad\alpha, \beta=0, 1, 2, 3.
\end{equation}
Next recall $K_\ast$ from (\ref{R0def}). If $|\bar{x}|\le \lambda K_\ast$ and $|x|\ge 2\lambda K_\ast$, then 
\[ |x-\bar{x}|\ge |x|-|\bar{x}|\ge\frac{1}{2}\,|x|. \] 
Therefore (\ref{mdue30}) and $\hat{T}^{\alpha\beta}
=\eta^{\alpha\alpha'}\eta^{\beta\beta'} T_{\alpha'\beta'}$ from (\ref{That}) 
together with the boundedness condition~\eqref{fuvor0} shows that 
\begin{eqnarray}\label{mdue32}  
   |\tensor{({\rm rem}_{32})}{^\alpha^\beta^k}(t, x)|
   & \le & C\,\frac{G}{c^4}\,|x|\int_{|\bar{x}|\le\lambda K_\ast}
   \frac{1}{|x-\bar{x}|^3}\,|\hat{T}^{\alpha\beta}(t-c^{-1}|x-\bar{x}|, \bar{x})|\,d\bar{x}
   +\frac{1}{\lambda}\,\frac{C\delta_h^2}{(\lambda^{-1}|x|)^{\gamma_0-\delta_h}}
   \nonumber
   \\ & \le & C\,\frac{\lambda}{|x|^2}\,\delta_T^{(0)}
   +\frac{1}{\lambda}\,\frac{C\delta_h^2}{(\lambda^{-1}|x|)^{\gamma_0-\delta_h}}
\end{eqnarray} 
for $\alpha, \beta=0, 1, 2, 3$ and $k=1, 2, 3$. We also remark that, since $|ct-|x-\bar{x}||\le ||x-\bar{x}|-|x||+|ct-|x||\le\lambda(K_\ast+1)$, there would be no further gain from this term if some decay of the energy-momentum tensor in terms of $|x^0|$ were assumed in \eqref{fuvor0}.

Next we estimate $\rem{33}$. An inspection of the algebra prior to its definition in (\ref{eq_rem33}) implies that this term is given by
\begin{eqnarray*} 
   \lefteqn{{({\rm rem}_{33})}^{\alpha\beta}(t, x)}
   \\ & = & {({\rm rem}_{31})}^{\alpha\beta}(t, x)
   \\ & & +\,\frac{4G}{c^5}\int_{|\bar{x}|\le\lambda K_\ast}\bigg(\frac{1}{|x-\bar{x}|}
   \,\partial_t\hat{T}^{\alpha\beta}(t-c^{-1}|x-\bar{x}|, \bar{x})\,d\bar{x} 
   -\frac{1}{r}\,\partial_t\hat{T}^{\alpha\beta}(t-c^{-1}r, \bar{x})
   \\ & & \hspace{7em} 
   -\,\frac{(x\cdot\bar{x})}{cr^2}\,\partial^2_t\hat{T}^{\alpha\beta}(t-c^{-1}r, \bar{x})
   -\frac{(x\cdot\bar{x})^2}{2c^2r^3}\,\partial^3_t\hat{T}^{\alpha\beta}(t-c^{-1}r, \bar{x})
   \bigg)\,d\bar{x}, 
\end{eqnarray*} 
where $|u|\le\lambda c^{-1}$ and $|x|\ge 5\lambda K_\ast$. Hence using (\ref{mdue31}), 
and by Lemma \ref{mupol}(b) applied pointwise 
to $\psi=\partial_t\hat{T}^{\alpha\beta}(\cdot, \bar{x})$ 
(and the $\lambda$'s inserted), we obtain for $\alpha, \beta=0, 1, 2, 3$ with the additional help of the boundedness condition~\eqref{fuvor0}
\begin{eqnarray}\label{mdue33} 
   \lefteqn{|{({\rm rem}_{33})}^{\alpha\beta}(t, x)|}
   \nonumber
   \\ & \le & \frac{1}{\lambda}\,\frac{C\delta_h^2}{(\lambda^{-1}|x|)^{\gamma_0-\delta_h}}
   +C\,\frac{G\lambda^3}{c^5}
   \,r^{-2}\sum_{j=0}^2 c^{-j}\lambda^{j+1}{\|\partial_t^{j+1}\hat{T}^{\alpha\beta}\|}_{L^\infty_{tx}}
   +C\,\frac{G\lambda^3}{c^5}
   \,r^{-1}c^{-3}\lambda^3\,{\|\partial_t^4\hat{T}^{\alpha\beta}\|}_{L^\infty_{tx}}
   \nonumber 
   \\ & \le & \frac{1}{\lambda}\,\frac{C\delta_h^2}{(\lambda^{-1}|x|)^{\gamma_0-\delta_h}}
   +C\,\frac{G\lambda^3}{c^4 r^2}
   \Big(\lambda {\|\partial_{x^0} T_{\alpha\beta}\|}_{L^\infty_{tx}}
   +\lambda^2 {\|\partial_{x^0}^2 T_{\alpha\beta}\|}_{L^\infty_{tx}}
   +\lambda^3 {\|\partial_{x^0}^3 T_{\alpha\beta}\|}_{L^\infty_{tx}}\Big)
   \nonumber   
   \\ & & +\,C\,\frac{G\lambda^6}{c^4 r}
   \,{\|\partial_{x^0}^4 T_{\alpha\beta}\|}_{L^\infty_{tx}}
   \nonumber 
   \\ & \le & \frac{1}{\lambda}\,\frac{C\delta_h^2}{(\lambda^{-1}|x|)^{\gamma_0-\delta_h}}
   +C\,\frac{G\lambda^3}{c^4 r^2}
   \Big(\lambda\frac{c^4}{G\lambda^3}\,\delta_{T,\,\alpha\beta}^{(1), x^0}
   +\lambda^2\frac{c^4}{G\lambda^4}\,\delta_{T,\,\alpha\beta}^{(2), x^0}
   +\lambda^3\frac{c^4}{G\lambda^5}\,\delta_{T,\,\alpha\beta}^{(3), x^0}\Big)
   +C\frac{G\lambda^6}{c^4 r}\,\frac{c^4}{G\lambda^6}\,\delta_{T,\,\alpha\beta}^{(4), x^0}
   \nonumber 
   \\ & \le & \frac{1}{\lambda}\,\frac{C\delta_h^2}{(\lambda^{-1}|x|)^{\gamma_0-\delta_h}}
   +C\,\frac{\lambda}{r^2}\,(\delta_T^{(1)}+\delta_T^{(2)}+\delta_T^{(3)})
   +C\,\frac{1}{r}\,\delta_{T,\,\alpha\beta}^{(4), x^0}  
\end{eqnarray}

Similarly, the full expression for $\rem{34}$ given in \eqref{eq_rem34} is 
\begin{eqnarray*} 
   \lefteqn{{({\rm rem}_{34})}^{\alpha\beta k}(t, x)} 
   \\ & = & {({\rm rem}_{32})}^{\alpha\beta k}(t, x)
   -\frac{4G}{c^5}\int_{|\bar{x}|\le\lambda K_\ast} x^k\bigg(\frac{1}{|x-\bar{x}|^2}
   \,\partial_t\hat{T}^{\alpha\beta}(t-c^{-1}|x-\bar{x}|, \bar{x})
   -\frac{1}{r^2}\,\partial_t\hat{T}^{\alpha\beta}(t-c^{-1}r, \bar{x})
   \\ & & \hspace{13em} -\,\frac{(x\cdot\bar{x})}{cr^3}
   \,\partial^2_t\hat{T}^{\alpha\beta}(t-c^{-1}r, \bar{x})
   -\,\frac{(x\cdot\bar{x})^2}{2c^2 r^4}
   \,\partial^3_t\hat{T}^{\alpha\beta}(t-c^{-1}r, \bar{x})
   \bigg)\,d\bar{x}. 
\end{eqnarray*} 
In conjunction with (\ref{mdue32}) and Lemma \ref{mupol}(c), this leads to the following bound for $\alpha, \beta=0, 1, 2, 3$ and $k=1, 2, 3$:
\begin{eqnarray}\label{mdue34}  
   |{({\rm rem}_{34})}^{\alpha\beta k}(t, x)| 
   & \le & C\,\frac{\lambda}{|x|^2}\,\delta_T^{(0)}
   +\frac{1}{\lambda}\,\frac{C\delta_h^2}{(\lambda^{-1}|x|)^{\gamma_0-\delta_h}}
   \nonumber   
   \\ & & +\,C\,\frac{G\lambda^3}{c^5}\,|x|\,r^{-3}
   \sum_{j=0}^2 c^{-j}\lambda^{j+1}{\|\partial^{j+1}_t\hat{T}^{\alpha\beta}\|}_{L^\infty_{tx}}
   +C\,\frac{G\lambda^3}{c^5}\,|x|\,r^{-2}c^{-3}\lambda^3
   \,{\|\partial^4_t\hat{T}^{\alpha\beta}\|}_{L^\infty_{tx}}
   \nonumber
   \\ & \le & C\,\frac{\lambda}{|x|^2}\,\delta_T^{(0)}
   +\frac{1}{\lambda}\,\frac{C\delta_h^2}{(\lambda^{-1}|x|)^{\gamma_0-\delta_h}}
   \nonumber   
   \\ & & +\,C\,\frac{G\lambda^3}{c^4}\,r^{-2}
   \Big(\lambda {\|\partial_{x^0} T_{\alpha\beta}\|}_{L^\infty_{tx}}
   +\lambda^2 {\|\partial_{x^0}^2 T_{\alpha\beta}\|}_{L^\infty_{tx}}
   +\lambda^3 {\|\partial_{x^0}^3 T_{\alpha\beta}\|}_{L^\infty_{tx}}\Big)   
   \nonumber 
   \\ & & +\,C\,\frac{G\lambda^6}{c^4}\,r^{-1}
   \,{\|\partial_{x^0}^4\hat{T}^{\alpha\beta}\|}_{L^\infty_{tx}}
   \nonumber
   \\ & \le & C\,\frac{\lambda}{|x|^2}\,\delta_T^{(0)}
   +\frac{1}{\lambda}\,\frac{C\delta_h^2}{(\lambda^{-1}|x|)^{\gamma_0-\delta_h}}
   +C\,\frac{\lambda}{|x|^2}\,
   \,(\delta_T^{(1)}+\delta_T^{(2)}+\delta_T^{(3)})   
   +C\,\frac{1}{|x|}\,\delta_{T,\,\alpha\beta}^{(4), x^0}
   \nonumber
   \\ & \le & \frac{1}{\lambda}\,\frac{C\delta_h^2}{(\lambda^{-1}|x|)^{\gamma_0-\delta_h}}
   +C\,\frac{\lambda}{|x|^2}\,\Big(\sum_{j=0}^3\delta_T^{(j)}\Big)
   +C\,\frac{1}{|x|}\,\delta_{T,\,\alpha\beta}^{(4), x^0}
\end{eqnarray} 

Remainders 35 through 37, defined in (\ref{eq_rem35}), (\ref{eq_rem36}) and (\ref{eq_rem37}), 
can now be directly estimated using pointwise bounds for their integrands. As an example, for each $i, j=1, 2, 3$
\begin{eqnarray} \label{mdue35}
   \lefteqn{|{({\rm rem}_{35})}^{ij}(t, x)|}
   \nonumber
   \\ & \le & \frac{1}{\lambda}\,\frac{C\delta_h^2}{(\lambda^{-1}|x|)^{\gamma_0-\delta_h}}
   +C\,\frac{\lambda}{|x|^2}\,(\delta_T^{(1)}+\delta_T^{(2)}+\delta_T^{(3)})
   +C\,\frac{1}{|x|}\,\delta_{T,\,ij}^{(4), x^0}
   \nonumber 
   \\ & & +\,C\,\frac{G}{c^5}\,\lambda^3\Big(\frac{\lambda}{c|x|}
   \,{\|\partial_t^2 T_{ij}\|}_{L^\infty_{tx}}
   +\frac{\lambda^2}{c^2 |x|}\,{\|\partial_t^3 T_{ij}\|}_{L^\infty_{tx}}
   \Big)
   \nonumber
   \\ & \le & \frac{1}{\lambda}\,\frac{C\delta_h^2}{(\lambda^{-1}|x|)^{\gamma_0-\delta_h}}
   +C\,\frac{\lambda}{|x|^2}\,(\delta_T^{(1)}+\delta_T^{(2)}+\delta_T^{(3)})
   +C\,\frac{1}{|x|}\,\delta_{T,\,ij}^{(4), x^0}
   \nonumber
   \\ & & +\,C\,\frac{G}{c^4}\,\frac{\lambda^4}{|x|}\Big(
   {\|\partial_{x^0}^2 T_{ij}\|}_{L^\infty_{tx}}
   +\lambda\,{\|\partial_{x^0}^3 T_{ij}\|}_{L^\infty_{tx}}\Big)  
   \nonumber
   \\ & \le & \frac{1}{\lambda}\,\frac{C\delta_h^2}{(\lambda^{-1}|x|)^{\gamma_0-\delta_h}}
   +C\,\frac{\lambda}{|x|^2}\,(\delta_T^{(1)}+\delta_T^{(2)}+\delta_T^{(3)})
   +C\,\frac{1}{|x|}\,(\delta_{T,\,ij}^{(4), x^0}+\delta_{T,\,ij}^{(2), x^0}+\delta_{T,\,ij}^{(3), x^0}).
\end{eqnarray}  
Similarly we find for $i, j, k=1, 2, 3$
\begin{equation} \label{mdue36}
   |\tensor{({\rm rem}_{36})}{^0^j^k}(t, x)| \leq  \frac{1}{\lambda}\,\frac{C\delta_h^2}{(\lambda^{-1}|x|)^{\gamma_0-\delta_h}}
   +C\,\frac{\lambda}{|x|^2}\,\Big(\sum_{j=0}^3\delta_T^{(j)}\Big)
   +C\,\frac{1}{|x|}\,(\delta_{T,\,0j}^{(4), x^0}+\delta_{T,\,0j}^{(3), x^0})
\end{equation}
and
\begin{equation} \label{mdue37}
    |\tensor{({\rm rem}_{37})}{^i^j^k}(t, x)| \leq \frac{1}{\lambda}\,\frac{C\delta_h^2}{(\lambda^{-1}|x|)^{\gamma_0-\delta_h}}
   +C\,\frac{\lambda}{|x|^2}\,\Big(\sum_{j=0}^3\delta_T^{(j)}\Big)
   +\,C\,\frac{1}{|x|}\,(\delta_{T,\,ij}^{(4), x^0}+\delta_{T,\,ij}^{(2), x^0}+\delta_{T,\,ij}^{(3), x^0})
\end{equation}

\subsection{Remainders 38 through 42: (\ref{eq_rem38})--(\ref{eq_rem42})} 

These remainders were defined in Subsection~\ref{eotm_sect} as algebraic expressions involving remainders 18, 20, 21, 24, 28 and 33--37, as well as $x^k$ terms (which are directly bounded by $r$). Estimates for them are immediate. Using (\ref{maumau20}), (\ref{maumau24}) and (\ref{mdue33}) as well as (\ref{hinfty3}) we get 
\begin{eqnarray}\label{maumau38}  
   |{({\rm rem}_{38})}^{00}(t, x)|
   & \le & C\,\frac{G}{c^5}\,\frac{1}{r}\,\frac{c^5}{G}\,(\delta_T^{(0)}+\delta_T^{(1)}+\delta_T^{(2)})
   \Big({\|\hat{h}(u)\|}_{L^\infty_x}+\lambda {\|\partial_{x^\gamma}\hat{h}(u)\|}_{L^\infty_x}
   +\lambda^2 {\|\partial^2_{x^\beta x^\gamma}\hat{h}(u)\|}_{L^\infty_x}\Big)
   \nonumber 
   \\ & & +\,\frac{1}{\lambda}\,\frac{C\delta_h^2}{(\lambda^{-1}|x|)^{\gamma_0-\delta_h}}
   +C\,\frac{\lambda}{r^2}\,(\delta_T^{(1)}+\delta_T^{(2)}+\delta_T^{(3)})
   +C\,\frac{1}{r}\,\delta_{T,\,00}^{(4), x^0}
   \nonumber
   \\ & \le & \frac{1}{\lambda}\,\frac{C\delta_h^2}{(\lambda^{-1}|x|)^{\gamma_0-\delta_h}}
   +C\,\frac{\lambda}{|x|^2}\,(\delta_T^{(1)}+\delta_T^{(2)}+\delta_T^{(3)})
   +C\,\frac{1}{|x|}\,\Big(\delta_{T,\,00}^{(4), x^0}+(\delta_T^{(0)}+\delta_T^{(1)}+\delta_T^{(2)})\delta_h\Big). 
   \nonumber 
   \\ & & 
\end{eqnarray} 
Using (\ref{maumau18}) and (\ref{mdue35}) 
together with (\ref{hinfty3}) and (\ref{coe}), we get for $i, j=1, 2, 3$
\begin{eqnarray}\label{maumau39}  
   |{({\rm rem}_{39})}^{ij}(t, x)|
   & \le & C\,\frac{G}{c^5}\,\frac{1}{|x|}\,
   \frac{c^5}{G}\,\Big(\sum_{j=0}^3\delta_T^{(j)}\Big)\,
   \Big(\sum_{|\alpha|\le 3}\lambda^{|\alpha|}{\|\partial^\alpha\hat{h}(u)\|}_{L^\infty_x}
   +\sum_{|\alpha|\le 3}\lambda^{|\alpha|}{\|\partial^\alpha\hat{H}(u)\|}_{L^\infty_x}\Big)
   \nonumber 
   \\ & & +\,\frac{1}{\lambda}\,\frac{C\delta_h^2}{(\lambda^{-1}|x|)^{\gamma_0-\delta_h}}
   +C\,\frac{\lambda}{|x|^2}\,(\delta_T^{(1)}+\delta_T^{(2)}+\delta_T^{(3)})
   +C\,\frac{1}{|x|}\,(\delta_{T,\,ij}^{(4), x^0}+\delta_{T,\,ij}^{(2), x^0}+\delta_{T,\,ij}^{(3), x^0})  
   \nonumber 
   \\ & \le & \frac{1}{\lambda}\,\frac{C\delta_h^2}{(\lambda^{-1}|x|)^{\gamma_0-\delta_h}}
   +C\,\frac{\lambda}{|x|^2}\,(\delta_T^{(1)}+\delta_T^{(2)}+\delta_T^{(3)})
   \nonumber
   \\ & & +\,C\,\frac{1}{|x|}\,\Big(\delta_{T,\,ij}^{(4), x^0}+\delta_{T,\,ij}^{(2), x^0}+\delta_{T,\,ij}^{(3), x^0}
   +\Big(\sum_{j=0}^3\delta_T^{(j)}\Big)\delta_h\Big)
\end{eqnarray} 
From (\ref{maumau20}), (\ref{maumau24}) and (\ref{mdue34}) we see that, for $k=1, 2, 3$,
\begin{eqnarray}\label{maumau40}  
   |\tensor{({\rm rem}_{40})}{^0^0^k}(t, x)|
   & \le & C\,\frac{G}{c^5}\,\frac{1}{r}\,\frac{c^5}{G}
   \,(\delta_T^{(0)}+\delta_T^{(1)}+\delta_T^{(2)})\,
   \Big({\|\hat{h}(u)\|}_{L^\infty_x}+\lambda {\|\partial_{x^\gamma}\hat{h}(u)\|}_{L^\infty_x}
   +\lambda^2 {\|\partial^2_{x^\beta x^\gamma}\hat{h}(u)\|}_{L^\infty_x}\Big)
   \nonumber 
   \\ & & +\,\frac{1}{\lambda}\,\frac{C\delta_h^2}{(\lambda^{-1}|x|)^{\gamma_0-\delta_h}}
   +C\,\frac{\lambda}{|x|^2}\,\Big(\sum_{j=0}^3\delta_T^{(j)}\Big)
   +C\,\frac{1}{|x|}\,\delta_{T,\,00}^{(4), x^0} 
   \nonumber 
   \\ & \le & \frac{1}{\lambda}\,\frac{C\delta_h^2}{(\lambda^{-1}|x|)^{\gamma_0-\delta_h}}
   +C\,\frac{\lambda}{|x|^2}\,\Big(\sum_{j=0}^3\delta_T^{(j)}\Big)
   +C\,\frac{1}{|x|}\,\Big(\delta_{T,\,00}^{(4), x^0}+(\delta_T^{(0)}+\delta_T^{(1)}+\delta_T^{(2)})\delta_h\Big)
   \nonumber 
   \\ & & 
\end{eqnarray} 
We can use (\ref{eq_rem21c}), (\ref{maumau28}) and (\ref{mdue36}) to obtain, for $j, k=1, 2, 3$,
\begin{eqnarray}\label{maumau41}  
   |\tensor{({\rm rem}_{41})}{^0^j^k}(t, x)|
   & \le & C\,\frac{G}{c^5}\,\frac{1}{r}\,\frac{c^5}{G}
   \,\Big(\sum_{j=0}^3\delta_T^{(j)}\Big)\,
   \Big(\sum_{|\alpha|\le 3}\lambda^{|\alpha|}{\|\partial^\alpha\hat{h}(u)\|}_{L^\infty_x}
   +\sum_{|\alpha|\le 3}\lambda^{|\alpha|}{\|\partial^\alpha\hat{H}(u)\|}_{L^\infty_x}\Big)
   \nonumber 
   \\ & & +\,\frac{1}{\lambda}\,\frac{C\delta_h^2}{(\lambda^{-1}|x|)^{\gamma_0-\delta_h}}
   +C\,\frac{\lambda}{|x|^2}\,\Big(\sum_{j=0}^3\delta_T^{(j)}\Big)
   +C\,\frac{1}{|x|}\,(\delta_{T,\,0j}^{(4), x^0}+\delta_{T,\,0j}^{(3), x^0})
   \nonumber
   \\ & \le & \frac{1}{\lambda}\,\frac{C\delta_h^2}{(\lambda^{-1}|x|)^{\gamma_0-\delta_h}}
   +C\,\frac{\lambda}{|x|^2}\,\Big(\sum_{j=0}^3\delta_T^{(j)}\Big)
   +C\,\frac{1}{|x|}\,\Big(\delta_{T,\,0j}^{(4), x^0}+\delta_{T,\,0j}^{(3), x^0}
   +\Big(\sum_{j=0}^3\delta_T^{(j)}\Big)\delta_h\Big)
   \nonumber 
   \\ & & 
\end{eqnarray}
And finally, owing to (\ref{maumau18}) and (\ref{mdue37}), we have for $i, j, k=1, 2, 3$
\begin{eqnarray}\label{maumau42}  
   |\tensor{({\rm rem}_{42})}{^i^j^k}(t, x)|
   & \le & C\,\frac{G}{c^5}\,\frac{1}{|x|}\,
   \frac{c^5}{G}\,\Big(\sum_{j=0}^3\delta_T^{(j)}\Big)\,
   \Big(\sum_{|\alpha|\le 3}\lambda^{|\alpha|}{\|\partial^\alpha\hat{h}(u)\|}_{L^\infty_x}
   +\sum_{|\alpha|\le 3}\lambda^{|\alpha|}{\|\partial^\alpha\hat{H}(u)\|}_{L^\infty_x}\Big)
   \nonumber 
   \\ & & +\,\frac{1}{\lambda}\,\frac{C\delta_h^2}{(\lambda^{-1}|x|)^{\gamma_0-\delta_h}}
   +C\,\frac{\lambda}{|x|^2}\,\Big(\sum_{j=0}^3\delta_T^{(j)}\Big)
   \nonumber
   \\ & & +\,C\,\frac{1}{|x|}\,(\delta_{T,\,ij}^{(4), x^0}+\delta_{T,\,ij}^{(2), x^0}+\delta_{T,\,ij}^{(3), x^0}) 
   \nonumber 
   \\ & \le & \frac{1}{\lambda}\,\frac{C\delta_h^2}{(\lambda^{-1}|x|)^{\gamma_0-\delta_h}}
   +C\,\frac{\lambda}{|x|^2}\,\Big(\sum_{j=0}^3\delta_T^{(j)}\Big)
   \nonumber
   \\ & & +\,C\,\frac{1}{|x|}\,\Big(\delta_{T,\,ij}^{(4), x^0}+\delta_{T,\,ij}^{(2), x^0}+\delta_{T,\,ij}^{(3), x^0}
   +\Big(\sum_{j=0}^3\delta_T^{(j)}\Big)\delta_h\Big)
\end{eqnarray} 

\subsection{Remainders 43 through 50}

These remainders, introduced in Subsection~\ref{expansion_of_energy_flux_sect}, ultimately depend on remainders 10 and 38--42, as well as on the quadrupole moment $D$. The latter was defined in (\ref{ito}) as
\[ D^{\alpha\beta}(u)=\frac{1}{c^2}\int_{\R^3} T_{00}(u, \bar{x})\,\bar{x}^\alpha
   \,\bar{x}^\beta\,d\bar{x}. \] 
We will need to know a bound for its triple derivative:
\[ \dddot{D}^{\alpha\beta}(u)=c\int_{\bbR^3} \partial^3_{x^0}T_{00}(u, \bar{x})\,\bar{x}^\alpha
   \,\bar{x}^\beta\,d\bar{x}, \] 
where $u=t-c^{-1}r$ satisfies by hypothesis $|u|\le\lambda c^{-1}$. As always, the integration domain may be reduced to
\[ \dddot{D}^{\alpha\beta}(u)=c\int_{|\bar{x}|\leq\lambda K_\ast}
   \partial^3_{x^0}T_{00}(u, \bar{x})\,\bar{x}^\alpha
   \,\bar{x}^\beta\,d\bar{x} \] 
and from the boundedness condition~\eqref{fuvor0} we immediately have, for $\alpha, \beta=0, 1, 2, 3$,
\begin{equation}\label{Dtriple} 
   |\dddot{D}^{\alpha\beta}(u)|\le C\,c\lambda^5
   \,\frac{c^4}{G\lambda^5}\,\delta_{T,\,00}^{(3), x^0}
   =C\,\frac{c^5}{G}\,\delta_{T,\,00}^{(3), x^0}
\end{equation} 

Using (\ref{maumau38}) and (\ref{maumau39}) we deduce from~\eqref{eq_rem43} that
\begin{eqnarray}\label{maumau43} 
   |({\rm rem}_{43})(t,x)|
   & \le & \frac{1}{\lambda}\,\frac{C\delta_h^2}{(\lambda^{-1}|x|)^{\gamma_0-\delta_h}}
   +C\,\frac{\lambda}{|x|^2}\,(\delta_T^{(1)}+\delta_T^{(2)}+\delta_T^{(3)})
   \nonumber
   \\ & & +\,C\,\frac{1}{|x|}\,\Big(\delta_{T,\,00}^{(4), x^0}+\sum_{j=1}^3\delta_{T,\,jj}^{(2), x^0}
   +\sum_{j=1}^3\delta_{T,\,jj}^{(3), x^0}
   +\Big(\sum_{j=0}^3\delta_T^{(j)}\Big)\delta_h\Big).  
\end{eqnarray} 
Then the bounds in (\ref{maumau40}) and (\ref{maumau42}) imply for~\eqref{eq_rem44} and for $i=1, 2, 3$
\begin{eqnarray}\label{maumau44} 
   |({\rm rem}_{44})^i(t, x)|
   & \le & \frac{1}{\lambda}\,\frac{C\delta_h^2}{(\lambda^{-1}|x|)^{\gamma_0-\delta_h}}
   +C\,\frac{\lambda}{|x|^2}\,\Big(\sum_{j=0}^3\delta_T^{(j)}\Big)
   \nonumber
   \\ & & +\,C\,\frac{1}{|x|}\,\Big(\delta_{T,\,00}^{(4), x^0}
   +\sum_{j=1}^3\delta_{T,\,jj}^{(4), x^0}
   +\sum_{j=1}^3\delta_{T,\,jj}^{(2), x^0}
   +\sum_{j=1}^3\delta_{T,\,jj}^{(3), x^0}
   +\Big(\sum_{j=0}^3\delta_T^{(j)}\Big)\delta_h\Big)\quad 
\end{eqnarray}  
We also find from (\ref{maumau44}) that~\eqref{eq_rem45} is bounded by
\begin{eqnarray}\label{maumau45} 
   |({\rm rem}_{45})(t, x)|
   & \le & \frac{1}{\lambda}\,\frac{C\delta_h^2}{(\lambda^{-1}|x|)^{\gamma_0-\delta_h}}
   +C\,\frac{\lambda}{|x|^2}\,\Big(\sum_{j=0}^3\delta_T^{(j)}\Big)
   \nonumber
   \\ & & +\,C\,\frac{1}{|x|}\,\Big(\delta_{T,\,00}^{(4), x^0}
   +\sum_{j=1}^3\delta_{T,\,jj}^{(4), x^0}
   +\sum_{j=1}^3\delta_{T,\,jj}^{(2), x^0}
   +\sum_{j=1}^3\delta_{T,\,jj}^{(3), x^0}
   +\Big(\sum_{j=0}^3\delta_T^{(j)}\Big)\delta_h\Big).
   \qquad 
\end{eqnarray} 
Next, from (\ref{Dtriple}), (\ref{maumau45}) and (\ref{maumau43}) we deduce for the formula~\eqref{eq_rem46} (where one needs to remember that the argument of all $D$ terms is the retarded time $u = t - c^{-1}r$)
\begin{eqnarray}\label{maumau46}  
   \lefteqn{|({\rm rem}_{46})(t,x)|}
   \nonumber 
   \\ & \le & C\,\frac{G}{c^5|x|}\,\frac{c^5}{G}\,\delta_{T,\,00}^{(3), x^0}
   \,\Big[\frac{1}{\lambda}\,\frac{\delta_h^2}{(\lambda^{-1}|x|)^{\gamma_0-\delta_h}}
   +\frac{\lambda}{|x|^2}\,\Big(\sum_{j=0}^3\delta_T^{(j)}\Big)
   \nonumber
   \\ & & \hspace{8em} +\,\frac{1}{|x|}\,\Big(\delta_{T,\,00}^{(4), x^0}
   +\sum_{j=1}^3\delta_{T,\,jj}^{(4), x^0}
   +\sum_{j=1}^3\delta_{T,\,jj}^{(2), x^0}
   +\sum_{j=1}^3\delta_{T,\,jj}^{(3), x^0}
   +\Big(\sum_{j=0}^3\delta_T^{(j)}\Big)\delta_h\Big)\Big]
   \nonumber 
   \\ & & +\,C\Big[ 
   \frac{1}{\lambda}\,\frac{\delta_h^2}{(\lambda^{-1}|x|)^{\gamma_0-\delta_h}}
   +\frac{\lambda}{|x|^2}\,(\delta_T^{(1)}+\delta_T^{(2)}+\delta_T^{(3)})
   \nonumber
   \\ & & \hspace{3em} +\,\frac{1}{|x|}\,\Big(\delta_{T,\,00}^{(4), x^0}
   +\sum_{j=1}^3\delta_{T,\,jj}^{(2), x^0}
   +\sum_{j=1}^3\delta_{T,\,jj}^{(3), x^0}
   +\Big(\sum_{j=0}^3\delta_T^{(j)}\Big)\delta_h\Big)\Big]
   \nonumber 
   \\ & & \hspace{2em}\times\Big[\frac{G}{c^5 |x|}\,\frac{c^5}{G}\,\delta_{T,\,00}^{(3), x^0}
   +\frac{1}{\lambda}\,\frac{\delta_h^2}{(\lambda^{-1}|x|)^{\gamma_0-\delta_h}}
   +\frac{\lambda}{|x|^2}\,\Big(\sum_{j=0}^3\delta_T^{(j)}\Big)
   \nonumber
   \\ & & \hspace{3em} +\,\frac{1}{|x|}\,\Big(\delta_{T,\,00}^{(4), x^0}
   +\sum_{j=1}^3\delta_{T,\,jj}^{(4), x^0}
   +\sum_{j=1}^3\delta_{T,\,jj}^{(2), x^0}
   +\sum_{j=1}^3\delta_{T,\,jj}^{(3), x^0}
   +\Big(\sum_{j=0}^3\delta_T^{(j)}\Big)\delta_h\Big)\Big]
   \nonumber
   \\ & \le & C\Big[ 
   \frac{1}{\lambda}\,\frac{\delta_h^2}{(\lambda^{-1}|x|)^{\gamma_0-\delta_h}}
   +\frac{\lambda}{|x|^2}\,\Big(\sum_{j=0}^3\delta_T^{(j)}\Big)
   \nonumber
   \\ & & \hspace{3em} +\,\frac{1}{|x|}
   \,\Big(\delta_{T,\,00}^{(3), x^0}+\delta_{T,\,00}^{(4), x^0}
   +\sum_{j=1}^3\delta_{T,\,jj}^{(4), x^0}
   +\sum_{j=1}^3\delta_{T,\,jj}^{(2), x^0}
   +\sum_{j=1}^3\delta_{T,\,jj}^{(3), x^0}
   +\Big(\sum_{j=0}^3\delta_T^{(j)}\Big)\delta_h\Big)\Big]
   \nonumber 
   \\ & & \hspace{2em}\times\Big[
   \frac{1}{\lambda}\,\frac{\delta_h^2}{(\lambda^{-1}|x|)^{\gamma_0-\delta_h}}
   +\frac{\lambda}{|x|^2}\,\Big(\sum_{j=0}^3\delta_T^{(j)}\Big)
   \nonumber
   \\ & & \hspace{3em} +\,\frac{1}{|x|}\,
   \Big(\delta_{T,\,00}^{(4), x^0}
   +\sum_{j=1}^3\delta_{T,\,jj}^{(4), x^0}
   +\sum_{j=1}^3\delta_{T,\,jj}^{(2), x^0}
   +\sum_{j=1}^3\delta_{T,\,jj}^{(3), x^0}
   +\Big(\sum_{j=0}^3\delta_T^{(j)}\Big)\delta_h\Big)\Big]. 
\end{eqnarray} 
The lengthy expression (\ref{bigrem1}) can be bounded using (\ref{Dtriple}), (\ref{maumau38}), (\ref{maumau39}), (\ref{maumau40}), (\ref{maumau41}), (\ref{maumau42}) and (\ref{maumau10}). Note that the structure of (\ref{bigrem1}) is simple: In self-explanatory symbolical notation, it comes down to
\[ \rem{47}\sim
   \frac{\dddot{D}}{cr}\sum_{j=38}^{42} \rem{j} 
   +\frac{c^4}{G}\sum_{i,j=38}^{42} \rem{i}\rem{j} 
   +\frac{c^4}{G}\rem{10}. \]
After a straightforward estimate we therefore obtain for $i=1, 2, 3$ 
\begin{eqnarray}\label{maumau47} 
   \lefteqn{|({\rm rem}_{47})^i (t, x)|}
   \nonumber 
   \\ & \le & C\,\frac{1}{cr}\,
   \frac{c^5}{G}\,\delta_{T,\,00}^{(3), x^0}
   \,\Big[\frac{1}{\lambda}\,\frac{\delta_h^2}{(\lambda^{-1}|x|)^{\gamma_0-\delta_h}}
   +\frac{\lambda}{|x|^2}\,\Big(\sum_{j=0}^3\delta_T^{(j)}\Big)
   \nonumber
   \\ & & +\,\frac{1}{|x|}\,\Big(\delta_{T,\,00}^{(4), x^0}
   +\sum_{i, j=1}^3\delta_{T,\,ij}^{(4), x^0}
   +\sum_{j=1}^3\delta_{T,\,0j}^{(4), x^0}   
   +\sum_{i, j=1}^3\delta_{T,\,ij}^{(3), x^0}
   +\sum_{j=1}^3\delta_{T,\,0j}^{(3), x^0}
   +\sum_{i, j=1}^3\delta_{T,\,ij}^{(2), x^0}   
   +\Big(\sum_{j=0}^3\delta_T^{(j)}\Big)\delta_h\Big)\Big]
   \nonumber
   \\ & & +\,C\,\frac{c^4}{G}\,
   \Big[\frac{1}{\lambda}\,\frac{\delta_h^2}{(\lambda^{-1}|x|)^{\gamma_0-\delta_h}}
   +\frac{\lambda}{|x|^2}\,\Big(\sum_{j=0}^3\delta_T^{(j)}\Big)
   \nonumber
   \\ & & +\,\frac{1}{|x|}\,\Big(\delta_{T,\,00}^{(4), x^0}
   +\sum_{i, j=1}^3\delta_{T,\,ij}^{(4), x^0}
   +\sum_{j=1}^3\delta_{T,\,0j}^{(4), x^0}   
   +\sum_{i, j=1}^3\delta_{T,\,ij}^{(3), x^0}
   +\sum_{j=1}^3\delta_{T,\,0j}^{(3), x^0}
   +\sum_{i, j=1}^3\delta_{T,\,ij}^{(2), x^0}   
   +\Big(\sum_{j=0}^3\delta_T^{(j)}\Big)\delta_h\Big)\Big]^2
   \nonumber   
   \\ & & +\,C\,\frac{c^4}{G}
   \,(|\hat{h}(t, x)|+|\hat{H}(t, x)|)\,
   \Big(|\partial_{x^\gamma}\hat{h}(t, x)|^2+|\partial_{x^\gamma}\hat{H}(t, x)|^2\Big)
   \nonumber 
   \\ & \le & C\,\frac{c^4}{G}\,
   \Big[\frac{1}{\lambda}\,\frac{\delta_h^2}{(\lambda^{-1}|x|)^{\gamma_0-\delta_h}}
   +\frac{\lambda}{|x|^2}\,\Big(\sum_{j=0}^3\delta_T^{(j)}\Big)
   \nonumber
   \\ & & \hspace{3em}+\,\frac{1}{|x|}\,\Big(\delta_{T,\,00}^{(3), x^0}+\delta_{T,\,00}^{(4), x^0}
   +\sum_{i, j=1}^3\delta_{T,\,ij}^{(4), x^0}
   +\sum_{j=1}^3\delta_{T,\,0j}^{(4), x^0}
   \nonumber   
   \\ & & \hspace{6em} +\,\sum_{i, j=1}^3\delta_{T,\,ij}^{(3), x^0}
   +\sum_{j=1}^3\delta_{T,\,0j}^{(3), x^0}
   +\sum_{i, j=1}^3\delta_{T,\,ij}^{(2), x^0}   
   +\Big(\sum_{j=0}^3\delta_T^{(j)}\Big)\delta_h\Big)\Big]
   \nonumber 
   \\ & & \hspace{3em}\times\Big[\frac{1}{\lambda}\,\frac{\delta_h^2}{(\lambda^{-1}|x|)^{\gamma_0-\delta_h}}
   +\frac{\lambda}{|x|^2}\,\Big(\sum_{j=0}^3\delta_T^{(j)}\Big)
   \nonumber
   \\ & & \hspace{5em} +\,\frac{1}{|x|}\,\Big(\delta_{T,\,00}^{(4), x^0}
   +\sum_{i, j=1}^3\delta_{T,\,ij}^{(4), x^0}
   +\sum_{j=1}^3\delta_{T,\,0j}^{(4), x^0}   
   \nonumber 
   \\ & & \hspace{8em}
   +\,\sum_{i, j=1}^3\delta_{T,\,ij}^{(3), x^0}
   +\sum_{j=1}^3\delta_{T,\,0j}^{(3), x^0}
   +\sum_{i, j=1}^3\delta_{T,\,ij}^{(2), x^0}   
   +\Big(\sum_{j=0}^3\delta_T^{(j)}\Big)\delta_h\Big)\Big]
   \nonumber   
   \\ & & +\,C\,\frac{c^4}{G}
   \,(|\hat{h}(t, x)|+|\hat{H}(t, x)|)\,
   \Big(|\partial_{x^\gamma}\hat{h}(t, x)|^2+|\partial_{x^\gamma}\hat{H}(t, x)|^2\Big).
\end{eqnarray} 
This directly implies for $\rem{48}$ as defined in~\eqref{eq_rem48}:
\begin{eqnarray}\label{maumau48} 
   \lefteqn{|({\rm rem}_{48})(t, x)|}
   \nonumber 
   \\ & \le & C\,\frac{c^4}{G}\,
   \Big[\frac{1}{\lambda}\,\frac{\delta_h^2}{(\lambda^{-1}|x|)^{\gamma_0-\delta_h}}
   +\frac{\lambda}{|x|^2}\,\Big(\sum_{j=0}^3\delta_T^{(j)}\Big)
   \nonumber
   \\ & & \hspace{3em}+\,\frac{1}{|x|}\,\Big(\delta_{T,\,00}^{(3), x^0}+\delta_{T,\,00}^{(4), x^0}
   +\sum_{i, j=1}^3\delta_{T,\,ij}^{(4), x^0}
   +\sum_{j=1}^3\delta_{T,\,0j}^{(4), x^0}
   \nonumber   
   \\ & & \hspace{6em} +\,\sum_{i, j=1}^3\delta_{T,\,ij}^{(3), x^0}
   +\sum_{j=1}^3\delta_{T,\,0j}^{(3), x^0}
   +\sum_{i, j=1}^3\delta_{T,\,ij}^{(2), x^0}   
   +\Big(\sum_{j=0}^3\delta_T^{(j)}\Big)\delta_h\Big)\Big]
   \nonumber 
   \\ & & \hspace{3em}\times\Big[\frac{1}{\lambda}\,\frac{\delta_h^2}{(\lambda^{-1}|x|)^{\gamma_0-\delta_h}}
   +\frac{\lambda}{|x|^2}\,\Big(\sum_{j=0}^3\delta_T^{(j)}\Big)
   \nonumber
   \\ & & \hspace{5em} +\,\frac{1}{|x|}\,\Big(\delta_{T,\,00}^{(4), x^0}
   +\sum_{i, j=1}^3\delta_{T,\,ij}^{(4), x^0}
   +\sum_{j=1}^3\delta_{T,\,0j}^{(4), x^0}   
   \nonumber 
   \\ & & \hspace{8em}
   +\,\sum_{i, j=1}^3\delta_{T,\,ij}^{(3), x^0}
   +\sum_{j=1}^3\delta_{T,\,0j}^{(3), x^0}
   +\sum_{i, j=1}^3\delta_{T,\,ij}^{(2), x^0}   
   +\Big(\sum_{j=0}^3\delta_T^{(j)}\Big)\delta_h\Big)\Big]
   \nonumber   
   \\ & & +\,C\,\frac{c^4}{G}
   \,(|\hat{h}(t, x)|+|\hat{H}(t, x)|)\,
   \Big(|\partial_{x^\gamma}\hat{h}(t, x)|^2+|\partial_{x^\gamma}\hat{H}(t, x)|^2\Big). 
\end{eqnarray} 

To treat $\rem{49}$ defined in~\eqref{eq_rem49}, we first need (\ref{h49}) and (\ref{H49}) in Corollary \ref{H49cor} to find
\begin{eqnarray*} 
   & & (|\hat{h}(t, x)|+|\hat{H}(t, x)|)\,
   \Big(|\partial_{x^\gamma}\hat{h}(t, x)|^2+|\partial_{x^\gamma}\hat{H}(t, x)|^2\Big)
   \\ & & \hspace{3em}\le 
   C\,\frac{\delta_h^3}{\lambda^2}
   \,\frac{(1+\lambda^{-1}|x^0|)^{3\delta_h}}{(1+\lambda^{-1}|x^0|+\lambda^{-1}|x|)^3}
   \\ & & \hspace{3em}\le 
   C\,\frac{\delta_h^3}{\lambda^2}
   \,\frac{1}{(1+\lambda^{-1}|x|)^{3(1-\delta_h)}}, 
\end{eqnarray*}
the latter for $|t-c^{-1}|x||\le\lambda c^{-1}$.
Hence (\ref{maumau46}) and (\ref{maumau48}) implies that 
\begin{eqnarray}\label{maumau49} 
\lefteqn{|({\rm rem}_{49})(t, x)|}
   \nonumber 
   \\ & \le & C\,\frac{c^4}{G}\,
   \Big[\frac{1}{\lambda}\,\frac{\delta_h^2}{(\lambda^{-1}|x|)^{\gamma_0-\delta_h}}
   +\frac{\lambda}{|x|^2}\,\Big(\sum_{j=0}^3\delta_T^{(j)}\Big)
   \nonumber
   \\ & & \hspace{3em}+\,\frac{1}{|x|}\,\Big(\delta_{T,\,00}^{(3), x^0}+\delta_{T,\,00}^{(4), x^0}
   +\sum_{i, j=1}^3\delta_{T,\,ij}^{(4), x^0}
   +\sum_{j=1}^3\delta_{T,\,0j}^{(4), x^0}
   \nonumber   
   \\ & & \hspace{6em} +\,\sum_{i, j=1}^3\delta_{T,\,ij}^{(3), x^0}
   +\sum_{j=1}^3\delta_{T,\,0j}^{(3), x^0}
   +\sum_{i, j=1}^3\delta_{T,\,ij}^{(2), x^0}   
   +\Big(\sum_{j=0}^3\delta_T^{(j)}\Big)\delta_h\Big)\Big]
   \nonumber 
   \\ & & \hspace{3em}\times\Big[\frac{1}{\lambda}\,\frac{\delta_h^2}{(\lambda^{-1}|x|)^{\gamma_0-\delta_h}}
   +\frac{\lambda}{|x|^2}\,\Big(\sum_{j=0}^3\delta_T^{(j)}\Big)
   \nonumber
   \\ & & \hspace{5em} +\,\frac{1}{|x|}\,\Big(\delta_{T,\,00}^{(4), x^0}
   +\sum_{i, j=1}^3\delta_{T,\,ij}^{(4), x^0}
   +\sum_{j=1}^3\delta_{T,\,0j}^{(4), x^0}   
   \nonumber 
   \\ & & \hspace{8em}
   +\,\sum_{i, j=1}^3\delta_{T,\,ij}^{(3), x^0}
   +\sum_{j=1}^3\delta_{T,\,0j}^{(3), x^0}
   +\sum_{i, j=1}^3\delta_{T,\,ij}^{(2), x^0}   
   +\Big(\sum_{j=0}^3\delta_T^{(j)}\Big)\delta_h\Big)\Big]
   \nonumber
   \\ & & +\,C\,\frac{c^4}{G}
   \,\frac{\delta_h^3}{\lambda^2}\,\frac{1}{(1+\lambda^{-1}|x|)^{3(1-\delta_h)}}. 
\end{eqnarray} 

Lastly, we recall the definition~\eqref{eq_rem50} of the last remainder term. By integration of (\ref{maumau49}) on the spherical surface $\partial_r(0)$ (which just amounts to a multiplication by $r^2$), we find 
\begin{eqnarray}\label{maumau50} 
\lefteqn{|({\rm rem}_{50})(t, x)|}
   \nonumber 
   \\ & \le & C\,\frac{c^5}{G}\,
   \Big[(\lambda^{-1}|x|)^{1-\gamma_0+\delta_h}\,\delta_h^2
   +\frac{\lambda}{|x|}\,\Big(\sum_{j=0}^3\delta_T^{(j)}\Big)
   \nonumber
   \\ & & \hspace{4em} +\,\Big(\delta_{T,\,00}^{(3), x^0}+\delta_{T,\,00}^{(4), x^0}
   +\sum_{i, j=1}^3\delta_{T,\,ij}^{(4), x^0}
   +\sum_{j=1}^3\delta_{T,\,0j}^{(4), x^0}
   \nonumber   
   \\ & & \hspace{6em} +\,\sum_{i, j=1}^3\delta_{T,\,ij}^{(3), x^0}
   +\sum_{j=1}^3\delta_{T,\,0j}^{(3), x^0}
   +\sum_{i, j=1}^3\delta_{T,\,ij}^{(2), x^0}   
   +\Big(\sum_{j=0}^3\delta_T^{(j)}\Big)\delta_h\Big)\Big]
   \nonumber 
   \\ & & \hspace{3em}\times\Big[
   (\lambda^{-1}|x|)^{1-\gamma_0+\delta_h}\,\delta_h^2
   +\frac{\lambda}{|x|}\,\Big(\sum_{j=0}^3\delta_T^{(j)}\Big)
   \nonumber
   \\ & & \hspace{5em} +\,\Big(\delta_{T,\,00}^{(4), x^0}
   +\sum_{i, j=1}^3\delta_{T,\,ij}^{(4), x^0}
   +\sum_{j=1}^3\delta_{T,\,0j}^{(4), x^0}   
   \nonumber 
   \\ & & \hspace{7em}
   +\,\sum_{i, j=1}^3\delta_{T,\,ij}^{(3), x^0}
   +\sum_{j=1}^3\delta_{T,\,0j}^{(3), x^0}
   +\sum_{i, j=1}^3\delta_{T,\,ij}^{(2), x^0}   
   +\Big(\sum_{j=0}^3\delta_T^{(j)}\Big)\delta_h\Big)\Big]
   \nonumber
   \\ & & +\,C\,\frac{c^5}{G}
   \,\frac{1}{(\lambda^{-1}|x|)^{1-3\delta_h}}\,\delta_h^3. 
\end{eqnarray}
This concludes the proof of Theorem \ref{mainthm}.

\appendix 

\section{Appendix: The energy-momentum pseudo tensor}
\label{sec_en_mom_ps_tensor}
\setcounter{equation}{0}

The Landau-Lifschitz energy-momentum pseudo tensor $t^{\alpha\beta}$, 
see \cite[Section 28.2]{stephani} or \cite[equ.~(20.22)]{MTW}, is given by 
\begin{eqnarray}\label{LaLi} 
   \frac{16\pi G}{c^4}\,|g|\,t^{\alpha\beta} 
   & = & \partial_{x^\gamma}\mathfrak{g}^{\alpha\beta}\,\partial_{x^\delta}\mathfrak{g}^{\delta\gamma}
   -\partial_{x^\gamma}\mathfrak{g}^{\alpha\gamma}\,\partial_{x^\delta}\mathfrak{g}^{\beta\delta}
   +\frac{1}{2}\,g^{\alpha\beta}\,g_{\delta\eps}\,\partial_{x^\sigma}\mathfrak{g}^{\delta\gamma}
   \,\partial_{x^\gamma}\mathfrak{g}^{\sigma\eps}
   \nonumber
   \\ & & -g_{\eps\gamma}\,\partial_{x^\delta}\mathfrak{g}^{\eps\sigma}\,
   (g^{\alpha\delta}\,\partial_{x^\sigma}\mathfrak{g}^{\beta\gamma}
   +g^{\beta\delta}\,\partial_{x^\sigma}\mathfrak{g}^{\alpha\gamma})
   +g_{\delta\eps}\,g^{\gamma\sigma}\,\partial_{x^\gamma}\mathfrak{g}^{\beta\delta}
   \,\partial_{x^\sigma}\mathfrak{g}^{\alpha\eps}
   \nonumber
   \\ & & +\frac{1}{8}\,(2g^{\alpha\delta}\,g^{\beta\gamma}-g^{\alpha\beta}\,g^{\delta\gamma})
   (2g_{\eps\sigma}\,g_{\kappa\nu}-g_{\sigma\kappa}\,g_{\eps\nu})
   \,\partial_{x^\delta}\mathfrak{g}^{\eps\nu}\,\partial_{x^\gamma}\mathfrak{g}^{\sigma\kappa}, 
\end{eqnarray} 
where 
\begin{equation}\label{mfg} 
   \mathfrak{g}^{\alpha\beta}=|g|^{1/2}\,g^{\alpha\beta}.
\end{equation}  
It is not a tensor; cf.~for instance \cite[p.~233]{stephani}. Defining also the quantities 
\[ U^{\alpha\beta\gamma\delta}
   =\frac{1}{2}\,|g|\,(g^{\alpha\beta}\,g^{\gamma\delta}-g^{\alpha\gamma}\,g^{\beta\delta})
   \quad\mbox{and}\quad 
   U^{\alpha\beta\gamma}=\partial_{x^\delta} U^{\alpha\beta\gamma\delta}, \]   
(the latter being sometimes called a superpotential), we immediately have
\begin{equation}\label{macar} 
   |g|\,\Big(G^{\alpha\beta}+\frac{8\pi G}{c^4}\,t^{\alpha\beta}\Big)
   =\partial_{x^\gamma} U^{\alpha\beta\gamma},
   \quad U^{\alpha\beta\gamma}=-U^{\alpha\gamma\beta},
   \quad \partial_{x^\gamma} U^{\alpha\beta\gamma}=\partial_{x^\gamma} U^{\beta\alpha\gamma},
\end{equation}  
from which it follows that 
\begin{equation}\label{conecti} 
   \partial^2_{x^\beta x^\gamma} U^{\alpha\beta\gamma}=0.
\end{equation}  
Consequently,
\[ \partial_{x^\beta}\Big[|g|\,\Big(G^{\alpha\beta}
   +\frac{8\pi G}{c^4}\,t^{\alpha\beta}\Big)\Big]=0, \] 
which together with the Einstein equations (\ref{einst}) implies the conservation law
\begin{equation} \label{cons_law}
    \partial_{x^\beta}[\,|g|\,(T^{\alpha\beta}+t^{\alpha\beta})]=0.
\end{equation}
This has to be contrasted to the identity $\nabla_\beta T^{\alpha\beta}=0$ 
in terms of covariant derivatives. Equation~\eqref{cons_law} implies that the quantities 
$|g|(T^{\alpha\beta}+t^{\alpha\beta})$ can be viewed as components of a fictitious energy-momentum tensor 
of the matter system coupled to gravity, and in particular the quantities
\[ \mathfrak{p}^\alpha(t,x) = \int_{B_r(0)} |g(t, \bar{x})|\,(T^{0\alpha}(t, \bar{x})
   +t^{0\alpha}(t, \bar{x}))\,d\bar{x} , \quad \alpha=0,1,2,3 , \] 
represent the total energy (for $\alpha=0$) or $j$-momentum (for $\alpha=j$) present inside 
a ball of radius $r = |x|$ at time $t$. Here $d\bar{x} = d\bar{x}^1\, d\bar{x}^2\, d\bar{x}^3$.
\medskip 

Now note that (\ref{conecti}) also implies
\[ 0=\partial^2_{x^\beta x^\gamma} U^{\alpha\beta\gamma}
   =\partial^2_{x^0 x^\gamma} U^{\alpha 0\gamma}+\partial^2_{x^i x^\gamma} U^{\alpha i\gamma}, \] 
which upon integration yields
\begin{eqnarray}\label{gtls}  
   \partial_{x^0}\int_{B_r(0)} \partial_{x^\gamma} U^{\alpha 0\gamma}\,d\bar{x}
   & = & -\int_{B_r(0)}\partial^2_{x^i x^\gamma} U^{\alpha i\gamma}\,d\bar{x}
   \nonumber
   \\ & = & -\sum_{i=1}^3\int_{\partial B_r(0)}\frac{\bar{x}^i}{r}\,\partial_{x^\gamma} 
   U^{\alpha i\gamma}\,dS(\bar{x}),
\end{eqnarray}
where we have used Gauss's Theorem in the last step, with $dS(\bar{x})
=dS(\bar{x}^1, \bar{x}^2, \bar{x}^3)$ representing the surface measure on $\partial B_r(0)$. 
We have suppressed the function arguments $(t, \bar{x})$ inside the integrals, but we remark that 
$|x| = |\bar{x}| = r$ for any integral performed over $\partial B_r(0)$. From the first and third relation 
in (\ref{macar}) it similarly follows that 
\begin{equation}\label{wehi} 
   \mathfrak{p}^\alpha
   =\frac{c^4}{8\pi G}\int_{B_r(0)}\partial_{x^\gamma} U^{0\alpha\gamma}\,d\bar{x}
   =\frac{c^4}{8\pi G}\int_{B_r(0)}\partial_{x^\gamma} U^{\alpha 0\gamma}\,d\bar{x}.
\end{equation}  
Also (\ref{macar}) yields $U^{\alpha\beta\beta}=0$ for all $\alpha, \beta$, which we use to simplify (\ref{wehi}) into
\[ \mathfrak{p}^\alpha
   =\frac{c^4}{8\pi G}\int_{B_r(0)}\partial_{x^i} U^{\alpha 0 i} \,d\bar{x}
   =\frac{c^4}{8\pi G}\sum_{i=1}^3\int_{\partial B_r(0)}
   \frac{\bar{x}^i}{r}\,U^{\alpha 0 i} \,dS(\bar{x}), \]    
using Gauss's Theorem one more time. Because of $\partial_{x^0}=c^{-1}\partial_t$, 
equations (\ref{wehi}), (\ref{gtls}) and (\ref{macar}) imply the following expression for the rate of change 
of the total energy inside $B_r(0)$:
\begin{eqnarray}\label{morgo} 
   \frac{d}{dt}\,\mathfrak{p}^\alpha 
   & = & c\,\partial_{x^0}\,\mathfrak{p}^\alpha
   \nonumber
   \\ & = & \frac{c^5}{8\pi G}\,\partial_{x^0}
   \int_{B_r(0)}\partial_{x^\gamma} U^{\alpha 0\gamma} \,d\bar{x}
   \nonumber
   \\ & = & -\frac{c^5}{8\pi G}\sum_{i=1}^3
   \int_{\partial B_r(0)}\frac{\bar{x}^i}{r}\,\partial_{x^\gamma} 
   U^{\alpha i\gamma}\,dS(\bar{x})
   \nonumber
   \\ & = & -c\sum_{i=1}^3\int_{\partial B_r(0)} |g|\,\frac{\bar{x}^i}{r}\,
   (T^{\alpha i}+t^{\alpha i})\,dS(\bar{x}). 
\end{eqnarray}
Thus~\eqref{ogin} in the introduction is justified.


\section{Appendix: The harmonic gauge} 
\label{hargau} 
\setcounter{equation}{0} 

In this section, for the sake of making this work self contained, we provide some well known facts related to the harmonic gauge. We follow \cite[p.~162]{weinb}.

The harmonic-gauge condition is 
\begin{equation}\label{hgc} 
   \Gamma^\lambda:=g^{\mu\nu}\Gamma^{\lambda}_{\mu\nu}=0.
\end{equation} 
 
\begin{lemma} (a) One has $\Gamma^\lambda
=-|g|^{-1/2}\,\partial_{x^\kappa}(|g|^{1/2} g^{\lambda\kappa})$. 
In particular, (\ref{hgc}) is equivalent to 
the relation $\partial_{x^\kappa}(|g|^{1/2} g^{\lambda\kappa})=0$. 
\smallskip 

\noindent 
(b) One has $\Gamma^\lambda=-\Box_g x^\lambda$, where $\Box_g\phi=\nabla_\mu\nabla^\mu\phi
=\nabla_\mu (g^{\nu\mu}\nabla_\nu\phi)$. In particular, (\ref{hgc}) is equivalent 
to the relation $\Box_g x^\lambda=0$. 
\smallskip 

\noindent 
(c) The condition (\ref{hgc}) is also equivalent to 
\begin{equation}\label{nijm} 
   g^{\mu\nu}\,\partial_{x^\mu} g_{\nu\lambda}
   =\frac{1}{2}\,g^{\mu\nu}\,\partial_{x^\lambda} g_{\mu\nu}
\end{equation} 
and furthermore to 
\begin{equation}\label{nijm2} 
   \partial_{x^\mu} g^{\mu\lambda}=\frac{1}{2}\,g_{\mu\nu}\,g^{\lambda\kappa}\,\partial_{x^\kappa} g^{\mu\nu}. 
\end{equation}
\end{lemma} 
{\bf Proof\,:} (a) Since 
\[ \Gamma^{\lambda}_{\mu\nu}=\frac{1}{2}\,g^{\lambda\kappa}(\partial_{x^\nu} g_{\kappa\mu}
   +\partial_{x^\mu} g_{\kappa\nu}-\partial_{x^\kappa} g_{\mu\nu}), \] 
one has 
\begin{equation}\label{hgc2} 
   \Gamma^\lambda=\frac{1}{2}\,g^{\mu\nu} g^{\lambda\kappa}(\partial_{x^\nu} g_{\kappa\mu}
   +\partial_{x^\mu} g_{\kappa\nu}-\partial_{x^\kappa} g_{\mu\nu}).
\end{equation} 
First note that $g^{\lambda\kappa} g_{\kappa\mu}=\delta^\lambda_\nu$ yields upon differentiation 
$\partial_{x^\nu}$ that 
\begin{equation}\label{capa} 
   g^{\lambda\kappa}\partial_{x^\nu} g_{\kappa\mu}=-g_{\kappa\mu}\partial_{x^\nu} g^{\lambda\kappa}.
\end{equation} 
Next, in general the determinant function $\det: \R^{n\times n}\to\R$ is differentiable 
at any invertible matrix $M$ and such that $D\det(M)(A)=\det M\,{\rm tr}\,(M^{-1} A)$. Thus if $M=M(x)$ is $x$-dependent, we get 
\[ \partial_{x^\mu}\det M(x)=\det M(x)\,{\rm tr}\,(M^{-1}(x)\partial_{x^\mu} M(x)). \] 
Taking for $M=(g_{\mu\nu})$ the metric tensor, this yields for $g=\det\,(g_{\mu\nu})$ the relation 
\[ g^{-1}\partial_{x^\kappa} g={\rm tr}\,((g^{\mu'\nu'})\,(\partial_{x^\kappa} g_{\mu\nu}))
   =g^{\mu\nu}\,\partial_{x^\kappa} g_{\mu\nu}. \] 
Since $g<0$ this may be rewritten as 
\begin{equation}\label{gartba} 
  |g|^{-1}\partial_{x^\kappa} |g|
  =g^{\mu\nu}\,\partial_{x^\kappa} g_{\mu\nu}. 
\end{equation}   
Together with (\ref{capa}) we arrive at 
\begin{eqnarray}\label{hilt}  
   \Gamma^\lambda & = & 
   \frac{1}{2}\,g^{\mu\nu} g^{\lambda\kappa}\partial_{x^\nu} g_{\kappa\mu}
   +\frac{1}{2}\,g^{\mu\nu} g^{\lambda\kappa}\partial_{x^\mu} g_{\kappa\nu}
   -\frac{1}{2}\,g^{\mu\nu} g^{\lambda\kappa}\partial_{x^\kappa} g_{\mu\nu}
   \nonumber
   \\ & = & -\frac{1}{2}\,g^{\mu\nu} g_{\kappa\mu}\partial_{x^\nu} g^{\lambda\kappa}
   -\frac{1}{2}\,g^{\mu\nu} g_{\kappa\nu}\partial_{x^\mu} g^{\lambda\kappa} 
   -\frac{1}{2}\,g^{\lambda\kappa} |g|^{-1}\partial_{x^\kappa} |g|
   \nonumber
   \\ & = & -\partial_{x^\kappa} g^{\lambda\kappa}
   -\frac{1}{2}\,g^{\lambda\kappa} |g|^{-1}\partial_{x^\kappa} |g|
   \nonumber
   \\ & = & -|g|^{-1/2}\,\partial_{x^\kappa}(|g|^{1/2} g^{\lambda\kappa}), 
\end{eqnarray} 
as claimed. (b) To make a connection to $\Box_g$, we assert that in general 
\begin{equation}\label{boxg} 
   \Box_g\phi=g^{\nu\mu}\,\partial^2_{x^\nu x^\mu}\phi
   -\Gamma^\nu\partial_{x^\nu}\phi.
\end{equation} 
To establish (\ref{boxg}), the first thing to observe is that $\nabla_\nu\phi=\partial_{x^\nu}\phi$, 
since $\phi$ is a scalar. Let $(V^\mu)$ be a contravariant vector. From 
\[ \Gamma^{\mu}_{\mu\nu}=\frac{1}{2}\,g^{\mu\kappa}(\partial_{x^\mu} g_{\kappa\nu}
   +\partial_{x^\nu} g_{\kappa\mu}-\partial_{x^\kappa} g_{\mu\nu})
   =\frac{1}{2}\,g^{\mu\kappa} \partial_{x^\nu} g_{\mu\kappa} \]   
and (\ref{gartba}) we obtain 
\[ \Gamma^{\mu}_{\mu\nu}=\frac{1}{2}\,|g|^{-1}\partial_{x^\nu} |g|
   =|g|^{-1/2}\,\partial_{x^\nu} |g|^{1/2}. \] 
This in turn yields 
\[ \nabla_\mu V^\mu=\partial_{x^\mu} V^\mu+\Gamma^{\mu}_{\mu\nu} V^\nu
   =|g|^{-1/2}\,\partial_{x^\mu} (|g|^{1/2}\,V^\mu) \] 
for the covariant divergence of a contravariant vector. 
Altogether, and using once again (\ref{hilt}), 
\begin{eqnarray*} 
   \Box_g\phi & = & \nabla_\mu (g^{\nu\mu}\nabla_\nu\phi)
   \\ & = & \nabla_\mu (g^{\nu\mu}\,\partial_{x^\nu}\phi)
   \\ & = & |g|^{-1/2}\,\partial_{x^\mu} (|g|^{1/2}\,g^{\nu\mu}\,\partial_{x^\nu}\phi)
   \\ & = & g^{\nu\mu}\,\partial^2_{x^\nu x^\mu}\phi
   +|g|^{-1/2}\,\partial_{x^\mu} (|g|^{1/2}\,g^{\nu\mu})\,\partial_{x^\nu}\phi
   \\ & = & g^{\nu\mu}\,\partial^2_{x^\nu x^\mu}\phi-\Gamma^\nu\partial_{x^\nu}\phi,
\end{eqnarray*} 
which completes the proof of (\ref{boxg}). Since $\partial^2_{x^\nu x^\mu} x^\lambda=0$ 
for all $\nu, \mu, \lambda$, this specializes to $\Box_g x^\lambda=-\Gamma^\lambda$, 
so that the claim follows. 
(c) First note that by definition  
\begin{eqnarray*} 
   2\Gamma^\lambda & = & g^{\mu\nu} g^{\lambda\kappa}(\partial_{x^\nu} g_{\kappa\mu}
   +\partial_{x^\mu} g_{\kappa\nu}-\partial_{x^\kappa} g_{\mu\nu})
   \\ & = & g^{\lambda\kappa}\partial^\mu g_{\kappa\mu}
   +g^{\lambda\kappa}\partial^\nu g_{\kappa\nu}
   -g^{\mu\nu} g^{\lambda\kappa}\partial_{x^\kappa} g_{\mu\nu}
   \\ & = & 2g^{\lambda\kappa}\partial^\mu g_{\kappa\mu}
   -g^{\mu\nu} g^{\lambda\kappa}\partial_{x^\kappa} g_{\mu\nu}, 
\end{eqnarray*} 
which means that 
\begin{equation}\label{wmt} 
   \frac{1}{2}\,g^{\mu\nu} g^{\lambda\kappa}\partial_{x^\kappa} g_{\mu\nu}
   =g^{\lambda\kappa}\partial^\mu g_{\kappa\mu}-\Gamma^\lambda.
\end{equation} 
Hence 	   
\begin{eqnarray}\label{utry}  
   \frac{1}{2}\,g^{\mu\nu}\,\partial_{x^\lambda} g_{\mu\nu}
   & = & \frac{1}{2}\,g^{\mu\nu}\,\delta^\kappa_\lambda\,\partial_{x^\kappa} g_{\mu\nu}
   =\frac{1}{2}\,g^{\mu\nu}\,g_{\lambda\sigma}\,g^{\sigma\kappa}\,\partial_{x^\kappa} g_{\mu\nu}
   =g_{\lambda\sigma}\,(g^{\sigma\kappa}\partial^\mu g_{\kappa\mu}-\Gamma^\sigma)
   \nonumber
   \\ & = & \delta_\lambda^\kappa\,\partial^\mu g_{\kappa\mu}-g_{\lambda\sigma}\Gamma^\sigma
   =\partial^\mu g_{\lambda\mu}-\Gamma_\lambda
   =g^{\mu\nu}\,\partial_{x^\mu} g_{\nu\lambda}-\Gamma_\lambda. 
\end{eqnarray} 
To prove the equivalence of (\ref{hgc}), (\ref{nijm}) and (\ref{nijm2}), 
suppose first that (\ref{hgc}) holds. Then, due to (\ref{utry}),  
\[ \frac{1}{2}\,g^{\mu\nu}\,\partial_{x^\lambda} g_{\mu\nu}
   =g^{\mu\nu}\,\partial_{x^\mu} g_{\nu\lambda}, \] 
which is (\ref{nijm}). If (\ref{nijm}) is verified, 
then we may use (\ref{capa}) twice to obtain 
\begin{eqnarray*} 
   \frac{1}{2}\,g_{\mu\nu}\,g^{\lambda\kappa}\,\partial_{x^\kappa} g^{\mu\nu}
   & = & -\frac{1}{2}\,g^{\lambda\kappa}\,g^{\mu\nu}\,\partial_{x^\kappa} g_{\mu\nu} 
   =-g^{\lambda\kappa}\,g^{\mu\nu}\,\partial_{x^\mu} g_{\nu\kappa}
   \\ & = & g^{\lambda\kappa}\,g_{\nu\kappa}\,\partial_{x^\mu} g^{\mu\nu}
   =\delta_\nu^\lambda\,\partial_{x^\mu} g^{\mu\nu}=\partial_{x^\mu} g^{\mu\lambda}, 
\end{eqnarray*} 
and hence (\ref{nijm2}) is valid. Lastly, suppose that (\ref{nijm2}) holds. 
Then, once again by (\ref{capa}), 
\begin{eqnarray*} 
   \frac{1}{2}\,g^{\mu\nu} g^{\lambda\kappa}\partial_{x^\kappa} g_{\mu\nu}
   & = & \frac{1}{2}\,g^{\lambda\kappa} g^{\mu\nu}\partial_{x^\kappa} g_{\mu\nu}
   =-\frac{1}{2}\,g^{\lambda\kappa} g_{\mu\nu}\partial_{x^\kappa} g^{\mu\nu}
   =-\frac{1}{2}\,g_{\mu\nu} g^{\lambda\kappa}\partial_{x^\kappa} g^{\mu\nu}
   \\ & = & -\partial_{x^\mu} g^{\mu\lambda} = -g_{\kappa\mu}\,\partial^\kappa g^{\mu\lambda}
   = g^{\mu\lambda}\,\partial^\kappa g_{\kappa\mu}, 
\end{eqnarray*} 
but according to (\ref{wmt}) this implies (\ref{hgc}). 
{\hfill$\Box$}\bigskip  

Next we rewrite the Ricci tensor, following \cite[Lemma 3.1]{LR}. First introduce 
\[ \Gamma_{\mu\alpha\nu}=g_{\alpha\lambda}\Gamma_{\mu\nu}^\lambda. \] 
Then 
\begin{eqnarray}\label{alic} 
   \Gamma_{\mu\alpha\nu}
   & = & \frac{1}{2}\,g_{\alpha\lambda}
   \,g^{\lambda\delta}(\partial_{x^\mu} g_{\nu\delta}
   +\partial_{x^\nu} g_{\mu\delta}-\partial_{x^\delta} g_{\mu\nu})
   \nonumber
   \\ & = & \frac{1}{2}\,(\partial_{x^\mu} g_{\nu\alpha}
   +\partial_{x^\nu} g_{\mu\alpha}-\partial_{x^\alpha} g_{\mu\nu}), 
\end{eqnarray}  
and in particular 
\[ \Gamma_{\beta\lambda\alpha}
   =\frac{1}{2}\,(\partial_{x^\beta} g_{\alpha\lambda}
   +\partial_{x^\alpha} g_{\beta\lambda}-\partial_{x^\lambda} g_{\beta\alpha})
   =\frac{1}{2}\,(\partial_{x^\alpha} g_{\beta\lambda}
   +\partial_{x^\beta} g_{\alpha\lambda}-\partial_{x^\lambda} g_{\alpha\beta})
   =\Gamma_{\alpha\lambda\beta}. \] 
Furthermore, 
\[ \partial_{x^\alpha} g_{\beta\mu}
   =\Gamma_{\alpha\beta\mu}+\Gamma_{\alpha\mu\beta}. \]   
In fact, 
\begin{eqnarray*} 
   \Gamma_{\alpha\beta\mu}+\Gamma_{\alpha\mu\beta}
   =\frac{1}{2}\,(\partial_{x^\alpha} g_{\mu\beta}
   +\partial_{x^\mu} g_{\alpha\beta}-\partial_{x^\beta} g_{\alpha\mu})
   +\frac{1}{2}\,(\partial_{x^\alpha} g_{\beta\mu}
   +\partial_{x^\beta} g_{\alpha\mu}-\partial_{x^\mu} g_{\alpha\beta}) 
   =\partial_{x^\alpha} g_{\beta\mu}. 
\end{eqnarray*}  
This in turn implies that 
\begin{eqnarray*} 
   \partial_{x^\beta}\Gamma_{\mu\alpha\nu}
   & = & \partial_{x^\beta}(g_{\alpha\lambda}\Gamma_{\mu\nu}^\lambda)
   =g_{\alpha\lambda}(\partial_{x^\beta}\Gamma_{\mu\nu}^\lambda)
   +(\partial_{x^\beta} g_{\alpha\lambda})\Gamma_{\mu\nu}^\lambda
   \\ & = & g_{\alpha\lambda}(\partial_{x^\beta}\Gamma_{\mu\nu}^\lambda)
   +(\Gamma_{\beta\alpha\lambda}+\Gamma_{\beta\lambda\alpha})\Gamma_{\mu\nu}^\lambda.
\end{eqnarray*} 
The Riemann curvature tensor we will use in the form 
\begin{eqnarray*}
   \tensor{R}{_\mu_\alpha_\nu_\beta}
   & = & g_{\alpha\lambda}\,\tensor{R}{_\mu^\lambda_\nu_\beta}
   \\ & = & g_{\alpha\lambda}\,(\partial_{x^\beta}\Gamma^{\lambda}_{\mu\nu}
   -\partial_{x^\nu}\Gamma^{\lambda}_{\mu\beta}
   +\Gamma^{\lambda}_{\gamma\beta}\,\Gamma^{\gamma}_{\mu\nu}
   -\Gamma^{\lambda}_{\gamma\nu}\,\Gamma^{\gamma}_{\mu\beta})
   \\ & = & \partial_{x^\beta}\Gamma_{\mu\alpha\nu}
    -(\Gamma_{\beta\alpha\lambda}+\Gamma_{\beta\lambda\alpha})\Gamma_{\mu\nu}^\lambda
   -\partial_{x^\nu}\Gamma_{\mu\alpha\beta}
    +(\Gamma_{\nu\alpha\lambda}+\Gamma_{\nu\lambda\alpha})\Gamma_{\mu\beta}^\lambda
   \\ & & +\,\Gamma_{\gamma\alpha\beta}\,\Gamma^{\gamma}_{\mu\nu}
   -\Gamma_{\gamma\alpha\nu}\,\Gamma^{\gamma}_{\mu\beta}
   \\ & = & \partial_{x^\beta}\Gamma_{\mu\alpha\nu}
    -\partial_{x^\nu}\Gamma_{\mu\alpha\beta}
    +\Gamma_{\nu\lambda\alpha}\Gamma_{\mu\beta}^\lambda
    -\Gamma_{\alpha\lambda\beta}\Gamma_{\mu\nu}^\lambda. 
\end{eqnarray*}  
For the Ricci tensor, we consequently obtain 
\begin{equation}\label{bersa} 
   R_{\mu\nu}=\tensor{R}{_\mu^\alpha_\nu_\alpha}
   =g^{\alpha\beta}\,\tensor{R}{_\mu_\beta_\nu_\alpha}
    =g^{\alpha\beta}\,(\partial_{x^\alpha}\Gamma_{\mu\beta\nu}
    -\partial_{x^\nu}\Gamma_{\mu\beta\alpha})
    +g^{\alpha\beta}\,(\Gamma_{\nu\lambda\beta}\Gamma_{\mu\alpha}^\lambda
    -\Gamma_{\beta\lambda\alpha}\Gamma_{\mu\nu}^\lambda).
\end{equation}  
We look at the different parts on the right-hand side of (\ref{bersa}) separately. 
To begin with, 
\begin{equation}\label{aland} 
   g^{\alpha\beta}\,\Gamma_{\beta\lambda\alpha}\Gamma_{\mu\nu}^\lambda
   =g^{\alpha\beta}\,\Gamma_{\alpha\lambda\beta}\Gamma_{\mu\nu}^\lambda
   =g^{\alpha\beta}\,g_{\lambda\gamma}\,\Gamma_{\alpha\beta}^\gamma
   \,\Gamma_{\mu\nu}^\lambda
   =g_{\lambda\gamma}\,\Gamma_{\mu\nu}^\lambda\,\Gamma^\gamma
   =\Gamma_{\mu\nu}^\lambda\,\Gamma_\lambda.
\end{equation} 
Next, by (\ref{alic}) and through a renaming of indices 
$\lambda\leftrightarrow\alpha$, and thereafter $\lambda\leftrightarrow\beta'$, 
$\gamma\leftrightarrow\alpha'$,  
\begin{eqnarray}\label{dafta} 
   g^{\alpha\beta}\,\Gamma_{\nu\lambda\beta}\Gamma_{\mu\alpha}^\lambda
   & = & \frac{1}{4}\,g^{\alpha\beta}\,(\partial_{x^\nu} g_{\beta\lambda}
   +\partial_{x^\beta} g_{\nu\lambda}-\partial_{x^\lambda} g_{\nu\beta})
   \,g^{\lambda\gamma}(\partial_{x^\mu} g_{\alpha\gamma}
   +\partial_{x^\alpha} g_{\mu\gamma}-\partial_{x^\gamma} g_{\mu\alpha})
   \nonumber
   \\ & = & \frac{1}{4}\,g^{\beta\beta'} g^{\alpha\alpha'}
   \,(\partial_{x^\nu} g_{\beta\alpha}
   +\partial_{x^\beta} g_{\alpha\nu}-\partial_{x^\alpha} g_{\beta\nu})
   \,(\partial_{x^\mu} g_{\beta'\alpha'}
   +\partial_{x^{\beta'}} g_{\alpha'\mu}-\partial_{x^{\alpha'}} g_{\beta'\mu}).
   \qquad 
\end{eqnarray} 
To proceed further, note that by exchanging 
$\alpha\leftrightarrow\beta'$, $\beta\leftrightarrow\alpha'$,   
\begin{equation}\label{ta1} 
   g^{\beta\beta'} g^{\alpha\alpha'}\,(\partial_{x^\beta} g_{\alpha\nu})
   \,\partial_{x^{\beta'}} g_{\alpha'\mu}
   =g^{\alpha'\alpha} g^{\beta'\beta}\,\partial_{x^{\alpha'}} g_{\beta'\nu}
   \,\partial_{x^\alpha} g_{\beta\mu},
\end{equation}  
and moreover from $\beta\leftrightarrow\beta'$, $\alpha\leftrightarrow\alpha'$,
\begin{equation}\label{ta2} 
   g^{\beta\beta'} g^{\alpha\alpha'}\,\partial_{x^\alpha} g_{\beta\nu}
   \,\partial_{x^{\alpha'}} g_{\beta'\mu}
   =g^{\beta'\beta} g^{\alpha'\alpha}\,\partial_{x^{\alpha'}} g_{\beta'\nu}
   \,\partial_{x^\alpha} g_{\beta\mu}.
\end{equation}
Similarly, $\beta\leftrightarrow\beta'$, $\alpha\leftrightarrow\alpha'$, yields 
\begin{equation}\label{ta3} 
   g^{\beta\beta'} g^{\alpha\alpha'}\,\partial_{x^\beta} g_{\alpha\nu}
   \,\partial_{x^{\alpha'}} g_{\beta'\mu}
   =g^{\beta'\beta} g^{\alpha'\alpha}\,\partial_{x^{\beta'}} g_{\alpha'\nu}
   \,\partial_{x^\alpha} g_{\beta\mu}
\end{equation}
and $\beta\leftrightarrow\alpha'$, $\alpha\leftrightarrow\beta'$, leads to  
\begin{equation}\label{ta4} 
   g^{\beta\beta'} g^{\alpha\alpha'}\,\partial_{x^\alpha} g_{\beta\nu}
   \,\partial_{x^{\beta'}} g_{\alpha'\mu}
   =g^{\alpha'\alpha} g^{\beta'\beta}\,\partial_{x^{\beta'}} g_{\alpha'\nu}
   \,\partial_{x^\alpha} g_{\beta\mu}.
\end{equation} 
Furthermore, $\alpha\leftrightarrow\beta$, $\alpha'\leftrightarrow\beta'$ implies 
\begin{equation}\label{ta5} 
   g^{\beta\beta'} g^{\alpha\alpha'}\,\partial_{x^\nu} g_{\alpha\beta}
   \,\partial_{x^{\beta'}} g_{\alpha'\mu}
   =g^{\alpha\alpha'} g^{\beta\beta'}\,\partial_{x^\nu} g_{\beta\alpha}
   \,\partial_{x^{\alpha'}} g_{\beta'\mu}
\end{equation}
as well as 
\begin{equation}\label{ta6} 
   g^{\beta\beta'} g^{\alpha\alpha'}\,\partial_{x^\beta} g_{\alpha\nu}
   \,\partial_{x^\mu} g_{\alpha'\beta'}
   =g^{\alpha\alpha'} g^{\beta\beta'}\,\partial_{x^\alpha} g_{\beta\nu}
   \,\partial_{x^\mu} g_{\beta'\alpha'}.
\end{equation} 
Using (\ref{ta1})-(\ref{ta6}) in (\ref{dafta}), it follows that 
\begin{eqnarray*} 
   \lefteqn{g^{\alpha\beta}\,\Gamma_{\nu\lambda\beta}\Gamma_{\mu\alpha}^\lambda}
   \\ & = & \frac{1}{4}\,g^{\beta\beta'} g^{\alpha\alpha'}\,
   \partial_{x^\nu} g_{\beta\alpha}\,\partial_{x^\mu} g_{\beta'\alpha'}
   +\frac{1}{4}\,g^{\beta\beta'} g^{\alpha\alpha'}\,
   \partial_{x^\nu} g_{\beta\alpha}\,\partial_{x^{\beta'}} g_{\alpha'\mu}
   -\frac{1}{4}\,g^{\beta\beta'} g^{\alpha\alpha'}\,
   \partial_{x^\nu} g_{\beta\alpha}\,\partial_{x^{\alpha'}} g_{\beta'\mu}
   \\ & & +\,\frac{1}{4}\,g^{\beta\beta'} g^{\alpha\alpha'}\,
   \partial_{x^\beta} g_{\alpha\nu}\,\partial_{x^\mu} g_{\beta'\alpha'}
   +\frac{1}{4}\,g^{\beta\beta'} g^{\alpha\alpha'}\,
   \partial_{x^\beta} g_{\alpha\nu}\,\partial_{x^{\beta'}} g_{\alpha'\mu}
   -\frac{1}{4}\,g^{\beta\beta'} g^{\alpha\alpha'}\,
   \partial_{x^\beta} g_{\alpha\nu}\,\partial_{x^{\alpha'}} g_{\beta'\mu}
   \\ & & -\,\frac{1}{4}\,g^{\beta\beta'} g^{\alpha\alpha'}\,
   \partial_{x^\alpha} g_{\beta\nu}\,\partial_{x^\mu} g_{\beta'\alpha'}
   -\frac{1}{4}\,g^{\beta\beta'} g^{\alpha\alpha'}\,
   \partial_{x^\alpha} g_{\beta\nu}\,\partial_{x^{\beta'}} g_{\alpha'\mu}
   +\frac{1}{4}\,g^{\beta\beta'} g^{\alpha\alpha'}\,
   \partial_{x^\alpha} g_{\beta\nu}\,\partial_{x^{\alpha'}} g_{\beta'\mu}
   \\ & = & \frac{1}{4}\,g^{\beta\beta'} g^{\alpha\alpha'}\,
   \partial_{x^\nu} g_{\beta\alpha}\,\partial_{x^\mu} g_{\beta'\alpha'}
   +\frac{1}{4}\,g^{\alpha\alpha'} g^{\beta\beta'}\,\partial_{x^\nu} g_{\beta\alpha}
   \,\partial_{x^{\alpha'}} g_{\beta'\mu}
   -\frac{1}{4}\,g^{\beta\beta'} g^{\alpha\alpha'}\,
   \partial_{x^\nu} g_{\beta\alpha}\,\partial_{x^{\alpha'}} g_{\beta'\mu}
   \\ & & +\,\frac{1}{4}\,g^{\alpha\alpha'} g^{\beta\beta'}\,\partial_{x^\alpha} g_{\beta\nu}
   \,\partial_{x^\mu} g_{\beta'\alpha'}
   +\frac{1}{4}\,g^{\alpha'\alpha} g^{\beta'\beta}\,\partial_{x^{\alpha'}} g_{\beta'\nu}
   \,\partial_{x^\alpha} g_{\beta\mu}
   -\frac{1}{4}\,g^{\beta'\beta} g^{\alpha'\alpha}\,\partial_{x^{\beta'}} g_{\alpha'\nu}
   \,\partial_{x^\alpha} g_{\beta\mu}
   \\ & & -\,\frac{1}{4}\,g^{\beta\beta'} g^{\alpha\alpha'}\,
   \partial_{x^\alpha} g_{\beta\nu}\,\partial_{x^\mu} g_{\beta'\alpha'}
   -\frac{1}{4}\,g^{\alpha'\alpha} g^{\beta'\beta}\,\partial_{x^{\beta'}} g_{\alpha'\nu}
   \,\partial_{x^\alpha} g_{\beta\mu}
   +\frac{1}{4}\,g^{\beta'\beta} g^{\alpha'\alpha}\,\partial_{x^{\alpha'}} g_{\beta'\nu}
   \,\partial_{x^\alpha} g_{\beta\mu}
   \\ & = & g^{\alpha\alpha'} g^{\beta\beta'}\,\bigg(
   \frac{1}{4}\,\partial_{x^\nu} g_{\alpha\beta}\,\partial_{x^\mu} g_{\alpha'\beta'}
   +\frac{1}{2}\,\partial_{x^\alpha} g_{\beta\mu}\,\partial_{x^{\alpha'}} g_{\beta'\nu}
   -\frac{1}{2}\,\partial_{x^\alpha} g_{\beta\mu}\,\partial_{x^{\beta'}} g_{\alpha'\nu}\bigg). 
\end{eqnarray*} 
This is further rewritten as 
\begin{eqnarray}\label{fgwaltp} 
   \lefteqn{g^{\alpha\beta}\,\Gamma_{\nu\lambda\beta}\Gamma_{\mu\alpha}^\lambda}
   \nonumber
   \\ & = & g^{\alpha\alpha'} g^{\beta\beta'}\,\bigg(
   \frac{1}{4}\,\partial_{x^\nu} g_{\alpha\beta}\,\partial_{x^\mu} g_{\alpha'\beta'}
   +\frac{1}{2}\,\partial_{x^\alpha} g_{\beta\mu}\,\partial_{x^{\alpha'}} g_{\beta'\nu}
   -\frac{1}{2}\,\partial_{x^{\beta'}} g_{\beta\mu}\,\partial_{x^\alpha} g_{\alpha'\nu}\bigg)
   \nonumber
   \\ & & -\frac{1}{2}\,g^{\alpha\alpha'} g^{\beta\beta'}\,
   \bigg(\partial_{x^\alpha} g_{\beta\mu}\,\partial_{x^{\beta'}} g_{\alpha'\nu}
   -\partial_{x^{\beta'}} g_{\beta\mu}\,\partial_{x^\alpha} g_{\alpha'\nu}\bigg). 
\end{eqnarray}  
The next step is to observe that, by (\ref{utry}),   
\begin{eqnarray*} 
   g^{\alpha\alpha'} g^{\beta\beta'}
   \,\partial_{x^{\beta'}} g_{\beta\mu}\,\partial_{x^\alpha} g_{\alpha'\nu} 
   & = & (g^{\beta\beta'}\,\partial_{x^{\beta'}} g_{\beta\mu})\,
   (g^{\alpha\alpha'}\partial_{x^\alpha} g_{\alpha'\nu})
   \\ & = & \bigg(\frac{1}{2}\,g^{\beta\beta'}\,\partial_{x^\mu} g_{\beta\beta'}+\Gamma_\mu\bigg)\,
   \bigg(\frac{1}{2}\,g^{\alpha\alpha'}\partial_{x^\nu} g_{\alpha\alpha'}+\Gamma_\nu\bigg).
\end{eqnarray*}  
Going back to (\ref{fgwaltp}), we arrive at
\begin{eqnarray}\label{fgwalt}
   g^{\alpha\beta}\,\Gamma_{\nu\lambda\beta}\Gamma_{\mu\alpha}^\lambda
   & = & g^{\alpha\alpha'} g^{\beta\beta'}\,\bigg(
   \frac{1}{4}\,\partial_{x^\nu} g_{\alpha\beta}\,\partial_{x^\mu} g_{\alpha'\beta'}
   +\frac{1}{2}\,\partial_{x^\alpha} g_{\beta\mu}\,\partial_{x^{\alpha'}} g_{\beta'\nu}\bigg)
   \nonumber
   \\ & & -\,\frac{1}{2}\,\bigg(\frac{1}{2}\,g^{\beta\beta'}
   \,\partial_{x^\mu} g_{\beta\beta'}+\Gamma_\mu\bigg)\,
   \bigg(\frac{1}{2}\,g^{\alpha\alpha'}\partial_{x^\nu} g_{\alpha\alpha'}+\Gamma_\nu\bigg)
   \nonumber
   \\ & & -\frac{1}{2}\,g^{\alpha\alpha'} g^{\beta\beta'}\,
   \bigg(\partial_{x^\alpha} g_{\beta\mu}\,\partial_{x^{\beta'}} g_{\alpha'\nu}
   -\partial_{x^{\beta'}} g_{\beta\mu}\,\partial_{x^\alpha} g_{\alpha'\nu}\bigg). 
\end{eqnarray} 
Lastly, we look at the contribution of $g^{\alpha\beta}
(\partial_{x^\alpha}\Gamma_{\mu\beta\nu}-\partial_{x^\nu}\Gamma_{\mu\beta\alpha})$ 
to (\ref{bersa}), which contains the `principal part'. From (\ref{alic}) we find 
\begin{eqnarray}\label{peterso}  
   g^{\alpha\beta}(\partial_{x^\alpha}\Gamma_{\mu\beta\nu}
   -\partial_{x^\nu}\Gamma_{\mu\beta\alpha})
   & = & \frac{1}{2}\,g^{\alpha\beta}(\partial^2_{x^\alpha x^\mu} g_{\nu\beta}
   +\partial^2_{x^\alpha x^\nu} g_{\mu\beta}-\partial^2_{x^\alpha x^\beta} g_{\mu\nu})
   \nonumber
   \\ & & -\,\frac{1}{2}\,g^{\alpha\beta}
   (\partial^2_{x^\nu x^\mu} g_{\alpha\beta}
   +\partial^2_{x^\nu x^\alpha} g_{\mu\beta}-\partial^2_{x^\nu x^\beta} g_{\mu\alpha})
   \nonumber
   \\ & = & -\frac{1}{2}\,\Box'_g\,g_{\mu\nu}
   +\frac{1}{2}\,g^{\alpha\beta}(\partial^2_{x^\alpha x^\mu} g_{\beta\nu}
   +\partial^2_{x^\nu x^\beta} g_{\mu\alpha}-\partial^2_{x^\nu x^\mu} g_{\alpha\beta})
   \nonumber
   \\ & = & -\frac{1}{2}\,\Box'_g\,g_{\mu\nu}
   +\frac{1}{2}\,g^{\alpha\beta}\bigg(\partial^2_{x^\alpha x^\mu} g_{\beta\nu}
   -\frac{1}{2}\,\partial^2_{x^\nu x^\mu} g_{\alpha\beta}\bigg)
   \nonumber
   \\ & & +\,\frac{1}{2}\,g^{\alpha\beta}\bigg(\partial^2_{x^\nu x^\beta} g_{\mu\alpha}
   -\frac{1}{2}\,\partial^2_{x^\nu x^\mu} g_{\alpha\beta}\bigg), 
\end{eqnarray} 
with the operator $\Box'_g=g^{\alpha\beta}\,\partial^2_{x^\alpha x^\beta}$; 
note that $\Box'_g=\Box_g$ in the harmonic gauge due to (\ref{boxg}). 
From (\ref{utry}) we get 
\[ g^{\alpha\beta}\,\partial_{x^\alpha} g_{\beta\nu}
   -\frac{1}{2}\,g^{\alpha\beta}\,\partial_{x^\nu} g_{\alpha\beta}-\Gamma_\nu=0, \] 
and differentiation $\partial_{x^\mu}$ yields 
\[ g^{\alpha\beta}\bigg(\partial^2_{x^\alpha x^\mu} g_{\beta\nu}
   -\frac{1}{2}\,\partial^2_{x^\nu x^\mu} g_{\alpha\beta}\bigg)
   =-\partial_{x^\mu} g^{\alpha\beta}\bigg(\partial_{x^\alpha} g_{\beta\nu}
   -\frac{1}{2}\,\partial_{x^\nu} g_{\alpha\beta}\bigg)
   +\partial_{x^\mu}\Gamma_\nu. \]    
Using this, and the same relation for 
$\alpha\leftrightarrow\beta$, $\mu\leftrightarrow\nu$, 
(\ref{peterso}) leads to    
\begin{eqnarray}\label{scao}  
   \lefteqn{g^{\alpha\beta}(\partial_{x^\alpha}\Gamma_{\mu\beta\nu}
   -\partial_{x^\nu}\Gamma_{\mu\beta\alpha})}
   \nonumber
   \\ & = & -\frac{1}{2}\,\Box'_g\,g_{\mu\nu}
   -\frac{1}{2}\,\partial_{x^\mu} g^{\alpha\beta}\bigg(\partial_{x^\alpha} g_{\beta\nu}
   -\frac{1}{2}\,\partial_{x^\nu} g_{\alpha\beta}\bigg)
   +\frac{1}{2}\,\partial_{x^\mu}\Gamma_\nu
   \nonumber
   \\ & & -\,\frac{1}{2}\,\partial_{x^\nu} g^{\alpha\beta}\bigg(\partial_{x^\beta} g_{\alpha\mu}
   -\frac{1}{2}\,\partial_{x^\mu} g_{\alpha\beta}\bigg)
   +\frac{1}{2}\,\partial_{x^\nu}\Gamma_\mu. 
\end{eqnarray}   
Furthermore, 
\begin{eqnarray*}
   \partial_{x^\mu} g^{\alpha\beta}
   & = & \partial_{x^\mu} (g^{\alpha\alpha'} g^{\beta\beta'} g_{\alpha'\beta'})
   \\ & = & (\partial_{x^\mu} g^{\alpha\alpha'}) g^{\beta\beta'} g_{\alpha'\beta'}
   +g^{\alpha\alpha'} (\partial_{x^\mu} g^{\beta\beta'}) g_{\alpha'\beta'}  
   +g^{\alpha\alpha'} g^{\beta\beta'} (\partial_{x^\mu}g_{\alpha'\beta'})
   \\ & = & 2\partial_{x^\mu} g^{\alpha\beta}
   +g^{\alpha\alpha'} g^{\beta\beta'} (\partial_{x^\mu}g_{\alpha'\beta'})      
\end{eqnarray*} 
shows that 
\[ \partial_{x^\mu} g^{\alpha\beta}
   = -g^{\alpha\alpha'} g^{\beta\beta'} (\partial_{x^\mu}g_{\alpha'\beta'}). \]    
Using this, and the same relation for 
$\alpha\leftrightarrow\beta$, $\mu\leftrightarrow\nu$, 
we can continue (\ref{scao}) as 
\begin{eqnarray}\label{madeir} 
   \lefteqn{g^{\alpha\beta}(\partial_{x^\alpha}\Gamma_{\mu\beta\nu}
   -\partial_{x^\nu}\Gamma_{\mu\beta\alpha})}
   \nonumber
   \\ & = & -\frac{1}{2}\,\Box'_g\,g_{\mu\nu}
   +\frac{1}{2}\,g^{\alpha\alpha'} g^{\beta\beta'} (\partial_{x^\mu}g_{\alpha'\beta'})   
   \bigg(\partial_{x^\alpha} g_{\beta\nu}
   -\frac{1}{2}\,\partial_{x^\nu} g_{\alpha\beta}\bigg)
   \nonumber
   \\ & & +\,\frac{1}{2}\,g^{\beta\alpha'} g^{\alpha\beta'} (\partial_{x^\nu}g_{\alpha'\beta'})
   \bigg(\partial_{x^\beta} g_{\alpha\mu}
   -\frac{1}{2}\,\partial_{x^\mu} g_{\alpha\beta}\bigg)
   +\frac{1}{2}\,(\partial_{x^\mu}\Gamma_\nu+\partial_{x^\nu}\Gamma_\mu). 
\end{eqnarray} 
Due to $\alpha\leftrightarrow\beta$ we get 
\begin{equation}\label{trep1} 
   g^{\beta\alpha'} g^{\alpha\beta'} (\partial_{x^\nu}g_{\alpha'\beta'})
   \,(\partial_{x^\beta} g_{\alpha\mu})
   =g^{\alpha\alpha'} g^{\beta\beta'} (\partial_{x^\nu}g_{\alpha'\beta'})
   \,(\partial_{x^\alpha} g_{\beta\mu})
\end{equation} 
as well as 
\begin{equation}\label{trep2} 
   g^{\beta\alpha'} g^{\alpha\beta'} (\partial_{x^\nu}g_{\alpha'\beta'})
   \,(\partial_{x^\mu} g_{\alpha\beta})
   =g^{\alpha\alpha'} g^{\beta\beta'} (\partial_{x^\nu}g_{\alpha'\beta'})
   \,(\partial_{x^\mu} g_{\alpha\beta}),
\end{equation}   
and $\alpha\leftrightarrow\alpha'$, $\beta\leftrightarrow\beta'$ gives 
\begin{equation}\label{trep3} 
   g^{\alpha\alpha'} g^{\beta\beta'} (\partial_{x^\mu}g_{\alpha'\beta'})   
   \,(\partial_{x^\nu} g_{\alpha\beta})
   =g^{\alpha\alpha'} g^{\beta\beta'} (\partial_{x^\mu}g_{\alpha\beta})   
   \,(\partial_{x^\nu} g_{\alpha'\beta'}).
\end{equation}  
From (\ref{trep1})-(\ref{trep3}) applied in (\ref{madeir}) we deduce 
\begin{eqnarray}\label{walde} 
   \lefteqn{g^{\alpha\beta}(\partial_{x^\alpha}\Gamma_{\mu\beta\nu}
   -\partial_{x^\nu}\Gamma_{\mu\beta\alpha})}
   \nonumber
   \\ & = & -\frac{1}{2}\,\Box'_g\,g_{\mu\nu}
   \nonumber
   \\ & & +\,\frac{1}{2}\,g^{\alpha\alpha'} g^{\beta\beta'}
   \Big(\partial_{x^\mu}g_{\alpha'\beta'}\,\partial_{x^\alpha} g_{\beta\nu}
   +\partial_{x^\nu}g_{\alpha'\beta'}\,\partial_{x^\alpha} g_{\beta\mu}
   -\partial_{x^\nu} g_{\alpha'\beta'}\,\partial_{x^\mu}g_{\alpha\beta}\Big)
   \nonumber
   \\ & & +\,\frac{1}{2}\,(\partial_{x^\mu}\Gamma_\nu+\partial_{x^\nu}\Gamma_\mu). 
\end{eqnarray} 
To re-express the middle term on the right-hand side, 
we invoke (\ref{utry}) twice to obtain 
\begin{eqnarray*} 
   \lefteqn{g^{\alpha\alpha'} g^{\beta\beta'}
   \partial_{x^\mu}g_{\alpha'\beta'}\,\partial_{x^\alpha} g_{\beta\nu}} 
   \\ & = & g^{\alpha\alpha'} g^{\beta\beta'}
   \partial_{x^\alpha} g_{\alpha'\beta'}\,\partial_{x^\mu} g_{\beta\nu}
   +g^{\alpha\alpha'} g^{\beta\beta'}
   \Big(\partial_{x^\mu}g_{\alpha'\beta'}\,\partial_{x^\alpha} g_{\beta\nu}   
   -\partial_{x^\alpha} g_{\alpha'\beta'}\,\partial_{x^\mu} g_{\beta\nu}\Big)
   \\ & = & \frac{1}{2}\,g^{\alpha\alpha'} g^{\beta\beta'}
   \,\partial_{x^{\beta'}} g_{\alpha\alpha'}\,\partial_{x^\mu} g_{\beta\nu}   
   +\Gamma^{\beta}\partial_{x^\mu} g_{\beta\nu}
   \\ & & +\,g^{\alpha\alpha'} g^{\beta\beta'}
   \Big(\partial_{x^\mu}g_{\alpha'\beta'}\,\partial_{x^\alpha} g_{\beta\nu}   
   -\partial_{x^\alpha} g_{\alpha'\beta'}\,\partial_{x^\mu} g_{\beta\nu}\Big)
   \\ & = & \frac{1}{2}\,g^{\alpha\alpha'} g^{\beta\beta'}
   \,\partial_{x^\mu} g_{\alpha'\alpha}\,\partial_{x^{\beta'}} g_{\beta\nu}   
   \\ & & +\,g^{\alpha\alpha'} g^{\beta\beta'}
   \Big(\frac{1}{2}\,(\partial_{x^{\beta'}} g_{\alpha\alpha'}\,\partial_{x^\mu} g_{\beta\nu}
   -\partial_{x^\mu} g_{\alpha'\alpha}\,\partial_{x^{\beta'}} g_{\beta\nu})
   +(\partial_{x^\mu}g_{\alpha'\beta'}\,\partial_{x^\alpha} g_{\beta\nu}   
   -\partial_{x^\alpha} g_{\alpha'\beta'}\,\partial_{x^\mu} g_{\beta\nu})\Big)
   \\ & & +\,\Gamma^{\beta}\partial_{x^\mu} g_{\beta\nu}
   \\ & = & \frac{1}{4}\,g^{\alpha\alpha'} g^{\beta\beta'}
   \,\partial_{x^\mu} g_{\alpha\alpha'}\,\partial_{x^\nu} g_{\beta\beta'}
   \\ & & +\,g^{\alpha\alpha'} g^{\beta\beta'}
   \Big(\frac{1}{2}\,(\partial_{x^{\beta'}} g_{\alpha\alpha'}\,\partial_{x^\mu} g_{\beta\nu}
   -\partial_{x^\mu} g_{\alpha'\alpha}\,\partial_{x^{\beta'}} g_{\beta\nu})
   +(\partial_{x^\mu}g_{\alpha'\beta'}\,\partial_{x^\alpha} g_{\beta\nu}   
   -\partial_{x^\alpha} g_{\alpha'\beta'}\,\partial_{x^\mu} g_{\beta\nu})\Big)
   \\ & & +\,\Gamma^{\beta}\partial_{x^\mu} g_{\beta\nu}
   +\frac{1}{2}\,g^{\alpha\alpha'} 
   \,\partial_{x^\mu} g_{\alpha'\alpha}\,\Gamma_\nu.               
\end{eqnarray*}
From this expression, and the same with $\mu\leftrightarrow\nu$, 
in (\ref{walde}), we get
\begin{eqnarray}\label{johans}
   \lefteqn{g^{\alpha\beta}(\partial_{x^\alpha}\Gamma_{\mu\beta\nu}
   -\partial_{x^\nu}\Gamma_{\mu\beta\alpha})}
   \nonumber
   \\ & = & -\frac{1}{2}\,\Box'_g\,g_{\mu\nu}
   \nonumber
   \\ & & +\,\frac{1}{8}\,g^{\alpha\alpha'} g^{\beta\beta'}
   \,\partial_{x^\mu} g_{\alpha\alpha'}\,\partial_{x^\nu} g_{\beta\beta'}
   \nonumber
   \\ & & +\,\frac{1}{2}\,g^{\alpha\alpha'} g^{\beta\beta'}
   \Big(\frac{1}{2}\,(\partial_{x^{\beta'}} g_{\alpha\alpha'}\,\partial_{x^\mu} g_{\beta\nu}
   -\partial_{x^\mu} g_{\alpha'\alpha}\,\partial_{x^{\beta'}} g_{\beta\nu})
   +(\partial_{x^\mu}g_{\alpha'\beta'}\,\partial_{x^\alpha} g_{\beta\nu}   
   -\partial_{x^\alpha} g_{\alpha'\beta'}\,\partial_{x^\mu} g_{\beta\nu})\Big)
   \nonumber
   \\ & & +\frac{1}{2}\,\Gamma^{\beta}\partial_{x^\mu} g_{\beta\nu}
   +\frac{1}{4}\,g^{\alpha\alpha'} 
   \,\partial_{x^\mu} g_{\alpha'\alpha}\,\Gamma_\nu
   \nonumber
   \\ & & +\frac{1}{8}\,g^{\alpha\alpha'} g^{\beta\beta'}
   \,\partial_{x^\nu} g_{\alpha\alpha'}\,\partial_{x^\mu} g_{\beta\beta'}
   \nonumber
   \\ & & +\,\frac{1}{2}\,g^{\alpha\alpha'} g^{\beta\beta'}
   \Big(\frac{1}{2}\,(\partial_{x^{\beta'}} g_{\alpha\alpha'}\,\partial_{x^\nu} g_{\beta\mu}
   -\partial_{x^\nu} g_{\alpha'\alpha}\,\partial_{x^{\beta'}} g_{\beta\mu})
   +(\partial_{x^\nu} g_{\alpha'\beta'}\,\partial_{x^\alpha} g_{\beta\mu}   
   -\partial_{x^\alpha} g_{\alpha'\beta'}\,\partial_{x^\nu} g_{\beta\mu})\Big)
   \nonumber
   \\ & & +\frac{1}{2}\,\Gamma^{\beta}\partial_{x^\nu} g_{\beta\mu}
   +\frac{1}{4}\,g^{\alpha\alpha'} 
   \,\partial_{x^\nu} g_{\alpha'\alpha}\,\Gamma_\mu
   \nonumber
   \\ & & -\,\frac{1}{2}\,g^{\alpha\alpha'} g^{\beta\beta'}
   \partial_{x^\nu} g_{\alpha'\beta'}\,\partial_{x^\mu}g_{\alpha\beta}
   \nonumber
   \\ & & +\,\frac{1}{2}\,(\partial_{x^\mu}\Gamma_\nu+\partial_{x^\nu}\Gamma_\mu)
   \nonumber
   \\ & = & -\frac{1}{2}\,\Box'_g\,g_{\mu\nu}
   \nonumber
   \\ & & +\,g^{\alpha\alpha'} g^{\beta\beta'}\bigg(
   \frac{1}{4}\,\partial_{x^\mu} g_{\alpha\alpha'}\,\partial_{x^\nu} g_{\beta\beta'}
   -\frac{1}{2}\,\partial_{x^\nu} g_{\alpha'\beta'}\,\partial_{x^\mu} g_{\alpha\beta}\bigg)
   \nonumber
   \\ & & +\,\frac{1}{4}\,g^{\alpha\alpha'} g^{\beta\beta'} 
   \Big((\partial_{x^{\beta'}} g_{\alpha\alpha'}\,\partial_{x^\mu} g_{\beta\nu}
   -\partial_{x^\mu} g_{\alpha'\alpha}\,\partial_{x^{\beta'}} g_{\beta\nu})
   +(\partial_{x^{\beta'}} g_{\alpha\alpha'}\,\partial_{x^\nu} g_{\beta\mu}
   -\partial_{x^\nu} g_{\alpha'\alpha}\,\partial_{x^{\beta'}} g_{\beta\mu})\Big)
   \nonumber
   \\ & & +\,\frac{1}{2}\,g^{\alpha\alpha'} g^{\beta\beta'} 
   \Big((\partial_{x^\mu}g_{\alpha'\beta'}\,\partial_{x^\alpha} g_{\beta\nu}   
   -\partial_{x^\alpha} g_{\alpha'\beta'}\,\partial_{x^\mu} g_{\beta\nu})
   +(\partial_{x^\nu} g_{\alpha'\beta'}\,\partial_{x^\alpha} g_{\beta\mu}   
   -\partial_{x^\alpha} g_{\alpha'\beta'}\,\partial_{x^\nu} g_{\beta\mu})\Big)
   \nonumber
   \\ & & +\,\frac{1}{2}\,(\partial_{x^\mu}\Gamma_\nu+\partial_{x^\nu}\Gamma_\mu)
   +\frac{1}{2}\,\Gamma^{\beta}(\partial_{x^\mu} g_{\beta\nu}
   +\partial_{x^\nu} g_{\beta\mu})
   +\frac{1}{4}\,g^{\alpha\alpha'}(\partial_{x^\mu} g_{\alpha\alpha'}\,\Gamma_\nu
   +\partial_{x^\nu} g_{\alpha\alpha'}\,\Gamma_\mu).  
\end{eqnarray} 
Summarizing (\ref{bersa}), (\ref{johans}), (\ref{fgwalt}) and (\ref{aland}), 
we finally obtain 
\begin{eqnarray*} 
   R_{\mu\nu} & = & g^{\alpha\beta}\,(\partial_{x^\alpha}\Gamma_{\mu\beta\nu}
   -\partial_{x^\nu}\Gamma_{\mu\beta\alpha})
   +g^{\alpha\beta}\,\Gamma_{\nu\lambda\beta}\Gamma_{\mu\alpha}^\lambda
   -g^{\alpha\beta}\,\Gamma_{\beta\lambda\alpha}\Gamma_{\mu\nu}^\lambda
   \\ & = &  -\frac{1}{2}\,\Box'_g\,g_{\mu\nu}
   \\ & & +\,g^{\alpha\alpha'} g^{\beta\beta'}\bigg(
   \frac{1}{4}\,\partial_{x^\mu} g_{\alpha\alpha'}\,\partial_{x^\nu} g_{\beta\beta'}
   -\frac{1}{2}\,\partial_{x^\nu} g_{\alpha'\beta'}\,\partial_{x^\mu} g_{\alpha\beta}\bigg)
   \\ & & +\,\frac{1}{4}\,g^{\alpha\alpha'} g^{\beta\beta'} 
   \Big((\partial_{x^{\beta'}} g_{\alpha\alpha'}\,\partial_{x^\mu} g_{\beta\nu}
   -\partial_{x^\mu} g_{\alpha'\alpha}\,\partial_{x^{\beta'}} g_{\beta\nu})
   +(\partial_{x^{\beta'}} g_{\alpha\alpha'}\,\partial_{x^\nu} g_{\beta\mu}
   -\partial_{x^\nu} g_{\alpha'\alpha}\,\partial_{x^{\beta'}} g_{\beta\mu})\Big)
   \\ & & +\,\frac{1}{2}\,g^{\alpha\alpha'} g^{\beta\beta'} 
   \Big((\partial_{x^\mu}g_{\alpha'\beta'}\,\partial_{x^\alpha} g_{\beta\nu}   
   -\partial_{x^\alpha} g_{\alpha'\beta'}\,\partial_{x^\mu} g_{\beta\nu})
   +(\partial_{x^\nu} g_{\alpha'\beta'}\,\partial_{x^\alpha} g_{\beta\mu}   
   -\partial_{x^\alpha} g_{\alpha'\beta'}\,\partial_{x^\nu} g_{\beta\mu})\Big)
   \\ & & +\,\frac{1}{2}\,(\partial_{x^\mu}\Gamma_\nu+\partial_{x^\nu}\Gamma_\mu)
   +\frac{1}{2}\,\Gamma^{\beta}(\partial_{x^\mu} g_{\beta\nu}
   +\partial_{x^\nu} g_{\beta\mu})
   +\frac{1}{4}\,g^{\alpha\alpha'}(\partial_{x^\mu} g_{\alpha\alpha'}\,\Gamma_\nu
   +\partial_{x^\nu} g_{\alpha\alpha'}\,\Gamma_\mu) 
   \\ & & +\,g^{\alpha\alpha'} g^{\beta\beta'}\,\bigg(
   \frac{1}{4}\,\partial_{x^\nu} g_{\alpha\beta}\,\partial_{x^\mu} g_{\alpha'\beta'}
   +\frac{1}{2}\,\partial_{x^\alpha} g_{\beta\mu}\,\partial_{x^{\alpha'}} g_{\beta'\nu}\bigg)
   \\ & & -\,\frac{1}{2}\,\bigg(\frac{1}{2}\,g^{\beta\beta'}
   \,\partial_{x^\mu} g_{\beta\beta'}+\Gamma_\mu\bigg)\,
   \bigg(\frac{1}{2}\,g^{\alpha\alpha'}\partial_{x^\nu} g_{\alpha\alpha'}+\Gamma_\nu\bigg)
   \\ & & -\frac{1}{2}\,g^{\alpha\alpha'} g^{\beta\beta'}\,
   \bigg(\partial_{x^\alpha} g_{\beta\mu}\,\partial_{x^{\beta'}} g_{\alpha'\nu}
   -\partial_{x^{\beta'}} g_{\beta\mu}\,\partial_{x^\alpha} g_{\alpha'\nu}\bigg)
   \\ & & -\,\Gamma_{\mu\nu}^\lambda\,\Gamma_\lambda
   \\ & = & -\frac{1}{2}\,\Box'_g\,g_{\mu\nu}
   \\ & & +\,\frac{1}{4}\,g^{\alpha\alpha'} g^{\beta\beta'}
   \partial_{x^\mu} g_{\alpha\alpha'}\,\partial_{x^\nu} g_{\beta\beta'}
   -\frac{1}{2}\,g^{\alpha\alpha'} g^{\beta\beta'}
   \partial_{x^\nu} g_{\alpha'\beta'}\,\partial_{x^\mu} g_{\alpha\beta}
   \\ & & +\,\frac{1}{4}\,g^{\alpha\alpha'} g^{\beta\beta'}
   \,\partial_{x^\nu} g_{\alpha\beta}\,\partial_{x^\mu} g_{\alpha'\beta'}
   -\frac{1}{8}\,g^{\alpha\alpha'} g^{\beta\beta'}
   \partial_{x^\mu} g_{\beta\beta'}\,\partial_{x^\nu} g_{\alpha\alpha'}
   \\ & & +\,\frac{1}{2}\,g^{\alpha\alpha'} g^{\beta\beta'}
   \partial_{x^\alpha} g_{\beta\mu}\,\partial_{x^{\alpha'}} g_{\beta'\nu}
   -\frac{1}{2}\,g^{\alpha\alpha'} g^{\beta\beta'}\,
   \bigg(\partial_{x^\alpha} g_{\beta\mu}\,\partial_{x^{\beta'}} g_{\alpha'\nu}
   -\partial_{x^{\beta'}} g_{\beta\mu}\,\partial_{x^\alpha} g_{\alpha'\nu}\bigg)
   \\ & & +\,\frac{1}{2}\,g^{\alpha\alpha'} g^{\beta\beta'} 
   \Big((\partial_{x^\mu}g_{\alpha'\beta'}\,\partial_{x^\alpha} g_{\beta\nu}   
   -\partial_{x^\alpha} g_{\alpha'\beta'}\,\partial_{x^\mu} g_{\beta\nu})
   +(\partial_{x^\nu} g_{\alpha'\beta'}\,\partial_{x^\alpha} g_{\beta\mu}   
   -\partial_{x^\alpha} g_{\alpha'\beta'}\,\partial_{x^\nu} g_{\beta\mu})\Big)
   \\ & & +\,\frac{1}{4}\,g^{\alpha\alpha'} g^{\beta\beta'} 
   \Big((\partial_{x^{\beta'}} g_{\alpha\alpha'}\,\partial_{x^\mu} g_{\beta\nu}
   -\partial_{x^\mu} g_{\alpha'\alpha}\,\partial_{x^{\beta'}} g_{\beta\nu})
   +(\partial_{x^{\beta'}} g_{\alpha\alpha'}\,\partial_{x^\nu} g_{\beta\mu}
   -\partial_{x^\nu} g_{\alpha'\alpha}\,\partial_{x^{\beta'}} g_{\beta\mu})\Big)   
   \\ & & +\,\frac{1}{2}\,(\partial_{x^\mu}\Gamma_\nu+\partial_{x^\nu}\Gamma_\mu)
   +\frac{1}{2}\,\Gamma^{\beta}(\partial_{x^\mu} g_{\beta\nu}
   +\partial_{x^\nu} g_{\beta\mu})
   +\frac{1}{4}\,g^{\alpha\alpha'}(\partial_{x^\mu} g_{\alpha\alpha'}\,\Gamma_\nu
   +\partial_{x^\nu} g_{\alpha\alpha'}\,\Gamma_\mu)    
   \\ & & -\,\frac{1}{4}\,g^{\beta\beta'}\partial_{x^\mu} g_{\beta\beta'}\,\Gamma_\nu
   -\frac{1}{4}\,g^{\alpha\alpha'}\partial_{x^\nu} g_{\alpha\alpha'}\,\Gamma_\mu
   -\frac{1}{2}\,\Gamma_\mu\Gamma_\nu-\Gamma_{\mu\nu}^\lambda\,\Gamma_\lambda
   \\ & = & -\frac{1}{2}\,\Box'_g\,g_{\mu\nu}
   +g^{\alpha\alpha'} g^{\beta\beta'}\bigg(
   -\frac{1}{4}\,\partial_{x^\nu} g_{\alpha'\beta'}\,\partial_{x^\mu} g_{\alpha\beta}
   +\frac{1}{8}\,\partial_{x^\mu} g_{\alpha\alpha'}\,\partial_{x^\nu} g_{\beta\beta'}\bigg)
   \\ & & +\,\frac{1}{2}\,g^{\alpha\alpha'} g^{\beta\beta'}
   \partial_{x^\alpha} g_{\beta\mu}\,\partial_{x^{\alpha'}} g_{\beta'\nu}
   -\frac{1}{2}\,g^{\alpha\alpha'} g^{\beta\beta'}\,
   \bigg(\partial_{x^\alpha} g_{\beta\mu}\,\partial_{x^{\beta'}} g_{\alpha'\nu}
   -\partial_{x^{\beta'}} g_{\beta\mu}\,\partial_{x^\alpha} g_{\alpha'\nu}\bigg)
   \\ & & +\,\frac{1}{2}\,g^{\alpha\alpha'} g^{\beta\beta'} 
   \Big((\partial_{x^\mu}g_{\alpha'\beta'}\,\partial_{x^\alpha} g_{\beta\nu}   
   -\partial_{x^\alpha} g_{\alpha'\beta'}\,\partial_{x^\mu} g_{\beta\nu})
   +(\partial_{x^\nu} g_{\alpha'\beta'}\,\partial_{x^\alpha} g_{\beta\mu}   
   -\partial_{x^\alpha} g_{\alpha'\beta'}\,\partial_{x^\nu} g_{\beta\mu})\Big)
   \\ & & +\,\frac{1}{4}\,g^{\alpha\alpha'} g^{\beta\beta'} 
   \Big((\partial_{x^{\beta'}} g_{\alpha\alpha'}\,\partial_{x^\mu} g_{\beta\nu}
   -\partial_{x^\mu} g_{\alpha'\alpha}\,\partial_{x^{\beta'}} g_{\beta\nu})
   +(\partial_{x^{\beta'}} g_{\alpha\alpha'}\,\partial_{x^\nu} g_{\beta\mu}
   -\partial_{x^\nu} g_{\alpha'\alpha}\,\partial_{x^{\beta'}} g_{\beta\mu})\Big)   
   \\ & & +\,\frac{1}{2}\,(\partial_{x^\mu}\Gamma_\nu+\partial_{x^\nu}\Gamma_\mu)
   +\frac{1}{2}\,\Gamma^{\beta}(\partial_{x^\mu} g_{\beta\nu}
   +\partial_{x^\nu} g_{\beta\mu})    
   -\frac{1}{2}\,\Gamma_\mu\Gamma_\nu-\Gamma_{\mu\nu}^\lambda\,\Gamma_\lambda, 
\end{eqnarray*} 
where we did some index renaming in the last step. 
Therefore we have shown the following result. 

\begin{lemma}\label{Ric_expr} 
Let $\Box'_g=g^{\alpha\beta}\,\partial^2_{x^\alpha x^\beta}$. 
The Ricci tensor may be expressed as 
\begin{eqnarray*} 
   R_{\mu\nu} & = & -\frac{1}{2}\,\Box'_g\,g_{\mu\nu}
   +\frac{1}{2}\,\tilde{P}(g; \partial_{x^\mu} g, \partial_{x^\nu} g)
   +\frac{1}{2}\,\tilde{Q}_{\mu\nu}(g; \partial g, \partial g)   
   \\ & & +\,\frac{1}{2}\,(\partial_{x^\mu}\Gamma_\nu+\partial_{x^\nu}\Gamma_\mu)
   +\frac{1}{2}\,\Gamma^{\beta}(\partial_{x^\mu} g_{\beta\nu}
   +\partial_{x^\nu} g_{\beta\mu})    
   -\frac{1}{2}\,\Gamma_\mu\Gamma_\nu-\Gamma_{\mu\nu}^\lambda\,\Gamma_\lambda, 
\end{eqnarray*}
where 
\begin{eqnarray} 
   \tilde{P}(g; \partial_{x^\mu} h, \partial_{x^\nu} k) 
   & = & \frac{1}{4}\,g^{\alpha\alpha'}\partial_{x^\mu} h_{\alpha\alpha'}
   \,g^{\beta\beta'}\partial_{x^\nu} k_{\beta\beta'}
   -\frac{1}{2}\,g^{\alpha\alpha'} g^{\beta\beta'}
   \,\partial_{x^\mu} h_{\alpha\beta}\,\partial_{x^\nu} k_{\alpha'\beta'}, 
   \label{tildP} 
   \\ \tilde{Q}_{\mu\nu}(g; \partial h, \partial k) 
   & = & g^{\alpha\alpha'} g^{\beta\beta'}\,\partial_{x^\alpha} h_{\beta\mu}
   \,\partial_{x^{\alpha'}} k_{\beta'\nu}
   \nonumber
   \\ & & -\,g^{\alpha\alpha'} g^{\beta\beta'} 
   \,(\partial_{x^\alpha} h_{\beta\mu}\,\partial_{x^{\beta'}} k_{\alpha'\nu}
   -\partial_{x^{\beta'}} h_{\beta\mu}\,\partial_{x^\alpha} k_{\alpha'\nu})
   \nonumber
   \\ & & +\,g^{\alpha\alpha'} g^{\beta\beta'} 
   \,(\partial_{x^\mu} h_{\alpha'\beta'}\,\partial_{x^\alpha} k_{\beta\nu}
   -\partial_{x^\alpha} h_{\alpha'\beta'}\,\partial_{x^\mu} k_{\beta\nu})
   \nonumber
   \\ & & +\,g^{\alpha\alpha'} g^{\beta\beta'} 
   \,(\partial_{x^\nu} h_{\alpha'\beta'}\,\partial_{x^\alpha} k_{\beta\mu}
   -\partial_{x^\alpha} h_{\alpha'\beta'}\,\partial_{x^\nu} k_{\beta\mu})
   \nonumber
   \\ & & +\,\frac{1}{2}\,g^{\alpha\alpha'} g^{\beta\beta'} 
   \,(\partial_{x^{\beta'}} h_{\alpha\alpha'}\,\partial_{x^\mu} k_{\beta\nu}
   -\partial_{x^\mu} h_{\alpha\alpha'}\,\partial_{x^{\beta'}} h_{\beta\nu})
   \nonumber
   \\ & & +\,\frac{1}{2}\,g^{\alpha\alpha'} g^{\beta\beta'} 
   \,(\partial_{x^{\beta'}} h_{\alpha\alpha'}\,\partial_{x^\nu} k_{\beta\mu}
   -\partial_{x^\nu} h_{\alpha\alpha'}\,\partial_{x^{\beta'}} k_{\beta\mu}).
   \label{tildQ} 
\end{eqnarray}  
In particular, in the harmonic gauge one has 
\begin{equation}\label{Ric_harmo} 
   R_{\mu\nu}=-\frac{1}{2}\,\Box'_g\,g_{\mu\nu}
   +\frac{1}{2}\,\tilde{P}(g; \partial_{x^\mu} g, \partial_{x^\nu} g)
   +\frac{1}{2}\,\tilde{Q}_{\mu\nu}(g; \partial g, \partial g).
\end{equation} 
\end{lemma} 

Since $g^{\mu\nu} g_{\mu\nu}=4$ and hence $R=-\frac{8\pi}{c^4}\,{\rm tr}_g(T)$, 
we also have the following consequence. 

\begin{cor}\label{EE_harmonic} 
Let $\Box'_g=g^{\alpha\beta}\,\partial^2_{x^\alpha x^\beta}$. 
In the harmonic gauge, the Einstein equations become
\[ \Box'_g\,g_{\alpha\beta}=\tilde{P}(g; \partial_{x^\alpha} g, \partial_{x^\beta} g)
   +\tilde{Q}_{\alpha\beta}(g; \partial g, \partial g)
   -\frac{16\pi G}{c^4}\bigg(T_{\alpha\beta}-\frac{1}{2}\,{\rm tr}_g(T)\,g_{\alpha\beta}\bigg). \] 
\end{cor}


\section{Appendix: Discussion on Blanchet's condition}
\label{Blanch_discuss} 
\setcounter{equation}{0} 

When solving the wave equation~\eqref{jret2} that results from the choice 
of the harmonic gauge, a term of the form 
\begin{equation}\label{Ralph} 
   S_{\alpha\beta}(t, x)   
   =\int_{\R^3}\frac{1}{|x-\bar{x}|}
   \,\phi_{\alpha\beta}(t-c^{-1}|x-\bar{x}|, \bar{x})\,d\bar{x}
\end{equation}  
occurs in~\eqref{eq_rem29}, where $\phi_{\alpha\beta}$ is quadratic in $h$ and/or its derivatives. 
The quantity $S_{\alpha\beta}$ describes the gravitational part of the source term 
of the wave equation, in contrast to what happens in electrodynamics, where this source term 
is due to matter only and has compact spatial support. In most physics textbooks, (\ref{Ralph}) 
is swept under the rug as a term of ``higher order in the weak-field approximation 
and therefore dispensable''.

However, to those working in gravitational radiation (cf.~\cite[p.~54]{WiWi} for instance), 
it is well-known that (\ref{Ralph}) will generally be infinite without further assumptions. 
Let us briefly explain why this could happen, from a mathematics viewpoint. For simplicity 
we drop $\lambda$, $c$ and other constants in the short argument to follow. Let $(t, x)$ 
be fixed such that $t\ge 0$, $|t-|x||\le 1$ and $|x|\ge 1$, and define the set
\[ A = \{z\in\R^3: |z|\ge |x|\,\,{\rm and}\,\,\langle z, x\rangle\le 0\}, \]
where $\langle \cdot,\cdot\rangle$ denotes the Euclidean inner product. 
The relation $|x-z|^2\ge |x|^2+|z|^2$ holds for $z\in A$, and together with $|x|\ge 1$ 
it leads to $|x-z|^2\ge (|z|-|x|+1)^2$ and thus $|x-z|\ge ||z|-|x|+1|=|z|-|x|+1\ge |z|-t$. 
Furthermore, we also have $|x-z|\ge |z|-|x|+1\ge |x|-|z|+1 \ge t-|z|$. 
In conclusion, $|x-z|\ge |t-|z||$ is verified for all $z\in A$ 
and from what we know about the decay of $h$ (or its derivatives), 
one can therefore expect an estimate like
\begin{eqnarray*} 
   |S_{\alpha\beta}(t, x)|   
   & \sim & \int_{\R^3}\frac{1}{|x-\bar{x}|}
   \,|h(t-|x-\bar{x}|, \bar{x})|^2\,d\bar{x}
   =\int_{\R^3}\frac{1}{|z|}\,|h(t-|z|, x-z)|^2\,dz
   \\ & \gtrsim & \int_A\frac{1}{|z|}
   \,\dfrac{(1+|t-|z||)^{2\delta}}{(1+|x-z|+|t-|z||)^2(1+|x-z|-|t-|z||)^{2\gamma}}\,dz 
\end{eqnarray*} 
to hold. Now $1+|x-z|-|t-|z||\le 1+|x|+|z|+t-|z|=1+|x|+t$ yields 
\[ |S_{\alpha\beta}(t, x)|   
   \gtrsim\frac{1}{(1+|x|+t)^{2\gamma}}\int_A\frac{1}{|z|}
   \,\dfrac{(1+|t-|z||)^{2\delta}}{(1+|x-z|+|t-|z||)^2}\,dz. \] 
However, for points $z\in A$ with $|z|$ large, this integrand behaves 
like $|z|^{2\delta-3}$, causing the integral to diverge even if $\delta=0$, due to the fact that $A$ is a half space 
minus a bounded half ball.
\smallskip 

Blanchet \cite[equ.~(19)]{B_LRR} defines a \textit{no-incoming-radiation condition} 
on the metric of a spacetime.  First note that the expression 
that he calls the \textit{(inverse) metric perturbation},
\[ h_{\text{Blanchet}}^{\alpha\beta} := \sqrt{|g|} g^{\alpha\beta} - \eta^{\alpha\beta} \ , \]
actually differs from our inverse metric perturbation (which is the same as for \cite{LT}):
\[ h^{\alpha\beta}=\eta^{\alpha\alpha'}\eta^{\beta\beta'} h_{\alpha'\beta'}
   =\eta^{\alpha\alpha'}\eta^{\beta\beta'}(g_{\alpha'\beta'}-\eta_{\alpha'\beta'}); \] 
see Section \ref{swnir}. Blanchet's condition is written as follows: 
Some large enough $\calT>0$ exists such that, 
for all $\alpha,\beta=0, 1, 2, 3$,
\begin{equation}\label{Blanch_orig} 
   \partial_t(h_{\text{Blanchet}}^{\alpha\beta}) = \partial_t\left( \sqrt{|g|}g^{\alpha\beta} 
   - \eta^{\alpha\beta} \right) = 0 \quad \text{at all points } (t, x) \text{ with } t \leq -\calT \ .
\end{equation} 
We can rephrase this as: For each $\alpha,\beta=0, 1, 2, 3$ there exists a function 
$q^{\alpha\beta} = q^{\alpha\beta}(x)$ such that
\begin{equation}
\label{eq_blanchet_condition}
    q^{\alpha\beta}(x) + \eta^{\alpha\beta} = (\sqrt{|g|}g^{\alpha\beta})(t, x) 
    \quad \text{for all } t\leq -\calT \ .
\end{equation}
Physically meaningful asymptotically flat spacetimes will have $q^{\alpha\beta}$ bounded over the whole domain $\bbR^3$. 
More than that, $q^{\alpha\beta}$ should decay as $|x|\to\infty$, 
along with its derivatives, with decay rates given as
\[ |\partial^\kappa q^{\alpha\beta}(x)|\sim |x|^{-(1+|\kappa|)} \] 
for $|x|\ge {\cal S}$ and $|\kappa|\le 3$ in the case of an isolated body. 

\medskip
In order to compare this to the NIR condition~\eqref{Blanch}, all that is needed 
is the following lemma that relates the $q$-terms above to the $p$-terms 
that appear in NIR.

\begin{lemma}
Under the hypothesis of \eqref{eq_blanchet_condition}, we have
\begin{equation} \label{h_and_p}
   h_{\alpha\beta}(t,x) = -q_{\alpha\beta}(x) + \frac{1}{2}\,q(x)\,\eta_{\alpha\beta} 
   + {\cal O}(q^2,hq) \ ,
\end{equation}
where we defined
\[ q_{\alpha\beta} = \eta_{\alpha\alpha'}
   \eta_{\beta\beta'} q^{\alpha'\beta'} \ , 
   \quad q = q_{\alpha\beta} q^{\alpha\beta}, \]
and where ${\cal O}(q^2)$ represents remainder terms that are at least quadratic 
in $q^{\alpha\beta}$ terms or proportional to $h_{\alpha\beta} q^{\mu\nu}$ terms.
\end{lemma}
{\bf Proof\,:} Consider the matrices of functions
\[ \hat{q} = (q^{\alpha\beta}) \ , 
   \quad\hat{m}^{-1} = \hat{m} = (\eta^{\alpha\beta}) = (\eta_{\alpha\beta}) \ , 
   \quad\hat{g} = (g_{\alpha\beta}) \ , 
   \quad\hat{H} = (H^{\alpha\beta}) \ , \]
where $H^{\alpha\beta} = g^{\alpha\beta} - \eta^{\alpha\beta}$ 
is as in (\ref{grossH}). Then~\eqref{eq_blanchet_condition} is written as
\[ \hat{q} + \hat{m}^{-1} = \sqrt{|\det\hat{g}|} \ \hat{g}^{-1} \ . \]
In particular
\begin{eqnarray*}
   \det(\hat{q}+\hat{m}^{-1}) & = & \left(\sqrt{|\det\hat{g}|}\right)^4 (\det\hat{g})^{-1} 
   \\ & = & \left(\sqrt{|\det\hat{g}|}\right)^4\left(-\frac{1}{\sqrt{|\det\hat{g}|}
   \sqrt{|\det \hat{g}|}}\right) 
   = -\left(\sqrt{|\det\hat{g}|}\right)^2, 
\end{eqnarray*} 
so that $\sqrt{|\det\hat{g}|} = \sqrt{-\det(\hat{q}+\hat{m}^{-1})}$ and hence 
\[ \hat{g}^{-1} = \frac{\hat{q}+\hat{m}^{-1}}{\sqrt{-\det(\hat{q}+\hat{m}^{-1})}}, 
   \quad\hat{H} = -\hat{m}^{-1} + \frac{\hat{q}+\hat{m}^{-1}}
   {\sqrt{-\det(\hat{q}+\hat{m}^{-1})}}, \] 
or, what is the same,
\[ H^{\alpha\beta} = -\eta^{\alpha\beta} + (-\det(\eta^{\mu\nu} 
   + q^{\mu\nu}))^{-1/2}(\eta^{\alpha\beta} + q^{\alpha\beta}) \ . \]
The determinant appearing on the right side, to first order in $q$ terms, is
\[ -\det(\eta^{\mu\nu}+q^{\mu\nu}) = -\det(\hat{m}^{-1}+\hat{q}) 
   = -\det(\hat{m}^{-1})\det(1+\hat{m}\hat{q}) 
   \approx 1+\mathrm{tr}(\hat{m}\hat{q}) 
   = 1+\eta_{\mu\nu}q^{\mu\nu} \ . \]
With this, the formula for $H^{\alpha\beta}$ to first order in small quantities becomes
\begin{equation}\label{eq_H_q} 
    H^{\alpha\beta} \approx -\eta^{\alpha\beta} 
    + \left(1-\frac{\eta_{\mu\nu}q^{\mu\nu}}{2}\right)(\eta^{\alpha\beta}+q^{\alpha\beta})
    \approx q^{\alpha\beta} - \frac{1}{2}q\eta^{\alpha\beta} \ . 
\end{equation} 
Now we can find $h_{\alpha\beta}$ in terms of $H^{\mu\nu}$. The relation
\[ \delta^\alpha_\beta = g^{\alpha\mu}g_{\mu\beta} 
   = (\eta^{\alpha\mu}+H^{\alpha\mu})(\eta_{\mu\beta}+h_{\mu\beta}) = \delta^\alpha_\beta 
   + g^{\alpha\mu}h_{\mu\beta} + H^{\alpha\mu}\eta_{\mu\beta} \] 
yields $g^{\alpha\mu}h_{\mu\beta} = -H^{\alpha\mu}\eta_{\mu\beta}$ and therefore 
\[ -g_{\nu\alpha}H^{\alpha\mu}\eta_{\mu\beta} = g_{\nu\alpha}g^{\alpha\mu}h_{\mu\beta} 
   = \delta^\mu_\nu h_{\mu\beta} = h_{\nu\beta},\quad 
   h_{\alpha\beta} = -g_{\alpha\mu}\eta_{\beta\nu}H^{\mu\nu} \ . \] 
To first order, we can replace the $g$ term by an $\eta$ term. Then, 
by plugging in (\ref{eq_H_q}), we find
\[ h_{\alpha\beta} \approx -\eta_{\alpha\mu}\eta_{\beta\nu}
   \left(q^{\mu\nu}-\frac{1}{2}q\eta^{\mu\nu}\right) = -q_{\alpha\beta} 
   + \frac{1}{2}q\eta_{\alpha\beta} \ , \]
which yields the claim. 
{\hfill$\Box$}\bigskip 

Thus, to summarize, the assumption (\ref{Blanch}) that we use as a \textit{no-incoming-radiation condition} 
agrees to first order in $h$ (which is the relevant order for the quadrupole formula) 
with Blanchet's original condition (\ref{Blanch_orig}), with some natural decay assumptions added 
on the functions $p_{\alpha\beta}(x)$ for large $|x|$. We also remark that a factor of $\eps_h$ has been added 
to the right-hand side of (\ref{Blanch}). This accounts for the fact that the smallness 
of the perturbation $h$ is measured in terms of $\eps_h$.


\section{Appendix: Bounds derived from \cite{LT}}
\label{LTbounds_sect} 
\setcounter{equation}{0} 

This section describes how to adapt the relevant estimates from \cite{LT}, where nothing 
has a physical dimension, to our desired context where factors of $c$ and $G$ are kept. 
These estimates are only stated for reference, as a justification for our assumptions 
in Subsection~\ref{subsec_apriori}, but they are not further used in this paper 
(apart from the method of \cite[Prop.~2.1]{LT} to control the support of $f(t)$, 
which will appear in Subsection~\ref{subsubsec_support} below).

\subsection{Description of relevant results from \cite{LT}}
\label{subsec_LT}

Here we summarize and paraphrase the results from \cite{LT} that we need. Let $({\cal M}, g)$ 
be a Lorentzian $4$-manifold with signature $(-, +, +, +)$ and a coordinate system 
$(x^0, x^1, x^2, x^3)$ in the harmonic gauge. Let $(x^\mu, p^\nu)
=(x^0,x^1,x^2,x^3,p^0,p^1,p^2,p^3)$ denote the natural coordinates on $T{\cal M}$. Given $(x^\mu)\in M$, 
we let $r^2 = (x^1)^2 + (x^2)^2 + (x^3)^2$. For a tensor $U_{\alpha\beta}$, 
the notation $|U(x^\mu)|$ means $\sum_{\alpha,\beta}|U_{\alpha\beta}(x^\mu)|$, 
while $\lVert U(x^0,\cdot) \rVert_{L^2}$ 
means $\sum_{\alpha,\beta}\lVert U_{\alpha\beta}(x^0,\cdot) \rVert_{L^2}$ 
(and similarly for tensors of any rank). Define the mass shell
\[ P{\cal M} = \{ (x^\mu, p^\nu) \in T{\cal M} \ : \ (p^\nu) \text{ is future-directed, } 
   g_{\alpha\beta} p^\alpha p^\beta = -1\} \ . \]
For a fixed $x=(x^\mu)\in {\cal M}$ define also
\[ P{\cal M}(x) = \{ (p^\nu)\in T_x {\cal M} \ : \ (p^\nu) \text{ is future-directed, } 
   g_{\alpha\beta} p^\alpha p^\beta = -1\} \ . \]
Note that the dependence of $P{\cal M}(x)$ on $x$ enters through the $g_{\alpha\beta}$ terms. 
Let $f: P{\cal M}\longrightarrow [0,\infty)$ be a mass-density scalar function 
and define the energy-momentum tensor
\begin{equation} \label{LT_T}
    T^{\mu\nu}(x) = |g(x)|^{1/2}\int_{P{\cal M}(x)} p^\mu p^\nu\,f(x, p^\beta)
    \,\frac{1}{-p_0}\,dp^1 dp^2 dp^3 \ ,
\end{equation}
where $g<0$ denotes the metric determinant, $x$ denotes a spacetime 4-vector $(x^\alpha)$, 
the symbols $p^1,p^2,p^3$ are to be regarded as mere integration variables, 
and $p^0$ and its lowered counterpart $p_0 = g_{\alpha 0}\,p^\alpha$ 
are functions of $p^1,p^2,p^3$ implicitly defined by the relation 
$g_{\alpha\beta}\,p^\alpha p^\beta = -1$. [We remark that the formula for $T^{\mu\nu}$ 
actually presented in \cite{LT} differs from (\ref{LT_T}) above in that it doesn't have 
the minus sign and it shows $p^0$ instead of $p_0$ in the denominator, 
but this details doesn't change any of the estimates that we need. 
See Section \ref{EV_system} with $c=G=m=1$ and \cite{AndrLRR} for a consistent set of equations.] 
Finally, letting $X$ denote the generator of geodesic flow on ${\cal M}$, 
which is the vector field on $T{\cal M}$ given by
\[ X(x^\mu,p^\nu) = p^\mu\frac{\partial}{\partial x^\mu} 
   - p^\alpha p^\beta \Gamma^\mu_{\alpha\beta}\frac{\partial}{\partial p^\mu} \ , \]
where $(x^\mu, p^\nu)$ refer to the canonical coordinates of the tangent bundle as above, then the Vlasov equation for $f$ reads:
\begin{equation}\label{LT_X}
   X(f) = 0 \ .
\end{equation}
The so-called \textit{reduced Einstein-Vlasov system} is composed of equation (\ref{LT_T}) and (\ref{LT_X}) 
together with the Einstein equations, which \cite{LT} write in the equivalent trace-reduced formulation:
\begin{equation}\label{LT_E}
   \mathrm{Ric}(g)_{\alpha\beta} = T_{\alpha\beta} 
   -\frac{1}{2}\,g_{\alpha\beta}\,\mathrm{tr}_g T \ .
\end{equation}
In \cite{LT} this equation is expanded in the following form:
\[ \Box'_g\,g_{\alpha\beta}=F_{\alpha\beta}(g)(\partial g,\partial g)
   +T_{\alpha\beta}-\frac{1}{2}\,{\rm tr}_g(T)\,g_{\alpha\beta} \] 
for $\Box'_g=g^{\alpha\beta}\,\partial_{x^\alpha}\partial_{x^\beta}$. 
Here a factor of $-16\pi$ is missing on the $T$-terms (see Corollary \ref{EE_harmonic}), 
and the function $F_{\alpha\beta}$ turns out to be the following quadratic expression:
\[ F_{\alpha\beta}(g)(\partial g,\partial g)
   =\tilde{P}(g; \partial_{x^\alpha} g, \partial_{x^\beta} g)
   +\tilde{Q}_{\alpha\beta}(g; \partial g, \partial g) \] 
for $\tilde{P}$ and $\tilde{Q}_{\alpha\beta}$ defined as in (\ref{tildP}) and (\ref{tildQ}), respectively. 

Now let $\chi: [0,\infty)\longrightarrow\mathbb{R}$ be a smooth cutoff function such that $\chi(\xi)=1$ 
for $\xi\geq 3/4$, $\chi(\xi) = 0$ for $\xi\leq 1/2$ and $0\leq\chi(\xi´)\leq 1$ for all $\xi$.  
For a solution $g$ of the reduced Einstein-Vlasov system having ADM mass $M$, define $h^0$ and $h^1$ by
\begin{equation}\label{gsplit} 
   g = \eta + h = \eta + h^0 + h^1 \ , \quad 
   h^0_{\mu\nu}(x^\alpha) = \chi\left(\frac{r}{1+x^0}\right)\frac{M}{r}\,\delta_{\mu\nu} \ ,
\end{equation} 
where $\eta$ is the Minkowski metric; see \cite[(1.13)]{LT}. 
For fixed $0<\gamma<1$ and $0<\mu<1-\gamma$, let also
\[ w(x^\alpha) = \left\{\begin{array}{rcl}
    (1+|r-x^0|)^{1+2\gamma} & \text{ if } & r > x^0  \\
    1+(1+|r-x^0|)^{-2\mu} & \text{ if } & r \leq x^0 
    \end{array}\right. . \]
The \textit{energy} at time $x^0$ (depending on a natural number $N$) is
\[ E_N(x^0) = \sum_{|I|\leq N} \lVert \sqrt{w(x^0,\cdot)}
   \,\partial Z^I h^1(x^0,\cdot) \rVert_{L^2_x}^2 \ , \]
where $I$ is a multi-index and $Z^I$ runs over all possible compositions of $|I|$ of the vector fields
\begin{equation}\label{vefi} 
   \Omega_{ij} = x^i\frac{\partial}{\partial x^j} - x^j\frac{\partial}{\partial x^i} \ , 
   \quad B_i = x^i \frac{\partial}{\partial x^0} + x^0\frac{\partial}{\partial x^i} \ , 
   \quad S = x^0\frac{\partial}{\partial x^0} + x^i\frac{\partial}{\partial x^i} \ , 
   \quad \frac{\partial}{\partial x^\alpha} \ .
\end{equation} 
Given an initial particle distribution $f_0 = f_0(x^i,p^\beta)$ at time $x^0=0$, 
denote also, for a given $N$,
\begin{equation}\label{calVN} 
   \mathcal{V}_N = \sum_{k+l\leq N} \lVert \partial_x^k \partial_p^l f_0 \rVert_{L^2_x L^2_p} \ .
\end{equation} 
One of the main theorems in \cite{LT}, Theorem 1.2, together with estimate \cite[equ.~(7.3)]{LT} 
for the case $I=J=\emptyset$ in their notation, can be quoted as follows:

\begin{theorem}[\cite{LT}]\label{LT_main} 
For any $0<\gamma<1$, $0<\mu<1-\gamma$, $0<\delta<\frac{1}{8}\min\{\gamma, 1-\gamma\}$, 
$K>0$, $K'>0$ and $N\geq 11$, there are $\eps_0>0$ and constants $C_N, C'_N>0$ 
such that, for all $\eps \leq \eps_0$, given any initial data 
$(g_{\alpha\beta},\partial_{x^0} g_{\alpha\beta},f_0)\big|_{x^0=0}$ 
for the reduced Einstein-Vlasov system satisfying
\[ E_N(0)^{1/2} + \mathcal{V}_N + M < \eps \]
and
\[ \mathrm{supp}(f_0) \subset\{ (x,p): |x|\leq K , |p|\leq K' \}, \]
there exists a corresponding global solution satisfying, for all $x^0\geq 0$,
\begin{equation} \label{LT_L2}
    \sqrt{E_N(x^0)} + \sum_{|I|\leq N} (1+x^0)\lVert Z^I T^{\mu\nu}(x^0,\cdot) \rVert_{L^2_x} 
   \leq C_N\eps(1+x^0)^{C_N'\eps}
\end{equation}
along with the pointwise decay estimates
\[ |Z^I h^1_{\mu\nu}(x^\alpha)| \leq 
   \frac{C'_N\eps(1+x^0)^{C'_N\eps}}{(1+x^0+r)(1+(r-x^0)_+)^\gamma},
   \quad |I|\le N-3, \] 
and 
\[ |\partial h_{\mu\nu}(x^\alpha)|
   \le\frac{C'_N\eps (1+x^0)^{2\delta}}{(1+x^0+r)(1+|r-x^0|){(1+(r-x^0)_+)}^{2\delta}}, \]  
where $(r-x^0)_+$ stands for $r-x^0$ if $r \geq x^0$ and 0 otherwise.
\end{theorem}

Estimate~\eqref{LT_L2} immediately yields an $L^2$ estimate for $T^{\mu\nu}$, 
which we can transform into an $L^\infty$ estimate using the Sobolev inequality
\[ \lVert u\rVert_{L^\infty(\bbR^n)} \leq D_s \lVert u \rVert_{W^{s,2}(\bbR^n)} 
    \quad \text{for all } u\in W^{s,2}(\bbR^n) \ , \]
which is valid for $s>n/2$ and where $D_s>0$ is a universal constant 
(we need $n=3$ and $s=2$). By letting $N=11$ in Theorem \ref{LT_main} 
and by only taking into account the standard derivatives among the vector fields (\ref{vefi}), 
we arrive at the following

\begin{cor}\label{LT_cor} 
For any $0<\gamma<1$, $0<\mu<1-\gamma$, $0<\delta<\frac{1}{8}\min\{\gamma, 1-\gamma\}$, 
$K>0$ and $K'>0$, there is $\eps_0>0$ and a constant $C_0>0$ such that, 
for all $\eps \leq \eps_0$, given any initial data 
$(g_{\alpha\beta},\partial_{x^0} g_{\alpha\beta},f_0)\big|_{x^0=0}$ 
for the reduced Einstein-Vlasov system satisfying
\[ E_{11}(0)^{1/2} + \mathcal{V}_{11} + M < \eps \]
and
\[ \mathrm{supp}(f_0) \subset\{ (x,p): |x|\leq K , |p|\leq K' \}, \]
there exists a corresponding global solution satisfying the following pointwise 
decay estimates, for all $x^0\geq 0$ 
and multi-indices $\kappa$ with $|\kappa|\le 8$,
\begin{align}
    |\partial^\kappa T^{\mu\nu}(x^0, x)| &\leq \frac{C_0\eps}{(1+x^0)^{1-C_0\eps}} ,  
    \label{LT_cor_conclusion_T} \\
    |\partial h_{\mu\nu}(x^0, x)| &\leq \frac{C_0\eps(1+x^0)^{2\delta}}{(1+|x|+x^0)(1+||x|-x^0|)
   {(1+(|x|-x^0)_+)}^{2\delta}} , \nonumber \\
    |\partial^\kappa (h^1)_{\mu\nu}(x^0, x)| &\leq \dfrac{C_0\eps (1+x^0)^{C_0\eps}}{(1+|x|+x^0)
    (1+(|x|-x^0)_+)^\gamma} . \label{LT_cor_conclusion_h}
\end{align}
\end{cor}

It should be noted that we do not make use of the smallness of 
$E_{11}(0)^{1/2} + \mathcal{V}_{11}$ to derive the quadrupole formula; 
this condition is only needed in order for the above corollary to hold. 
Also note that estimates (\ref{fuvor0}) and (\ref{fuvor}) in our boundedness 
and decay conditions have been much relaxed as compared to what the above corollary gives. 
However, there would have been no gain in our estimates had we used the 
above better bound~\eqref{LT_cor_conclusion_T} for the energy-momentum tensor, 
since such bounds only get applied at points satisfying $x^0={\cal O}(1)$ 
and $t-|x| = {\cal O}(1)$, for $x$ in a ball of a fixed finite radius.

\subsection{Estimates with factors of $c$ and $G$ included}

We now translate the bounds from Corollary \ref{LT_cor} into the framework of equations 
that has been set up in our Section \ref{EV_system}. From now on $(x^\alpha, p^\beta)$ 
denote the coordinates on the tangent bundle $T{\cal M}$ as we used 
in Section \ref{EV_system}. Our coordinates $(x^0,x^1,x^2,x^3)$ have a dimension of \textit{length}, 
for which the constant $\lambda$ represents a fundamental scale. The second half $(p^0,p^1,p^2,p^3)$ 
of our canonical tangent-bundle coordinates have a dimension of \textit{momentum}, which in terms 
of our fundamental constants has units of $G^{-1}\lambda c^{3}$. Also let $m = G^{-1}\lambda c^2$ 
be the Vlasov particles' mass expressed in the correct physical units. Starting from this context, 
we define new, dimensionless objects denoted with an overbar, which are themselves functions 
of new dimensionless coordinates $\overline{x}$ and $\overline{p}$ 
on the tangent bundle:
\begin{align}
    \overline{x}^\alpha & = \lambda^{-1} \ x^\alpha \ , 
    \nonumber \\
    \overline{p}^\beta & = m^{-1} c^{-1} \ p^\beta \ , 
    \nonumber \\
    \overline{g}_{\mu\nu}(\overline{x}^\alpha) & = g_{\mu\nu}(\lambda\overline{x}^\alpha) \ , 
    \nonumber \\
    \overline{T}^{\mu\nu}(\overline{x}^\alpha) & = 
    8\pi G\lambda^2 c^{-4} \ T^{\mu\nu}(\lambda \overline{x}^\alpha) \ , 
    \label{aaaa_new_T} \\
    \overline{f}(\overline{x}^\alpha,\overline{p}^\beta) & = 8\pi G\lambda^2c \ m^5 \ 
    f(\lambda \overline{x}^\alpha, mc\overline{p}^\beta) \ . \label{aa_new_f}
\end{align}
We check now that, if our objects satisfy our version of the Einstein-Vlasov system 
as stated in Section \ref{EV_system}, then the equations of the Einstein-Vlasov system 
as described in subsection \ref{subsec_LT} are valid for the objects with an overbar. 
To begin, since Christoffel symbols involve taking first-order derivatives of the metric, 
the Christoffel symbols of the overbar system are 
\[ \overline{\Gamma}^\rho_{\mu\nu}(\overline{x}^\alpha) 
   = \lambda\Gamma^\rho_{\mu\nu}(\lambda \overline{x}^\alpha) \ . \]
Likewise, the components of the Einstein tensor in the new coordinates are
\[ \overline{G}_{\mu\nu}(\overline{x}^\alpha) 
   = \lambda^{2}G_{\mu\nu}(\lambda\overline{x}^\alpha) \ . \]
Then the Einstein equation $G_{\alpha\beta} = 8\pi G c^{-4} T_{\alpha\beta}$ 
from (\ref{einst}) implies
\[ \overline{G}_{\mu\nu}(\overline{x}^\alpha) 
    = \lambda^2\,\frac{8\pi G}{c^4}\,T_{\mu\nu}(\lambda\overline{x}^\alpha)
    = \overline{T}_{\mu\nu}(\overline{x}^\alpha) \ , \] 
and thus the Einstein equation~\eqref{einst} as in subsection \ref{subsec_LT} is verified. 
To check the validity of the Vlasov equation~\eqref{vlas3}, let us abbreviate 
the unimportant constant in front of $f$ as $C = 8\pi G \lambda^2 c m^5$. Note the derivative relations
\[ \frac{\partial \overline{f}}{\partial \overline{x}^\mu}(\overline{x}^\alpha,\overline{p}^\beta) 
   =  C\lambda\frac{\partial f}{\partial x^\mu}(\lambda\overline{x}^\alpha,mc\lambda\overline{p}^\beta) 
   \ , \quad \frac{\partial \overline{f}}{\partial \overline{p}^\mu}
   (\overline{x}^\alpha,\overline{p}^\beta) 
   =  Cmc\frac{\partial f}{\partial p^\mu}(\lambda\overline{x}^\alpha,mc\lambda\overline{p}^\beta) \ . \]
Thus, from the assumption that our version of the Vlasov equation (\ref{vlas3}) holds:
\[ p^\mu\frac{\partial f}{\partial x^\mu}(x^\alpha, b^\beta)
   -\Gamma_{\mu\nu}^\sigma\,p^\mu p^\nu\,\frac{\partial f}{\partial p^\sigma}(x^\alpha, p^\beta)=0, \] 
we conclude that, at all points $(\overline{x}^\alpha,\overline{p}^\beta)$ of the tangent bundle,
\begin{eqnarray*} 
   \lefteqn{\overline{p}^\mu\frac{\partial\overline{f}}{\partial\overline{x}^\mu}
   (\overline{x}^\alpha,\overline{p}^\beta) 
   -\overline{\Gamma}^\sigma_{\mu\nu}\,\overline{p}^\mu\overline{p}^{\nu}
   \frac{\partial\overline{f}}{\partial\overline{p}^\sigma}
   (\overline{x}^\alpha,\overline{p}^\beta)}
   \\ & = & (mc)^{-1} p^\mu C\lambda\,\frac{\partial f}{\partial x^\mu}
   (\lambda\overline{x}^\alpha,mc\overline{p}^\beta)
   - (mc)^{-2}p^\mu p^\nu \lambda\,\Gamma^\sigma_{\mu\nu} 
   Cmc\,\frac{\partial f}{\partial p^\sigma}(\lambda\overline{x}^\alpha,mc\overline{p}^\beta) 
   \\ & = &  C(mc)^{-1}\lambda \left( p^\mu\frac{\partial f}{\partial x^\mu}(x^\alpha,p^\beta) 
   -\Gamma^\sigma_{\mu\nu} p^\mu p^\nu\,\frac{\partial f}{\partial p^\sigma}(x^\alpha,p^\beta)\right) 
   =0,
\end{eqnarray*} 
as needed. Finally we check the relation~\eqref{tab3} between $T$ and $f$. Consider $x=(x^\alpha)$ fixed 
and denote by $\overline{x} = \lambda^{-1}x$ the corresponding point in the new manifold coordinates. 
We raise the indices and write the constant $m$ where it belongs in our definition~\eqref{tab3} 
of $T_{\alpha\beta}$:
\begin{equation} \label{aa_newT}
    T^{\mu\nu}(x)=mc\,|g(x)|^{1/2}\int_{P{\cal M}(x)} p^\mu p^\nu\,f(x, p^\beta)\,
   \frac{dp^1\,dp^2\,dp^3}{-p_0}.
\end{equation}
Then we perform the change of integration variables
\[ \overline{p}^i = \frac{1}{mc}\,p^i \quad , \ i=1,2,3 \ . \]
The $p_0=g_{\alpha 0}(x)p^\alpha$ inside the integral is a function of $(p^j)$ defined
by the relation $g_{\alpha\beta}(x)\,p^\alpha p^\beta = -m^2c^2$ from (\ref{aa_mass_shell}); 
use it to define a new function $\overline{p}^0$ of the new variables $\overline{p}^j$ as
\[ \overline{p}^0(\overline{p}^j) = \frac{1}{mc}\,p^0\left(mc\overline{p}^j\right) \ . \]
Then the implicit relation that defines $\overline{p}^0$ in terms of $\overline{p}^j$ is
\[ \overline{g}_{\mu\nu}(\overline{x})\,\overline{p}^\mu\overline{p}^\nu
   =g_{\mu\nu}(x)\,\overline{p}^\mu\overline{p}^\nu = -1 \ . \]
(We also define the function $\overline{p}_0$ by lowering the index from $\overline{p}^0$). 
Under this change of variables, the domain of integration
\[ P{\cal M}(x) = \{ (p^\nu) \in T_x {\cal M} \ : \ (p^\nu) \text{ is future-directed, } 
  g_{\alpha\beta}(x^\mu)p^\alpha p^\beta = -m^2c^2\} \]
is mapped onto 
\[ \{ (\overline{p}^\nu) \in T_{\bar{x}} {\cal M} 
   \ : \ (\overline{p}^\nu) \text{ is future-directed, } 
   \overline{g}_{\alpha\beta}(\overline{x})\overline{p}^\alpha \overline{p}^\beta = -1 \} 
   =: \overline{P{\cal M}}(\overline{x}) \ , \] 
which is exactly the definition of the mass shell found in subsection \ref{subsec_LT}. 
Hence the integral above for $T^{\mu\nu}(x)$, \eqref{aa_newT}, becomes
\begin{align*}
    T^{\mu\nu}(x) &= mc\,|g(x)|^{1/2}\int_{\overline{P{\cal M}}(\overline{x})} 
    m^2c^2 \overline{p}^\mu\overline{p}^\nu\,f(x,mc\overline{p}^\beta) 
    \frac{1}{-mc\overline{p}_0} m^3c^3 \ 
    \mathrm{d}\overline{p}^1 \mathrm{d}\overline{p}^2 \mathrm{d}\overline{p}^3 
    \\ &= m^5 c^5\,|\det\overline{g}(\overline{x})|^{1/2} \int_{\overline{P{\cal M}}(\overline{x})} 
    \overline{p}^\mu\overline{p}^\nu\,f(x,mc\overline{p}^\beta) 
    \,\frac{1}{-\overline{p}_0} 
    \ \mathrm{d}\overline{p}^1 \mathrm{d}\overline{p}^2 \mathrm{d}\overline{p}^3 \ . 
\end{align*}
Now we use the definition~\eqref{aa_new_f} of $\overline{f}$ to turn this into
\[ T^{\mu\nu}(x) = \frac{1}{8\pi}\,G^{-1}\lambda^{-2}c^4 
   |\det \overline{g}(\overline{x})|^{1/2}
   \int_{\overline{P{\cal M}}(\overline{x})}
   \overline{p}^\mu\overline{p}^\nu\,\overline{f}(\overline{x},\overline{p}) 
   \,\frac{1}{-\overline{p}_0} 
   \ \mathrm{d}\overline{p}^1 \mathrm{d}\overline{p}^2 \mathrm{d}\overline{p}^3 \ . \]
Moving the constants to the left and using the definition~\eqref{aaaa_new_T} of $\overline{T}$, 
we conclude that the energy-momentum-tensor formula~\eqref{tab3} from subsection \ref{subsec_LT} 
is satisfied.

In particular, Corollary \ref{LT_cor} applies to the objects with an overbar. To state what 
it implies in terms of our original objects, note the relation between higher-order derivatives 
of barred and unbarred quantities: For all multi-indices $\alpha$, $\beta$, we have
\begin{align*}
   \partial^\alpha_{\bar{x}}\overline{T}^{\mu\nu}(\overline{x}) 
   &= 8\pi G\lambda^{2+|\alpha|}c^{-4}\,\partial^\alpha_x T^{\mu\nu}(x) \ , 
   \\[1ex] \partial^\alpha_{\bar{x}}\overline{g}_{\mu\nu}(\overline{x}) 
    &= \lambda^{|\alpha|}\partial^\alpha_x g_{\mu\nu}(x) \ , 
    \\[1ex] \partial^\alpha_{\bar{x}}\partial^\beta_{\bar{p}}\overline{f}
    (\overline{x},\overline{p}) &= 8\pi G \lambda^{2+|\alpha|} c^{|\beta|} m^{5+|\beta|} 
    \,\partial^\alpha_x\partial^\beta_p f(x,p) \ .
\end{align*}
Also observe that, due to $\overline{r}=\lambda^{-1}r$ and $\overline{M}=\frac{GM}{c^2\lambda}$, 
the splitting of the metric (\ref{gsplit}) now translates into 
\begin{equation}\label{gsplit2} 
   g(x) = \eta + h^0(x) + h^1(x) \ , \quad 
   h^0_{\mu\nu}(x)=\chi\left(\frac{r}{\lambda+x^0}\right)
   \frac{GM}{c^2 r}\,\delta_{\mu\nu} \ ,
\end{equation}  
where $\chi$ is the same cutoff function used in~\eqref{gsplit}. In particular, 
$h^1(x)=\overline{h^1}(\overline{x})$ and the same for $h$. Therefore we obtain the following result 
(with a modified constant $C_0$): 

\begin{cor}\label{LT_cor_cG} 
For any $0<\gamma<1$, $0<\mu<1-\gamma$, $0<\delta<\frac{1}{8}\min\{\gamma, 1-\gamma\}$, 
$K>0$ and $K'>0$, there is $\eps_0>0$ and a constant $C_0>0$ such that, 
for all $\eps \leq \eps_0$, given any initial data 
$(g_{\alpha\beta},\partial_{x^0} g_{\alpha\beta}, f_0)\big|_{x^0=0}$ 
for the Einstein-Vlasov system as in Section \ref{EV_system} so that
\[ {\cal E}(0) + {\cal V}(0) + \frac{GM}{c^2\lambda} < \eps \]
and
\[ \mathrm{supp}(f_0) \subset\{ (x, p): 
   |x|\leq\lambda K, \ |p|\leq mcK' \}, \]
there exists a corresponding global solution satisfying, for all $x^0\geq 0$ 
and multi-indices $\kappa$ with $|\kappa|\le 8$, 
\begin{equation}\label{conc_T}
   |\partial^\kappa T^{\mu\nu}(x^0, x)| 
   \leq\frac{c^4}{G\lambda^{2+|\kappa|}}
   \,\frac{C_0\eps}{(1+\lambda^{-1}x^0)^{1-C_0\eps}} ,
\end{equation}
and furthermore the weak-field coefficients satisfy
\begin{equation}\label{conc_h} 
   |\partial h_{\mu\nu}(x^0, x)|
   \le\frac{1}{\lambda}
   \,\frac{C_0\eps (1+\lambda^{-1}x^0)^{2\delta}}{(1+\lambda^{-1}(|x|+x^0))
   (1+\lambda^{-1}||x|-x^0|){(1+\lambda^{-1}(|x|-x^0)_+)}^{2\delta}}
\end{equation}    
as well as 
\begin{equation}\label{conc_h1}
   |\partial^\kappa (h^1)_{\mu\nu}(x^0, x)|
   \le\dfrac{1}{\lambda^{|\kappa|}}\,
   \dfrac{C_0\eps (1+\lambda^{-1}x^0)^{C_0\eps}}
   {(1+\lambda^{-1}(|x|+x^0))(1+\lambda^{-1}(|x|-x^0)_+)^\gamma}. 
\end{equation}
The quantities ${\cal E}(0)$ and ${\cal V}(0)$ are made explicit in Remark \ref{balc}(a) immediately below.
\end{cor}

\begin{remark}\label{balc}{\rm (a) We have 
${\cal E}(0)=\overline{E}_{11}(0)$ and ${\cal V}(0)=\overline{\mathcal{V}}_{11}$ 
for the solution of the barred system resulting from the initial data 
$(\overline{g}_{\alpha\beta},\partial_{\overline{x}^0} \overline{g}_{\alpha\beta}, 
\overline{f}_0)\big|_{\overline{x}^0=0}$ as obtained from 
$(g_{\alpha\beta},\partial_{x^0} g_{\alpha\beta}, f_0)\big|_{x^0=0}$. 
Therefore, by (\ref{calVN}) for $N=11$, 
\[ {\cal V}(0)=\overline{\mathcal{V}}_{11}
   =\sum_{k+l\leq 11} 
   \|\partial^k_{\overline{x}}\,\partial^l_{\overline{p}}\,\overline{f}_0
   \|_{L^2_{\overline{x}} L^2_{\overline{p}}}. \]  
Both ${\cal V}(0)$ and ${\cal E}(0)$ could now be expressed in the new variables, 
but, since we will not need their explicit forms, we don't expand this any further.
\smallskip 

\noindent 
(b) All of the above inequalities for $T$ and $h^1$ concern only the time domain $x^0\geq 0$. 
In our conditions in the introduction and in our estimates, however, the same inequalities 
are needed also at negative values of time. We remark to this end that the exact same results 
(possibly with a new constant $C_0$) still hold with $x^0$ replaced by $|x^0|$, and the reason 
for this is that one can evolve the initial data for the Einstein-Vlasov system backwards in time as well.
\smallskip 

\noindent 
(c) Recall that in Section \ref{EV_system} we had (using the notation from there) 
$\tilde{T}_{\alpha\beta}(t, x^a)=T_{\alpha\beta}(ct, x^a)$, see (\ref{tab3b}), 
and the former quantity was called $T$ again. Hence, if we pass to the variables $(t,x^a)$ 
instead of $(x^0, x^a)=(ct, x^a)$, then $x^0$ has to be replaced by $ct$ in (\ref{conc_T}), 
and it has to be used that $\partial_t^n=c^n\partial_{x^0}^n$.  
\smallskip 

\noindent 
(d) Since $g$ and its derivatives turn out to be uniformly bounded in $L^\infty$, 
the bounds in (\ref{conc_T}) for $|\partial^\kappa T^{\mu\nu}|$ and in (\ref{conc_h}), 
(\ref{conc_h1}) for $|\partial h_{\mu\nu}|$, $|\partial^\kappa (h^1)_{\mu\nu}|$ 
could equally well be stated 
for their index-raised/lowered counterparts $|\partial^\kappa T_{\mu\nu}|$ 
and $|\partial h^{\mu\nu}|$, $|\partial^\kappa (h^1)^{\mu\nu}|$, for a possibly different constant $C_0>0$. 

}
\end{remark}


\section{Appendix: Some technical lemmas}
\label{sec_app_technical}
\setcounter{equation}{0}

This last section collects the statements and proofs of technical lemmas 
that have been used in our estimates.

\subsection{Bounds from our \emph{a priori} conditions} 
\setcounter{equation}{0} 

Here we derive further bounds that follow from the conditions imposed in 
Subsection~\ref{subsec_apriori}. Let $(g_{\alpha\beta},f)$ be a solution 
to the Einstein-Vlasov system satisfying those conditions for constants 
$\mathcal{T}$, $\mathcal{S}$, $\gamma$, $\delta$ etc. as specified in the statement of Theorem~\ref{mainthm}.

First we recall from (\ref{grossH}) and (\ref{def_h0h1}) the splitting of the metric, 
\[ g(x) = \eta + h^0(x) + h^1(x),\quad 
   h^0_{\mu\nu}(x)=\chi\left(\frac{r}{\lambda+|x_0|}\right)
   \frac{q}{\lambda^{-1}r}\,\delta_{\mu\nu},\quad x=(x^\alpha), \] 
where $0\le q\le C\eps_h\le C\delta_h$ by \eqref{ADM_boundbound}; the constants
\begin{equation}\label{deep} 
\delta_h=C_0\eps_h\quad\mbox{and}\quad\delta_{T,\,\mu\nu}^\kappa=C_0\eps_{T,\,\mu\nu}^\kappa,
\end{equation}  
as defined in Theorem~\ref{mainthm}, will appear later, and, as in~\eqref{eq_chichi}, 
$\chi: [0,\infty)\longrightarrow\mathbb{R}$ 
is a smooth cutoff function such that $\chi(\xi)=1$ 
for $\xi\geq 3/4$, $\chi(\xi) = 0$ for $\xi\leq 1/2$ and $0\leq\chi(\xi´)\leq 1$ for all $\xi$.  

\subsubsection{Bounds on $h^0$} 
\label{faten} 

Due to the support properties of $\chi$ we have 
\begin{equation}\label{supp_chi} 
   r\le\frac{1}{2}(\lambda+|x^0|)\,\,\Rightarrow
   \,\,\chi\Big(\frac{r}{\lambda+|x^0|}\Big)
   =\chi'\Big(\frac{r}{\lambda+|x^0|}\Big)=0
   \quad\mbox{and}\quad r\ge\frac{3}{4}(\lambda+|x^0|)
   \,\,\Rightarrow\,\,\chi'\Big(\frac{r}{\lambda+|x^0|}\Big)=0.
\end{equation} 
If $\mu\neq\nu$, then $h^0_{\mu\nu}=0$. 
Denoting $\xi=\frac{r}{\lambda+|x^0|}$, we calculate for $x^0\neq 0$ 
and equal indices 
\begin{eqnarray*} 
   & & \partial_{x^j}h^0_{\mu\mu}=-\lambda q\chi(\xi)\,\frac{x^j}{|x|^3}
   +\lambda q\chi'(\xi)\,\frac{1}{\lambda+|x^0|}\,\frac{x^j}{|x|^2},
   \quad\partial_{x^0} h^0_{\mu\mu}=-\lambda q\chi'(\xi)\frac{1}{(\lambda+|x^0|)^2}\frac{x_0}{|x_0|},
   \\ & & \partial_{x^k}\partial_{x^j}h^0_{\mu\mu}
   =-3\lambda q\chi'(\xi)\frac{1}{\lambda+|x^0|}\frac{x^k x^j}{|x|^4}
   -\lambda q\chi(\xi)\frac{\delta^{jk}}{|x|^3}+3\lambda q\chi(\xi)\frac{x^k x^j}{|x|^5}
   \\ & & \hspace{5.8em} +\,\lambda q\chi''(\xi)\frac{1}{(\lambda+|x^0|)^2}\frac{x^k x^j}{|x|^3}
   +\lambda q\chi'(\xi)\frac{1}{\lambda+|x^0|}\frac{\delta^{jk}}{|x|^2}, 
   \\ & & \partial_{x^0}\partial_{x^j} h^0_{\mu\mu}=-\lambda q\chi''(\xi)\frac{1}{(\lambda+|x^0|)^3}
   \frac{x^0}{|x^0|}\frac{x^j}{|x|},
   \\ & & \pm\partial_{x^0}^2 h^0_{\mu\mu}=\lambda q\chi''(\xi)\frac{|x|}{(\lambda+|x^0|)^4}
   +2\lambda q\chi'(\xi)\frac{1}{(\lambda+|x^0|)^3}\frac{x^0}{|x^0|},    
\end{eqnarray*} 
and similarly for the third-order and fourth-order derivatives. 
Hence for $\lambda=0, 1, 2, 3$, also using (\ref{supp_chi}), 
\begin{eqnarray*} 
   |\partial_{x^\lambda} h^0_{\mu\mu}|
   & \le & C\lambda\delta_h |\chi'(\xi)|\frac{1}{(\lambda+|x^0|)^2}
   +C\lambda\delta_h\chi(\xi)\,\frac{1}{|x|^2}
   +C\lambda\delta_h |\chi'(\xi)|\,\frac{1}{\lambda+|x^0|}\,\frac{1}{|x|}
   \\ & \le & C\lambda\delta_h\,{\bf 1}_{\{r\ge\frac{1}{2}(\lambda+|x^0|)\}}\,\frac{1}{|x|^2}
   +C\lambda\delta_h\,{\bf 1}_{\{\frac{1}{2}(\lambda+|x^0|)\le r\le\frac{3}{4}(\lambda+|x^0|)\}}\,
   \bigg(\frac{1}{\lambda+|x^0|}\,\frac{1}{|x|}+\frac{1}{(\lambda+|x^0|)^2}\bigg) 
   \\ & \le & C\lambda\delta_h\,{\bf 1}_{\{r\ge\frac{1}{2}(\lambda+|x^0|)\}}\,\frac{1}{|x|^2}; 
\end{eqnarray*}  
the last estimate is due to $||x^0|-|x||=|ct-|x||\le\lambda$, which implies 
$\lambda+|x^0|\ge |x|=r$. In an analogous manner it is verified that 
\[ |\partial^\alpha h^0_{\mu\mu}|
   \le C\lambda\delta_h\,{\bf 1}_{\{r\ge\frac{1}{2}(\lambda+|x^0|)\}}\,\frac{1}{|x|^3},
   \quad |\alpha|=2, \] 
and 
\[ |\partial^\alpha h^0_{\mu\mu}|
   \le\lambda C\delta_h
   \,{\bf 1}_{\{r\ge\frac{1}{2}(\lambda+|x^0|)\}}\,\frac{1}{|x|^4},
   \quad |\alpha|=3, \]
so that we get 
\begin{equation}\label{h0compl} 
   |\partial^\alpha h^0_{\mu\nu}|
   \le C\lambda\delta_h\,{\bf 1}_{\{r\ge\frac{1}{2}(\lambda+|x^0|)\}}\,\frac{1}{|x|^{1+|\alpha|}},
   \quad |\alpha|\le 3.
\end{equation} 
Therefore we can also state the following bounds for later use: 
\begin{eqnarray} 
   & & |(\partial_{x^\alpha} h^0_{\mu\nu}\partial_{x^\beta} h^0_{\mu'\nu'})(x^0, x)|
   \le C\lambda^2\delta_h^2\,{\bf 1}_{\{r\ge\frac{1}{2}(\lambda+|x^0|)\}}\,\frac{1}{|x|^4},
   \label{00bth} 
   \\ & & |\partial_{x^\alpha}(\partial_{x^\beta} h^0_{\mu\nu}
   \partial_{x^\gamma} h^0_{\mu'\nu'})(x^0, x)|
   \le C\lambda^2\delta_h^2\,{\bf 1}_{\{r\ge\frac{1}{2}(\lambda+|x^0|)\}}\,\frac{1}{|x|^5},
   \label{00cth} 
   \\ & & |(h^0_{\mu\nu}\partial^2_{x^\alpha x^\beta} h^0_{\mu'\nu'})(x^0, x)|
   \le C\lambda^2\delta_h^2\,{\bf 1}_{\{r\ge\frac{1}{2}(\lambda+|x^0|)\}}\,\frac{1}{|x|^4},
   \label{00bco} 
   \\ & & |\partial_{x^\alpha}(h^0_{\mu\nu}\partial^2_{x^\beta x^\gamma} 
   h^0_{\mu'\nu'})(x^0, x)|
   \le C\lambda^2\delta_h^2\,{\bf 1}_{\{r\ge\frac{1}{2}(\lambda+|x^0|)\}}\,\frac{1}{|x|^5}. 
   \label{00cco} 
\end{eqnarray}

\subsubsection{Bounds on $h^1$} 

Due to the decay condition~\eqref{ils} we have 
\begin{equation}\label{LT116}  
   |\partial^\kappa h^1_{\mu\nu}(x^0, x)|
   \le C\lambda^{-|\kappa|}\delta_h\,\frac{(1+\lambda^{-1}|x^0|)^{\delta_h}}
   {{(1+\lambda^{-1}|x^0|+\lambda^{-1}|x|)(1+\lambda^{-1}(|x|-|x^0|)_+)}^{\gamma_0}} 
\end{equation} 
for $|\kappa|\le 4$ and with $\delta_h=C_0\eps_h$. In particular, 
\begin{eqnarray} 
   & & \lambda^2 |(\partial_{x^\alpha} h^1_{\mu\nu}\partial_{x^\beta} h^1_{\mu'\nu'})(x^0, x)|
   +\lambda^3 |\partial_{x^\alpha}(\partial_{x^\beta} h^1_{\mu\nu}
   \partial_{x^\gamma} h^1_{\mu'\nu'})(x^0, x)|
   \nonumber
   \\ & & \hspace{5em}\le\,\,C\delta_h^2\,\frac{(1+\lambda^{-1}|x^0|)^{2\delta_h}}
   {{(1+\lambda^{-1}|x^0|+\lambda^{-1}|x|)^2(1+\lambda^{-1}(|x|-|x^0|)_+)}^{2\gamma_0}},
   \qquad
   \label{11acpt}  
   \\[1ex] & & \lambda^2 |(h^1_{\mu\nu}\partial^2_{x^\alpha x^\beta} h^1_{\mu'\nu'})(x^0, x)|
   +\lambda^3 |\partial_{x^\alpha}(h^1_{\mu\nu}\partial^2_{x^\beta x^\gamma} h^1_{\mu'\nu'})(x^0, x)|
   \nonumber 
   \\ & & \hspace{5em}\le\,\,C\delta_h^2\frac{(1+\lambda^{-1}|x^0|)^{2\delta_h}}
   {{(1+\lambda^{-1}|x^0|+\lambda^{-1}|x|)^2(1+\lambda^{-1}(|x|-|x^0|)_+)}^{2\gamma_0}},
   \label{11ast}   
\end{eqnarray} 
is verified. Note that there is no gain for $h^1_{\mu\nu}$ 
if one takes a derivative, and further that the term ${(1+\lambda^{-1}(|x|-|x^0|)_+)}^{\gamma_0}$ 
will not help for $||x^0|-|x||\le\lambda$. 

\subsubsection{Bounds on combinations of $h^0$ and $h^1$} 

Combining (\ref{h0compl}) with (\ref{LT116}), 
we get in particular the following estimates:  
\begin{eqnarray} 
   \lefteqn{
   \Big(\lambda |\partial_{x^\alpha} h^0_{\mu\nu}\partial_{x^\beta} h^1_{\mu'\nu'}|
   +\lambda^2 |\partial_{x^\alpha}(\partial_{x^\beta} h^0_{\mu\nu}
   \partial_{x^\gamma} h^1_{\mu'\nu'})|\Big)(x^0, x)}
   \nonumber
   \\ & \le & C\delta_h^2\,{\bf 1}_{\{r\ge\frac{1}{2}(\lambda+|x^0|)\}}
   \,\frac{(1+\lambda^{-1}|x^0|)^{\delta_h}}
   {(1+\lambda^{-1}|x^0|+\lambda^{-1}|x|){(1+\lambda^{-1}(|x|-|x^0|)_+)}^{\gamma_0}}
   \,\frac{1}{|x|},
   \label{0110bkb} 
   \\ \lefteqn{
   \Big(\lambda |h^0_{\mu\nu}\partial^2_{x^\alpha x^\beta} h^1_{\mu'\nu'}|
   +\lambda^2 |\partial_{x^\alpha}(h^0_{\mu\nu}
   \partial^2_{x^\beta x^\gamma} h^1_{\mu'\nu'})|\Big)(x^0, x)}
   \nonumber
   \\ & \le & C\delta_h^2\,{\bf 1}_{\{r\ge\frac{1}{2}(\lambda+|x^0|)\}}
   \,\frac{(1+\lambda^{-1}|x^0|)^{\delta_h}}
   {(1+\lambda^{-1}|x^0|+\lambda^{-1}|x|){(1+\lambda^{-1}(|x|-|x^0|)_+)}^{\gamma_0}}
   \,\frac{1}{|x|},
   \label{0110brb} 
   \\ \lefteqn{
   \Big(\lambda |\partial_{x^\alpha} h^1_{\mu\nu}\partial_{x^\beta} h^0_{\mu'\nu'}|
   +\lambda^2 |\partial_{x^\alpha}(\partial_{x^\beta} h^1_{\mu\nu}
   \partial_{x^\gamma} h^0_{\mu'\nu'})|\Big)(x^0, x)}
   \nonumber
   \\ & \le & C\delta_h^2\,{\bf 1}_{\{r\ge\frac{1}{2}(\lambda+|x^0|)\}}
   \,\frac{(1+\lambda^{-1}|x^0|)^{\delta_h}}
   {(1+\lambda^{-1}|x^0|+\lambda^{-1}|x|){(1+\lambda^{-1}(|x|-|x^0|)_+)}^{\gamma_0}}
   \,\frac{1}{|x|},
   \label{0110bsb} 
   \\ \lefteqn{
   \Big(\lambda |h^1_{\mu\nu}\partial^2_{x^\alpha x^\beta} h^0_{\mu'\nu'}|
   +\lambda^2 |\partial_{x^\alpha}(h^1_{\mu\nu}
   \partial^2_{x^\beta x^\gamma} h^0_{\mu'\nu'})|\Big)(x^0, x)}
   \nonumber
   \\ & \le & C\delta_h^2\,{\bf 1}_{\{r\ge\frac{1}{2}(\lambda+|x^0|)\}}
   \,\frac{(1+\lambda^{-1}|x^0|)^{\delta_h}}
   {(1+\lambda^{-1}|x^0|+\lambda^{-1}|x|){(1+\lambda^{-1}(|x|-|x^0|)_+)}^{\gamma_0}}
   \,\frac{1}{|x|}.
   \label{0110blb} 
\end{eqnarray} 
Also, for $h=h^0+h^1$ we have 
\begin{equation}\label{hinfty3} 
   \sum_{|\alpha|\le 4}\lambda^{|\alpha|} |\partial^\alpha h_{\mu\nu}(x^0, x)|\le C\delta_h,  
\end{equation} 
which in turn yields 
\begin{equation}\label{ginfty3} 
   \sum_{|\alpha|\le 4}\lambda^{|\alpha|} |\partial^\alpha g_{\mu\nu}(x^0, x)|\le C. 
\end{equation} 
In addition, 
\begin{equation}\label{h49} 
   |h_{\mu\nu}(x^0, x)|+\lambda |\partial_{x^\alpha} h_{\mu\nu}(x^0, x)|
   \le C\delta_h\,\frac{(1+\lambda^{-1}|x^0|)^{\delta_h}}{1+\lambda^{-1}|x^0|+\lambda^{-1}|x|};
\end{equation} 
this follows from (\ref{h0compl}) and (\ref{LT116}). 

\subsubsection{Bounds on $H$}
\label{boH}  

Consider the matrix equations for the metric and its inverse:
\begin{equation}\label{eq_names_gh}
    (g_{\alpha\beta}) = \hat{g} 
    = \hat{m} + \hat{h} 
    = (\eta_{\alpha\beta}) + (h_{\alpha\beta}) \ , 
    \quad (g^{\alpha\beta}) = \hat{g}^{-1} 
    = \hat{m}^{-1} + \hat{H} = (\eta^{\alpha\beta}) + (H^{\alpha\beta}) \ .
\end{equation}
If $\eps_h>0$ (and hence $\delta_h>0$) is sufficiently small, 
then (\ref{hinfty3}) in particular yields $|\hat{h}|\le 1/2$, 
which shows that the inverse $(g^{\alpha\beta})$ exists.   
We can find $\hat{H}$ in terms of the other matrices above as follows:
\[ I = (\hat{m}+\hat{h})(\hat{m}^{-1}+\hat{H}) 
   = I + \hat{h}\hat{m}^{-1} + \hat{g}\hat{H}, \] 
which implies that 
\[ \hat{H} = -\hat{g}^{-1}\hat{h}\hat{m}^{-1} 
   = -(\hat{m}+\hat{h})^{-1}\hat{h}\hat{m}^{-1} 
   = -(I+\hat{m}^{-1}\hat{h})^{-1}\hat{m}^{-1}\hat{h}\hat{m}^{-1} \ . \] 
Define
\[ Z = \hat{H}\hat{m} 
   \ , \quad X = \hat{m}^{-1}\hat{h} 
   \ , \quad Y = (I + X)^{-1} \ . \]
Then we have the relations
\begin{align*}
    Z &= -YX \ , \\
    \dot{Z} &= -\dot{Y}X - Y\dot{X} \ , \\
    \ddot{Z} &= -\ddot{Y}X - \dot{Y}\dot{X} - Y\ddot{X} \ , \\
    \dddot{Z} &= -\dddot{Y}X - \ddot{Y}\dot{X} - \dot{Y}\ddot{X} - Y\dddot{X} \ ,
\end{align*}
where a dot represents a derivative $\partial_{x^\gamma}$ for an arbitrary $\gamma = 0,1,2,3$. 
Note that these formulas for higher-order derivatives only work if all derivatives 
are with respect to the same index $\gamma$; if differentiating with respect to different indices, 
the formulas would be longer. However, the natural sup bound obtained from them 
would be the same as the one we will obtain now for the formulas above, so there is 
no loss in doing this calculation for one fixed $\gamma$ only.

From $Y(I + X) = I$ the derivative of $Y$ is found to be $\dot{Y} = -Y\dot{X}Y$. 
This in turn yields  
\begin{align*}
    \ddot{Y} &= -\dot{Y}\dot{X}Y - Y\ddot{X}Y - Y\dot{X}\dot{Y} \\
    &= Y\dot{X}Y\dot{X}Y - Y\ddot{X}Y + Y\dot{X}Y\dot{X}Y \\
    &= Y\big( 2\dot{X}Y\dot{X} - \ddot{X} \big)Y
\end{align*}
and 
\begin{align*}
    \dddot{Y} &= \dot{Y}\big( 2\dot{X}Y\dot{X} - \ddot{X} \big)Y 
    + Y\big( 2\ddot{X}Y\dot{X} + 2\dot{X}\dot{Y}\dot{X} + 2\dot{X}Y\ddot{X} 
    - \dddot{X} \big)Y + Y\big( 2\dot{X}Y\dot{X} - \ddot{X} \big)\dot{Y} \\
    &= -Y\dot{X}Y( 2\dot{X}Y\dot{X} - \ddot{X} )Y + Y\big( 2\ddot{X}Y\dot{X} 
    - 2\dot{X}Y\dot{X}Y\dot{X} + 2\dot{X}Y\ddot{X} - \dddot{X} \big)Y 
    - Y\big( 2\dot{X}Y\dot{X} - \ddot{X} \big)Y\dot{X}Y \\
    &= Y\bigg( -\dot{X}Y( 2\dot{X}Y\dot{X} - \ddot{X} ) + \big( 2\ddot{X}Y\dot{X} 
    - 2\dot{X}Y\dot{X}Y\dot{X} + 2\dot{X}Y\ddot{X} - \dddot{X} \big) 
    - \big( 2\dot{X}Y\dot{X} - \ddot{X} \big)Y\dot{X} \bigg)Y. 
\end{align*}
From (\ref{hinfty3}) and the constancy of $\hat{m}^{-1}$ we get 
\[ \sum_{|\alpha|\le 3}\lambda^{|\alpha|} |\partial^\alpha X|
   \le C\sum_{|\alpha|\le 3}\lambda^{|\alpha|} |\partial^\alpha\hat{h}|\le C\delta_h. \]  
As for $|Y|$, 
\[ |Y| = |(I+X)^{-1}| = \Big|\sum_{j=0}^\infty (-1)^j X^j\Big|
   \leq \sum_{j=0}^\infty |X|^j = \frac{1}{1-|X|} \leq 2, \]
since we may assume that $|X| \le C\delta_h\leq 1/2$.
Therefore, it follows from the above formulas that
\[ \sum_{1\le |\alpha|\le 3}\lambda^{|\alpha|} |\partial^\alpha Y|\le C\delta_h. \] 
Going back to the derivatives of $Z$ in terms of $X$ and $Y$, we deduce
\[ \sum_{|\alpha|\le 3}\lambda^{|\alpha|} |\partial^\alpha Z|\le C\delta_h. \] 
Finally, note that these same bounds apply for the matrix 
$\hat{H}=Z\hat{m}^{-1}$. As a consequence,  
\begin{equation}\label{coe} 
   \sum_{|\alpha|\le 3}\lambda^{|\alpha|} |\partial^\alpha H^{\mu\nu}|\le C\delta_h.
\end{equation}  

\subsubsection{Supplementary bounds}

Since $(g^{\alpha\beta})=(\eta^{\alpha\beta})+(H^{\alpha\beta})$ by (\ref{eq_names_gh}), 
it follows from (\ref{coe}) that 
\begin{equation}\label{gm1infty} 
   \sum_{|\alpha|\le 3}\lambda^{|\alpha|} |\partial^\alpha g^{\mu\nu}|\le C.   
\end{equation}  
For the Christoffel symbols, 
\begin{equation}\label{wasa} 
   \Gamma^{\sigma}_{\mu\nu}
   =\frac{1}{2}\,g^{\sigma\kappa}\,(\partial_{x^\mu} g_{\nu\kappa}
   +\partial_{x^\nu} g_{\mu\kappa}-\partial_{x^\kappa} g_{\mu\nu})
   =\frac{1}{2}\,g^{\sigma\kappa}\,(\partial_{x^\mu} h_{\nu\kappa}
   +\partial_{x^\nu} h_{\mu\kappa}-\partial_{x^\kappa} h_{\mu\nu}),
\end{equation}  
the last estimate and (\ref{hinfty3}) leads to 
\[ \sum_{|\alpha|\le 2}\lambda^{|\alpha|+1} 
   |\partial^\alpha\Gamma^{\sigma}_{\mu\nu}|\le C\delta_h. \] 
From (\ref{wasa}), (\ref{gm1infty}), $h=h^0+h^1$ and the decay condition~\eqref{fuvor0} we also find 
\begin{equation}\label{studun}  
   \lambda\,|\Gamma^{\sigma}_{\mu\nu}(t, x)|
   \le\frac{C\delta_h}{{(1+\lambda^{-1}|x^0|)}^{1-2\delta}(1+\lambda^{-1}||x|-|x^0||)}.    
\end{equation} 
As a reminder, the bounds in \eqref{fuvor0} are 
\[ \lambda^{|\kappa|}\,|\partial^\kappa T_{\mu\nu}(x^0, x)|
   \le C\frac{c^4}{G\lambda^2}\,\delta_{T,\,\mu\nu}^\kappa 
   \quad\mbox{for}\quad |\kappa|\le 4,\,\,(t, x)\in\R^4. \] 
Due to 
\[ T^{\mu\nu}=g^{\mu\mu'} g^{\nu\nu'} T_{\mu'\nu'} \] 
and (\ref{gm1infty}), this yields 
\begin{equation}\label{Tupper} 
   \lambda^{|\kappa|}\,|\partial^\kappa T^{\mu\nu}(x^0, x)|
   \le C\frac{c^4}{G\lambda^2}\,\Big(\sum_{j=0}^{|\kappa|}\delta_T^{(j)}\Big)
   \quad\mbox{for}\quad |\kappa|\le 4,\,\,(x^0, x)\in\R^4,
\end{equation} 
where 
\[ \delta_T^{(j)}=\max\{\delta_{T,\,\mu\nu}^\kappa: |\kappa|=j, \mu, \nu=0, 1, 2, 3\}. \]

\subsubsection{Bounds on the support of $f$ and $T_{\mu\nu}$}
\label{subsubsec_support}

Here we denote by $f(t)$ the function $f(t,\cdot,\cdot)$, where $t$ is fixed. The Vlasov equation
\[ c^{-1}\partial_t f+\frac{p^a}{p^0}\,\partial_{x^a}f
   -\frac{1}{p^0}\,\Gamma_{\beta\gamma}^a\,p^\beta p^\gamma\,\partial_{p^a} f=0 \]
from (\ref{vlas}) with $p^0>0$ being defined by the mass-shell condition 
$g_{\alpha\beta}\,p^\alpha p^\beta=-c^2$ implies that 
\[ \frac{d}{dt} f(t, x^a(t), p^b(t))=0 \] 
for the solutions $(x^a(t), p^b(t))$ of the characteristic equations 
\begin{equation}\label{c2w} 
   \frac{dx^a}{dt}=\frac{c}{p^0}\,p^a,
   \quad\frac{dp^b}{dt}
   =-\frac{c}{p^0}\,\Gamma_{\beta\gamma}^b\,p^\beta p^\gamma,
\end{equation}  
see (\ref{adr}) and (\ref{chart}). As observed in \cite[Prop.~2.1]{LT}, 
which we basically follow here, the main issue to control $\mathrm{supp}(f(t))$ 
is to show that after a finite time 
the characteristic has to enter the ``far zone'' where $cs-|X(s)|\succsim cs$, 
which makes it possible to make use of the additional factor in the denominator 
of (\ref{studun}). On the mass shell we have 
\begin{equation}\label{casm} 
   |-(p^0)^2+|p|^2+c^2|=|(\eta_{\alpha\beta}-g_{\alpha\beta}) p^\alpha p^\beta|
   \le C\delta_h\sum_{\alpha\beta}|p^\alpha p^\beta|
   \le C\delta_h ((p^0)^2+|p|^2)
\end{equation}   
by (\ref{hinfty3}) and H\"older's inequality. Thus 
\[ \frac{|p|^2}{(p^0)^2}\le C_h\delta\Big(1+\frac{|p|^2}{(p^0)^2}\Big)
   +1-\frac{c^2}{(p^0)^2} \] 
and hence 
\begin{equation}\label{lisb} 
   \frac{|p|^2}{(p^0)^2}
   \le\frac{1}{1-C\delta_h}\Big(C\delta_h+1-\frac{c^2}{(p^0)^2}\Big)
\end{equation}  
for $\delta_h\sim\eps_h$, cf.~(\ref{deep}), sufficiently small. 
But (\ref{casm}) also yields $(p^0)^2\le C\delta_h ((p^0)^2+|p|^2)+|p|^2+c^2$, so that 
\begin{equation}\label{sica} 
   (p^0)^2\le 4|p|^2+2c^2
\end{equation}  
for $\delta_h$ small enough. 

For $K'_0>0$ appearing in the compact-support condition~\eqref{CS_cond_f}, we may assume that $K'_0\ge 2$. 

\begin{lemma}\label{suppft} 
Under the assumption
\[ \mathrm{supp}(f(0))\subset\{(x, p): |x|\leq\lambda K_0, \ |p|\leq cK'_0 \} , \]
let $\beta_0\in (0, 1)$ be defined through  
\[ 1-\beta_0^2=\frac{1}{4(8K'_0+1)}. \] 
If $\eps>0$ is small enough, then
\[ \mathrm{supp}(f(t))\subset\Big\{(x, p): |x|\le \lambda K_0+\beta_0 c|t|, \ 
   |p|\le c(K'_0+1)\Big\},\quad t\in\R. \]
\end{lemma} 
{\bf Proof\,:} We consider non-negative $t$ only. 
If $(x, p)\in\mathrm{supp}(f(t))$, 
then $(x, p)=(X(t),P(t))$ for a unique characteristic 
such that $(\tilde{x}, \tilde{p})=(X(0), P(0))\in\mathrm{supp}(f_0)$. 
Therefore $|\tilde{x}|\le\lambda K_0$ and $|\tilde{p}|\le cK'_0$ by the compact-support condition~\eqref{CS_cond_f}. 
Let $S\in ]0, \infty]$ denote the largest time such that $|P(s)|\le 2cK'_0$ 
for all $s\in [0, S)$. 
Since the characteristic stays on the mass shell, (\ref{sica}) yields 
\[ (P^0(s))^2\le 4|P(s)|^2+2c^2\le (16K'_0+2)c^2=\frac{c^2}{2(1-\beta_0^2)},
   \quad s\in [0, S). \] 
Thus we have 
\[ (P^0(s))^2\le\frac{c^2}{C\delta_h(1+\beta_0^2)+1-\beta_0^2},
   \quad s\in [0, S), \] 
if $\delta_h\sim\eps_h$ is sufficiently small. 
From (\ref{lisb}) we hence obtain 
\begin{equation}\label{PP0} 
   \frac{|P(s)|^2}{(P^0(s))^2}
   \le\frac{1}{1-C\delta_h}\Big(C\delta_h+1-\frac{c^2}{(P^0(s))^2}\Big)\le\beta_0^2,
   \quad s\in [0, S).
\end{equation}  
Owing to (\ref{c2w}) this leads to 
\[ \frac{d}{ds}\,(cs-|X(s)|)=c-c\,\frac{X(s)}{|X(s)|}\cdot\frac{P(s)}{P^0(s)}
   \ge (1-\beta_0)c,\quad s\in [0, S),  \]   
and it follows that 
\begin{equation}\label{kupop} 
   cs-|X(s)|\ge -|\tilde{x}|+(1-\beta_0)cs\ge (1-\beta_0)cs-K_0,
   \quad s\in [0, S).
\end{equation}  
From the second equation in (\ref{c2w}) and (\ref{studun}) we deduce 
\begin{eqnarray*} 
    |P(s)| & \le & |\tilde{p}|+\int_0^s |\dot{P}(\tau)|\,d\tau
    \le c K'_0+c\int_0^s\frac{1}{P^0(\tau)}\,
    |\Gamma^{\cdot}_{\mu\nu}(\tau, X(\tau))\,P^\mu(\tau) P^\nu(\tau)|\,d\tau
    \\ & \le & c K'_0+2c^2\,C\beta_0 K'_0\delta_h\int_0^s\,
    \frac{d\tau}{{(1+c\tau)}^{1-2\delta}(1+|c\tau-|X(\tau)||)}.  
\end{eqnarray*} 
As a consequence, if $S\le\frac{K_0}{(1-\beta_0)c}$, then 
\[ |P(s)|\le c K'_0+c^2\,C\delta_h\int_0^{\frac{K_0}{(1-\beta_0)c}}\,
    \frac{d\tau}{{(1+c\tau)}^{1-2\delta}}
    \le c K'_0+c\,\frac{C\delta_h}{\delta}\Big(1+\frac{K_0}{1-\beta_0}\Big)^{2\delta}
    \le c(K'_0+1),\quad s\in [0, S), \] 
if $\delta_h\sim\eps_h$ is small enough. On the other hand, 
if $S\ge\frac{K_0}{(1-\beta_0)c}$, then using (\ref{kupop}) 
\begin{eqnarray*} 
   |P(s)| & \le & c K'_0+c\,C\delta_h+c^2\,C\delta_h\int_{\frac{K_0}{(1-\beta_0)c}}^\infty\,
   \frac{d\tau}{{(1+c\tau)}^{1-2\delta}(1+(1-\beta_0)c\tau-K_0)}
   \\ & = & c K'_0+c\,C\delta_h+c\,C\delta_h\int_{\frac{K_0}{1-\beta_0}}^\infty\,
   \frac{d\sigma}{{(1+\sigma)}^{1-2\delta}(1+(1-\beta_0)\sigma-K_0)}
   \\ & \le & c K'_0+c\,C\delta_h,\quad s\in [0, S),
\end{eqnarray*}  
since $1-2\delta+1=2-2\delta>1$ (recall that $\delta < \frac{1}{2}$). Therefore if $\delta_h\sim\eps_h$ 
is sufficiently small, then we can achieve the bound $|P(s)|\le c(K'_0+1)$ for $s\in [0, S)$ in both cases. 
Observing that $K'_0+1\le 2K'_0-1$, the definition of $S$ shows that we must have $S=\infty$, and in particular 
we get $|P(s)|\le c(K'_0+1)$ for all $s\in [0, \infty)$. Furthermore, due to (\ref{c2w}) and (\ref{PP0}) we find that 
\[ |X(s)|\le |\tilde{x}|+c\int_0^s\frac{|P(\tau)|}{P^0(\tau)}\,d\tau
   \le\lambda K_0+c\beta_0 s,\quad s\in [0, \infty), \] 
and this completes the argument. 
{\hfill$\Box$}\bigskip   

\begin{cor}\label{suppon} Let $u=t-c^{-1}r$ for $r=|x|$ and put 
\[ K_\ast=\frac{K_0+\beta_0}{1-\beta_0}. \]  
\begin{itemize} 
\item[(a)] If $|\bar{x}|\ge\lambda K_\ast$ and $|u|\le\lambda c^{-1}$, 
then $T_{\alpha\beta}(t-c^{-1}|x-\bar{x}|, \bar{x})
=T^{\alpha\beta}(t-c^{-1}|x-\bar{x}|, \bar{x})=0$. 
\item[(b)] If $|x|\ge\lambda K_\ast$ and $|u|\le\lambda c^{-1}$, 
then $T_{\alpha\beta}(t, x)=T^{\alpha\beta}(t, x)=0$. 
\end{itemize} 
The same is true for derivatives of $T_{\alpha\beta}$ or $T^{\alpha\beta}$. 
\end{cor} 
{\bf Proof\,:} (a) From (\ref{tab3}), more precisely (\ref{tab3b}), 
we obtain
\[ T^{\alpha\beta}(t, x)=c\,|g(t, x)|^{1/2}\int_{\R^3} p^\alpha p^\beta
   \,f(t, x, p)\,\frac{dp}{-p_0}. \] 
If $|\bar{x}|\ge\lambda K_\ast$ and $|u|\le\lambda c^{-1}$, then 
\[ \lambda K_0+\beta_0 c(t-c^{-1}|x-\bar{x}|)
   =\lambda K_0+\beta_0 cu+\beta_0 |x|-\beta_0|x-\bar{x}|
   \le \lambda K_0+\lambda \beta_0+\beta_0 |\bar{x}|\le |\bar{x}|. \]
Hence the claim follows from Lemma \ref{suppft}. 
(b) This is shown in the same way. 
{\hfill$\Box$}\bigskip   

\subsection{Bounds for some integrals}

Here we prove general bounds regarding integrals of the type appearing in Lemma~\eqref{hinteg}. 
As a preparation, for suitable functions $\psi$ and $\tilde{\psi}$ we wish to consider $\partial_{x^\alpha} I$, where
\[ I(t, x)=\int_{\R^3}\frac{(\psi\partial_{x^\lambda}\tilde{\psi})(t-|x-y|, y)}{|x-y|}\,dy, \quad\alpha, \lambda=0, 1, 2, 3. \]  
We split up the integral $I=I_1+I_2+I_3$ according to 
\begin{eqnarray*} 
   I_1(t, x) & = & \int_{|y|\le {\cal S}}\frac{(\psi\partial_{x^\lambda}\tilde{\psi})(t-|x-y|, y)}{|x-y|}\,dy, 
   \\ I_2(t, x) & = & \int_{|x-y|\ge t+{\cal T},\,|y|\ge {\cal S}}
   \frac{(\psi\partial_{x^\lambda}\tilde{\psi})(t-|x-y|, y)}{|x-y|}\,dy, 
   \\ I_3(t, x) & = & \int_{|x-y|\le t+{\cal T},\,|y|\ge {\cal S}}
   \frac{(\psi\partial_{x^\lambda}\tilde{\psi})(t-|x-y|, y)}{|x-y|}\,dy, 
\end{eqnarray*}
and study these integrals separately, under various assumptions 
on $\psi$ and $\tilde{\psi}$. We will assume throughout this section that 
\[ |t-|x||\le 1,\quad |x|\ge\max\{12, {\cal T}, 2{\cal S}\},
   \quad {\cal T}\ge 2,\quad {\cal S}\ge\max\{2{\cal T}, 6\}. \] 
For conciseness, no physical constants such as $c$ or $\lambda$ are included. 
In addition, $\rho>0$ will simply be a parameter that is carried along. 

\subsubsection{$I_3$} 

First we are going to deal with
\[ \partial_{x^\alpha} I_3,
   \quad I_3(t, x)=\int_{|x-y|\le t+{\cal T},\,|y|\ge {\cal S}}
   \frac{(\psi\partial_{x^\lambda}\tilde{\psi})(t-|x-y|, y)}{|x-y|}\,dy,
   \quad\alpha, \lambda=0, 1, 2, 3 \]  
Note that 
\begin{eqnarray*}
   I_3 & = & \int_{|z|\le t+{\cal T},\,|x+z|\ge {\cal S}}
   \frac{(\psi\partial_{x^\lambda}\tilde{\psi})(t-|z|, x+z)}{|z|}\,dz
   \\ & = & \int_0^{t+{\cal T}} d\sigma\int_{|z|=\sigma} dS(z)
   \,{\bf 1}_{\{|x+z|\ge {\cal S}\}}
   \frac{(\psi\partial_{x^\lambda}\tilde{\psi})(t-\sigma, x+z)}{\sigma}. 
\end{eqnarray*} 
This yields 
\begin{eqnarray}\label{dtIab}  
   \partial_t I_3(t, x) 
   & = & \int_{|z|=t+{\cal T}} dS(z)
   \,{\bf 1}_{\{|x+z|\ge {\cal S}\}}\frac{(\psi\partial_{x^\lambda}\tilde{\psi})(-{\cal T}, x+z)}{t+{\cal T}}
   \nonumber
   \\ & & +\,\int_0^{t+{\cal T}} d\sigma\int_{|z|=\sigma} dS(z)
   \,{\bf 1}_{\{|x+z|\ge {\cal S}\}}\frac{\partial_t(\psi\partial_{x^\lambda}
   \tilde{\psi})(t-\sigma, x+z)}{\sigma}
   \nonumber
   \\ & = & \frac{1}{t+{\cal T}}\int_{|z|=t+{\cal T}} dS(z)
   \,{\bf 1}_{\{|x+z|\ge {\cal S}\}} (\psi\partial_{x^\lambda}\tilde{\psi})(-{\cal T}, x+z)
   \nonumber
   \\ & & +\,\int_{|x-y|\le t+{\cal T},\,|y|\ge {\cal S}}
   \frac{\partial_t(\psi\partial_{x^\lambda}\tilde{\psi})(t-|x-y|, y)}{|x-y|}\,dy. 
\end{eqnarray} 
For the $x^k$-derivatives of $I_3$, $k=1, 2, 3$, Lemma \ref{xkabl} implies that
\begin{eqnarray*} 
   \lefteqn{\partial_{x^k} I_3(t, x)}
   \\ & = & -\frac{1}{t+{\cal T}}\int_{|z|=t+{\cal T}} dS(z)\,\frac{z_k}{|z|}
   \,{\bf 1}_{\{|x+z|\le {\cal S}\}}\,(\psi\partial_{x^\lambda}\tilde{\psi})(-{\cal T}, x+z)
   \\ & & +\,\int_{|x-y|\le t+{\cal T}}
   \frac{\partial_{x^k}(\psi\partial_{x^\lambda}\tilde{\psi})(t-|x-y|, y)}{|x-y|}\,dy
   \\ & & +\,\int_{|x-y|\le t+{\cal T}, |y|\le {\cal S}}
   \bigg(\frac{(\psi\partial_{x^\lambda}\tilde{\psi})(t-|x-y|, y)}{|x-y|}
   +\partial_t (\psi\partial_{x^\lambda}\tilde{\psi})(t-|x-y|, y)\bigg)\,\frac{x_k-y_k}{|x-y|^2}\,dy
   \\ & =: & \partial_{x^k} I_{3,1}(t, x)
   +\partial_{x^k} I_{3,2}(t, x)+\partial_{x^k} I_{3,3}(t, x). 
\end{eqnarray*} 
We consider three sets of assumptions (cases I-III in what follows) to bound these integrals. 
\medskip 

\noindent 
{\bf Case I.} We assume that 
\begin{eqnarray} 
   & & |(\psi\partial_{x^\lambda}\tilde{\psi})(t, x)|
   \le C\rho^2\,{\bf 1}_{\{r\ge\frac{1}{2}(1+|t|)\}}\,\frac{1}{|x|^3},
   \quad\lambda=0, 1, 2, 3.   
   \label{00b} 
   \\ & & |\partial_{x^\alpha}(\psi\partial_{x^\lambda}\tilde{\psi})(t, x)|
   \le C\rho^2\,{\bf 1}_{\{r\ge\frac{1}{2}(1+|t|)\}}\,\frac{1}{|x|^4},
   \quad\alpha, \lambda=0, 1, 2, 3. 
   \label{00c} 
\end{eqnarray} 
From (\ref{dtIab}) we recall that 
\begin{eqnarray*}  
   \partial_t I_3(t, x) 
   & = & \frac{1}{t+{\cal T}}\int_{|z|=t+{\cal T}} dS(z)
   \,{\bf 1}_{\{|x+z|\ge {\cal S}\}} (\psi\partial_{x^\lambda}\tilde{\psi})(-{\cal T}, x+z)
   \\ & & +\,\int_{|x-y|\le t+{\cal T},\,|y|\ge {\cal S}}
   \frac{\partial_t(\psi\partial_{x^\lambda}\tilde{\psi})(t-|x-y|, y)}{|x-y|}\,dy. 
\end{eqnarray*} 
For the boundary term, as a consequence of (\ref{00b}), we get  
\[ \frac{1}{t+{\cal T}}\int_{|z|=t+{\cal T}} dS(z)
   \,{\bf 1}_{\{|x+z|\ge {\cal S}\}} |(\psi\partial_{x^\lambda}\tilde{\psi})(-{\cal T}, x+z)|
   \le\frac{C\rho^2}{t+{\cal T}}\int_{|z|=t+{\cal T}}\frac{dS(z)}{|x+z|^3}
   \le\frac{C\rho^2}{|x|}, \] 
where we have used Lemma \ref{aved} for $\sigma=0$ in the last estimate. 
For the volume term, owing to (\ref{00c}) and Lemma \ref{fastverg} we obtain 
\begin{eqnarray*}
   \lefteqn{\int_{|x-y|\le t+{\cal T},\,|y|\ge {\cal S}}
   \frac{|\partial_t(\psi\partial_{x^\lambda}\tilde{\psi})(t-|x-y|, y)|}{|x-y|}\,dy}  
   \\ & \le & C\rho^2\int_{|x-y|\le t+{\cal T},\,|y|\ge {\cal S}}
   {\bf 1}_{\{|y|\ge\frac{1}{2}(1+|t-|x-y||)\}}\,\frac{dy}{|x-y||y|^4}
   \le C(1+{\cal T})\,\frac{\rho^2}{|x|}\le\frac{C\rho^2}{|x|}. 
\end{eqnarray*} 
Thus we have shown that 
\begin{equation}\label{earlier}  
   |\partial_t I_3|\le\frac{C\rho^2}{|x|}. 
\end{equation}  

For the $x^k$-derivatives, we first consider 
\[ \partial_{x^k} I_{3,1}
   =-\frac{1}{t+{\cal T}}\int_{|z|=t+{\cal T}} dS(z)\,\frac{z_k}{|z|}\,{\bf 1}_{\{|x+z|\le {\cal S}\}}
   \,(\psi\partial_{x^\lambda}\tilde{\psi})(-{\cal T}, x+z). \] 
Then (\ref{00b}) yields 
\[ |\partial_{x^k} I_{3,1}|
   \le\frac{C\rho^2}{t+{\cal T}}\int_{|z|=t+{\cal T}}\frac{dS(z)}{|x+z|^3}
   \le\frac{C\rho^2}{|x|}. \] 
Next, for 
\begin{eqnarray*} 
  \partial_{x^k} I_{3,2} 
  & = & \int_{|x-y|\le t+{\cal T}}
  \frac{\partial_{x^k}(\psi\partial_{x^\lambda}\tilde{\psi})(t-|x-y|, y)}{|x-y|}\,dy
\end{eqnarray*} 
we can use the bound (\ref{00c}). Therefore we find, as above,  
\[ |\partial_{x^k} I_{3,2}| 
   \le C\rho^2\int_{|x-y|\le t+{\cal T}}
  {\bf 1}_{\{|y|\ge\frac{1}{2}(1+|t-|x-y||)\}}\,\frac{dy}{|x-y||y|^4}
  \le C\rho^2(1+{\cal T})\,\frac{1}{|x|}\le\frac{C\rho^2}{|x|}, \]  
by means of Lemma \ref{fastverg}. 

Lastly, 
\begin{eqnarray*} 
   \lefteqn{\partial_{x^k} I_{3,3}}
   \\ & = & \int_{|x-y|\le t+{\cal T}, |y|\le {\cal S}}
   \bigg(\frac{(\psi\partial_{x^\lambda}\tilde{\psi})(t-|x-y|, y)}{|x-y|}
   +\partial_t (\psi\partial_{x^\lambda}\tilde{\psi})(t-|x-y|, y)\bigg)\,\frac{x_k-y_k}{|x-y|^2}\,dy. 
\end{eqnarray*} 
If $|y|\le {\cal S}$, then $|x-y|\ge |x|-|y|\ge |x|/2$. In addition, 
$|\psi\partial_{x^\lambda}\tilde{\psi}|\le C\rho^2$ by (\ref{00b}) 
and $|\partial_t(\psi\partial_{x^\lambda}\tilde{\psi})|\le C\rho^2$ by (\ref{00c}). 
Therefore 
\begin{eqnarray*} 
   |\partial_{x^k} I_{3,3}| & \le & C\rho^2\int_{|y|\le {\cal S}}
   \bigg(\frac{1}{|x-y|}+1\bigg)\,\frac{1}{|x-y|}\,dy\le\frac{C\rho^2}{|x|}.  
\end{eqnarray*} 
So if we summarize the bounds on $I_3$, we have shown 
that (\ref{00b})-(\ref{00c}) implies that
\begin{equation}\label{I00bd} 
   |\partial_{x^\alpha} I_3|\le\frac{C\rho^2}{|x|},
   \quad\alpha=0, 1, 2, 3.
\end{equation}  
\medskip 

\noindent 
{\bf Case II.} We assume that 
\begin{equation}\label{11a} 
   |(\psi\partial_{x^\lambda}\tilde{\psi})(t, x)|
   +|\partial_{x^\alpha}(\psi\partial_{x^\lambda}\tilde{\psi})(t, x)|
   \le C\rho^2\frac{(1+|t|)^{2\delta}}{(1+|t|+|x|)^2(1+(|x|-|t|)_+)^{2\gamma}},
   \quad\alpha, \lambda=0, 1, 2, 3.  
\end{equation} 

By (\ref{dtIab}) we have 
\begin{eqnarray}\label{weist}   
   \partial_t I_3 
   & = & \frac{1}{t+{\cal T}}\int_{|z|=t+{\cal T}} dS(z)
   \,{\bf 1}_{\{|x+z|\ge {\cal S}\}} (\psi\partial_{x^\lambda}\tilde{\psi})(-{\cal T}, x+z)
   \nonumber   
   \\ & & +\,\int_{|x-y|\le t+{\cal T},\,|y|\ge {\cal S}}
   \frac{\partial_t(\psi\partial_{x^\lambda}\tilde{\psi})(t-|x-y|, y)}{|x-y|}\,dy. 
\end{eqnarray} 
For the boundary term, (\ref{11a}) yields 
\begin{eqnarray}\label{thhue} 
   \lefteqn{\frac{1}{t+{\cal T}}\int_{|z|=t+{\cal T}} dS(z)
   \,{\bf 1}_{\{|x+z|\ge {\cal S}\}} |(\psi\partial_{x^\lambda}\tilde{\psi})(-{\cal T}, x+z)|}
   \nonumber 
   \\ & \le & C\rho^2\,\frac{(1+{\cal T})^{2\delta}}{t+{\cal T}}\int_{|z|=t+{\cal T}}
   \,\frac{dS(z)}{(1+{\cal T}+|x+z|)^2(1+(|x+z|-{\cal T})_+)^{2\gamma}}
   \,{\bf 1}_{\{|x+z|\ge {\cal S}\}}. 
\end{eqnarray} 
On the domain of integration we have $|x+z|\ge {\cal S}\ge 2{\cal T}$, 
and hence $|x+z|-{\cal T}\ge |x+z|/2$. Therefore taking $\gamma\ge 1/2$, 
the right-hand side of (\ref{thhue}) is bounded by 
\begin{eqnarray*} 
   C\rho^2\,\frac{1}{t+{\cal T}}\int_{|z|=t+{\cal T}}
   \,\frac{dS(z)}{(1+{\cal T}+|x+z|)^2(1+|x+z|/2)^{2\gamma}} 
   & \le & C\rho^2\,\frac{1}{t+{\cal T}}\int_{|z|=t+{\cal T}}
   \,\frac{dS(z)}{|x+z|^3}\le\frac{C\rho^2}{|x|},   
\end{eqnarray*} 
where Lemma \ref{aved} for $\sigma=0$ entered in the last step. 
For the main term in (\ref{weist}), we invoke again (\ref{11a}). 
This gives 
\begin{eqnarray}  
   \lefteqn{\int_{|x-y|\le t+{\cal T},\,|y|\ge {\cal S}}
   \frac{|\partial_t(\psi\partial_{x^\lambda}\tilde{\psi})(t-|x-y|, y)|}{|x-y|}\,dy} 
   \nonumber
   \\ & \le & C\rho^2\int_{|x-y|\le t+{\cal T},\,|y|\ge {\cal S}}
   \frac{(1+|t-|x-y||)^{2\delta}}{(1+|t-|x-y||+|y|)^2(1+(|y|-|t-|x-y||)_+)^{2\gamma}|x-y|}\,dy
   \nonumber
   \\ & \le & C\rho^2\int_{|x-y|\le t+{\cal T}}\frac{dy}{|x-y|}\,
   \frac{(1+|\tau|)^{2\delta}}{(1+|\tau|+\lambda)^2(1+(\lambda-|\tau|)_+)^{2\gamma}}
   \label{thgi} 
   \\ & \le & \frac{C\rho^2}{|x|}\int_{-{\cal T}}^t d\tau\,(1+|\tau|)^{2\delta}
   \nonumber
   \int_{||x|-t+\tau|}^{|x|+t-\tau} d\lambda	    
   \,\frac{\lambda}{(1+|\tau|+\lambda)^2(1+(\lambda-|\tau|)_+)^{2\gamma}}, 
\end{eqnarray} 
where we wrote $\tau=t-|x-y|$ and $\lambda=|y|$ 
and used (\ref{CalA1}) in the last step. 
The right-hand side is split up according to 
\begin{eqnarray*}
   & & \frac{C\rho^2}{|x|}\int_{-{\cal T}}^0 d\tau\,(1+|\tau|)^{2\delta}
   \int_{||x|-t+\tau|}^{|x|+t-\tau} d\lambda 
   \,\frac{\lambda}{(1+|\tau|+\lambda)^2(1+(\lambda-|\tau|)_+)^{2\gamma}}
   \\ & & +\,\frac{C\rho^2}{|x|}\int_0^t d\tau\,(1+\tau)^{2\delta}
   \int_{||x|-t+\tau|}^{|x|+t-\tau} d\lambda\,{\bf 1}	_{\{\lambda\ge\tau\}}    
   \,\frac{\lambda}{(1+\tau+\lambda)^2(1+\lambda-\tau)^{2\gamma}}
   \\ & & +\,\frac{C\rho^2}{|x|}\int_0^t d\tau\,(1+\tau)^{2\delta}
   \int_{||x|-t+\tau|}^{|x|+t-\tau} d\lambda\,{\bf 1}	_{\{\lambda\le\tau\}}    
   \,\frac{\lambda}{(1+\tau+\lambda)^2}
   \\ & =: & A_1(t, x)+A_2(t, x)+A_3(t, x).
\end{eqnarray*} 
For the first term, 
\[ A_1(t, x)|\le\frac{C\rho^2}{|x|}\,(1+{\cal T})^{2\delta}\int_{-{\cal T}}^0 d\tau
   \int_{||x|-t+\tau|}^{|x|+t-\tau}\frac{\lambda}{(1+\lambda)^2}\,d\lambda. \]  
Now 
\begin{eqnarray*} 
   \int_{||x|-t+\tau|}^{|x|+t-\tau}\frac{\lambda}{(1+\lambda)^2}\,d\lambda
   & \le & \int_{||x|-t+\tau|}^{|x|+t-\tau}\frac{\lambda}{(1+\lambda^2)}\,d\lambda
   \\ & = & C\Big(\ln(1+(|x|+t-\tau)^2)-\ln(1+(||x|-t+\tau|)^2)\Big)
   \\ & \le & C\ln(1+(|x|+t-\tau)^2)
   \le C\ln |x|, 
\end{eqnarray*} 
due to $\tau\in [-{\cal T}, 0]$. Therefore we arrive at 
\[ A_1(t, x)\le C\rho^2\,\frac{\ln |x|}{|x|}. \] 
For the second term, if we take $\gamma\ge 3/4$ for instance, then 
\begin{eqnarray*} 
   A_2(t, x) & \le & \frac{C\rho^2}{|x|}\,\int_0^t\,d\tau\,(1+\tau)^{2\delta}
   \int_{\tau}^\infty d\lambda\,
   \,\frac{\lambda}{(1+\tau+\lambda)^2(1+\lambda-\tau)^{3/2}} 
   \\ & \le & \frac{C\rho^2}{|x|}\,\int_0^t\,d\tau\,\frac{(1+\tau)^{2\delta}}{1+\tau}
   \int_{\tau}^\infty d\lambda\,
   \,\frac{1}{(1+\lambda-\tau)^{3/2}} 
   \\ & \le & \frac{C\rho^2}{|x|}\,\int_0^t\,d\tau\,\frac{(1+\tau)^{2\delta}}{1+\tau}
   \le\frac{C\rho^2}{|x|}\,(1+t)^{2\delta}
   \le\frac{C\rho^2}{|x|^{1-2\delta}}, 
\end{eqnarray*} 
the latter since $|t-|x||\le 1$ and $|x|\ge 10$. For the third term, 	
\[ A_3(t, x)\le\frac{C\rho^2}{|x|}\,\int_0^t\,d\tau\,\frac{(1+\tau)^{2\delta}}{(1+\tau)^2}
   \int_{||x|-t+\tau|}^{|x|+t-\tau} d\lambda\,{\bf 1}_{\{\lambda\le\tau\}}
   \,\lambda. \] 
On the domain of $\lambda$-integration we have $\lambda\le\tau$. 
Moreover, $\lambda\ge ||x|-t+\tau|\ge\tau-||x|-t|\ge\tau-1$. 
Hence the domain of $\lambda$-integration is an interval of length at most $1$.  
This yields 
\[ A_3(t, x)\le\frac{C\rho^2}{|x|}\,\int_0^t\,d\tau\,\frac{(1+\tau)^{2\delta}\tau}{(1+\tau)^2}
   \le\frac{C\rho^2}{|x|^{1-2\delta}}. \] 
Taking together all the estimates in this section so far, 
we have shown that 
\[ |\partial_t I_3|\le\frac{C\rho^2}{|x|^{1-2\delta}}. \]  

What concerns the $\partial_{x^k}$-derivatives, we first consider 
\[ \partial_{x^k} I_{3,1}
   =-\frac{1}{t+{\cal T}}\int_{|z|=t+{\cal T}} dS(z)\,\frac{z_k}{|z|}\,{\bf 1}_{\{|x+z|\le {\cal S}\}}
   \,(\psi\partial_{x^\lambda}\tilde{\psi})(-{\cal T}, x+z). \] 
Here (\ref{11a}) results in 
\begin{eqnarray*} 
   \lefteqn{\frac{1}{t+{\cal T}}\int_{|z|=t+{\cal T}} dS(z)
   \,{\bf 1}_{\{|x+z|\le {\cal S}\}} |(\psi\partial_{x^\lambda}\tilde{\psi})(-{\cal T}, x+z)|}
   \\ & \le & C\rho^2\,\frac{(1+{\cal T})^{2\delta}}{t+{\cal T}}\int_{|z|=t+{\cal T}}
   \,\frac{dS(z)}{(1+{\cal T}+|x+z|)^2(1+(|x+z|-{\cal T})_+)^{2\gamma}}
   \,{\bf 1}_{\{|x+z|\le {\cal S}\}}
   \\ & \le & C\rho^2\,\frac{1}{t+{\cal T}}\int_{|z|=t+{\cal T}}
   \,\frac{dS(z)}{|x+z|^2}\,\frac{{\cal S}}{|x+z|}
   \le\frac{C\rho^2}{|x|}, 
\end{eqnarray*} 
the latter due to Lemma \ref{aved} for $\sigma=0$. Next, for 
\begin{eqnarray*} 
  \partial_{x^k} I_{3,2} 
  & = & \int_{|x-y|\le t+{\cal T}}
  \frac{\partial_{x^k}(\psi\partial_{x^\lambda}\tilde{\psi})(t-|x-y|, y)}{|x-y|}\,dy
\end{eqnarray*} 
we can apply (\ref{11a}) once more to bound 
\begin{eqnarray*} 
  |\partial_{x^k} I_{3,2}| 
  & \le & C\rho^2\int_{|x-y|\le t+{\cal T}}\frac{dy}{|x-y|}\,
  \frac{(1+|t-|x-y||)^{2\delta}}{(1+|t-|x-y||+|y|)^2(1+(|y|-|t-|x-y||)_+)^{2\gamma}}. 
\end{eqnarray*} 
However, this is the same term as in (\ref{thgi}). Hence we obtain 
\[ |\partial_{x^k} I_{3,2}|\le\frac{C\rho^2}{|x|^{1-2\delta}}. \]  
Lastly, 
\begin{eqnarray*} 
   \lefteqn{\partial_{x^k} I_{3,3}}
   \\ & = & \int_{|x-y|\le t+{\cal T}, |y|\le {\cal S}}
   \bigg(\frac{(\psi\partial_{x^\lambda}\tilde{\psi})(t-|x-y|, y)}{|x-y|}
   +\partial_t (\psi\partial_{x^\lambda}\tilde{\psi})(t-|x-y|, y)\bigg)\,\frac{x_k-y_k}{|x-y|^2}\,dy. 
\end{eqnarray*} 
If $|y|\le {\cal S}$, then $|x-y|\ge |x|-|y|\ge |x|/2$.
Furthermore, $|\psi\partial_{x^\lambda}\tilde{\psi}|
+|\partial_t(\psi\partial_{x^\lambda}\tilde{\psi})|\le C\rho^2$ 
by (\ref{11a}). Thus
\begin{eqnarray*} 
   |\partial_{x^k} I_{3,3}| & \le & C\rho^2\int_{|y|\le {\cal S}}
   \bigg(\frac{1}{|x-y|}+1\bigg)\,\frac{1}{|x-y|}\,dy\le\frac{C\rho^2}{|x|}.  
\end{eqnarray*} 
So if we summarize the bounds on $I_3$, 
we have shown that if $\gamma\ge 3/4$ and (\ref{11a}) holds, then 
\begin{equation}\label{I11bd} 
   |\partial_{x^\alpha} I_3|
   \le\frac{C\rho^2}{|x|^{1-2\delta}},\quad\alpha=0, 1, 2, 3.
\end{equation}   
\medskip 

\noindent 
{\bf Case III.} Here we assume that 
\begin{eqnarray} 
   \lefteqn{
   \Big(|\psi\partial_{x^\lambda}\tilde{\psi}|
   +|\partial_{x^\alpha}(\psi\partial_{x^\lambda}\tilde{\psi})|\Big)(t, x)}
   \nonumber
   \\ & \le & C\rho^2\,{\bf 1}_{\{r\ge\frac{1}{2}(1+|t|)\}}
   \,\frac{(1+|t|)^\delta}{(1+|t|+|x|)(1+(|x|-|t|)_+)^{\gamma}}
   \,\frac{1}{|x|},\quad\alpha, \lambda=0, 1, 2, 3.  
   \label{0110b} 
\end{eqnarray} 

From (\ref{dtIab}) we once again have 
\begin{eqnarray}\label{nss}   
   \partial_t I_3 
   & = & \frac{1}{t+{\cal T}}\int_{|z|=t+{\cal T}} dS(z)
   \,{\bf 1}_{\{|x+z|\ge {\cal S}\}} (\psi\partial_{x^\lambda}\tilde{\psi})(-{\cal T}, x+z)
   \nonumber
   \\ & & +\,\int_{|x-y|\le t+{\cal T},\,|y|\ge {\cal S}}
   \frac{\partial_t(\psi\partial_{x^\lambda}\tilde{\psi})(t-|x-y|, y)}{|x-y|}\,dy. 
\end{eqnarray} 
For the boundary term, (\ref{0110b}) yields 
\begin{eqnarray}\label{thhue01}  
   \lefteqn{\frac{1}{t+{\cal T}}\int_{|z|=t+{\cal T}} dS(z)
   \,{\bf 1}_{\{|x+z|\ge {\cal S}\}} |(\psi\partial_{x^\lambda}\tilde{\psi})(-{\cal T}, x+z)|}
   \nonumber
   \\ & \le & C\rho^2\,\frac{(1+{\cal T})^\delta}{t+{\cal T}}\int_{|z|=t+{\cal T}}
   \,dS(z)\,\frac{1}{|x+z|}\,\frac{1}{(1+{\cal T}+|x+z|)(1+(|x+z|-{\cal T})_+)^\gamma}
   \,{\bf 1}_{\{|x+z|\ge {\cal S}\}}.\qquad 
\end{eqnarray} 
On the domain of integration we have $|x+z|\ge {\cal S}\ge 2{\cal T}$, 
and hence $|x+z|-{\cal T}\ge |x+z|/2$. 
Therefore the right-hand side of (\ref{thhue01}) is bounded by 
\begin{eqnarray*} 
   C\rho^2\,\frac{1}{t+{\cal T}}\int_{|z|=t+{\cal T}}
   \,\frac{dS(z)}{|x+z|(1+{\cal T}+|x+z|)(1+|x+z|/2)^\gamma} 
   & \le & C\rho^2\,\frac{1}{t+T}\int_{|z|=t+{\cal T}}
   \,\frac{dS(z)}{|x+z|^{2+\gamma}}\le\frac{C\rho^2}{|x|^\gamma},   
\end{eqnarray*} 
where in the last step Lemma \ref{aved} has been used.   
For the main term in (\ref{nss}), once again (\ref{0110b}) is helpful. 
Since $g(|y|)\le\frac{C}{|y|}$ for $|y|\ge {\cal S}$, we get 
\begin{eqnarray} 
   \lefteqn{\int_{|x-y|\le t+{\cal T},\,|y|\ge {\cal S}}
   \frac{|\partial_t(\psi\partial_{x^\lambda}\tilde{\psi})(t-|x-y|, y)|}{|x-y|}\,dy}
   \nonumber 
   \\ & \le & C\rho^2\int_{|x-y|\le t+{\cal T},\,|y|\ge {\cal S}}
   \frac{1}{|y|}\,\frac{(1+|t-|x-y||)^\delta}{(1+|t-|x-y||+|y|)(1+(|y|-|t-|x-y||)_+)^{\gamma}|x-y|}\,dy
   \nonumber   
   \\ & \le & C\rho^2\int_{|x-y|\le t+{\cal T}}\frac{dy}{|x-y|}
   \,\frac{(1+|\tau|)^\delta}{\lambda(1+|\tau|+\lambda)(1+(\lambda-|\tau|)_+)^{\gamma}}
   \label{fahren}    
   \\ & \le & \frac{C\rho^2}{|x|}\int_{-{\cal T}}^t d\tau\int_{||x|-t+\tau|}^{|x|+t-\tau} d\lambda
   \,\frac{(1+|\tau|)^\delta}{(1+|\tau|+\lambda)(1+(\lambda-|\tau|)_+)^{\gamma}}, 
   \nonumber
\end{eqnarray} 
where we wrote $\tau=t-|x-y|$ and $\lambda=|y|$, and simplified by means of (\ref{CalA1}) 
in the last step. The right-hand side is split up according to 
\begin{eqnarray*}
   & & \frac{C\rho^2}{|x|}\int_{-{\cal T}}^0 d\tau\,(1+|\tau|)^{\delta}
   \int_{||x|-t+\tau|}^{|x|+t-\tau} d\lambda 
   \,\frac{1}{(1+|\tau|+\lambda)(1+(\lambda-|\tau|)_+)^{\gamma}}
   \\ & & +\,\frac{C\rho^2}{|x|}\int_0^t d\tau\,(1+\tau)^{\delta}
   \int_{||x|-t+\tau|}^{|x|+t-\tau} d\lambda\,{\bf 1}	_{\{\lambda\ge\tau\}}    
   \,\frac{1}{(1+\tau+\lambda)(1+\lambda-\tau)^{\gamma}}
   \\ & & +\,\frac{C\rho^2}{|x|}\int_0^t d\tau\,(1+\tau)^{\delta}
   \int_{||x|-t+\tau|}^{|x|+t-\tau} d\lambda\,{\bf 1}	_{\{\lambda\le\tau\}}    
   \,\frac{1}{1+\tau+\lambda}
   \\ & =: & B_1(t, x)+B_2(t, x)+B_3(t, x).
\end{eqnarray*} 
For the first term, 
\[ B_1(t, x)|\le\frac{C\rho^2}{|x|}\,(1+{\cal T})^{\delta}\int_{-{\cal T}}^0 d\tau
   \int_{||x|-t+\tau|}^{|x|+t-\tau}\frac{d\lambda}{1+\lambda}. \]  
Now 
\begin{eqnarray*} 
   \int_{||x|-t+\tau|}^{|x|+t-\tau}\frac{d\lambda}{1+\lambda}
   & = & \ln(1+(|x|+t-\tau))-\ln(1+(||x|-t+\tau|))
   \\ & \le & C\ln(1+(|x|+t-\tau))
   \le C\ln |x|, 
\end{eqnarray*} 
due to $\tau\in [-{\cal T}, 0]$. Therefore we see that
\[ B_1(t, x)\le C\rho^2\,\frac{\ln |x|}{|x|}. \] 
The second term is 
\begin{eqnarray*} 
   B_2(t, x) & = & \frac{C\rho^2}{|x|}\int_0^{\min\{t, \frac{1}{2}(|x|+t)\}} d\tau\,(1+\tau)^{\delta}
   \int_{||x|-t+\tau|}^{|x|+t-\tau} d\lambda\,{\bf 1}	_{\{\lambda\ge\tau\}}    
   \,\frac{1}{(1+\tau+\lambda)(1+\lambda-\tau)^{\gamma}}
   \\ & \le & \frac{C\rho^2}{|x|}\int_0^{\min\{t, \frac{1}{2}(|x|+t)\}} d\tau
   \,\frac{(1+\tau)^{\delta}}{1+\tau}
   \int_{\tau}^{|x|+t-\tau}\frac{d\lambda}{(1+\lambda-\tau)^{\gamma}}
   \\ & \le & \frac{C\rho^2}{|x|}\int_0^{\min\{t, \frac{1}{2}(|x|+t)\}} d\tau
   \,\frac{(1+\tau)^{\delta}}{1+\tau}
   \,(1+|x|+t-2\tau)^{1-\gamma}
   \\ & \le & \frac{C\rho^2}{|x|}\,|x|^{1-\gamma}
   \int_0^{C|x|} d\tau\,\frac{(1+\tau)^{\delta}}{1+\tau}
   \le\frac{C\rho^2}{|x|^\gamma}\,|x|^\delta. 
\end{eqnarray*} 
For the third term, 
\begin{eqnarray*} 
   B_3(t, x) & = & \frac{C\rho^2}{|x|}\int_0^t d\tau\,(1+\tau)^{\delta}
   \int_{||x|-t+\tau|}^{|x|+t-\tau} d\lambda\,{\bf 1}	_{\{\lambda\le\tau\}}    
   \,\frac{1}{1+\tau+\lambda}
   \\ & \le & \frac{C\rho^2}{|x|}\,\int_0^t\,d\tau\,\frac{(1+\tau)^\delta}{1+\tau}
   \int_{||x|-t+\tau|}^{|x|+t-\tau} d\lambda\,{\bf 1}_{\{\lambda\le\tau\}}. 
\end{eqnarray*} 
On the domain of $\lambda$-integration we have $\lambda\le\tau$. 
Moreover, $\lambda\ge ||x|-t+\tau|\ge\tau-||x|-t|\ge\tau-1$. 
Hence the domain of $\lambda$-integration is an interval of length at most $1$.  
This yields 
\[ B_3(t, x)\le\frac{C\rho^2}{|x|}\,\int_0^t\frac{d\tau}{(1+\tau)^{1-\delta}}
   \le\frac{C\rho^2}{|x|}\,(1+t)^\delta\le\frac{C\rho^2}{|x|^{1-\delta}}.   \] 
In summary, we have shown that 
\[ |\partial_t I_3|\le\frac{C\rho^2}{|x|^{\gamma-\delta}}. \] 

We still need to bound the $\partial_{x^k}$-derivatives. 
For, we first consider 
\[ \partial_{x^k} I_{3,1}
   =-\frac{1}{t+{\cal T}}\int_{|z|=t+{\cal T}} dS(z)\,\frac{z_k}{|z|}\,{\bf 1}_{\{|x+z|\le {\cal S}\}}
   \,(\psi\partial_{x^\lambda}\tilde{\psi})(-{\cal T}, x+z). \] 
Then by (\ref{0110b})
\begin{eqnarray*}  
   \lefteqn{|\partial_{x^k} I_{3,1}|}
   \\ & \le & C\rho^2\,\frac{(1+{\cal T})^\delta}{t+{\cal T}}\int_{|z|=t+{\cal T}}
   \,\frac{dS(z)}{|x+z|}\,\frac{1}{(1+{\cal T}+|x+z|)(1+(|x+z|-{\cal T})_+)^\gamma}
   \,{\bf 1}_{\{|x+z|\le {\cal S}\}} 
   \\ & \le & C\rho^2\,\frac{1}{t+{\cal T}}\int_{|z|=t+{\cal T}}\,dS(z)\,\frac{1}{|x+z|^2}
   \,\frac{{\cal S}}{|x+z|}
   \le\frac{C\rho^2}{|x|}, 
\end{eqnarray*} 
the latter by Lemma \ref{aved} for $\sigma=0$. Next, for 
\begin{eqnarray*} 
  \partial_{x^k} I_{3,2} 
  & = & \int_{|x-y|\le t+{\cal T}}
  \frac{\partial_{x^k}(\psi\partial_{x^\lambda}\tilde{\psi})(t-|x-y|, y)}{|x-y|}\,dy
\end{eqnarray*} 
we can use (\ref{0110b}) to bound 
\begin{eqnarray*} 
  |\partial_{x^k} I_{3,2}| 
  & \le & C\rho^2\int_{|x-y|\le t+{\cal T}}\frac{dy}{|x-y|}\,
  \frac{(1+|t-|x-y||)^\delta}{|y|(1+|t-|x-y||+|y|)(1+(|y|-|t-|x-y||)_+)^{\gamma}} 
\end{eqnarray*} 
This term agrees with (\ref{fahren}), so we obtain the estimate 
\[ |\partial_{x^k} I_{3,2}|\le\frac{C\rho^2}{|x|^{\gamma-\delta}}. \] 
Lastly, 
\begin{eqnarray*} 
   \partial_{x^k} I_{3,3}
   & = & \int_{|x-y|\le t+{\cal T}, |y|\le {\cal S}}
   \bigg(\frac{(\psi\partial_{x^\lambda}\tilde{\psi})(t-|x-y|, y)}{|x-y|}
   \\ & & \hspace{8em} +\,\partial_t (\psi\partial_{x^\lambda}\tilde{\psi})
   (t-|x-y|, y)\bigg)\,\frac{x_k-y_k}{|x-y|^2}\,dy. 
\end{eqnarray*} 
If $|y|\le {\cal S}$, then $|x-y|\ge |x|-|y|\ge |x|/2$. 
In addition, by (\ref{0110b}) in particular 
\[ |(\psi\partial_{x^\lambda}\tilde{\psi})(t, x)|
   +|\partial_t(\psi\partial_{x^\lambda}\tilde{\psi})(t, x)|
   \le C\rho^2\,{\bf 1}_{\{r\ge\frac{1}{2}(1+|t|)\}}\,(1+|t|)^\delta
   \,\frac{1}{|x|}\le C\rho^2, \]  
owing to $||x|-t|\le 1$. This leads to
\[ |\partial_{x^k} I_{3,3}|\le C\rho^2\int_{|y|\le {\cal S}}
   \bigg(\frac{1}{|x-y|}+1\bigg)\,\frac{1}{|x-y|}\,dy
   \le\frac{C\rho^2}{|x|}. \]  
So if we summarize the bounds on $I_3$, 
we have shown that (\ref{0110b}) implies that  
\begin{equation}\label{I0110bd} 
   |\partial_{x^\alpha} I_3|
   \le\frac{C\rho^2}{|x|^{\gamma-\delta}},\quad\alpha=0, 1, 2, 3.
\end{equation}   

Hence our results for $I_3$ can be summarized as follows. 

\begin{cor}\label{I3sum} 
Let $\gamma\in [\frac{3}{4}, 1)$ and $\delta>0$ 
be such that $\gamma+\delta\le 1$. If 
\begin{itemize} 
\item[(a)] (\ref{00b})-(\ref{00c}) holds, or  
\item[(b)] (\ref{11a}) holds, or 
\item[(c)] (\ref{0110b}) holds, 
\end{itemize} 
then
\[ \bigg|\,\partial_{x^\alpha}\int_{|x-y|\le t+{\cal T},\,|y|\ge {\cal S}}
   \frac{(\psi\partial_{x^\lambda}\tilde{\psi})(t-|x-y|, y)}{|x-y|}\,dy\bigg|
   \le\frac{C\rho^2}{|x|^{\gamma-\delta}} \] 
for $\alpha, \lambda=0, 1, 2, 3$. 
\end{cor} 
{\bf Proof\,:} This follows from (\ref{I00bd}), (\ref{I11bd}) and (\ref{I0110bd}). 
{\hfill$\Box$}\bigskip

\subsubsection{$I_2$} 

Now we are going to bound
\[ \partial_{x^\alpha} I_2,
   \quad I_2(t, x)=\int_{|x-y|\ge t+{\cal T},\,|y|\ge {\cal S}}
   \frac{(\psi\partial_{x^\lambda}\tilde{\psi})(t-|x-y|, y)}{|x-y|}\,dy,
   \quad\alpha, \lambda=0, 1, 2, 3. \]  
In this section we will assume that for some $l\in\{0, 1\}$  
\begin{eqnarray}\label{I2psi} 
   & & \psi(t, x)=q(x),\quad t\le -{\cal T},
   \nonumber 
   \\ & & |\partial^\alpha q(x)|\le\frac{C\rho}{|x|^{l+1+|\alpha|}},
   \quad |x|\ge {\cal S},\,\,|\alpha|\le 2,  
\end{eqnarray} 
and for some $m\in\{0, 1\}$ 
\begin{eqnarray}\label{I2psitil} 
   & & \tilde{\psi}(t, x)=\tilde{q}(x),\quad t\le -{\cal T},
   \nonumber 
   \\ & & |\partial^\alpha\tilde{q}(x)|\le\frac{C\rho}{|x|^{m+1+|\alpha|}},
   \quad |x|\ge {\cal S},\,\,|\alpha|\le 2,  
\end{eqnarray} 
is verified. Since $t-|x-y|\le -{\cal T}$ on the domain of integration, this yields 
\[ I_2(t, x)=\int_{|x-y|\ge t+{\cal T},\,|y|\ge {\cal S}}
   \frac{q(y)\partial_{x^\lambda}\tilde{q}(y)}{|x-y|}\,dy. \] 
For the $\partial_t$-derivative, we write 
\begin{eqnarray*} 
   I_2 & = & \int_{|z|\ge t+{\cal T},\,|x+z|\ge {\cal S}}
   \frac{q(x+z)\partial_{x^\lambda}\tilde{q}(x+z)}{|z|}\,dz
   \\ & = & \int_{t+{\cal T}}^\infty d\sigma\int_{|z|=\sigma} dS(z)
   \,{\bf 1}_{\{|x+z|\ge {\cal S}\}}
   \,\frac{q(x+z)\partial_{x^\lambda}\tilde{q}(x+z)}{\sigma}
\end{eqnarray*} 
to get 
\[ \partial_t I_2 = -\frac{1}{t+{\cal T}}\int_{|z|=t+{\cal T}} dS(z)\,{\bf 1}_{\{|x+z|\ge {\cal S}\}}
   \,q(x+z)\partial_{x^\lambda}\tilde{q}(x+z). \] 
Hence, by (\ref{I2psi}) and (\ref{I2psitil}),   
\[ |\partial_t I_2| \le C\rho^2\,\frac{1}{t+{\cal T}}\int_{|z|=t+{\cal T}} 
   {\bf 1}_{\{|x+z|\ge {\cal S}\}}
   \,\frac{dS(z)}{|x+z|^{3+l+m}}\le C\rho^2\,\frac{1}{t+{\cal T}}\int_{|z|=t+{\cal T}}
   \,\frac{dS(z)}{|x+z|^3}\le\frac{C\rho^2}{|x|}, \] 
see (\ref{earlier}). For the $x^k$-derivatives, 
we obtain, completely analogous to Lemma \ref{xkabl}, 
\begin{eqnarray*} 
   |\partial_{x^k} I_2| & \le & 
   \int_{|x-y|\ge t+{\cal T}}\frac{|\partial_{x^k}(q\partial_{x^\lambda}\tilde{q})(y)|}{|x-y|}\,dy
   +\int_{|x-y|\ge t+{\cal T}, |y|\ge {\cal S}}\frac{|(q\partial_{x^\lambda}\tilde{q})(y)|}{|x-y|^2}\,dy
   \\ & & +\,\frac{1}{t+{\cal T}}\int_{|z|=t+{\cal T}} {\bf 1}_{\{|x+z|\ge {\cal S}\}}
   \,|(q\partial_{x^\lambda}\tilde{q})(x+z)|\,dS(z). 
\end{eqnarray*} 
On the first domain of integration we have $|y|\ge |x-y|-t+t-|x|\ge {\cal T}-1\ge 1$.  
In addition, on the second domain we may use that 
$1+|t-|x-y||+|y|\le 1+|t-|x||+||x|-|x-y||+|y|\le 2+2|y|\le 3|y|$.  
Therefore by (\ref{I2psi}) and (\ref{I2psitil}) 
\begin{eqnarray*} 
   |\partial_{x^k} I_2| & \le & 
   \frac{C\rho^2}{t+{\cal T}}\int_{|y|\ge 1}\frac{dy}{|y|^{4+l+m}}
   +C\rho^2\int_{|x-y|\ge t+{\cal T}, |y|\ge {\cal S}}\frac{dy}{|y|^{3+l+m} |x-y|^2}
   \\ & & +\,\frac{C\rho^2}{t+{\cal T}}\int_{|z|=t+{\cal T}}{\bf 1}_{\{|x+z|\ge {\cal S}\}}
   \,\frac{dS(z)}{|x+z|^{3+l+m}}
   \\ & \le & \frac{C\rho^2}{|x|} 
   +C\rho^2\int_{\R^3}\frac{dy}{(1+|t-|x-y||+|y|)^3 |x-y|^2}
   \\ & \le & \frac{C\rho^2}{|x|}
   +C\rho^2\,\frac{1}{1+|t|+|x|}\,\frac{1}{1+|t-|x||}
   \le\frac{C\rho^2}{|x|}, 
\end{eqnarray*} 
see (\ref{earlier}) and (\ref{CalA1b}).  
Hence we have shown that (\ref{I2psi}) and (\ref{I2psitil}) imply that 
\begin{equation}\label{I2sum} 
   \bigg|\partial_{x^\alpha}\int_{|x-y|\ge t+{\cal T}, |y|\ge {\cal S}}
   \frac{(\psi\partial_{x^\lambda}\tilde{\psi})(t-|x-y|, y)}{|x-y|}\,dy\bigg|
   \le\frac{C\rho^2}{|x|}.
\end{equation} 

\subsubsection{$I_1$} 

It remains to deal with
\[ \partial_{x^\alpha} I_1,
   \quad I_1(t, x)=\int_{|y|\le {\cal S}}
   \frac{(\psi\partial_{x^\lambda}\tilde{\psi})(t-|x-y|, y)}{|x-y|}\,dy,
   \quad\alpha, \lambda=0, 1, 2, 3, \]  
which is the easiest part. Throughout this section we will assume that 
\begin{equation}\label{I1assum} 
   |(\psi\partial_{x^\lambda}\tilde{\psi})(t, x)|
   +|\partial_t(\psi\partial_{x^\lambda}\tilde{\psi})(t, x)|\le C\rho^2. 
\end{equation}  

Due to (\ref{I1assum}), and since $|x-y|\ge |x|-|y|\ge |x|/2$ for $|y|\le {\cal S}$, 
\[ |\partial_t I_1(t, x)|
   =\bigg|\int_{|y|\le {\cal S}}\frac{\partial_t(\psi\partial_{x^\lambda}\tilde{\psi})(t-|x-y|, y)}
   {|x-y|}\,dy\,\bigg|
   \le C\rho^2\int_{|y|\le {\cal S}}\frac{dy}{|x-y|}
   \le\frac{C\rho^2}{|x|}. \]     
For the $x^k$-derivatives, we calculate 
\begin{eqnarray*}
   \partial_{x^k} I_1(t, x) 
   & = & -\int_{|y|\le {\cal S}}
   \bigg(\frac{(\psi\partial_{x^\lambda}\tilde{\psi})(t-|x-y|, y)}{|x-y|^3}(x^k-y^k)
   \\ & & \hspace{5em} +\,\frac{\partial_t(\psi\partial_{x^\lambda}\tilde{\psi})(t-|x-y|, y)}{|x-y|^2}
   (x^k-y^k)\bigg)\,dy 
\end{eqnarray*}
and proceed analogously, once again using (\ref{I1assum}), to find
\[ |\partial_{x^k} I_1(t, x)|
   \le C\int_{|y|\le {\cal S}}\bigg(\frac{1}{|x-y|^2}+\frac{1}{|x-y|}\bigg)\,dy
   \le\frac{C\rho^2}{|x|}. \] 
Thus altogether we have seen that (\ref{I1assum}) implies that 
\begin{equation}\label{I1sum} 
   \bigg|\partial_{x^\alpha}\int_{|y|\le {\cal S}}
   \frac{(\psi\partial_{x^\lambda}\tilde{\psi})(t-|x-y|, y)}{|x-y|}\,dy\bigg|
   \le\frac{C\rho^2}{|x|}.
\end{equation}

\subsection{Miscellaneous lemmas}

\begin{lemma}\label{pients} 
For $j, k, n, m\in\{1, 2, 3\}$ we have 
\[ \int_{\partial B_r(0)} x^j x^k\,dS(x)=\frac{4\pi r^4}{3}\,\delta_{jk}
   \quad\mbox{and}\quad
   \int_{\partial B_r(0)} x^j x^k x^n x^m\,dS(x)=\frac{4\pi r^6}{15}
   \,(\delta_{jk}\delta_{nm}+\delta_{jn}\delta_{km}+\delta_{jm}\delta_{kn}). \] 
\end{lemma} 
{\bf Proof\,:} Consider the first integral. If $j\neq k$, the result is zero because then the change 
of variable $x^j\mapsto -x^j$ would cause the integrand to be multiplied by $-1$ 
while maintaining the domain of integration. So we only need to study the case $j=k$, 
and there are $3$ different integrals under this case:
\[ \int_{\partial B_r(0)} (x^1)^2\,dS(x) = \int_{\partial B_r(0)} (x^2)^2\,dS(x) 
   = \int_{\partial B_r(0)} (x^3)^2\,dS(x) \ . \]
They are all equal by symmetry. Therefore 
\[ \int_{\partial B_r(0)} \big((x^1)^2 + (x^2)^2+ (x^3)^2\big)\,dS(x) 
   = r^2 \int_{\partial B_r(0)} \,dS(x) = 4\pi r^4 \ . \]
Hence each of the $3$ integrals has value $4\pi r^4/3$.

Now consider the second integral. Similarly to the above, if there exists an index from $\{1,2,3\}$ 
that only appears once among $\{j,k,n,m\}$, then $\int_{\partial B_r(0)} x^j x^k x^n x^m\,dS(x)=0$. 
This means that the only such integrals that are nonzero are the following ones (and the equality among 
them is again due to symmetry):
\[ I_1 = \int_{\partial B_r(0)} (x^1)^4\,dS(x) = \int_{\partial B_r(0)} (x^2)^4\,dS(x) 
   = \int_{\partial B_r(0)} (x^3)^4\,dS(x) \]
and
\[ I_2 = \int_{\partial B_r(0)} (x^1x^2)^2\,dS(x) = \int_{\partial B_r(0)} (x^1x^3)^2\,dS(x) 
   = \int_{\partial B_r(0)} (x^2x^3)^2\,dS(x) \ . \]
Note that
\begin{align}\label{3I16I2}
   3I_1 + 6I_2 &= \int_{\partial B_r(0)} \big((x^1)^4+(x^2)^4
   +(x^3)^4+2(x^1x^2)^2+2(x^1x^3)^2+2(x^2x^3)^2\big)\,dS(x) 
   \nonumber \\
   &= \int_{\partial B_r(0)} \big((x^1)^2+(x^2)^2+(x^3)^2\big)^2\,dS(x)
   = r^4 \int_{\partial B_r(0)} \,dS(x)= 4\pi r^6. 
\end{align}
Now we will find the ratio between $I_1$ and $I_2$. In spherical coordinates, 
we can transform $I_1$ into a multiple of the integral $\int_0^\pi \int_0^{2\pi} 
\sin^4\theta \cos^4\phi \,d\phi\,d\theta$ and $I_2$ into that same multiple 
of the integral $\int_0^\pi \int_0^{2\pi} \sin^4\theta \cos^2\phi\sin^2\phi \,d\phi\,d\theta$, 
hence all we need is to find out the ratio between the following integrals:
\[ J_1 = \int_0^{2\pi} \cos^4\phi\,d\phi \ , \quad J_2 = \int_0^{2\pi} \cos^2\phi\sin^2\phi\,d\phi \ ; \]
they are evaluated as $J_1=\frac{3\pi}{4}$ and $J_2=\frac{\pi}{4}$. 
Therefore $J_1 = 3J_2$ and thus also $I_1=3I_2$. Together with (\ref{3I16I2}), we can conclude
\[ I_1 = \frac{4\pi r^6}{5} \ , \quad I_2 = \frac{4\pi r^6}{15} \ , \]
which can be written more compactly using the Dirac delta as in the statement of the lemma. 
{\hfill$\Box$}\bigskip  

\begin{lemma}\label{tele} (a) If $A\in\R^{n\times n}$, then 
\begin{equation}\label{2beh}
   |\det(I+tA)-1-t\,{\rm tr} A|\le (2^n+n!)(1+|A|)^n\,t^2
\end{equation}
for $|t|\le 1$. In particular, if $A=A(x)$ is a matrix-valued function 
such that $|A(x)|\le\varphi(x)$ for some function $0<\varphi(x)\le 1$, then 
\begin{equation}\label{2behc}
   |\det(I+A(x))-1-{\rm tr} A(x)|\le 2^n(2^n+n!)\,\varphi(x)^2.  
\end{equation}
\smallskip 

\noindent 
(b) We have 
\[ \sqrt{1+x}=1+\frac{1}{2}\,x-\frac{1}{8}\,x^2+{\cal O}_1(x^3) \] 
for $|x|\le 1/2$, where $|{\cal O}_1(x^3)|\le |x|^3$. 
\end{lemma} 
{\bf Proof\,:} (a) For (\ref{2beh}) let $C=I+tA$, so that $c_{ij}=\delta_{ij}+ta_{ij}$. Then
\[ \det(I+tA)=(1+ta_{11})\cdot\ldots\cdot (1+ta_{nn})
   +\sum_{\pi\in {\cal S}_n,\,\pi\neq {\rm id}}\sigma(\pi)
   c_{1\pi(1)}c_{2\pi(2)}\ldots c_{n\pi(n)}. \] 
Notice that $(1+ta_{11})\cdot\ldots\cdot (1+ta_{nn})$ can be evaluated to yield $2^n$ terms.
Among them is $1$, and there are $ta_{11}+\ldots+ta_{nn}=t\,{\rm tr} A$. All the other terms
have a factor $t^k$ for some $k\ge 2$ and a product of $k$ $a_{ii}$'s [at least two,
but at most $n$ of them]. Since $|t^k|\le t^2$ due to $|t|\le 1$
and $|a_{i_1 i_1}\ldots a_{i_k i_k}|\le\|A\|^k\le (1+\|A\|)^n$,
it follows that
\[ \Big|(1+ta_{11})\cdot\ldots\cdot (1+ta_{nn})-1-t\,{\rm tr}A\Big|\le 2^n (1+\|A\|)^n\,t^2. \]
Furthermore, if $\pi\in {\cal S}_n$ is such that $\pi\neq {\rm id}$, then $\pi(i)\neq i$
for at least two $i$, say $i_1\neq i_2$. But then
\[ |c_{1\pi(1)}c_{2\pi(2)}\ldots c_{n\pi(n)}|\le (1+|t|\|A\|)^{n-2}\,t^2
   \,|a_{i_1 \pi(i_1)}|\,|a_{i_2 \pi(i_2)}|\le (1+\|A\|)^n\,t^2, \]
which yields (\ref{2beh}). From this we get (\ref{2behc}) 
by replacing $t$ with $\varphi(x)$ and $A$ with $A(x)/\varphi(x)$. 
(b) The Taylor expansion of $\phi(x)=\sqrt{1+x}$ is 
$\phi(x)=\phi(0)+\phi'(0)\,x+\frac{1}{2}\,\phi''(0)\,x^2+\frac{1}{6}\,x^3\,\phi'''(\xi)$ 
for some $|\xi|\le 1/2$. Then 
\[  |{\cal O}_1(x^3)|\le \frac{1}{6}\,|x|^3\,\cdot\frac{3}{8}\,
    \max_{|\xi|\le 1/2}\frac{1}{(1+\xi)^{5/2}}
    =2^{-3/2}\,|x|^3\le |x|^3, \] 
as claimed. 
{\hfill$\Box$}\bigskip  

\begin{lemma}\label{aved} For $\sigma\in [0, 1[$ we have 
\begin{equation}\label{mesi} 
   \frac{1}{t+{\cal T}}\int_{|z|=t+{\cal T}}\frac{dS(z)}{|x+z|^{3-\sigma}}
   \le\frac{C}{|x|^{1-\sigma}}.
\end{equation}  
\end{lemma} 
{\bf Proof\,:} First we take $\sigma=0$. Then we can use a rotation of $x$ to $re_3$ to obtain 
\begin{eqnarray*}
   \lefteqn{\frac{1}{t+{\cal T}}\int_{|z|=t+T}\frac{dS(z)}{|x+z|^3}}
   \\ & = & \frac{1}{t+{\cal T}}\int_{|z|=t+T}\frac{dS(z)}{|re_3+z|^3}
   \\ & = & (t+{\cal T})\int_{|\omega|=1}\frac{dS(\omega)}{|re_3+(t+{\cal T})\omega|^3}
   \\ & = & C\,(t+{\cal T})\int_0^\pi d\theta\,\sin\theta\,\frac{1}{((t+{\cal T})^2
   +2(t+{\cal T})\cos\theta\,r+r^2)^{3/2}}
   \\ & = & C\,(t+{\cal T})\int_{-1}^1 ds\,\frac{1}{((t+{\cal T})^2+2(t+{\cal T})s\,r+r^2)^{3/2}}
   \\ & = & \frac{C}{r}\Big(\frac{1}{((t+{\cal T})^2-2(t+{\cal T})\,r+r^2)^{1/2}}
   -\frac{1}{((t+{\cal T})^2+2(t+{\cal T})s\,r+r^2)^{1/2}}\Big)
   \\ & = & \frac{C}{r}\Big(\frac{1}{t+{\cal T}-r}-\frac{1}{t+{\cal T}+r}\Big)
   =C\,\frac{1}{(t+{\cal T}-r)(t+{\cal T}+r)}.
\end{eqnarray*} 
Since $t+{\cal T}-r\ge 2+t-|x|\ge 1$ and $t+{\cal T}+r\ge r=|x|$, 
we obtain (\ref{mesi}) for $\sigma=0$. If $|z|=t+{\cal T}$, 
then $|z|\le 1+|x|+{\cal T}\le 3|x|$ yields
\[ \frac{1}{|x+z|^{3-\sigma}}=\frac{|x+z|^\sigma}{|x+z|^3}
   \le C\,\frac{|x|^\sigma}{|x+z|^3}, \] 
so that the general case follows from the case $\sigma=0$. 
{\hfill$\Box$}\bigskip 

\begin{lemma} We have 
\begin{equation}\label{CalA1} 
   \int_{a\le |x-y|\le b} \frac{dy}{|x-y|}\,g(t-|x-y|, |y|)
   =\frac{2\pi}{|x|}\,\int_{t-b}^{t-a}\,d\tau
   \int_{||x|-t+\tau|}^{|x|+t-\tau} d\lambda\,\lambda\,g(\tau, \lambda).
\end{equation} 
\end{lemma} 
{\bf Proof\,:} See \cite[Lemma 7]{glstr} or \cite[Lemma A.1]{calog}.    
{\hfill$\Box$}\bigskip 

\begin{lemma} We have 
\begin{equation}\label{CalA1b} 
   \int_{\R^3} \frac{dy}{|x-y|^2}\,\frac{1}{(1+|t-|x-y||+|y|)^3} 
   \le C\,\frac{1}{1+|t|+|x|}\,\frac{1}{1+|t-|x||}. 
\end{equation} 
\end{lemma} 
{\bf Proof\,:} See \cite[Lemma 3]{calog} for $n=2$ and $q=3$. 
{\hfill$\Box$}\bigskip 

\begin{lemma}\label{zweix} 
Let $h: \R^3\times\R^3\longrightarrow\R$ be a sufficiently smooth and decaying function. 
For $a, b>0$ define 
\[ H(x)=\int_{|y|\le a, |y-x|\le b} h(x, y)\,dy. \]
Then the (distributional) derivative of $H$ is 
\[ \nabla H(x)=\int_{|y|\le a, |x-y|\le b}\nabla_1 h(x, y)\,dy
   -\int_{|z|=b}\frac{z}{|z|}\,{\bf 1}_{\{|x-z|\le a\}}\,h(x, x-z)\,dS(z), \] 
where $\nabla_1$ denotes the gradient with respect to the first variable.  
\end{lemma} 
{\bf Proof\,:} For $\varphi\in C_0^\infty(\R^3)$ 
we calculate by means of the divergence theorem 
\begin{eqnarray*}
   \lefteqn{\int_{\R^3} H(x)\nabla\varphi(x)\,dx}
   \\ & = & \int_{\R^3} dx\,\int_{|y|\le a} dy\,{\bf 1}_{\{|y-x|\le b\}}\,h(x, y)\,\nabla\varphi(x)
   \\ & = & \int_{|y|\le a} dy\int_{|z|\le b} dz\,h(y+z, y)\,\nabla_z[\varphi(y+z)]
   \\ & = & \int_{|y|\le a} dy\,\bigg(\int_{|z|=b} dS(z)\,h(y+z, y)\varphi(y+z)\,\frac{z}{|z|}
   -\int_{|z|\le b} dz\,\nabla_1 h(y+z, y)\,\varphi(y+z)\bigg)
   \\ & = & \int_{|z|=b} dS(z)\,\frac{z}{|z|}\int_{\R^3} dx\,{\bf 1}_{\{|x-z|\le a\}}\,h(x, x-z)\varphi(x)
   -\int_{|z|\le b} dz\int_{\R^3} dx\,{\bf 1}_{\{|x-z|\le a\}}\nabla_1 h(x, x-z)\,\varphi(x)
   \\ & = & -\int_{\R^3} dx\bigg(\int_{|z|\le b} dz\,{\bf 1}_{\{|x-z|\le a\}}\nabla_1 h(x, x-z)
   -\int_{|z|=b} dS(z)\,\frac{z}{|z|}\,{\bf 1}_{\{|x-z|\le a\}}\,h(x, x-z)\bigg)\varphi(x)\,dx, 
\end{eqnarray*} 
and the assertion follows.  
{\hfill$\Box$}\bigskip 

\begin{lemma}\label{xkabl} 
For sufficiently smooth and decaying $g$, and $k=1, 2, 3$,  
\begin{eqnarray*} 
   \lefteqn{\partial_{x^k}\int_{|x-y|\le t+{\cal T},\,|y|\ge {\cal S}}\frac{g(t-|x-y|, y)}{|x-y|}\,dy}
   \\ & = & \int_{|x-y|\le t+{\cal T}}\frac{\partial_{x^k} g(t-|x-y|, y)}{|x-y|}\,dy
   \\ & & +\,\int_{|x-y|\le t+{\cal T}, |y|\le {\cal S}}
   \bigg(\frac{g(t-|x-y|, y)}{|x-y|}
   +\partial_t g(t-|x-y|, y)\bigg)\,\frac{x_k-y_k}{|x-y|^2}\,dy
   \\ & & -\,\frac{1}{t+{\cal T}}\int_{|z|=t+{\cal T}}\frac{z_k}{|z|}
   \,{\bf 1}_{\{|x+z|\le {\cal S}\}}\,g(-{\cal T}, x+z)\,dS(z).  
\end{eqnarray*} 
\end{lemma} 
{\bf Proof\,:} Let $I$ denote the integral in question. Then 
\begin{eqnarray*} 
   I & = & \int_{|x-y|\le t+{\cal T}}\frac{g(t-|x-y|, y)}{|x-y|}\,dy
   -\int_{|x-y|\le t+{\cal T},\,|y|\le {\cal S}}\frac{g(t-|x-y|, y)}{|x-y|}\,dy
   \\ & = & \int_{|z|\le t+{\cal T}}\frac{g(t-|z|, x-z)}{|z|}\,dz
   -\int_{|x-y|\le t+{\cal T},\,|y|\le {\cal S}}\frac{g(t-|x-y|, y)}{|x-y|}\,dy. 
\end{eqnarray*} 
Therefore 
\[ \partial_{x^k} I=\int_{|z|\le t+{\cal T}}\frac{\partial_{x^k} g(t-|z|, x-z)}{|z|}\,dz
   -\partial_{x^k}\int_{|x-y|\le t+{\cal T},\,|y|\le {\cal S}}\frac{g(t-|x-y|, y)}{|x-y|}\,dy. \] 
For the second term we are going to use Lemma \ref{zweix} 
with $a={\cal S}$, $b=t+{\cal T}$ and $h(x, y)=\frac{g(t-|x-y|, y)}{|x-y|}$. Then 
\[ \nabla_1 h(x, y)=-\bigg(\frac{g(t-|x-y|, y)}{|x-y|}
   +\partial_t g(t-|x-y|, y)\bigg)\,\frac{x-y}{|x-y|^2}, \] 
which yields the claim. 
{\hfill$\Box$}\bigskip

\begin{lemma}\label{fastverg} We have 
\[ \int_{|x-y|\le t+{\cal T}}\frac{dy}{|x-y||y|^4}\,{\bf 1}_{\{|y|\ge\frac{1}{2}(1+|t-|x-y||)\}}
   \le C(1+{\cal T})\,\frac{1}{|x|}. \] 
\end{lemma} 
{\bf Proof\,:} Since $|y|\ge 1/2$ on the domain of integration,  
also $|y|\ge\frac{1}{3}(1+|y|)$. Hence, by (\ref{CalA1}), 
the integral is bounded by  
\begin{eqnarray*} 
   C\int_{|x-y|\le t+{\cal T}}\frac{dy}{|x-y|(1+|y|)^4}
   & = & \frac{C}{|x|}\int_{-{\cal T}}^t\,d\tau
   \int_{||x|-t+\tau|}^{|x|+t-\tau} d\lambda\,\frac{\lambda}{(1+\lambda)^4}
   \\ & \le & \frac{C}{|x|}\int_{-{\cal T}}^t\,d\tau
   \int_{||x|-t+\tau|}^{|x|+t-\tau}\frac{d\lambda}{(1+\lambda)^3}.  
\end{eqnarray*} 
If $|x|-t+\tau\le 0$, then $\tau\le t-|x|\le 1$, so that   
\[ \frac{C}{|x|}\int_{-{\cal T}}^t\,d\tau\,{\bf 1}_{\{|x|-t+\tau\le 0\}}
   \int_{t-|x|-\tau}^{|x|+t-\tau}\frac{d\lambda}{(1+\lambda)^3}
   \le\frac{C}{|x|}\int_{-{\cal T}}^1\,d\tau
   \int_0^\infty\frac{d\lambda}{(1+\lambda)^3}
   \le\frac{C(1+{\cal T})}{|x|}. \]  
On the other hand, if $|x|-t+\tau\ge 0$, then 
\begin{eqnarray*}
   \lefteqn{\frac{C}{|x|}\int_{-{\cal T}}^t\,d\tau\,{\bf 1}_{\{|x|-t+\tau\ge 0\}}
   \int_{|x|-t+\tau}^{|x|+t-\tau}\frac{d\lambda}{(1+\lambda)^3}} 
   \\ & = & \frac{C}{|x|}\int_{-{\cal T}}^t\,d\tau\,{\bf 1}_{\{|x|-t+\tau\ge 0\}}
   \bigg(\frac{1}{(1+|x|-t+\tau)^2}-\frac{1}{(1+|x|+t-\tau)^2}\bigg)
   \\ & = & \frac{C}{|x|}\int_{-{\cal T}}^t\,d\tau\,{\bf 1}_{\{|x|-t+\tau\ge 0\}}
   \frac{(1+|x|+t-\tau)^2-(1+|x|-t+\tau)^2}{(1+|x|-t+\tau)^2(1+|x|+t-\tau)^2}
   \\ & = & \frac{C}{|x|}\int_{-{\cal T}}^t\,d\tau\,{\bf 1}_{\{|x|-t+\tau\ge 0\}}
   \frac{(1+|x|)(t-\tau)}{(1+|x|-t+\tau)^2(1+|x|+t-\tau)^2}
   \\ & \le & C\int_{-{\cal T}}^t\,d\tau\,{\bf 1}_{\{|x|-t+\tau\ge 0\}}
   \frac{t-\tau}{(1+|x|-t+\tau)^2(1+|x|+t-\tau)^2}
   \\ & = & C\int_0^{t+{\cal T}}\,ds\,{\bf 1}_{\{|x|\ge s\}}
   \frac{s}{(1+|x|-s)^2(1+|x|+s)^2}
   \\ & \le & \frac{C}{1+|x|}\int_0^{|x|}\,\frac{ds}{(1+|x|-s)^2} 
   =\frac{C}{1+|x|}\int_0^{|x|}\,\frac{d\sigma}{(1+\sigma)^2} 
   \le\frac{C}{1+|x|}.  
\end{eqnarray*} 
Taking together both estimates, we obtain the claim. 
{\hfill$\Box$}\bigskip 

The following multipole expansion corresponds 
to \cite[(27.19)-(27.21)]{stephani}, with the error terms included. 

\begin{lemma}\label{mupol} Let $x, \bar{x}\in\R^3$ be such that $\bar{r}\le r_\ast$ 
and $r\ge 5r_\ast$ for some $r_\ast\ge 1$, where $r=|x|$ and $\bar{r}=|\bar{x}|$. 
\smallskip

\noindent 
(a) We have the expansion
\begin{eqnarray} 
   |x-\bar{x}| & = & r-\frac{1}{r}\,(x\cdot\bar{x})
   +\frac{1}{2r}\bigg(\bar{r}^2-\frac{(x\cdot\bar{x})^2}{r^2}\bigg)
   +{\cal O}(r^{-2}), 
   \label{xdiff}
   \\ \frac{1}{|x-\bar{x}|} 
   & = & \frac{1}{r}+\frac{1}{r^3}(x\cdot\bar{x})
   +\frac{1}{2r^3}\,\bigg(\frac{3(x\cdot\bar{x})^2}{r^2}-\bar{r}^2\bigg)
   +{\cal O}(r^{-4}), 
   \label{xdiff1} 
\end{eqnarray} 
with the implied constants depending only on powers of $r_\ast$. 
\smallskip 

\noindent
(b) Let $\psi=\psi(t)$ be a suitable function. Then 
\begin{eqnarray*} 
   \lefteqn{\frac{\psi(t-c^{-1}|x-\bar{x}|)}{|x-\bar{x}|}} 
   \\ & = & \bigg[\frac{1}{r}+\frac{1}{r^3}(x\cdot\bar{x})
   +\frac{1}{2r^3}\,\bigg(\frac{3(x\cdot\bar{x})^2}{r^2}
   -\bar{r}^2\bigg)\bigg]\,\psi(t-c^{-1}r)
   \\ & & +\,\frac{1}{c}\,\bigg[\frac{1}{r^2}\,(x\cdot\bar{x})
   +\frac{1}{2r^2}\bigg(\frac{3(x\cdot\bar{x})^2}{r^2}-\bar{r}^2\bigg)\bigg]\,\psi'(t-c^{-1}r)
   +\frac{1}{2c^2}\,\frac{(x\cdot\bar{x})^2}{r^3}\,\psi''(t-c^{-1}r)
   \\ & & +\,{\cal O}\Big(\sum_{j=0}^3 c^{-j} r^{-(4-j)} {\|\psi^{(j)}\|}_\infty\Big), 
\end{eqnarray*} 
with the implied constants depending only on powers of $r_\ast$. In particular, 
\begin{eqnarray}\label{bergk}  
   \frac{\psi(t-c^{-1}|x-\bar{x}|)}{|x-\bar{x}|} 
   & = & \frac{1}{r}\,\psi(t-c^{-1}r)
   +\frac{(x\cdot\bar{x})}{cr^2}\,\psi'(t-c^{-1}r)
   +\frac{(x\cdot\bar{x})^2}{2c^2r^3}\,\psi''(t-c^{-1}r)
   \nonumber
   \\ & & +\,{\cal O}\Big(r^{-2}\sum_{j=0}^2 c^{-j} {\|\psi^{(j)}\|}_\infty
   +c^{-3} r^{-1} {\|\psi'''\|}_\infty\Big).  
\end{eqnarray}  
\smallskip 

\noindent
(c) Let $\psi=\psi(t)$ be a suitable function. Then  
\begin{eqnarray*} 
   \frac{\psi(t-c^{-1}|x-\bar{x}|)}{|x-\bar{x}|^2} 
   & = & \frac{1}{r^2}\,\psi(t-c^{-1}r)
   +\frac{(x\cdot\bar{x})}{cr^3}\,\psi'(t-c^{-1}r)
   +\frac{(x\cdot\bar{x})^2}{2c^2r^4}\,\psi''(t-c^{-1}r)
   \\ & & +\,{\cal O}\Big(r^{-3}\sum_{j=0}^2 c^{-j} {\|\psi^{(j)}\|}_\infty
   +c^{-3}r^{-2}{\|\psi'''\|}_\infty\Big), 
\end{eqnarray*} 
with the implied constants depending only on powers of $r_\ast$.
\end{lemma} 
{\bf Proof\,:} (a) For $|\eps|\le 1/2$ the function $\phi(\eps)=\sqrt{1-\eps}$ has the expansion 
$\phi(\eps)=\phi(0)+\phi'(0)\,\eps+\frac{1}{2}\,\phi''(0)\,\eps^2+\frac{1}{6}\,\eps^3\,\phi'''(\xi)$ 
for some $|\xi|\le 1/2$. It follows that 
\begin{equation}\label{einsmin} 
   \sqrt{1-\eps}=1-\frac{1}{2}\,\eps-\frac{1}{8}\,\eps^2+{\cal O}_1(\eps^3),
\end{equation}
where 
\[ |{\cal O}_1(\eps^3)|=\frac{1}{6}\,\eps^3\,\cdot\frac{3}{8}\,\max_{|\xi|\le 1/2}\frac{1}{(1-\xi)^{5/2}}
   =2^{-3/2}\,\eps^3\le\eps^3. \] 
Below we will take $\eps=\frac{2}{r^2}\,x\cdot\bar{x}
-\frac{\bar{r}^2}{r^2}$; note that 
$|\eps|=|\frac{2}{r^2}\,x\cdot\bar{x}
-\frac{\bar{r}^2}{r^2}|\le\frac{2\bar{r}}{r}+\frac{\bar{r}^2}{r^2}
\le\frac{2r_\ast}{5r_\ast}+\frac{r_\ast^2}{25 r_\ast^2}\le 1/2$ as well as 
$|\eps|\le\frac{2\bar{r}}{r}+\frac{\bar{r}^2}{r^2}\le\frac{3r_\ast}{r}$. 
Concerning (\ref{xdiff}), the expansion (\ref{einsmin}) yields
\begin{eqnarray*} 
   |x-\bar{x}| & = & \sqrt{r^2-2x\cdot\bar{x}+\bar{r}^2}
   =r\,\sqrt{1-\frac{2}{r^ 2}\,x\cdot\bar{x}+\frac{\bar{r}^2}{r^2}}
   \\ & = & r-\frac{1}{2r}\,(2\,x\cdot\bar{x}-\bar{r}^2)
   -\frac{1}{8r^3}\,(2\,x\cdot\bar{x}-\bar{r}^2)^2
   +r\,{\cal O}_1\bigg(\frac{27r_\ast^3}{r^3}\bigg)
   \\ & = & r-\frac{1}{r}\,(x\cdot\bar{x})
   +\frac{1}{2r}\bigg(\bar{r}^2-\frac{(x\cdot\bar{x})^2}{r^2}\bigg)
   +\frac{1}{2r^3}\,(x\cdot\bar{x})\,\bar{r}^2
   -\frac{\bar{r}^4}{8r^3}+r\,{\cal O}_1\bigg(\frac{27r_\ast^3}{r^3}\bigg)
   \\ & = & r-\frac{1}{r}\,(x\cdot\bar{x})
   +\frac{1}{2r}\bigg(\bar{r}^2-\frac{(x\cdot\bar{x})^2}{r^2}\bigg)
   +{\cal O}_2(r^{-2})
\end{eqnarray*}  
with 
\[ |{\cal O}_2(r^{-2})|=\bigg(\frac{r_\ast^3}{2}
   +\frac{r_\ast^4}{40}+27r_\ast^3\bigg) r^{-2}. \] 
Similarly, the constants in the order terms below will depend on powers $r_\ast$. 
To verify (\ref{xdiff1}), we write (\ref{xdiff}) as $|x-\bar{x}|=r+A_0+A_1+A_2$ according 
to the orders in $r^{-1}$ and expand $\frac{1}{M+R}=\frac{1}{M}-\frac{R}{(M+R)M}$ 
for a main term $M$ and a remainder $R$. Then $(A_0+A_1+A_2)^2=A_0^2+{\cal O}(r^{-1})$, so that  
\begin{eqnarray*}
   \frac{1}{|x-\bar{x}|} 
   & = & \frac{1}{r}-\frac{A_0+A_1+A_2}{r |x-\bar{x}|}
   =\frac{1}{r}-\frac{1}{r}\,(A_0+A_1+A_2)\,\bigg(\frac{1}{r}
   -\frac{A_0+A_1+A_2}{r |x-\bar{x}|}\bigg) 
   \\ & = & \frac{1}{r}-\frac{1}{r^2}\,(A_0+A_1+A_2)+\frac{(A_0+A_1+A_2)^2}{r^2 |x-\bar{x}|}
   \\ & = & \frac{1}{r}-\frac{1}{r^2}\,(A_0+A_1)+{\cal O}(r^{-4})
   +\frac{A_0^2+{\cal O}(r^{-1})}{r^2 |x-\bar{x}|}
   \\ & = & \frac{1}{r}-\frac{A_0}{r^2}-\frac{A_1}{r^2}+\frac{A_0^2}{r^3}+{\cal O}(r^{-4}),
\end{eqnarray*} 
which gives (\ref{xdiff1}), due $A_0=-\frac{1}{r}(x\cdot\bar{x})$, 
$A_1=\frac{1}{2r}(\bar{r}^2-\frac{(x\cdot\bar{x})^2}{r^2})$ and $A_2={\cal O}(r^{-2})$. 
(b) Let $T=t-c^{-1}|x-\bar{x}|$ and $u=t-c^{-1}r$. Then, owing to (\ref{xdiff}),  
\begin{equation}\label{Tminu} 
   T-u=\frac{1}{c}\,(|x|-|x-\bar{x}|)
   =\frac{1}{c}\,\bigg(\frac{1}{r}\,(x\cdot\bar{x})
   -\frac{1}{2r}\bigg(\bar{r}^2-\frac{(x\cdot\bar{x})^2}{r^2}\bigg)+{\cal O}(r^{-2})\bigg)
\end{equation} 
and in particular $T-u=c^{-1}\,{\cal O}(1)={\cal O}(c^{-1})$. By Taylor expansion, 
\begin{eqnarray}\label{milef} 
   \psi(T) & = & \psi(u)+\psi'(u)\,(T-u)+\frac{1}{2}\,\psi''(u)\,(T-u)^2
   +\frac{1}{6}\,\psi'''(\xi)\,(T-u)^3
   \nonumber
   \\ & = & \psi(u)+\psi'(u)\,(T-u)+\frac{1}{2}\,\psi''(u)\,(T-u)^2
   +{\cal O}({\|\psi'''\|}_\infty c^{-3}). 
\end{eqnarray} 
Using (\ref{xdiff1}), (\ref{milef}) and (\ref{Tminu}), 
\begin{eqnarray*} 
   \frac{\psi(T)}{|x-\bar{x}|}
   & = & \bigg[\frac{1}{r}+\frac{1}{r^3}(x\cdot\bar{x})
   +\frac{1}{2r^3}\,\bigg(\frac{3(x\cdot\bar{x})^2}{r^2}-\bar{r}^2\bigg)
   +{\cal O}(r^{-4})\bigg]
   \\ & & \,\times\bigg(\psi(u)+\psi'(u)\,(T-u)+\frac{1}{2}\,\psi''(u)\,(T-u)^2
   +{\cal O}({\|\psi'''\|}_\infty c^{-3})\bigg)
   \\ & = & \bigg[\frac{1}{r}+\frac{1}{r^3}(x\cdot\bar{x})
   +\frac{1}{2r^3}\,\bigg(\frac{3(x\cdot\bar{x})^2}{r^2}-\bar{r}^2\bigg)\bigg]
   \,\psi(u)+{\cal O}({\|\psi\|}_\infty r^{-4})
   \\ & & +\,\frac{1}{c}\,\bigg[\frac{1}{r}+\frac{1}{r^3}(x\cdot\bar{x})
   +\frac{1}{2r^3}\,\bigg(\frac{3(x\cdot\bar{x})^2}{r^2}-\bar{r}^2\bigg)
   +{\cal O}(r^{-4})\bigg]\,\psi'(u)
   \\ & & \quad\times\bigg(\frac{1}{r}\,(x\cdot\bar{x})
   -\frac{1}{2r}\bigg(\bar{r}^2-\frac{(x\cdot\bar{x})^2}{r^2}\bigg)+{\cal O}(r^{-2})\bigg)
   \\ & & +\,\frac{1}{2c^2}\,\bigg[\frac{1}{r}+\frac{1}{r^3}(x\cdot\bar{x})
   +\frac{1}{2r^3}\,\bigg(\frac{3(x\cdot\bar{x})^2}{r^2}-\bar{r}^2\bigg)
   +{\cal O}(r^{-4})\bigg]\,\psi''(u)\,\bigg(\frac{(x\cdot\bar{x})^2}{r^2}+{\cal O}(r^{-1})\bigg)
   \\ & & +\,{\cal O}({\|\psi'''\|}_\infty c^{-3}r^{-1})
   \\ & = & \bigg[\frac{1}{r}+\frac{1}{r^3}(x\cdot\bar{x})
   +\frac{1}{2r^3}\,\bigg(\frac{3(x\cdot\bar{x})^2}{r^2}-\bar{r}^2\bigg)\bigg]\,\psi(u)
   \\ & & +\,\frac{1}{c}\,\bigg(\frac{1}{r^2}\,(x\cdot\bar{x})
   -\frac{1}{2r^2}\bigg(\bar{r}^2-\frac{(x\cdot\bar{x})^2}
   {r^2}\bigg)+{\cal O}(r^{-3})\bigg)\psi'(u)
   \\ & & +\,\frac{1}{c}\,\bigg[\frac{1}{r^3}(x\cdot\bar{x})+{\cal O}(r^{-3})\bigg]\,\psi'(u)
   \,\bigg(\frac{1}{r}\,(x\cdot\bar{x})+{\cal O}(r^{-1})\bigg)
   \\ & & +\,\frac{1}{2c^2}\,\bigg[\frac{1}{r}+{\cal O}(r^{-2})\bigg]
   \,\psi''(u)\,\bigg(\frac{(x\cdot\bar{x})^2}{r^2}+{\cal O}(r^{-1})\bigg)
   \\ & & +\,{\cal O}({\|\psi\|}_\infty r^{-4})+{\cal O}({\|\psi'''\|}_\infty c^{-3}r^{-1})
   \\ & = & \bigg[\frac{1}{r}+\frac{1}{r^3}(x\cdot\bar{x})
   +\frac{1}{2r^3}\,\bigg(\frac{3(x\cdot\bar{x})^2}{r^2}-\bar{r}^2\bigg)\bigg]\,\psi(u)
   \\ & & +\,\frac{1}{c}\,\bigg(\frac{1}{r^2}\,(x\cdot\bar{x})
   -\frac{1}{2r^2}\bigg(\bar{r}^2-\frac{(x\cdot\bar{x})^2}{r^2}\bigg)\bigg)\psi'(u)
   +\frac{1}{c}\,\frac{(x\cdot\bar{x})^2}{r^4}\,\psi'(u)
   +\frac{1}{2c^2}\,\psi''(u)\,\frac{(x\cdot\bar{x})^2}{r^3}
   \\ & & +\,{\cal O}({\|\psi\|}_\infty r^{-4})+{\cal O}({\|\psi'''\|}_\infty c^{-3}r^{-1})
   +{\cal O}({\|\psi'\|}_\infty c^{-1}r^{-3})+{\cal O}({\|\psi''\|}_\infty c^{-2}r^{-2}), 
\end{eqnarray*} 
and this yields the claim. (c) This follows from (\ref{xdiff1}) and (b). 
{\hfill$\Box$}\bigskip 


\end{document}